\documentclass[prb,twocolumn,showpacs,amsmath,amssymb,superscriptaddress,floatfix]{revtex4-1}
\usepackage{amsmath}
\usepackage{amssymb}
\usepackage{amsthm}
\usepackage{amsfonts}
\usepackage{dsfont}
\usepackage{listings}
\usepackage{enumerate}
\usepackage{latexsym}

\usepackage{psfrag}
\usepackage{verbatim}

\usepackage{afterpage}

\usepackage{bm}
\usepackage{graphicx}
\usepackage{subfigure}
\usepackage{tikz}
\newcommand{\mm}{\text{-}1}
\newcommand{\mcP}{\mathcal{P}}

\newcommand{\beq}{\begin{equation}}
\newcommand{\eneq}{\end{equation}}

\newcommand{\prodal}[2]{\underset{#1}{\overset{#2}{\prod}}}
\newcommand{\sumal}[2]{\underset{#1}{\overset{#2}{\sum}}}

\newcommand{\braket}[2]{\left\langle #1 | #2 \right\rangle}
\newcommand{\bra}[1]{\left\langle#1\right|}
\newcommand{\ket}[1]{\left|#1\right\rangle}

\newcommand{\vac}{\ket{\theta}}

\usepackage[draft]{todonotes}

\def\eg{{\it e.g.}\ }
\def\ea{{\it et al.}}
\input{epsf}

\newcommand{\nn}{\nonumber}

\newcommand{\twopartdef}[4]
{
	\left\{
		\begin{array}{ll}
			#1 & \mbox{if } #2 \\
			#3 & \mbox{if } #4
		\end{array}
	\right.
}

\newcommand{\threepartdef}[6]
{
	\left\{
		\begin{array}{ll}
			#1 & \mbox{if } #2 \\
			#3 & \mbox{if } #4 \\
			#5 & \mbox{if } #6 
		\end{array}
	\right.
}

\usepackage{blkarray}

\begin{document}

\tolerance 10000

\newcommand{\vk}{{\bf k}}

\title{Exact Excited States of Non-Integrable Models}

\author{Sanjay Moudgalya}
\affiliation{Department of Physics, Princeton University, Princeton, NJ 08544, USA}
\author{Stephan Rachel}
\affiliation{Institut f\"{u}̈r Theoretische Physik, Technische Universit\"{a}̈t Dresden, 01062 Dresden, Germany}
\affiliation{Department of Physics, Princeton University, Princeton, NJ 08544, USA}
\author{B. Andrei Bernevig}
\affiliation{Department of Physics, Princeton University, Princeton, NJ 08544, USA}
\affiliation{Donostia International Physics Center, P. Manuel de Lardizabal 4, 20018 Donostia-San Sebasti\'{a}́n, Spain}
\affiliation{Laboratoire Pierre Aigrain, Ecole Normale Sup\'{e}rieure-PSL Research University, CNRS, Universit\'{e} Pierre et Marie Curie-Sorbonne Universit\'{e}s,
Universit\'{e} Paris Diderot-Sorbonne Paris Cit\'{e}, 24 rue Lhomond, 75231 Paris Cedex 05, France}
\affiliation{Sorbonne Universit\'{e}s, UPMC Univ Paris 06, UMR 7589, LPTHE, F-75005, Paris, France}
\author{Nicolas Regnault}
\affiliation{Laboratoire Pierre Aigrain, Ecole Normale Sup\'{e}rieure-PSL Research University, CNRS, Universit\'{e} Pierre et Marie Curie-Sorbonne Universit\'{e}s,
Universit\'{e} Paris Diderot-Sorbonne Paris Cit\'{e}, 24 rue Lhomond, 75231 Paris Cedex 05, France}
\affiliation{Department of Physics, Princeton University, Princeton, NJ 08544, USA}

\date{\today}

\begin{abstract}
We discuss a method of numerically identifying exact energy eigenstates for a finite system, whose form can then be obtained analytically. We demonstrate our method by identifying and deriving exact analytic expressions for several excited states, including an infinite tower, of the one dimensional spin-1 AKLT model, a celebrated non-integrable model. The states thus obtained for the AKLT model can be interpreted as one-to-an extensive number of quasiparticles on the ground state or on the highest excited state when written in terms of dimers. Included in these exact states is a tower of states spanning energies from the ground state to the highest excited state. To our knowledge, this is the first time that exact analytic expressions for a tower of excited states have been found in non-integrable models.  Some of the states of the tower appear to be in the bulk of the energy spectrum, allowing us to make conjectures on the strong Eigenstate Thermalization Hypothesis (ETH). We also generalize these exact states including the tower of states to the generalized integer spin AKLT models. Furthermore, we establish a correspondence between some of our states and those of the Majumdar-Ghosh model, yet another non-integrable model, and extend our construction to the generalized integer spin AKLT models.
\end{abstract}

\maketitle

\section{Introduction}
Many-body localization \cite{basko2006metal, pal2010many, imbrie2016many, nandkishore2014many, luitz2015many, vosk2013many, huse2013localization, canovi2011quantum, chandran2014many} has sparked a renewed interest in fundamental questions about thermalization in quantum systems. \cite{gogolin2011absence, rigol2009breakdown, deutsch1991quantum, srednicki1994chaos, rigol2008thermalization, de2013ergodicity}
The quests to protect exotic equilibrium phenomena from thermalization\cite{gring2012relaxation,kemp2017long,else2017prethermal} and realize them in a non-equilibrium setting\cite{huse2013localization, chandran2014many, bahri2015localization} call for a deeper understanding of quantum dynamics.
The dynamics of a quantum system is tied to the properties of {\it all} its energy eigenstates and not only to the ground state features. It is thus very important to have models where we know the analytical structure of the excited states in the bulk of the energy spectrum. Integrable models, including free systems, fall into this category but are unfortunately one of the two well known examples (along with the many-body localized states\cite{nandkishore2014many,bauer2013area}) where the Eigenstate Thermalization Hypothesis (ETH)\cite{deutsch1991quantum, srednicki1994chaos, rigol2008thermalization, beugeling2014finite, kim2014testing, garrison2015does,steinigeweg2013eigenstate,ikeda2013finite} breaks down.\cite{rigol2009breakdown,gogolin2011absence, pozsgay2014failure} Thus having some simple non-integrable models where a partial or complete analytical description beyond the low energy states is available, would be a perfect avenue to investigate the ETH.
For generic non-integrable systems, none of the energy eigenstates can be obtained analytically. However, the ground state is known exactly for some non-integrable models with local Hamiltonians. One such model is the one dimensional spin-1 AKLT chain,\cite{aklt1987rigorous, affleck1988valence} which was first introduced as a simple model to exemplify the Haldane gap in integer spin chains. Indeed, the ground state of the AKLT chain can be explicitly built and it belongs to the same universality class as that of the spin-1 Heisenberg model. Along with its generalizations to higher integer spin values, it is representative of the Haldane phase.\cite{haldane1983nonlinear, haldane1983continuum} The simplicity of the ground state of the AKLT model makes it one of the most elegant introductory examples for various concepts in condensed matter physics, including entanglement in spin-chains,\cite{fan2004entanglement, katsura2007exact, korepin2010entanglement, xu2008entanglement, santos2011negativity, santos2012entanglement} matrix product state representations of ground states,\cite{schollwock2011density,perez2006matrix,orus2014practical} bosonic symmetry protected topological (SPT) phases in one dimension\cite{pollmann2010entanglement, chen2011classification} and even some aspects of the fractional quantum Hall effect.\cite{arovas1988extended} Another example of a non-integrable model with a ground state whose expression is analytically known is the Majumdar-Ghosh model,\cite{majumdar1969next} a spin-$1/2$ Heisenberg chain with an extra fine-tuned next nearest neighbor coupling. 

Beyond the ground state, very little is known about excited states. Even more difficult is the question whether excited states with a closed-form expression, that we dub exact excited states, exist in these non-integrable quantum spin chains. Exact expressions for any of the excited states, even the ones close to the ground state or the highest excited states would help in testing predictions and conjectures made, on general grounds, about the nature of eigenstates at the edges of the energy spectrum.\cite{pizorn2012universality, thomale2015entanglement, zauner2015transfer, zauner2018topological} Caspers \ea \cite{caspers1982some} and Arovas\cite{arovas1989two} have derived three and two exact excited states in the Majumdar Ghosh model and the spin-1 AKLT model, respectively. An obstacle to the discovery of new exact excited states is the lack of physical intuition regarding their nature. In this article, we propose to find possible simple excited states by looking at the entanglement structure of eigenstates obtained in finite size systems by exact diagonalization. By looking at the reduced density matrix of each individual eigenstate and targeting those having a low rank, we are able to unveil new exact excited states whose analytic expressions we then obtain. Interestingly, their energy is, most of the time, an integer or a rational number, given a suitable choice of the Hamiltonian normalization.

The paper is organized as follows. In Sec.~\ref{sec:akltmodels} we review the spin-1 AKLT model and the construction of its ground state using the dimer basis. In Sec.~\ref{sec:whatarethey} we introduce the concept of exact states, i.e. eigenstates having an analytic closed-form expression. We discuss a numerical approach based on the rank of the reduced density matrix to track these states in exact diagonalization studies.  We show an extensive numerical study of the spectrum of the spin-1 AKLT chain, listing all the exact states up to $16$ spins. We then proceed to derive the analytical expressions for all the states. We first consider the low energy states in Sec.~\ref{sec:spin1aklt}, recovering the two Arovas states.\cite{arovas1989two} In Sec.~\ref{sec:ExactMiddle}, we derive the tower of states, a series of spin-2 magnon excitations on top of the ground state, ranging from the ground state to the highest energy state and present evidence that shows that their position in the bulk of the energy spectrum. In Sec.~\ref{sec:ExactHighEnergy}, we discuss the exact states situated close to the highest excited state. To show that our approach is valid beyond the spin-1 AKLT chain, we discuss its generalization to higher integer spin-$S$ in Sec.~\ref{sec:spinsaklt}, obtaining the analytical expression of all the exact states that we numerically observe to have a low entanglement rank. In Sec.~\ref{sec:projection}, we derive a correspondence between certain exact states of the Majumdar-Ghosh Model and the spin-1 AKLT model as well as between exact AKLT states with different spin-$S$.

\section{The Spin-1 AKLT Model}\label{sec:akltmodels}
\subsection{Hamiltonian}\label{sec:aklthamil}
The spin-1 AKLT Hamiltonian is defined as a sum projectors that projects two nearest neighbor spins onto spin 2.\cite{aklt1987rigorous, affleck1988valence} Denoting the projector of two spin-1s on sites $i$ and $j$ onto total spin 2 as  $P^{(2,1)}_{ij}$, the AKLT Hamiltonian for a chain of length $L$ with periodic boundary conditions (i.e., $L + 1 \equiv 1$) simply reads
\begin{equation}
    H = \sum_{i=1}^L{P^{(2,1)}_{i, i+1}}.
\label{S1Hamiltonian}
\end{equation}
The action of the projector on various configurations of nearest neighbor spins in given in Appendix~\ref{spins}. The projector can also be expressed in terms of the spin operators,
\begin{equation}\label{P2-projector}
    P^{(2,1)}_{ij} = \frac{1}{24}(\vec{S}_i + \vec{S}_j)^2((\vec{S}_i + \vec{S}_j)^2 - 2).
\end{equation}
Simplifying the expression Eq.~\eqref{P2-projector}, the AKLT Hamiltonian Eq.~(\ref{S1Hamiltonian}) can be written in a more familiar form as
\begin{equation}
    H = \sum_{i = 1}^L{\left(\frac{1}{3} + \frac{1}{2}\vec{S}_i\cdot\vec{S}_{i+1} + \frac{1}{6}(\vec{S}_i\cdot\vec{S}_{i+1})^2\right)}.
\label{S1HamiltonianSpin}
\end{equation}

The AKLT Hamiltonian Eq.~(\ref{S1HamiltonianSpin}) has many symmetries. In particular, it possesses $SU(2)$, translation, inversion (reflection about a bond $\{i,i+1\}$) and spin-flip ($S_z \rightarrow -S_z$) symmetries. Here, we associate the following quantum numbers to all the eigenstates of the Hamiltonian: $s$ for the total spin, $S_z$ for the projection of the total spin along the $z$ direction, and momentum $k$ (quantized in integer multiples of $2\pi/L$) for the translation symmetry. Furthermore, the eigenstates with momentum $k = 0, \pi$ can be labelled by a quantum number $I = \pm 1$ corresponding to inversion symmetry and the eigenstates with $S_z = 0$ can be labelled by another quantum number $P_z = \pm 1$ corresponding to the spin-flip symmetry.  

\begin{figure}[t!]
\hspace*{-0.2cm}\includegraphics[scale=0.45]{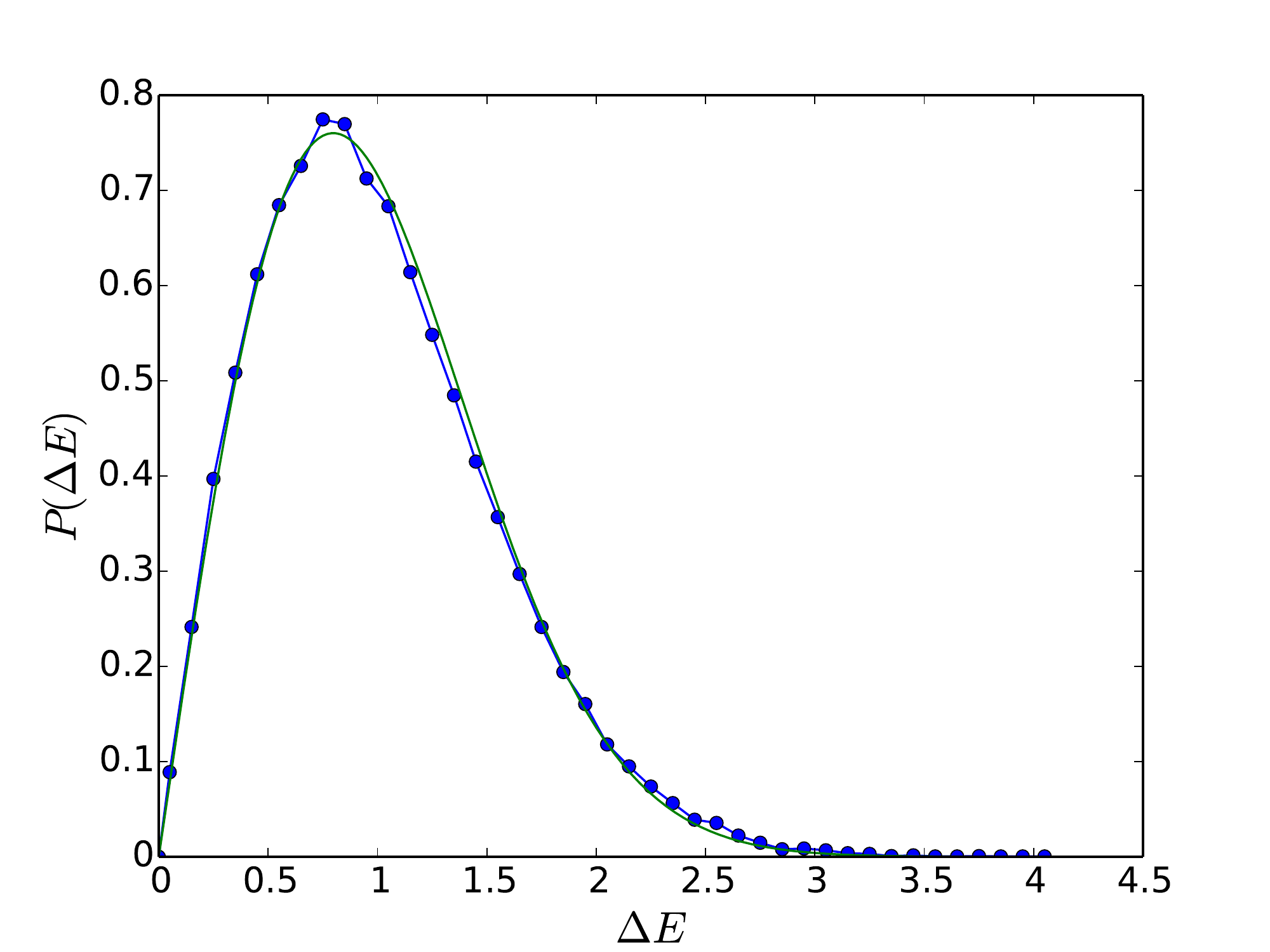}
\caption{(Color online) Energy level spacing statistics for $L = 16$ with periodic boundary conditions in a typical quantum number sector $(s, S_z, k, I, P_z) = (4, 0, 0, -1, 1)$ that has a Hilbert space dimension 26429.  $\Delta E$ is the level spacing between adjacent energy levels after a mapping of the energy spectrum to produce a constant density of states. $P(\Delta E)$ is the distribution of the level spacings. The peak of the distribution at non-zero $\Delta E$ indicates level repulsion. The green curve is the GOE distribution. The mean ratio of adjacent level spacings is $\langle r \rangle \approx 0.5316$, close to the GOE value of $\langle r\rangle \approx 0.5295$.\cite{oganesyan2007localization, atas2013distribution}}
\label{fig:goe}
\end{figure}

In spite of these symmetries, the AKLT Hamiltonian is non-integrable. Indeed the energy levels of eigenstates with a fixed set of quantum numbers corresponding to different symmetries do show level repulsion. In Fig.~\ref{fig:goe}, we plot the energy level spacing statistics of a typical quantum number sector of the AKLT model. We find that the level spacing distribution is close to that of a Gaussian Orthogonal Ensemble (GOE). Such a distribution is typical of non-integrable models.\cite{poilblanc1993poisson, brezin1993universality, nandkishore2014many} 

\subsection{Ground state}\label{sec:groundstate}
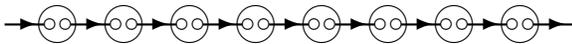
\begin{figure}[t!]
\centering
\setlength{\unitlength}{1pt}
\begin{picture}(180,20)(0,0)
\linethickness{0.8pt}
\multiput(0,10)(25,0){8}{\circle{4}} 
\multiput(6,10)(25,0){8}{\circle{4}} 
\multiput(3,10)(25,0){8}{\circle{14}}
\thicklines
\multiput(-17,10)(25,0){9}{\line(1,0){15}} 
\multiput(-5,10)(25,0){9}{\vector(1,0){0}} 
\end{picture}
\caption{AKLT Ground state. The big circles are physical spin-1\,s and the smaller circles within the spin-1\,s are spin-1/2 Schwinger bosons. Symmetric combinations of the Schwinger bosons on each site form the physical spin-1. The lines joining the Schwinger bosons represent singlets. $\ket{G}$ with periodic boundary conditions}
\label{fig:groundstate}
\end{figure}

The beauty of the AKLT model is that despite its lack of integrability, the ground state can be constructed explicitly.\cite{affleck1987rigorous} To do this, we write each spin-1 as two symmetrized spin-1/2 degrees of freedom that can either have $S_z = +1/2$ or $S_z = -1/2$. Thus, two nearest neighbor spins consist of four spin-1/2 degrees of freedom. If a singlet is formed between two of them as in Fig.~\ref{fig:groundstate}, the remaining two spin-1/2's can form at most a spin-1 configuration, meaning that the projector $P^{(2,1)}_{i,i+1}$ annihilates such a configuration. The cartoon picture of the ground state $\ket{G}$ of energy $E = 0$ with periodic boundary conditions is shown in Fig.~\ref{fig:groundstate}a.
With open boundary conditions, there are two spin-1/2 degrees of freedom at each edge that are not bound into singlets, so-called {\it dangling spins}. These fractionalized degrees of freedom, i.e.\ half-odd integer spins in a model that contained only integer spins, represent the topological nature of the spin-1 Haldane phase.\cite{haldane1983nonlinear} The ground state for open boundary conditions is shown in Appendix~\ref{sec:groundstateobc}.

In this paper, we mainly work with periodic boundary conditions (PBC), although we comment on open boundary conditions (OBC) in Appendix~\ref{sec:obc}. In this case, it has been shown that $\ket{G}$ is the unique ground state (with energy $0$) of the Hamiltonian Eq.~(\ref{S1HamiltonianSpin}).\cite{aklt1987rigorous, affleck1988valence} The groundstate $\ket{G}$ is separated from the excitation spectrum by an energy gap\cite{aklt1987rigorous} (the ``Haldane gap''), which for the AKLT model can be bounded from below. The AKLT ground state shown in Fig.~\ref{fig:groundstate}a can be described more rigorously using a dimer (a singlet) basis. Since the spin-1/2 degrees of freedom on each site are symmetrized, it is convenient to introduce Schwinger bosons, i.e.\ bosonic creation (annihilation) operators $a_i^\dagger$ ($a_i$) and $b_i^\dagger$ ($b_i$) for the $S_z = +1/2$ ($\uparrow$) and $S_z = -1/2$ ($\downarrow$) spin-1/2 degrees of freedom, respectively. Any wavefunction written in terms of Schwinger bosons on site $i$ can be converted to the normalized spin-1 basis on a site $i$ ($\ket{1}_i, \ket{0}_i , \ket{\mm}_i$ corresponding to $S_z = +1, 0, \mm$) using the dictionary:
\begin{equation}
    \ket{1}_i = \frac{(a^\dagger_i)^2}{\sqrt{2}}\vac_i, \,\,\, \ket{0}_i = a^\dagger_i b^\dagger_i \vac_i, \,\,\, \ket{\mm}_i = \frac{(b^\dagger_i)^2}{\sqrt{2}}\vac_i
\label{spindict}
\end{equation}
where $\vac_i$ is the local vacuum defined by the kernel of the boson annihilation operators of site $i$, i.e.\ $a_i\vac_i = 0, b_i\vac_i = 0$. 
Since there are two spin-1/2 degrees of freedom on each site, the number operator $N_i = a_i^\dagger a_i + b_i^\dagger b_i$ has the constraint $N_i \ket{\psi} = 2\ket{\psi}$ where $\ket{\psi}$ is any configuration of spin-1s. To describe dimers, one could then define a dimer creation operator that forms singlets between the bosons on different sites as
\begin{equation}
c^\dagger_{ij} = a_i^\dagger b_j^\dagger - a_j^\dagger b_i^\dagger.
\end{equation}
The complete algebra of dimers and Schwinger bosons is given in Appendix~\ref{dimers}. 

The spin-1/2 Schwinger boson creation and annihilation operators can be related to the spin-1 operators by
\begin{eqnarray}
    S^z &=& \frac{1}{2}(a_i^\dagger a_i - b^\dagger_i b_i) \nonumber \\
    S^+ &=& a_i^\dagger b_i \nonumber \\
    S^- &=& b_i^\dagger a_i.
\end{eqnarray}
In this notation, the operator $\vec{S_i}.\vec{S_j}$ can be written as 
\begin{eqnarray}
    \vec{S_i}\cdot\vec{S_j} &=& \frac{1}{2}(S_i^+ S_j^- + S_i^- S_j^+) + S_i^z S_j^z \nonumber \\
    &=& -\frac{1}{2}c_{i j}^\dagger c_{i j} + \frac{1}{4}(a_i^\dagger a_i + b_i^\dagger b_i)(a_j^\dagger a_j + b_j^\dagger b_j) \nonumber \\
    &=& 1 - \frac{1}{2}c_{i j}^\dagger c_{i j}
\label{Si.Sj}
\end{eqnarray}
where we have made use of the Schwinger boson number constraint on each site. Using Eq.~(\ref{Si.Sj}), Eq.~(\ref{S1HamiltonianSpin}) can be written in terms of dimer creation and annihilation operators. In particular, $P^{(2,1)}_{ij}$ can be written as 
\begin{eqnarray}
    P^{(2,1)}_{ij} &=& 1 - \frac{5}{12} c_{ij}^\dagger c_{ij} + \frac{1}{24}c_{ij}^\dagger c_{ij} c_{ij}^\dagger c_{ij} \nonumber \\
    &=& 1 - \frac{1}{4}c_{ij}^\dagger c_{ij} + \frac{1}{24}{c_{ij}^\dagger}^2 c_{ij}^2 
\label{P2dimer}
\end{eqnarray}
where the expression has been normal ordered using Eq.~(\ref{dimernormexpand}).

With this representation, the unnormalized AKLT state $\ket{G}$ of Fig.~\ref{fig:groundstate}a  for periodic boundary conditions can be written as
\begin{equation}
    \ket{G} = \prod_{i = 1}^L{{c_{i,i+1}^\dagger}}\vac.
\label{groundstate}
\end{equation}
Here the vacuum $\vac$ is the global vacuum defined as the kernel of all the annihilation operators, i.e. $c_{ij}\vac = a_i\vac = b_i\vac = 0$, $\forall\, i,j$. We also define the normalized ground state\cite{affleck1988valence} $\widetilde{\ket{G}}$ as
\begin{equation}
    \widetilde{\ket{G}} = \frac{\ket{G}}{\sqrt{3^L +  3 (-1)^L }}.
\end{equation} 

\section{Exact States}\label{sec:whatarethey}
\begin{table*}[ht]
\hspace{-38mm}
\begin{minipage}{0.35\textwidth}
\begin{tabular}{|c|c|c|c|c|c|c|}
    \hline
    \multicolumn{7}{|c|}{\boldsymbol{$L = 12$}} \\
    \hline
    \boldsymbol{$E$} & \boldsymbol{$D$} & \boldsymbol{$k$} & \boldsymbol{$s$} & \boldsymbol{$I$} & \boldsymbol{$P_z$} & $\ket{\boldsymbol{\psi}}$ \\
    \hline
    \hline
    0 & 1 & 0 & 0 & 1 & 1 & $\ket{G}$ \\
    \hline
    $\frac{5}{3}$ & 1 & $\pi$ & 0 & 1 & 1 & $\ket{A}$ \\
    \hline
    2 & 1 & $\pi$ & 0 & -1 & 1 & $\ket{B}$ \\
    \hline
    2 & 1 & $\pi$ & 2 & -1 & 1 & $\ket{S_2}$ \\
    \hline
    4 & 1 & 0 & 4 & 1 & 1 & $\ket{S_4}$ \\
    \hline
    6 & 1 & $\pi$ & 6 & -1 & 1 & $\ket{S_6}$ \\
    \hline
    8 & 1 & 0 & 8 & 1 & 1 & $\ket{S_8}$ \\
    \hline
    10 & 5 & $\pi$ & 10 & $(-1)^n$ & 1 & $\ket{2_n}$, $n = 1,3-6$ \\
    \hline
    10 & 1 & $\frac{(2n+1)\pi}{6}$ & 10 & - & 1 & $\ket{2_k}$ \\
    \hline
    10 & 1 & $\pi$ & 11 & 1 & -1 & $\ket{1_{k = \pi}}$ \\
    \hline
    $\frac{32}{3}$ & 1 & $\pi$ & 10 & 1 & 1 & $\ket{2_{n = 0}}$ \\
    \hline
    $\frac{21}{2}$ & 1 & $\frac{2\pi}{3}$, $\frac{4\pi}{3}$ & 11 & - & -1 & $\ket{1_k}$\\
    \hline
    11 & 1 & 0 & 10 & -1 & 1 & $\ket{6_0}$ \\
    \hline
    11 & 1 & $\frac{\pi}{2}$, $\frac{3\pi}{2}$ & 11 & - & -1 & $\ket{1_k}$ \\
    \hline
    $\frac{23}{2}$ & 1 & $\frac{\pi}{3}$, $\frac{5\pi}{3}$ & 11 & - & -1 & $\ket{1_k}$ \\
    \hline
    12 & 1 & 0 & 12 & 1 & 1 & $\ket{F}$\\
    \hline
    \end{tabular}
    \end{minipage}%
    \begin{minipage}{0.43\textwidth}
    \centering
    \begin{tabular}{|c|c|c|c|c|c|c|}
    \hline
    \multicolumn{7}{|c|}{\boldsymbol{$L = 14$}} \\
    \hline
    \boldsymbol{$E$} & \boldsymbol{$D$} & \boldsymbol{$k$} & \boldsymbol{$s$} & \boldsymbol{$I$} & \boldsymbol{$P_z$} & $\ket{\boldsymbol{\psi}}$ \\
    \hline 
    \hline
    0 & 1 & 0 & 0 & 1 & 1 & $\ket{G}$ \\
    \hline
    $\frac{5}{3}$ & 1 & $\pi$ & 0 & 1 & 1 & $\ket{A}$ \\
    \hline
    2 & 1 & $\pi$ & 0 & -1 & 1 & $\ket{B}$ \\
    \hline
    2 & 1 & $\pi$ & 2 & -1 & 1 & $\ket{S_2}$ \\
    \hline
    4 & 1 & 0 & 4 & 1 & 1 & $\ket{S_4}$ \\
    \hline
    6 & 1 & $\pi$ & 6 & -1 & 1 & $\ket{S_6}$ \\
    \hline
    8 & 1 & 0 & 8 & 1 & 1 & $\ket{S_8}$ \\
    \hline
    10 & 1 & $\pi$ & 10 & -1 & 1 & $\ket{S_{10}}$ \\
    \hline 
    12 & 5 & $\pi$ & 12 & $(-1)^n$ & 1 & $\ket{2_n}$, $n = 1,3-6$ \\
    \hline
    12 & 1 & $\frac{2n\pi}{7}$ & 12 & - & 1 & $\ket{2_k}$ \\
    \hline
    12 & 1 & $\pi$ & 13 & 1 & -1 & $\ket{1_{k = \pi}}$ \\
    \hline
    $\frac{38}{3}$ & 1 & $\pi$ & 12 & 1 & 1 & $\ket{2_{n = 0}}$ \\
    \hline
    14 & 1 & 0 & 14 & 1 & 1 & $\ket{F}$\\
    \hline
    \end{tabular}
    \end{minipage}%
    \begin{minipage}{0.1\textwidth}
    \hspace{-30mm}
    \begin{tabular}{|c|c|c|c|c|c|c|}
    \hline
    \multicolumn{7}{|c|}{\boldsymbol{$L = 16$}} \\
    \hline
    \boldsymbol{$E$} & \boldsymbol{$D$} & \boldsymbol{$k$} & \boldsymbol{$s$} & \boldsymbol{$I$} & \boldsymbol{$P_z$} & $\ket{\boldsymbol{\psi}}$ \\
    \hline 
    \hline
    0 & 1 & 0 & 0 & 1 & 1 & $\ket{G}$ \\
    \hline
    $\frac{5}{3}$ & 1 & $\pi$ & 0 & 1 & 1 & $\ket{A}$ \\
    \hline
    2 & 1 & $\pi$ & 0 & -1 & 1 & $\ket{B}$ \\
    \hline
    2 & 1 & $\pi$ & 2 & -1 & 1 & $\ket{S_2}$ \\
    \hline
    4 & 1 & 0 & 4 & 1 & 1 & $\ket{S_4}$ \\
    \hline
    6 & 1 & $\pi$ & 6 & -1 & 1 & $\ket{S_6}$ \\
    \hline
    8 & 1 & 0 & 8 & 1 & 1 & $\ket{S_8}$ \\
    \hline
    10 & 1 & $\pi$ & 10 & -1 & 1 & $\ket{S_{10}}$ \\
    \hline 
    12 & 1 & $\pi$ & 12 & -1 & 1 & $\ket{S_{12}}$ \\
    \hline
    14 & 7 & $\pi$ & 14 & $(-1)^n$ & 1 & $\ket{2_n}$, $n = 1,3-8$ \\
    \hline
    14 & 1 & $\pi$ & 15 & 1 & -1 & $\ket{1_{k = \pi}}$ \\
    \hline
    $\frac{44}{3}$ & 1 & $\pi$ & 14 & 1 & 1 & $\ket{2_{n = 0}}$ \\
    \hline
    16 & 1 & 0 & 16 & 1 & 1 & $\ket{F}$\\
    \hline
\end{tabular}
\end{minipage}
\caption{Table showing the highest weight states with rational energies for $L = 12$, $L = 14$ and $L = 16$ (sectors $k = 0, \pi$). $E$ is energy, $D$ is the degeneracy (excluding the $SU(2)$ multiplet degeneracy), $k$ momentum, $s$ the spin quantum number, $I$ the eigenvalue under bond inversion symmetry, $P_z$ is the eigenvalue under spin flips for the $S_z = 0$ state of the multiplet and $\ket{\psi}$ the state we identify it with. The states in this table with $s < L - 2$ have a very sparse entanglement spectrum compared to typical states in their quantum number sectors.}
\label{fig:tableofstates}
\end{table*}
The energy spectrum of the spin-1 (and, as we will see, integer spin-$S$) AKLT model for a finite size chain with periodic boundary conditions exhibits some remarkable features. Beyond the unique ground state whose energy is $0$ (for $H$ of Eq.~(\ref{S1HamiltonianSpin})), there are many other states with rational energies up to machine precision, some of them seemingly located in the bulk of the spectrum. Moreover, several of these states are even at integer energies. This observation holds for chains with a length up to $L=16$, the upper numerical limit where we can compute the full spectrum. In this paper, we show that states with such rational energies are not coincidences and we can derive analytical expressions for them akin to Eq.~(\ref{groundstate}) for the ground state. Being exact eigenstates for particular finite system sizes with a closed analytical expression and rational energy, we dub these states ``exact states".

One could argue that looking for exact states by targeting rational energies is ad-hoc. Indeed, rescaling the energy by a random positive number or shifting the ground state energy would scramble this information although simple algorithms could be devised to recover it. Moreover, in finite precision arithmetic, any number can be written as a rational number. To hunt for possible exact states, we propose another approach based on the entanglement spectrum.\cite{lihaldane2008} 

For any eigenstate $\ket{\psi}$ of a spin-$S$ chain, we consider the spatial partition into two continuous regions $A$ and $B$ with $L_A$ spins in $A$ and $L_B$ spins in $B$. We then construct the reduced density matrix $\rho_A = {\rm Tr}_B \ket{\psi}\bra{\psi}$. The entanglement spectrum is the eigenvalue spectrum of $-\log \rho_A$. Assuming that $L_A \le L_B$, the rank of $\rho_A$ (i.e. the number of levels in the entanglement spectrum) is bounded by $(2S+1)^{L_A}$. Unless $\ket{\psi}$ has a peculiar structure, this bound is usually saturated. Most eigenstates (including ground states) of local Hamiltonians saturate this bound. The fact that the entanglement entropy of the ground state of a gapped Hamiltonian is not volume law\cite{hastings2007area} merely means that the ground state of the system can be \emph{approximated} by a state with a sparse spectrum.\cite{verstraete2006matrix} However, states whose entanglement spectrum is truly sparse and whose number of levels in the entanglement spectrum do not saturate the bound are special. This includes the ground state of the AKLT model Eq.~(\ref{groundstate}), which has exactly $4$ out of $(2S + 1)^{L_A}$ levels in its entanglement spectrum irrespective of the length $L_A$. 
\begin{figure*}
\begin{minipage}{\textwidth}
    \begin{equation}
    \begin{split}
    &P^{(2,1)}_{ij}
    \begin{picture}(150,-30)(-10,7.5)
    \put(-8,10){\makebox(0,0)[c]{$\Big|$}}
    \put(139,10){\makebox(0,0)[c]{$\Big\rangle$}}
    \linethickness{0.8pt}
    \multiput(0,10)(25,0){2}{\circle*{4}}
    \multiput(106,10)(25,0){2}{\circle*{4}}
    \multiput(0,10)(25,0){6}{\circle{4}} 
    \multiput(6,10)(25,0){6}{\circle{4}} 
    \multiput(3,10)(25,0){6}{\circle{14}}
    \thicklines
    \qbezier(8,10)(31,35)(54,10)
    \multiput(34.5,23)(69,0){2}{\vector(1,0){0}}
    \qbezier(77,10)(100,35)(123,10)
    \multiput(33,10)(50,0){2}{\line(1,0){15}}
    \multiput(45,10)(50,0){2}{\vector(1,0){0}}
    \put(0,-5){m}
    \put(25,-5){n}
    \put(75,-5){j}
    \put(50,-5){i}
    \put(100,-5){p}
    \put(125,-5){r}
    \end{picture}
    = 
    \begin{picture}(150,-30)(-10,7.5)
    \put(-8,10){\makebox(0,0)[c]{$\Big|$}}
    \put(139,10){\makebox(0,0)[c]{$\Big\rangle$}}
    \linethickness{0.8pt}
    \multiput(0,10)(25,0){2}{\circle*{4}}
    \multiput(106,10)(25,0){2}{\circle*{4}}
    \multiput(0,10)(25,0){6}{\circle{4}} 
    \multiput(6,10)(25,0){6}{\circle{4}} 
    \multiput(3,10)(25,0){6}{\circle{14}}
    \thicklines
    \qbezier(8,10)(31,35)(54,10)
    \multiput(34.5,23)(69,0){2}{\vector(1,0){0}}
    \qbezier(77,10)(100,35)(123,10)
    \multiput(33,10)(50,0){2}{\line(1,0){15}}
    \multiput(45,10)(50,0){2}{\vector(1,0){0}}
    \put(0,-5){m}
    \put(25,-5){n}
    \put(75,-5){j}
    \put(50,-5){i}
    \put(100,-5){p}
    \put(125,-5){r}
    \end{picture}
    \\[5mm]
    & + \frac{1}{4}\;\;\;
    \begin{picture}(150,-30)(-10,7.5)
    \put(-12,10){\makebox(0,0)[c]{$\left(\,\Big|\right.$}}
    \put(139,10){\makebox(0,0)[c]{$\Big\rangle$}}
    \linethickness{0.8pt}
    \multiput(0,10)(25,0){6}{\circle{4}} 
    \multiput(0,10)(25,0){2}{\circle*{4}}
    \multiput(106,10)(25,0){2}{\circle*{4}}
    \multiput(6,10)(25,0){6}{\circle{4}} 
    \multiput(3,10)(25,0){6}{\circle{14}}
    \thicklines
    \qbezier(8,10)(28,35)(48,10)
    \multiput(31.5,23)(75,0){2}{\vector(1,0){0}}
    \qbezier(83,10)(103,35)(123,10)
    \qbezier(33,10)(65.5,35)(98,10)
    \put(69,23){\vector(1,0){0}}
    \put(58,10){\line(1,0){15}}
    \put(70,10){\vector(1,0){0}}
    \put(0,-5){m}
    \put(25,-5){n}
    \put(75,-5){j}
    \put(50,-5){i}
    \put(100,-5){p}
    \put(125,-5){r}
    \end{picture}
    +
    \begin{picture}(150,-30)(-10,7.5)
    \put(-8,10){\makebox(0,0)[c]{$\Big|$}}
    \put(139,10){\makebox(0,0)[c]{$\Big\rangle$}}
    \linethickness{0.8pt}
    \multiput(0,10)(25,0){2}{\circle*{4}}
    \multiput(106,10)(25,0){2}{\circle*{4}}
    \multiput(0,10)(25,0){6}{\circle{4}} 
    \multiput(6,10)(25,0){6}{\circle{4}} 
    \multiput(3,10)(25,0){6}{\circle{14}}
    \thicklines
    \multiput(106.5,23)(69,0){1}{\vector(1,0){0}}
    \qbezier(83,10)(103,35)(123,10)
    \multiput(33,10)(50,0){1}{\line(1,0){15}}
    \multiput(45,10)(50,0){1}{\vector(1,0){0}}
    \qbezier(8,10)(53,35)(98,10)
    \put(56.5,23){\vector(1,0){0}}
    \put(58,10){\line(1,0){15}}
    \put(70,10){\vector(1,0){0}}
    \put(0,-5){m}
    \put(25,-5){n}
    \put(75,-5){j}
    \put(50,-5){i}
    \put(100,-5){p}
    \put(125,-5){r}
    \end{picture}
    \\[7mm]
    &+
    \begin{picture}(150,-30)(-10,7.5)
    \put(-8,10){\makebox(0,0)[c]{$\Big|$}}
    \put(139,10){\makebox(0,0)[c]{$\Big\rangle$}}
    \linethickness{0.8pt}
    \multiput(0,10)(25,0){2}{\circle*{4}}
    \multiput(106,10)(25,0){2}{\circle*{4}}
    \multiput(0,10)(25,0){6}{\circle{4}} 
    \multiput(6,10)(25,0){6}{\circle{4}} 
    \multiput(3,10)(25,0){6}{\circle{14}}
    \thicklines
    \qbezier(8,10)(28,35)(48,10)
    \multiput(31.5,23)(75,0){1}{\vector(1,0){0}}
    \multiput(83,10)(50,0){1}{\line(1,0){15}}
    \multiput(95,10)(50,0){1}{\vector(1,0){0}}
    \qbezier(33,10)(88,35)(123,10)
    \put(91.5,23){\vector(1,0){0}}
    \put(58,10){\line(1,0){15}}
    \put(70,10){\vector(1,0){0}}
    \put(0,-5){m}
    \put(25,-5){n}
    \put(75,-5){j}
    \put(50,-5){i}
    \put(100,-5){p}
    \put(125,-5){r}
    \end{picture}
    +
    \begin{picture}(150,-30)(-10,7.5)
    \put(-8,10){\makebox(0,0)[c]{$\Big|$}}
    \put(139,10){\makebox(0,0)[c]{$\left.\Big\rangle\right)$}}
    \linethickness{0.8pt}
    \multiput(0,10)(25,0){2}{\circle*{4}}
    \multiput(106,10)(25,0){2}{\circle*{4}}
    \multiput(0,10)(25,0){6}{\circle{4}} 
    \multiput(6,10)(25,0){6}{\circle{4}} 
    \multiput(3,10)(25,0){6}{\circle{14}}
    \thicklines
    \multiput(33,10)(50,0){2}{\line(1,0){15}}
    \multiput(45,10)(50,0){2}{\vector(1,0){0}}
    \put(58,10){\line(1,0){15}}
    \put(70,10){\vector(1,0){0}}
    \qbezier(8,10)(65.5,38)(123,10)
    \put(69,24){\vector(1,0){0}}
    \put(0,-5){m}
    \put(25,-5){n}
    \put(75,-5){j}
    \put(50,-5){i}
    \put(100,-5){p}
    \put(125,-5){r}
    \end{picture}\\[7mm]
    &+ \frac{1}{12}\;\;\;
    \begin{picture}(150,-30)(-10,7.5)
    \put(-12,10){\makebox(0,0)[c]{$\left(\,\Big|\right.$}}
    \put(139,10){\makebox(0,0)[c]{$\Big\rangle$}}
    \linethickness{0.8pt}
    \multiput(0,10)(25,0){2}{\circle*{4}}
    \multiput(106,10)(25,0){2}{\circle*{4}}
    \multiput(0,10)(25,0){6}{\circle{4}} 
    \multiput(6,10)(25,0){6}{\circle{4}} 
    \multiput(3,10)(25,0){6}{\circle{14}}
    \thicklines
    \qbezier(52,10)(65.5,22)(79,10)
    \put(69,16){\vector(1,0){0}}
    \qbezier(33,10)(65.5,35)(98,10)
    \put(69,23){\vector(1,0){0}}
    \qbezier(8,10)(65.5,44)(123,10)
    \put(69,27){\vector(1,0){0}}
    \put(58,10){\line(1,0){15}}
    \put(70,10){\vector(1,0){0}}
    \put(0,-5){m}
    \put(25,-5){n}
    \put(75,-5){j}
    \put(50,-5){i}
    \put(100,-5){p}
    \put(125,-5){r}
    \end{picture}
    +
    \begin{picture}(150,-30)(-10,7.5)
    \put(-8,10){\makebox(0,0)[c]{$\Big|$}}
    \put(139,10){\makebox(0,0)[c]{$\left.\Big\rangle\right)$}}
    \linethickness{0.8pt}
    \multiput(0,10)(25,0){2}{\circle*{4}}
    \multiput(106,10)(25,0){2}{\circle*{4}}
    \multiput(0,10)(25,0){6}{\circle{4}} 
    \multiput(6,10)(25,0){6}{\circle{4}} 
    \multiput(3,10)(25,0){6}{\circle{14}}
    \thicklines
    \put(58,10){\line(1,0){15}}
    \put(70,10){\vector(1,0){0}}
    \qbezier(52,10)(65.5,22)(79,10)
    \put(69,16){\vector(1,0){0}}
    \qbezier(8,10)(53,35)(98,10)
    \put(56.5,23){\vector(1,0){0}}
    \qbezier(33,10)(88,35)(123,10)
    \put(91.5,23){\vector(1,0){0}}
    \put(0,-5){m}
    \put(25,-5){n}
    \put(75,-5){j}
    \put(50,-5){i}
    \put(100,-5){p}
    \put(125,-5){r}
    \end{picture}
    \end{split}
    \tag{\ref{S1singletrules}}
    \end{equation}
    \end{minipage}
\caption{An example of the action of $P^{(2,1)}_{ij}$ on a dimer configuration. The configurations of the filled small circles are not relevant for the scattering equation and are the same on all terms in the equation. The directions of the arrows are crucial; reversing an arrow contributes a factor of (-1).}
\label{fig:scatteringexample}
\end{figure*}

We propose to use the sparsity of entanglement spectrum, the ratio between the rank of $\rho_A$ and its dimension, as a probe to search for exact states. A brute force approach is thus to numerically compute all the eigenstates for a given system size and label them with their quantum numbers. We then focus on those exhibiting an entanglement spectrum sparsity at the largest possible value of $L_A$ (the integer part of $L/2$). For the spin-1 AKLT model, we observe that most of the states in the bulk of the full energy spectrum have a sparsity close or equal to $1$, as expected. However, there are a few eigenstates that have an entanglement spectrum sparsity less than $5\%$ for the largest system sizes we have computed. For reasons that are still not fully clear to us, most of these eigenstates coincide with those having a rational energy. Of course, some of these states are trivially exact states. An example is the highest excited state of the AKLT Hamiltonian that has all spin-1s with $S_z = 1$. It is a product state and its reduced density matrix has a single eigenvalue. But as we show below, many of these exact states have an interesting non-trivial structure.  We list these set of exact states for system sizes $L = 12$, $L = 14$ and $L = 16$ for the spin-1 case with all their useful quantum numbers and degeneracies in Table~\ref{fig:tableofstates}. The derivation of their analytic expressions will be detailed in the following sections.  
\section{Exact Low energy Excited states of the Spin-1 AKLT Model}\label{sec:spin1aklt}
To find and give an expression for the exact excited states of the AKLT model, we need to choose a convenient basis to work with. There are two bases that we use. The first is the usual spin basis. Since the Hamiltonian is simply a projector onto a particular total spin $J$, any configuration of nearest neighbors with $S_z = m_1$ and $S_z = m_2$ in a spin $S$ chain scatters to the state with $J = S$ and $S_z = m_1 + m_2$ with an amplitude given by a Clebsch-Gordan coefficient. A complete set of rules for scattering of configurations in the spin basis are presented in the Appendix~\ref{spins}.

The second is the dimer basis that we have introduced in Sec.~\ref{sec:akltmodels}. Here, the basis states are defined as linearly independent states of dimer or Schwinger boson creation operators acting on the vacuum $\vac$. Though this representation allows an elegant representation of (most of) the exact states we will discuss, the set of dimer basis states is highly overcomplete and non-orthonormal (see Appendix~\ref{ortho} for an example). To study the AKLT model in the dimer basis, we need to derive rules for the scattering of basis states upon the action of the Hamiltonian. Since the Hamiltonian in Eq.~(\ref{P2dimer}) is normal ordered, it is sufficient to compute the actions of $c_{ij}^m$ on the basis states. The actions of $c_{ij}$ and $P^{(2,1)}_{ij}$ on various configurations on dimers along with some useful identities are specified in Appendix~\ref{scattering} and Fig.~\ref{diagrep}. An example of such a scattering rule is diagrammatically depicted in Fig.~\ref{fig:scatteringexample}.

We have already exemplified the construction of the ground state in Sec.~\ref{sec:groundstate}. We now focus on two exact low energy excited states, namely the Arovas states.\cite{arovas1989two}
\subsection{{\bf Arovas A state}}
We now follow Arovas\cite{arovas1989two} to construct two exact excited states. Consider a configuration of dimers $\ket{A_n}$ defined in Eq.~(\ref{Acartoon}) with the cartoon picture of dimers as shown in Fig.~\ref{fig:arovasA}a.
\begin{eqnarray}
&\ket{A_n}=\left(\prod_{j=1}^{n-2} c_{j,j+1}^\dagger\right) c_{n-1, n+2}^\dagger (c_{n, n+1}^\dagger)^2 \nn \\
&\left(\prod_{j=n+2}^L c_{j, j+1}^\dagger\right) \vac.
\label{Acartoon}
\end{eqnarray}
The Arovas A state is a translation invariant linear superposition of these $\ket{A_n}$ with momentum $k = \pi$. Up to a global normalization factor, it is given by
\begin{equation}
    \ket{A} = \sum_n{(-1)^n \ket{A_n}}.
    \label{pisuparovas}
\end{equation} 
The system size is even, greater than 5 sites and we impose periodic boundary conditions. 

For pedagogical purposes, we now show the derivation of this first exact state beyond the ground state. This exemplifies the mechanism that underlies the derivation of all these exact states of the AKLT Hamiltonian. The proof relies on several properties of the dimer basis that are given in Appendix~\ref{scattering}. From the Hamiltonian Eq.~(\ref{S1Hamiltonian}) and the property Eq.~(\ref{projectorvanishing}), we deduce that the only terms in the Hamiltonian that give a non-vanishing contribution upon action on $\ket{A_n}$ are $P^{(2,1)}_{n-1,n}$ and $P^{(2,1)}_{n+1,n+2}$. That is, 
\begin{equation}
    H \ket{A_n} = (P^{(2,1)}_{n-1,n} + P^{(2,1)}_{n+1,n+2})\ket{A_n}.
\label{AOnlyP}
\end{equation}
Using the scattering rules of Eq.~(\ref{S1singletrules}) and the cartoon picture of $\ket{A_n}$, it is easy to see that
\begin{equation}
    P^{(2,1)}_{n-1,n}\ket{A_n} = \ket{A_n} + \frac{1}{6}\ket{A_{n-1}} + \frac{1}{2}\ket{G} + \ket{B_n}
\label{Anscattering}
\end{equation}
where $\ket{B_n}$ is shown in Fig.~\ref{fig:arovasA}b and defined as 
\begin{eqnarray}
    &\ket{B_n} = \left(\prod_{j=1}^{n-3}c_{j,j+1}^\dagger\right) c_{n-1,n}^\dagger c^\dagger_{n-2,n+1}  c^\dagger_{n-1, n+2} c_{n,n+1}^\dagger \nn \\
    &\left(\prod_{j=n+2}^L c_{j, j+1}^\dagger\right) \vac
\label{Bcartoon}
\end{eqnarray}
and $\ket{G}$ is the unnormalized ground state of Eq.~(\ref{groundstate}).
As seen in Fig.~\ref{fig:arovasA}a, $\ket{A_n}$ is symmetric under inversion about the mid bond $\{n, n+1\}$. Thus the scattering terms obtained by $P^{(2,1)}_{n+1,n+2}\ket{A_n}$ are the same as those in Eq.~(\ref{Anscattering}), but with all the terms inverted about the mid bond $\{n,n+1\}$. Under bond inversion, since $\ket{B_n} \rightarrow \ket{B_{n+1}}$, $\ket{A_{n-1}} \rightarrow \ket{A_{n+1}}$ and $\ket{G} \rightarrow \ket{G}$, we obtain
\begin{equation}
    P^{(2,1)}_{n+1,n+2}\ket{A_n} = \ket{A_n} + \frac{1}{6}\ket{A_{n+1}} + \frac{1}{2}\ket{G} + 
    \frac{1}{2}\ket{B_{n+1}}.
\label{Anreflect}
\end{equation}
It is important to note that $\ket{B_n}$ transforms in that way because it is symmetric about a site $n$, and not about any bond. Combining Eqs.~(\ref{Anscattering}) and (\ref{Anreflect}) with Eq.~(\ref{AOnlyP}), we obtain
\begin{eqnarray}
    &H\ket{A_n} = 2 \ket{A_n} + \frac{1}{6}(\ket{A_{n-1}} + \ket{A_{n+1}}) + \ket{G} \nonumber \\
    &+\frac{1}{2}(\ket{B_n} + \ket{B_{n+1}}).
\label{AnHamilaction}
\end{eqnarray}
It follows that
\begin{equation}
    H\ket{A} = 2\ket{A} - \frac{1}{3}\ket{A}
    =\frac{5}{3}\ket{A}.
\label{Arovas1}
\end{equation}
Thus, $\ket{A}$ is an exact eigenstate of the spin-1 AKLT Hamiltonian with energy $E = \frac{5}{3}$ and momentum $k = \pi$. The crucial part of the exactness of this state lies in the fact that $\ket{B_n}$ and $\ket{B_{n+1}}$ appeared in Eq.~(\ref{AnHamilaction}) with equal weight and they could be cancelled off by momentum superposition.  However, the cancellation would not hold if $L$ were not even (since we need $k = \pi$) or for open boundary conditions (because the edge scattering terms would not cancel). Moreover, $\ket{A}$ appears only for $L \geq 6$ because for $L = 4$, the scattering equation of $\ket{A_n}$ Eq.~(\ref{AnHamilaction}) would no longer be the same due to boundary conditions.

\begin{figure}[t!]
    \centering
    \begin{subfigure}{}
        \\[1mm]
        \begin{picture}(175,15)(0,0)
        \linethickness{0.8pt}
        \multiput(0,15)(25,0){3}{\circle{4}} 
        \multiput(6,15)(25,0){3}{\circle{4}} 
        \multiput(125,15)(25,0){3}{\circle{4}}
        \multiput(131,15)(25,0){3}{\circle{4}}
        \multiput(78,12.5)(25,0){2}{\circle{4}}
        \multiput(78,17.5)(25,0){2}{\circle{4}}
        \multiput(3,15)(25,0){8}{\circle{14}}
        \thicklines
        \multiput(-17,15)(25,0){3}{\line(1,0){15}} 
        \multiput(-5,15)(25,0){3}{\vector(1,0){0}} 
        \multiput(133,15)(25,0){3}{\line(1,0){15}}
        \multiput(145,15)(25,0){3}{\vector(1,0){0}} 
        \put(80,12.5){\line(1,0){21}}
        \put(95,12.5){\vector(1,0){0}} 
        \put(80,17.5){\line(1,0){21}}
        \put(95,17.5){\vector(1,0){0}}
        \qbezier(58,15)(90.5,40)(123,15)
        \put(95.5,28){\vector(1,0){0}}
        \put(75,0){n}
        \put(47,0){n-1}
        \put(95,0){n+1}
        \put(120,0){n+2}
        \put(-32,12.5){(a)}
        \end{picture}
    \end{subfigure}\\[-2mm]
    \begin{subfigure}{}
        \begin{picture}(175,45)(0,0)
        \linethickness{0.8pt}
        \multiput(0,15)(25,0){8}{\circle{4}} 
        \multiput(6,15)(25,0){8}{\circle{4}} 
        \multiput(3,15)(25,0){8}{\circle{14}}
        \thicklines
        \multiput(-17,15)(25,0){2}{\line(1,0){15}} 
        \multiput(-5,15)(25,0){2}{\vector(1,0){0}}
        \multiput(58,15)(25,0){2}{\line(1,0){15}}
        \multiput(70,15)(25,0){2}{\vector(1,0){0}}
        \multiput(133,15)(25,0){3}{\line(1,0){15}}
        \multiput(145,15)(25,0){3}{\vector(1,0){0}}
        \qbezier(33,15)(68.5,45)(104,15)
        \put(65.5, 30){\vector(1,0){0}}
        \qbezier(52,15)(87.5,45)(123,15)
        \put(90.5, 30){\vector(1,0){0}}
        \put(75,0){n}
        \put(47,0){n-1}
        \put(95,0){n+1}
        \put(120,0){n+2}
        \put(-32,12.5){(b)}
        \end{picture}
    \end{subfigure}
    \caption{(a) Arovas $A$ configuration $\ket{A_n}$ (b) Scattering term configuration $\ket{B_n}$}
    \label{fig:arovasA}
\end{figure}

Using Eqs.~(\ref{singletidentities1}) and (\ref{Si.Sj}), it can be seen that $\widetilde{\ket{A}}$, the normalized version of $\ket{A}$, can be rewritten as\cite{arovas1989two}
\begin{equation}
    \widetilde{\ket{A}} = \frac{3}{\sqrt{2L}}\sum_{n = 1}^L{(-1)^n \vec{S}_n\cdot\vec{S}_{n+1}} \widetilde{\ket{G}}.
\label{Arovas1Spin}
\end{equation}
Thus, we have an exact eigenstate with $\boldsymbol{E = \frac{5}{3}}$, $\boldsymbol{k = \pi}$ and $\boldsymbol{s = 0}$ (and $P_z = +1$, $I = +1$). This state is an exact example of the Single-Mode Approximation (SMA). This can be viewed as a magnon with form factor $\sin(k)$ as evident from the momentum space expression of the state
\begin{equation}
    \widetilde{\ket{A}} = \frac{3}{\sqrt{2L}}\sum_k{\sin(k) \vec{S}_k\cdot\vec{S}_{\pi-k}}\widetilde{\ket{G}}
\label{Arovas1Momentum}
\end{equation}
where $\vec{S}_k$ is the Fourier transform of $\vec{S}_n$.

\subsection{{\bf Arovas B state}}\label{sec:arovasb}
\begin{figure}
    \centering
    \begin{subfigure}{}
        \begin{picture}(200,45)(0,0)
        \linethickness{0.8pt}
        \multiput(0,35)(25,0){9}{\circle{4}} 
        \multiput(6,35)(25,0){9}{\circle{4}} 
        \multiput(3,35)(25,0){9}{\circle{14}}
        \thicklines
        \multiput(-17,35)(25,0){3}{\line(1,0){15}} 
        \multiput(-5,35)(25,0){3}{\vector(1,0){0}}
        \multiput(83,35)(25,0){2}{\line(1,0){15}}
        \multiput(95,35)(25,0){2}{\vector(1,0){0}}
        \multiput(158,35)(25,0){3}{\line(1,0){15}}
        \multiput(170,35)(25,0){3}{\vector(1,0){0}}
        \qbezier(58,35)(93.5,65)(129,35)
        \put(90.5, 50){\vector(1,0){0}}
        \qbezier(77,35)(112.5,65)(148,35)
        \put(122.5, 50){\vector(1,0){0}}
        \put(100,20){n}
        \put(72,20){n-1}
        \put(120,20){n+1}
        \put(145,20){n+2}
        \put(47,20){n-2}
        \put(-32,32.5){(a)}
        \end{picture}
    \end{subfigure}
    \begin{subfigure}{}
        \begin{picture}(200,10)(0,0)
        \linethickness{0.8pt}
        \multiput(0,15)(25,0){4}{\circle{4}} 
        \multiput(6,15)(25,0){4}{\circle{4}} 
        \multiput(150,15)(25,0){3}{\circle{4}}
        \multiput(156,15)(25,0){3}{\circle{4}}
        \multiput(103,12.5)(25,0){2}{\circle{4}}
        \multiput(103,17.5)(25,0){2}{\circle{4}}
        \multiput(3,15)(25,0){9}{\circle{14}}
        \thicklines
        \multiput(-17,15)(25,0){4}{\line(1,0){15}} 
        \multiput(-5,15)(25,0){4}{\vector(1,0){0}} 
        \multiput(158,15)(25,0){3}{\line(1,0){15}}
        \multiput(170,15)(25,0){3}{\vector(1,0){0}} 
        \put(105,12.5){\line(1,0){21}}
        \put(120,12.5){\vector(1,0){0}} 
        \put(105,17.5){\line(1,0){21}}
        \put(120,17.5){\vector(1,0){0}}
        \qbezier(83,15)(115.5,40)(148,15)
        \put(120.5,28){\vector(1,0){0}}
        \put(100,0){n}
        \put(47,0){n-2}
        \put(72,0){n-1}
        \put(120,0){n+1}
        \put(145,0){n+2}
        \put(-32,12.5){(b)}
        \end{picture}
    \end{subfigure}
    \begin{subfigure}{}
        \begin{picture}(147.5,28)(0,0)
        \linethickness{0.8pt}
        \multiput(-25,10)(25,0){9}{\circle{4}} 
        \multiput(-19,10)(25,0){9}{\circle{4}} 
        \multiput(-22,10)(25,0){9}{\circle{14}}
        \thicklines
        \multiput(-42,10)(25,0){3}{\line(1,0){15}} 
        \multiput(-30,10)(25,0){3}{\vector(1,0){0}}
        \put(58,10){\line(1,0){15}}
        \put(70,10){\vector(1,0){0}}
        \put(108,10){\line(1,0){15}}
        \put(120,10){\vector(1,0){0}}
        \multiput(158,10)(25,0){2}{\line(1,0){15}}
        \multiput(170,10)(25,0){2}{\vector(1,0){0}}
        \qbezier(52,10)(93.5,-20)(129,10)
        \put(95,-5){\vector(1,0){0}}
        \qbezier(83,10)(112.5,40)(148,10)
        \put(115,25){\vector(1,0){0}}
        \qbezier(33,10)(62.5,40)(98,10)
        \put(65,25){\vector(1,0){0}}
        \put(75,-11){n}
        \put(47,-11){n-1}
        \put(95,-11){n+1}
        \put(120,-11){n+2}
        \put(22,-11){n-2}
        \put(-57,7.5){(c)}
        \end{picture}
    \end{subfigure}
    \begin{subfigure}{}
        \begin{picture}(147.5,35)(0,0)
        \linethickness{0.8pt}
        \multiput(-25,10)(25,0){9}{\circle{4}} 
        \multiput(-19,10)(25,0){9}{\circle{4}} 
        \multiput(-22,10)(25,0){9}{\circle{14}}
        \thicklines
        \multiput(-42,10)(25,0){3}{\line(1,0){15}} 
        \multiput(-30,10)(25,0){3}{\vector(1,0){0}}
        \multiput(58,10)(25,0){3}{\line(1,0){15}}
        \multiput(70,10)(25,0){3}{\vector(1,0){0}}
        \multiput(158,10)(25,0){2}{\line(1,0){15}}
        \multiput(170,10)(25,0){2}{\vector(1,0){0}}
        \qbezier(52,10)(90.5,40)(129,10)
        \put(95,25){\vector(1,0){0}}
        \qbezier(33,10)(92.5,-22)(148,10)
        \put(95,-6){\vector(1,0){0}}
        \put(75,-13){n}
        \put(47,-13){n-1}
        \put(95,-13){n+1}
        \put(120,-13){n+2}
        \put(22,-13){n-2}
        \put(-57,7.5){(d)}
        \end{picture}
    \end{subfigure}
    \begin{subfigure}{}
        \begin{picture}(147.5,38)(0,0)
        \linethickness{0.8pt}
        \multiput(-25,10)(25,0){3}{\circle{4}} 
        \multiput(-19,10)(25,0){3}{\circle{4}} 
        \multiput(150,10)(25,0){2}{\circle{4}}
        \multiput(156,10)(25,0){2}{\circle{4}}
        \multiput(53,7.5)(25,0){4}{\circle{4}}
        \multiput(53,12.5)(25,0){4}{\circle{4}}
        \multiput(-22,10)(25,0){9}{\circle{14}}
        \thicklines
        \multiput(-42,10)(25,0){3}{\line(1,0){15}} 
        \multiput(-30,10)(25,0){3}{\vector(1,0){0}} 
        \multiput(158,10)(25,0){2}{\line(1,0){15}}
        \multiput(170,10)(25,0){2}{\vector(1,0){0}} 
        \multiput(55,7.5)(50,0){2}{\line(1,0){21}}
        \multiput(70,7.5)(50,0){2}{\vector(1,0){0}} 
        \multiput(55,12.5)(50,0){2}{\line(1,0){21}}
        \multiput(70,12.5)(50,0){2}{\vector(1,0){0}}
        \qbezier(33,10)(90.5,42.5)(148,10)
        \put(95.5,26.5){\vector(1,0){0}}
        \put(75,-5){n}
        \put(47,-5){n-1}
        \put(95,-5){n+1}
        \put(120,-5){n+2}
        \put(22,-5){n-2}
        \put(-57,7.5){(e)}
        \end{picture}
    \end{subfigure}
    \caption{(a) Arovas $B$ configuration $\ket{B_n}$. (b) Scattering term $\ket{A_n}$. (c) Scattering term $\ket{C_n}$. (d) Scattering term $\ket{D_n}$. (e) Scattering term $\ket{E_n}$.}
    \label{fig:ArovasBConfig}
\end{figure}
Similar to the Arovas A state built from the $\ket{A_n}$s, another exact state can be constructed from $\ket{B_n}$s of Eq.~(\ref{Bcartoon}). The Arovas B state then reads  
\begin{equation}
\ket{B} = \sum_n{(-1)^n \ket{B_n}}
\end{equation}
up to a global normalization factor.

We consider the configuration $\ket{B_n}$ defined in Eq.~(\ref{Bcartoon}) and also shown in Fig.~\ref{fig:ArovasBConfig}a. Since the only nearest neighbor bonds that do not have dimers between them are $\{n - 2, n -1\}$ and $\{n+1,n+2\}$, analogous to Eq.~(\ref{AOnlyP}), we can write
\begin{equation}
    H \ket{B_n} = (P^{(2,1)}_{n-2,n-1} + P^{(2,1)}_{n+1,n+2})\ket{B_n}.
\label{BOnlyP}
\end{equation}
The scattering terms of obtained upon action of $P^{(2,1)}_{n-2,n-1}$, by using rules of Eq.~(\ref{S1singletrules}) are
\begin{eqnarray}
    P^{(2,1)}_{n-2,n-1}\ket{B_n} &=& \ket{B_n} + \frac{1}{4} (-\ket{G} + \ket{A_{n-1}} \nonumber \\
    && + \ket{C_{n-1}} + \ket{D_{n-1}}) \nonumber \\
    &&+ \frac{1}{24}(-\ket{A_{n-2}} + \ket{E_{n-1}}) 
\label{Bnscattering}
\end{eqnarray}
where $\ket{A_n}$, $\ket{C_n}$, $\ket{D_n}$, $\ket{E_n}$ are defined according to the cartoon pictures in Figs.~\ref{fig:ArovasBConfig}b, \ref{fig:ArovasBConfig}c, \ref{fig:ArovasBConfig}d and \ref{fig:ArovasBConfig}e. A few typographical errors of Eq.~(8) in Ref.~[\onlinecite{arovas1989two}] including the omission of the scattering term $\ket{E_{n-1}}$ have been corrected here in Eq.~(\ref{BnHamilaction}). Since the scattering configurations are all symmetric terms that are symmetric under inversion about a bond $\{n, n+1\}$, whereas $\ket{B_n}$ is symmetric under inversion about site $n$, the action of $P^{(2,1)}_{n+1,n+2}$ changes $\ket{A_{n-1}} \rightarrow \ket{A_n}$, $\ket{C_{n-1}} \rightarrow \ket{C_n}$, $\ket{D_{n-1}} \rightarrow \ket{D_n}$ and $\ket{E_{n-1}} \rightarrow \ket{E_n}$. The action of the Hamiltonian on $\ket{B_n}$ thus reads
\begin{eqnarray}
    H \ket{B_n} &=& 2 \ket{B_n} - \frac{1}{2}\ket{G} + \frac{1}{4}(\ket{A_{n-1}} + \ket{A_n}) \nonumber \\
    &&+ \frac{1}{4}(\ket{C_{n-1}} + \ket{C_n}) + \frac{1}{4}(\ket{D_{n-1}} + \ket{D_n}) \label{BnHamilaction} \\
    &&-\frac{1}{24}(\ket{A_{n-2}} + \ket{A_{n+1}}) + \frac{1}{24}(\ket{E_{n-1}} + \ket{E_n}).\nn
\end{eqnarray}
With this property, for $L$ even with periodic boundary conditions, we obtain 
\begin{equation}
    H \ket{B} = 2 \ket{B}.
\label{Arovas2}
\end{equation}
Thus, we have an exact state with $\boldsymbol{E = 2}$, $\boldsymbol{k = \pi}$ and $\boldsymbol{s = 0}$ (and $P_z = +1$, $I = -1$). Again the key ingredient for this derivation was the fact that the scattering terms were symmetric and have the opposite (site/bond) symmetry. As for $\ket{A}$, it is not hard to see that this result would not hold for open boundary conditions or for odd $L$. Moreover, $\ket{B}$ appears only for $L \geq 8$: we need $L \geq 5$ to define $\ket{B_n}$ and for $L = 6$ with periodic boundary conditions, the state itself vanishes.
To formulate $\ket{B}$ within the SMA, we need to note that we obtain $\ket{B_n}$ by acting the $c^\dagger c$ term of the projector in Eq.~(\ref{P2dimer}) on $\ket{A_n}$. With this observation, along with identities of Eq.~(\ref{singletidentities}), we obtain the normalized eigenstate\cite{arovas1989two}
\begin{eqnarray}
    &\widetilde{\ket{B}} = \sqrt{\frac{27}{22L}}\sum_{n = 1}^L{(-1)^n \left(\vec{S}_{n}\cdot\vec{S}_{n+1}\right.} \nn \\
    &\left.- \frac{1}{2} (\vec{S}_{n-1}\cdot\vec{S}_{n})(\vec{S}_n\cdot\vec{S}_{n+1})\right)\widetilde{\ket{G}}
\label{Arovas2Spinorig}
\end{eqnarray} 
Eq.~(\ref{Arovas2Spinorig}) can also be written as a Hermitian operator on the ground state:
\begin{eqnarray}
    &\widetilde{\ket{B}} = \frac{1}{4}\sqrt{\frac{27}{22L}}\sum_{n = 1}^L{(-1)^n \{\vec{S}_{n-1}\cdot\vec{S}_n, \vec{S}_n\cdot\vec{S}_{n+1}\}}\widetilde{\ket{G}}\nn \\[0.1mm]
\label{Arovas2Spin}
\end{eqnarray}
where $\{,\}$ denotes the anti-commutator.

One might wonder if the pattern of exactness might continue for other dimer configurations such as $\ket{C_n}$, $\ket{D_n}$ or a combination of the two. However, in such cases, the scattering terms are no longer inversion symmetric, and hence do not appear in pairs, precluding cancellations at $k = \pi$. For the Arovas states considered here, the scattering terms appeared in pairs due to the presence of only two non-vanishing projectors on the state, forcing the exact state to have a momentum $k = \pi$. 
%

\section{Mid-Spectrum Exact States}\label{sec:ExactMiddle}
We now move on to the study of non-singlet states, i.e. states with a total spin $s \neq 0$. Since the AKLT Hamiltonian Eq.~(\ref{S1Hamiltonian}) is $SU(2)$ symmetric, it is sufficient to consider the highest weight state of each multiplet of spin $s$. The entire multiplet of $2s + 1$ states can be obtained by repeated application of $S^- = \sum_{n =1}^L{S_n^-}$ on the highest weight state.  
\subsection{{\bf Spin-2 magnon state}}\label{sec:spin2magnon}
\begin{figure}[t!]
    \centering
        \setlength{\unitlength}{1pt}
        \begin{picture}(150,20)(0,0)
        \linethickness{0.8pt}
        \multiput(0,10)(25,0){7}{\circle{4}} 
        \multiput(6,10)(25,0){7}{\circle{4}} 
        \multiput(3,10)(25,0){7}{\circle{14}}
        \thicklines
        \multiput(-17,10)(25,0){3}{\line(1,0){15}} 
        \multiput(-5,10)(25,0){3}{\vector(1,0){0}}
        \multiput(108,10)(25,0){3}{\line(1,0){15}}
        \multiput(120,10)(25,0){3}{\vector(1,0){0}}
        \multiput(56,5)(25,0){2}{\vector(0,1){12}} 
        \multiput(75,5)(25,0){2}{\vector(0,1){12}} 
        \put(75,-5){n}
        \put(47,-5){n-1}
        \put(95,-5){n+1}
        \put(-32, 7.5){(a)}
        \end{picture}\\[5mm]
    
        \setlength{\unitlength}{1pt}
        \begin{picture}(150,20)(0,0)
        \linethickness{0.8pt}
        \multiput(0,10)(25,0){7}{\circle{4}} 
        \multiput(6,10)(25,0){7}{\circle{4}} 
        \multiput(3,10)(25,0){7}{\circle{14}}
        \thicklines
        \multiput(-17,10)(25,0){3}{\line(1,0){15}} 
        \multiput(-5,10)(25,0){3}{\vector(1,0){0}}
        \multiput(133,10)(25,0){2}{\line(1,0){15}} 
        \multiput(145,10)(25,0){2}{\vector(1,0){0}}
        \put(83,10){\line(1,0){15}}
        \put(95,10){\vector(1,0){0}}
        \multiput(56,5)(50,0){2}{\vector(0,1){12}} 
        \multiput(75,5)(50,0){2}{\vector(0,1){12}} 
        \put(75,-5){n}
        \put(47,-5){n-1}
        \put(95,-5){n+1}
        \put(-32,7.5){(b)}
    \end{picture}\\[5mm]

    \caption{The two configurations that appear in the derivation of the spin-2 magnon state (a) The spin-2 magnon state $\ket{M_n}$. (b) Scattering state $\ket{N_n}$.}
    \label{fig:S2Magnon}
\end{figure}
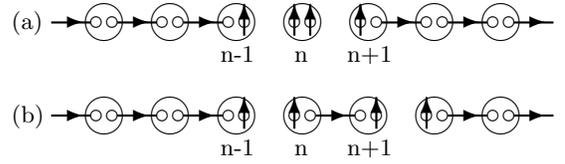
We start with a configuration with the cartoon picture as shown in Fig.~\ref{fig:S2Magnon}a. This particular configuration is used because of the rule Eq.~(\ref{singlethopping}) of Fig.~\ref{diagrep}, derived in Appendix~\ref{scattering}. It shows a fairly simply and symmetric scattering process. We define $\ket{M_n}$ as a quasiparticle with no dimers around site $n$ as
\begin{equation}
    \ket{M_n} = \prod_{j = 1}^{n-2}c_{j,j+1}^\dagger a_{n-1}^\dagger (a_n^\dagger)^2 a_{n+1}^\dagger \prod_{j = n+1}^{L} c_{j,j+1}^\dagger \vac.
\label{npnmdimer}
\end{equation}
From the cartoon picture Fig.~\ref{fig:S2Magnon}a, it is clear that the only projectors in the Hamiltonian that do not vanish on $\ket{M_n}$ are $P^{(2,1)}_{n-1, n}$ and $P^{(2,1)}_{n,n+1}$.
Using Eq.~(\ref{singlethopping}),
\begin{equation}
    P^{(2,1)}_{n-1, n} \ket{M_n} = \ket{M_n} + \frac{1}{2}\ket{N_{n-1}}.
\label{npnmscattering}
\end{equation}
where $\ket{N_n}$ is shown in Fig.~\ref{fig:S2Magnon}b.
Since $\ket{N_n}$ is bond inversion symmetric under  bond $\{n, n+1\}$ whereas $\ket{M_n}$ is site inversion symmetric about site $n$ (they have opposite types of symmetries), the action of the Hamiltonian on $\ket{M_n}$ reads
\begin{equation}
    H \ket{M_n} = 2 \ket{M_n} + \frac{1}{2}(\ket{N_{n-1}} + \ket{N_n})
\label{npnmhamilaction}
\end{equation}
and one can use the $k = \pi$ superposition to remove the $\ket{N_n}$ states. The translation invariant state is thus 
\begin{equation}
\ket{S_2} = \sum_{n = 1}^L{(-1)^n \ket{M_n}}.
\end{equation}
With $L$ even and periodic boundary conditions,
\begin{equation}
    H \ket{S_2} = 2 \ket{S_2}.
\label{HOnFirstess}
\end{equation}
Thus we have an exact multiplet of states with $\boldsymbol{s = 2}$ (4 $a_i^\dagger$s in the state), $\boldsymbol{E = 2}$, $\boldsymbol{k = \pi}$.
This state can again be written as an SMA with a spin 2 magnon. From Fig.~\ref{fig:S2Magnon}a, we immediately see that $\ket{M_n} = -({S_n^+}^2/2)\ket{G}$, that is, the spin on site $n$ is forced to have $S_z = 1$. Thus, including the normalization factor, the full state can be written as
\begin{equation}
    \widetilde{\ket{S_2}} = \sqrt{\frac{3}{4L}}\sum_{n = 1}^L{(-1)^n{S_n^+}^2}\widetilde{\ket{G}}.
\label{firstess}
\end{equation}
This state exists for all $L$ even and $L \geq 4$. In terms of momentum space operators, the state has the expression
\begin{equation}
    \widetilde{\ket{S_2}} = \sqrt{\frac{3}{4L}}\sum_k{S_k^+ S_{\pi-k}^+}\widetilde{\ket{G}}.
\label{firstesskspace}
\end{equation}
The entire multiplet of states with different $S_z$ can be obtained from $\ket{S_2}$ by applying the $S^-$ operator. 
An exact spin-2 magnon state can be constructed similarly for the AKLT Hamiltonian with {\bf open boundary conditions} too, see Appendix~\ref{sec:spin2magnonobc}.

\subsection{{\bf Tower of states}}\label{sec:towerofstates}
\begin{figure}[t!]
    \centering
    \begin{subfigure}{}
        \setlength{\unitlength}{1pt}
        \begin{picture}(150,20)(0,0)
        \linethickness{0.8pt}
        \multiput(-25,10)(25,0){9}{\circle{4}} 
        \multiput(-19,10)(25,0){9}{\circle{4}} 
        \multiput(-22,10)(25,0){9}{\circle{14}}
        \thicklines
        \multiput(-42,10)(25,0){2}{\line(1,0){15}} 
        \multiput(-30,10)(25,0){2}{\vector(1,0){0}}
        
        \multiput(58,10)(25,0){2}{\line(1,0){15}} 
        \multiput(70,10)(25,0){2}{\vector(1,0){0}}
        
        \multiput(106,5)(25,0){2}{\vector(0,1){12}} 
        \multiput(125,5)(25,0){2}{\vector(0,1){12}}
        
        \multiput(158,10)(25,0){2}{\line(1,0){15}}
        \multiput(170,10)(25,0){2}{\vector(1,0){0}}
        \multiput(6,5)(25,0){2}{\vector(0,1){12}} 
        \multiput(25,5)(25,0){2}{\vector(0,1){12}} 
        \put(25,-5){n}
        \put(-3,-5){n-1}
        \put(45,-5){n+1}
        \put(122,-5){n+4}
        \put(97,-5){n+3}
        \put(145,-5){n+5}
        \put(-57, 7.5){(a)}
        \end{picture}
    \end{subfigure}
    \begin{subfigure}{}
        \setlength{\unitlength}{1pt}
        \begin{picture}(150,30)(0,0)
        \linethickness{0.8pt}
        \multiput(-25,10)(25,0){9}{\circle{4}} 
        \multiput(-19,10)(25,0){9}{\circle{4}} 
        \multiput(-22,10)(25,0){9}{\circle{14}}
        \thicklines
        \multiput(-42,10)(25,0){2}{\line(1,0){15}} 
        \multiput(-30,10)(25,0){2}{\vector(1,0){0}}
        
        \multiput(83,10)(25,0){1}{\line(1,0){15}} 
        \multiput(95,10)(25,0){1}{\vector(1,0){0}}
        
        \multiput(106,5)(25,0){2}{\vector(0,1){12}} 
        \multiput(125,5)(25,0){2}{\vector(0,1){12}}
        
        \multiput(158,10)(25,0){2}{\line(1,0){15}}
        \multiput(170,10)(25,0){2}{\vector(1,0){0}}
        
        \multiput(6,5)(50,0){2}{\vector(0,1){12}} 
        \multiput(25,5)(50,0){2}{\vector(0,1){12}}
        
        \put(33,10){\line(1,0){15}}
        \put(45,10){\vector(1,0){0}}
        
        \put(25,-5){n}
        \put(-3,-5){n-1}
        \put(45,-5){n+1}
        \put(122,-5){n+4}
        \put(97,-5){n+3}
        \put(145,-5){n+5}
        \put(-57, 7.5){(b)}
    \end{picture}
    \end{subfigure}
    \begin{subfigure}{}
        \begin{picture}(150,30)(0,0)
        \linethickness{0.8pt}
        \multiput(-25,10)(25,0){9}{\circle{4}} 
        \multiput(-19,10)(25,0){9}{\circle{4}} 
        \multiput(-22,10)(25,0){9}{\circle{14}}
        \thicklines
        \multiput(-42,10)(25,0){2}{\line(1,0){15}} 
        \multiput(-30,10)(25,0){2}{\vector(1,0){0}}
        
        \multiput(108,10)(25,0){2}{\line(1,0){15}} 
        \multiput(120,10)(25,0){2}{\vector(1,0){0}}
        
        \multiput(56,5)(25,0){2}{\vector(0,1){12}} 
        \multiput(75,5)(25,0){2}{\vector(0,1){12}}
        
        \multiput(158,10)(25,0){2}{\line(1,0){15}}
        \multiput(170,10)(25,0){2}{\vector(1,0){0}}
        \multiput(6,5)(25,0){2}{\vector(0,1){12}} 
        \multiput(25,5)(25,0){2}{\vector(0,1){12}} 
        \put(25,-5){n}
        \put(-3,-5){n-1}
        \put(45,-5){n+1}
        \put(97,-5){n+3}
        \put(72,-5){n+2}
        \put(-57, 7.5){(c)}
        \end{picture}
    
        \begin{picture}(150,30)(0,0)
        \linethickness{0.8pt}
        \multiput(-25,10)(25,0){9}{\circle{4}} 
        \multiput(-19,10)(25,0){9}{\circle{4}} 
        \multiput(-22,10)(25,0){9}{\circle{14}}
        \thicklines
        \multiput(-42,10)(25,0){2}{\line(1,0){15}} 
        \multiput(-30,10)(25,0){2}{\vector(1,0){0}}
        \multiput(83,10)(50,0){2}{\line(1,0){15}} 
        \multiput(95,10)(50,0){2}{\vector(1,0){0}}
        \multiput(56,5)(50,0){2}{\vector(0,1){12}} 
        \multiput(75,5)(50,0){2}{\vector(0,1){12}}
        \multiput(158,10)(25,0){2}{\line(1,0){15}}
        \multiput(170,10)(25,0){2}{\vector(1,0){0}}
        \multiput(6,5)(25,0){2}{\vector(0,1){12}} 
        \multiput(25,5)(25,0){2}{\vector(0,1){12}} 
        \put(25,-5){n}
        \put(-3,-5){n-1}
        \put(45,-5){n+1}
        \put(97,-5){n+3}
        \put(72,-5){n+2}
        \put(-57, 7.5){(d)}
        \end{picture}
    
    \end{subfigure}
    \caption{(a) Two magnon configuraton  $\ket{M_n, M_{n+4}}$. (b) Scattering state $\ket{N_n,M_{n+4}}$. (c) Two magnon configuration $\ket{M_n, M_{n+2}}$. (d) Two magnon scattering configuration $\ket{M_n, N_{n+2}}$}
    \label{fig:towerofstates}
\end{figure}
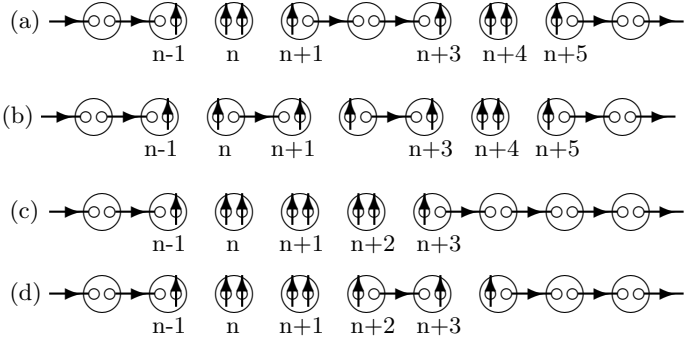

We now denote the spin-2 magnon ``creation" operator as $\mcP = \sum_{n = 1}^L{(-1)^n {S_n^+}^2}$. The state $\mcP^2 \ket{G}$ is a state with two of the $k = \pi$ spin-2 magnons dispersing on the chain. The correct basis state is proportional to ${S_n^+}^2 {S^+}^2_{n+m} \ket{G}$, with $n + m$ defined modulo $L$, containing two magnons (${S_n^+}^4\ket{G}$ = 0 for $m = 0$). It is convenient to denote this basis state $\ket{M_n,M_{n+m}}$, $n$ and $n+m$ denoting the position of the magnons. Similarly, basis states $\ket{M_n, N_{n+m}}$ can be defined as the configuration with the spin-2 magnon at position $m$ and the scattering magnon of Fig.~\ref{fig:S2Magnon}b at position $n+m$. Since ${S_n^+}^2\ket{G}$ annihilates all spin configurations of $\ket{G}$ that do not have $S_z = -1$ on site $n$, it follows that $S_z \neq -1$ on site $n + 1$ because the sites $n$ and $n + 1$ share a singlet in $\ket{G}$. Thus, $(S_n^+)^2 (S_{n+1}^+)^2 \ket{G} = 0$. For other $m$, there are two possibilities. If $3 \leq m \leq L - 3$, we can write down the action of the Hamiltonian on the two magnon basis states using Eq.~(\ref{npnmhamilaction}) as
\begin{equation}
\begin{split}
    &H \ket{M_n, M_{n+m}} = 4 \ket{M_n, M_{n+m}} + \frac{1}{2}\left(\ket{N_{n-1}, M_{n+m}}  \right.\\
     & \qquad \left. + \ket{N_n, M_{n+m}} +\ket{M_n, N_{n+m-1}} + \ket{M_n, N_{n+m}}\right).
\label{S4hamilaction}\end{split}
\end{equation}
An example of such a state and its scattering configuration are shown in Fig.~\ref{fig:towerofstates}a and Fig.~\ref{fig:towerofstates}b respectively.
If $m = 2$ or $m = L - 2$, the two magnons fuse to form one spin-4 magnon, as shown in Fig.~\ref{fig:towerofstates}c. Using Eq.~(\ref{singlethopping}), the action of the Hamiltonian can be written as 
\begin{equation}\begin{split}
    &H \ket{M_n, M_{n+2}} = 4 \ket{M_n, M_{n+2}}  \\ 
    &\qquad+ \frac{1}{2}(\ket{N_{n-1},M_{n+2}} + \ket{M_n, N_{n+2}}).
\label{S4fathamilaction}\end{split}
\end{equation}
In terms of these basis states, the translation invariant state comprising of two spin-2 magnons on the AKLT chain is
\begin{equation}
    \ket{S_4} = \sum_{n = 1}^L{\sum_{m = 2}^{L-2}{{(-1)^{m} \ket{M_n, M_{n+m}}}}}.
\label{S4ket}
\end{equation}
Using Eqs.~(\ref{S4hamilaction}) and (\ref{S4fathamilaction}) the action of the Hamiltonian on $\ket{S_4}$ can be written as
\begin{equation}
    H \ket{S_4} = 4\ket{S_4} + \frac{1}{2}(\ket{C_s} + \ket{C_t}).
\end{equation}
where we group the scattering terms into two: $\ket{C_s}$ arising from basis states in Eq.~(\ref{S4ket}) of the type Fig.~\ref{fig:towerofstates}a, (i.e. magnons are separated, $3 \leq m \leq L - 3$) and $\ket{C_t}$ arising from basis states of the type Fig.~\ref{fig:towerofstates}c (i.e. magnons are next to each other, $m = 2,\, L -2$). Using Eqs.~(\ref{S4hamilaction}), (\ref{S4fathamilaction}) and (\ref{S4ket}), $\ket{C_s}$ and $\ket{C_t}$ can be simplified to 
\begin{eqnarray}
    \ket{C_s} &=& \sum_{n=1}^L{\sum_{m = 3}^{L-3}{(-1)^m (\ket{N_{n-1}, M_{n+m}} + \ket{N_n, M_{n+m}}}} \nonumber \\ && \qquad + \ket{M_n, N_{n+m-1}} + \ket{M_n, N_{n+m}}) \label{S4scatteringterms} \\[5pt]
    \ket{C_t} &=& \sum_{n=1}^L{(\ket{N_n, M_{n+L-2}} + \ket{N_{n}, M_{n+3}}} \nonumber \\
    && \qquad + \ket{M_n, N_{n+2}} + \ket{M_n, N_{n+L-3})}. \label{S4scatteringterms2}
\end{eqnarray}
In Eq.~(\ref{S4scatteringterms}), considering the summation over just the first scattering term and relabelling the summation indices considering $L$ even, we obtain
\begin{eqnarray}
    &\sum_{n=1}^L{\sum_{m = 3}^{L-3}{(-1)^m \ket{N_{n-1}, M_{n+m}}}}  \nonumber \\
      =& -\sum_{n=1}^L{\sum_{m=4}^{L-2}{(-1)^m \ket{N_n, M_{n+m}}}}. 
\end{eqnarray}
Similarly for the third scattering term in Eq.~(\ref{S4scatteringterms}), we obtain
\begin{eqnarray}
    &\sum_{n=1}^L{\sum_{m = 3}^{L-3}{(-1)^m \ket{M_n, N_{n+m-1}}}}  \nonumber \\
    =& -\sum_{n=1}^L{\sum_{m=2}^{L-4}{(-1)^m \ket{M_n, N_{n+m}}}}. 
\end{eqnarray}
Adding all the terms in Eq.~(\ref{S4scatteringterms}) back, we obtain
\begin{eqnarray}
    \ket{C_s} &=& -\sum_{n=1}^L{(\ket{N_n, M_{n+L-2}} + \ket{N_n, M_{n+3}}} \nonumber \\
    &&+ \ket{M_n, N_{n+2}} + \ket{M_n, N_{n+L-3}}).
\label{S4scatteringtermssimplify}
\end{eqnarray}
Using Eqs.~(\ref{S4scatteringtermssimplify}) and (\ref{S4scatteringterms2}), $\ket{C_s} + \ket{C_t} = 0$. Thus,
\begin{equation}
    H\ket{S_4} = 4 \ket{S_4}.
\end{equation}
This is an exact state with $\boldsymbol{s = 4}$, $\boldsymbol{k = 0}$ and $\boldsymbol{E = 4}$. As with all the states we have presented, the cancellation here works only if $L$ is even.

\begin{figure}
        \setlength{\unitlength}{1pt}
        \begin{picture}(150,30)(0,0)
        \linethickness{0.8pt}
        \multiput(-25,10)(25,0){9}{\circle{4}} 
        \multiput(-19,10)(25,0){9}{\circle{4}} 
        \multiput(-22,10)(25,0){9}{\circle{14}}
        \thicklines
        \multiput(-42,10)(25,0){2}{\line(1,0){15}} 
        \multiput(-30,10)(25,0){2}{\vector(1,0){0}}
        
        \multiput(108,10)(25,0){2}{\line(1,0){15}} 
        \multiput(120,10)(25,0){2}{\vector(1,0){0}}
        
        \multiput(56,5)(25,0){2}{\vector(0,1){12}} 
        \multiput(75,5)(25,0){2}{\vector(0,1){12}}
        
        \multiput(158,10)(25,0){2}{\line(1,0){15}}
        \multiput(170,10)(25,0){2}{\vector(1,0){0}}
        \multiput(6,5)(25,0){2}{\vector(0,1){12}} 
        \multiput(25,5)(25,0){2}{\vector(0,1){12}} 
        \put(25,-5){n}
        \put(-3,-5){n-1}
        \put(45,-5){n+1}
        \put(97,-5){n+3}
        \put(72,-5){n+2}
        \put(-57, 7.5){(a)}
        \end{picture}\\[5mm]
        
        \begin{picture}(150,20)(0,0)
        \linethickness{0.8pt}
        \multiput(-25,10)(25,0){9}{\circle{4}} 
        \multiput(-19,10)(25,0){9}{\circle{4}} 
        \multiput(-22,10)(25,0){9}{\circle{14}}
        \thicklines
        \multiput(-42,10)(25,0){2}{\line(1,0){15}} 
        \multiput(-30,10)(25,0){2}{\vector(1,0){0}}
        
        \multiput(58,10)(25,0){1}{\line(1,0){15}} 
        \multiput(70,10)(25,0){1}{\vector(1,0){0}}
        
        \multiput(81,5)(25,0){2}{\vector(0,1){12}} 
        \multiput(100,5)(25,0){2}{\vector(0,1){12}}
        
        \multiput(133,10)(25,0){3}{\line(1,0){15}}
        \multiput(145,10)(25,0){3}{\vector(1,0){0}}
        \multiput(6,5)(25,0){2}{\vector(0,1){12}} 
        \multiput(25,5)(25,0){2}{\vector(0,1){12}} 
        \put(25,-5){n}
        \put(-3,-5){n-1}
        \put(45,-5){n+1}
        \put(122,-5){n+4}
        \put(97,-5){n+3}
        \put(70,-5){n+2}
        \put(-57, 7.5){(b)}
        \end{picture}\\[1mm]
    
        \begin{picture}(150,30)(0,0)
        \linethickness{0.8pt}
        \multiput(-25,10)(25,0){9}{\circle{4}} 
        \multiput(-19,10)(25,0){9}{\circle{4}} 
        \multiput(-22,10)(25,0){9}{\circle{14}}
        \thicklines
        \multiput(-42,10)(25,0){2}{\line(1,0){15}} 
        \multiput(-30,10)(25,0){2}{\vector(1,0){0}}
        \multiput(83,10)(50,0){2}{\line(1,0){15}} 
        \multiput(95,10)(50,0){2}{\vector(1,0){0}}
        \multiput(56,5)(50,0){2}{\vector(0,1){12}} 
        \multiput(75,5)(50,0){2}{\vector(0,1){12}}
        \multiput(158,10)(25,0){2}{\line(1,0){15}}
        \multiput(170,10)(25,0){2}{\vector(1,0){0}}
        \multiput(6,5)(25,0){2}{\vector(0,1){12}} 
        \multiput(25,5)(25,0){2}{\vector(0,1){12}} 
        \put(25,-5){n}
        \put(-3,-5){n-1}
        \put(45,-5){n+1}
        \put(97,-5){n+3}
        \put(72,-5){n+2}
        \put(-57, 7.5){(c)}
        \end{picture}
            
\caption{(a) A configuration with magnons on sites $n$ and $n + 2$ (b) A configuration with magnons on sites $n$ and $n+3$ (c) Common scattering configuration} 
\label{fig:towercancellation}
\end{figure}
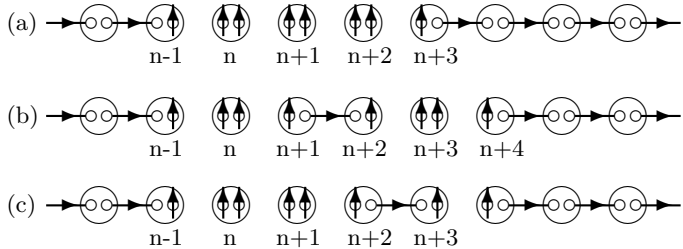

This construction can be easily generalized by noting that the spin-2 magnons on the spin chain behave as solitons.
A state with $N$ $k = \pi$ spin-2 magnons reads
\begin{equation}
    \ket{S_{2N}} = \sum_{\{l_j\}}{(-1)^{\sum_{j=1}^N{l_j}}\ket{M_{l_1}, M_{l_2} \dots M_{l_N}}}.
    \label{nmagnonstate} 
\end{equation}
As we have seen earlier, ${{S^+}^2_n}{{S^+}^2_{n+1}}\ket{G} = 0$. Hence, in Eq.~(\ref{nmagnonstate}), all configurations $\ket{\dots, M_n, M_{n + 1}, \dots}$ vanish. Thus the set $\{l_j\}$ satisfies the constraints $1 \leq j \leq N$, $l_{j+1} > l_j + 1$, $1 \leq l_j \leq L$ where addition is defined modulo $L$.
Upon the action of the Hamiltonian on Eq.~(\ref{nmagnonstate}), a term $\ket{M_{l_1}, \dots, M_{l_k} \dots, M_{l_N}}$ scatters to $\ket{M_{l_1}, \dots,N_{p_k},\dots,M_{l_N}}$ where $p_k = l_k - 1$ or $p_k = l_k$. 
From Eqs.~(\ref{npnmhamilaction}), (\ref{S4hamilaction}) and (\ref{S4fathamilaction}), observe that for each such scattering term, there always is a unique different term in Eq.~(\ref{nmagnonstate}) $\ket{M_{l_1}, \dots, M_{q_k} \dots, M_{l_N}}$ where $q_k = l_k-1$ or $q_k = l_k + 1$  that scatters to the same term. 
For example, a state with spin-2 magnons on several sites, including $n$ and $n + 2$ (Fig.~\ref{fig:towercancellation}a) and another state with spin-2 magnons on the same set of sites, except for $n + 2$ replaced by $n + 3$ (Fig.~\ref{fig:towercancellation}b), share a scattering term (Fig.~\ref{fig:towercancellation}c). 
However, for the scattering terms to cancel, the terms $\ket{M_{l_1}, \dots, M_{q_k}, \dots, M_{l_N}}$ and $\ket{M_{l_1}, \dots,M_{l_k},\dots,M_{l_N}}$ should have the opposite sign in Eq.~(\ref{nmagnonstate}). This is true for the case when $l_k = l_1 = 1$ and $q_k = l_N = L$ only if $L$ is even.
Thus, all the scattering terms arising from Eq.~(\ref{nmagnonstate}) cancel out and we have an exact state. $\ket{S_{2N}}$ can also be written as $\ket{S_{2N}} = \mcP^N \ket{G}$ and has a momentum $k = 0$ or $\pi$ depending on whether $N$ is even or odd. Its total spin is $\boldsymbol{s = 2N}$, and its energy is $\boldsymbol{E = 2N}$.
Since each spin-2 magnon annihilates two dimers from the ground state and $L$ is even, $\ket{S_{L}}$ is a state with $E = L$, $s = L$, the highest excited state of the model. 
Thus, $\{\ket{S_{2N}}\}$ is a tower of exact states from the ground state to the highest excited state. 
In terms of the spin operators, the highest weight states of this tower can be written as
\begin{equation}
    \widetilde{\ket{S_{2N}}} = \mathcal{N} \sum_{\{l_j\}}{(-1)^{\sum_{j=1}^N{l_j}}\prod_{j=1}^N{(S_{l_j}^+)^2}} \widetilde{\ket{G}}
    \label{evenspinsequence}
\end{equation}
where $\mathcal{N}$ is a normalization factor. Similar to the spin-2 magnon, the entire tower of states can be extended to the AKLT chain with {\bf open boundary conditions}, see Appendix~\ref{sec:obctowerofstates}.

\subsection{Position in the energy spectrum}\label{sec:position}
In this section we study the positions of the tower of states $\{\ket{S_{2N}}\}$ in the energy spectrum.
It is believed\cite{deutsch1991quantum, srednicki1994chaos, kim2014testing} that {\it all} energy eigenstates of non-integrable models that lie in a region of finite density of states satisfy ETH. This is commonly known {\it strong} ETH.\cite{kim2014testing, garrison2015does}
However, the local density of states can always be changed by tuning the Hamiltonian to allow level crossings from states with different quantum numbers.
To avoid this possibility, we study the position of the tower of states in the density of states of their own quantum number sector defined by $(s, S_z, k, I, P_z)$.

\begin{figure}[t!]
\hspace*{-0.2cm}\includegraphics[scale=0.45]{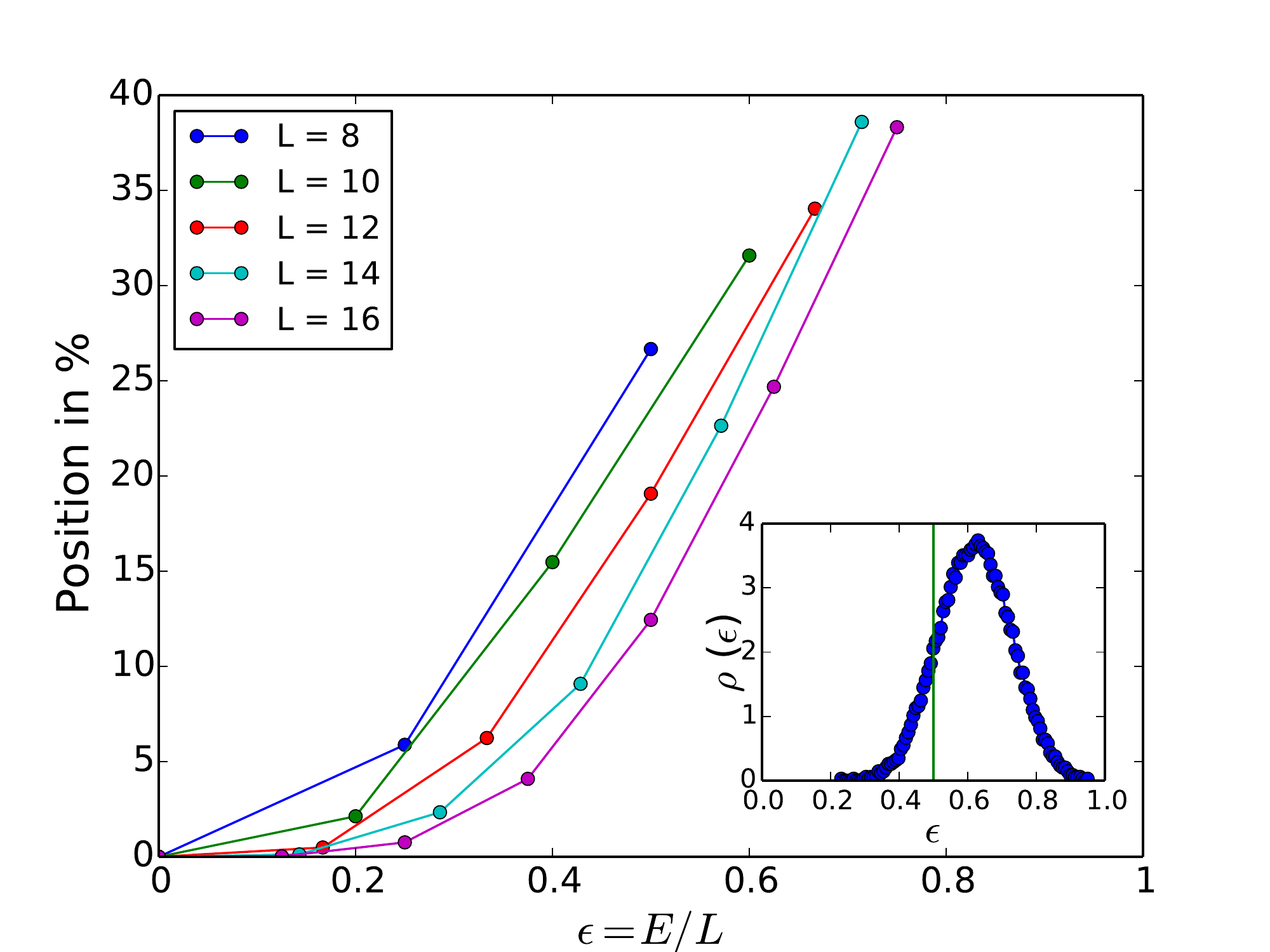}
\caption{(Color online) Positions of the $S_z = 0$ states of the tower of states within the energy spectrum of their own quantum number sector plotted against the energy density $\epsilon = E/L$ for $L = 8, 10, 12, 14,16$ with periodic boundary conditions. Each dot corresponds to a state with $N$ spin-2 magnons with $N = 0, 1, \dots, L/2 - 2$. The inset shows the density of states for $L = 16$ in the quantum number sector $(s, S_z, k, I, P_z) = (8, 0, 0, -1, 1)$. The vertical green line in the inset indicates the position of $\ket{S_8}$.}
\label{fig:tower}
\end{figure}

The inset of Fig.~\ref{fig:tower}, we show a typical example of the density of states for the spin-1 AKLT chain for periodic boundary conditions. We focus on system size $L = 16$ and the sector defined by the set of quantum numbers $(s, S_z, k, I, P_z) = (8, 0, 0, -1, 1)$. The quantum numbers are defined in Table~\ref{fig:tableofstates}. In this sector lies, for example, the state of the tower with $N = L/4$ magnons, i.e. $\ket{S_{2N}} = \ket{S_8}$. As can be observed, this state is located in a region with finite density of states.
Since any given state of the tower of states has a fixed energy as the thermodynamic limit ($L \rightarrow \infty$) is taken, it is natural to expect that such a state would eventually lie in a region of zero density of states.
However, for a fixed $L$, we have a tower of states of $E = 2N, s = 2N$, $N \in [0, L/2]$, and hence the number of states and their energies increase as the system size increases. One could look at the states at a finite energy density $\epsilon = E/L$ and then take the thermodynamic limit ($E \rightarrow \infty$, $L \rightarrow \infty$).
In this limit, we conjecture that some states of the tower lie in the bulk of the energy spectrum in the thermodynamic limit.

\section{Exact High Energy States}\label{sec:ExactHighEnergy}
For completeness, we now focus on the exact states in the upper part of the energy spectrum shown in Table~\ref{fig:tableofstates}. Not all the states presented here are specific to the AKLT model. Some are eigenstates merely as a consequence of having $SU(2)$ symmetry and translation invariance. Moreover, it has been shown that all the states with quantum numbers $S_z = L - 1$ and $S_z = L - 2$ (and hence $s = L - 1$ and $s = L - 2$) can be analytically obtained for a general $SU(2)$ symmetric spin-1 Hamiltonian in the thermodynamic limit\cite{bibikov2016three} and for any even system size,\cite{kiwata1994bethe} in spite of the fact that a general $SU(2)$ symmetric spin-1 Hamiltonian is non-integrable. This is due to the fact that the scattering equations of states with $S_z = L - 1$ and $S_z = L - 2$ correspond to one and two-body scattering problems respectively, which are integrable. Indeed, states in these quantum number sectors do not exhibit level repulsion. However, for $S_z = L - 2$, in most cases it is impossible to obtain a closed-form expression of the eigenstates for a finite system size. The states presented in this section are thus examples of high energy eigenstates with $s = L - 1$ and $s = L - 2$ that have a simple analytical expression for any finite system in an otherwise completely solvable quantum number sector. The scattering equation of states with $S_z = L - 3$ similarly corresponds to a three-magnon scattering problem, that has been solved partially.\cite{bibikov2016three} However, we have not found any exact states with $s = L - 3$.   
\subsection{{\bf Ferromagnetic and \boldsymbol{$1_k$} states}}
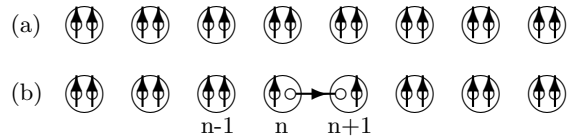
\begin{figure}[t!]
\begin{picture}(180,10)(0,0)
\linethickness{0.8pt}
\multiput(0,10)(25,0){8}{\circle{4}} 
\multiput(6,10)(25,0){8}{\circle{4}} 
\multiput(3,10)(25,0){8}{\circle{14}}
\thicklines
\multiput(0,5)(25,0){8}{\vector(0,1){12}} 
\multiput(6,5)(25,0){8}{\vector(0,1){12}} 
\put(-25,7.5){(a)}
\end{picture}

\begin{picture}(180,25)(0,0)
\linethickness{0.8pt}
\multiput(0,10)(25,0){8}{\circle{4}} 
\multiput(6,10)(25,0){8}{\circle{4}} 
\multiput(3,10)(25,0){8}{\circle{14}}
\thicklines
\multiput(0,5)(25,0){4}{\vector(0,1){12}} 
\multiput(6,5)(25,0){3}{\vector(0,1){12}} 
\put(83,10){\line(1,0){15}}
\put(95,10){\vector(1,0){0}}
\multiput(106,5)(25,0){4}{\vector(0,1){12}}
\multiput(125,5)(25,0){3}{\vector(0,1){12}}
\put(75,-5){n}
\put(47,-5){n-1}
\put(95,-5){n+1}
\put(-25,7.5){(b)}
\end{picture}
\caption{(a) Ferromagnetic State $\ket{F}$ (b) One dimer configuration $\ket{n}$. }
\label{fig:1kstates}
\end{figure}
As we have seen in Sec.~\ref{sec:towerofstates}, the highest excited state of the spin-1 AKLT model with periodic boundary conditions is the ferromagnetic multiplet that has $s = L$ and $E = L$. In the highest weight state, all of the spin-1s are ferromagnetic and are in the $S_z = 1$ state (Fig.~\ref{fig:1kstates}a). Since nearest neighbors already form a spin-2 configuration, each of the projectors in Eq.~(\ref{S1Hamiltonian}) contribute 1 unit to the energy of the state. In terms of Schwinger bosons, the state can be written as $\ket{F} = \ket{S_L} = \prod_{j = 1}^L{{a_i^\dagger}^2}\vac$. Using Eq.~(\ref{spindict}), the normalized state is given by
\begin{equation}
    \widetilde{\ket{F}} = \frac{\ket{F}}{2^{L/2}}.
\end{equation}
As also seen from Eq.~(\ref{allspin}), the total energy $E$ of the state is $L$.

Another trivial series of states that are exact for periodic boundary conditions are those with spin $s = L - 1$. There are $L - 1$ of them and they can be expressed by the action of $S_n^-$. Indeed the normalized states read
\begin{equation}
    \widetilde{\ket{1_k}} = \frac{1}{\sqrt{2L}}\sum_n{e^{ikn} S_n^-}\widetilde{\ket{F}}.
\label{1k}
\end{equation}
with momentum $k = 2\pi l/L$, where $l = 1, 2, \dots, L - 1$ ($l = 0$ belongs to the multiplet of $\ket{F}$). In the dimer basis, $\ket{1_k}$ is the translation invariant superposition of single dimer configurations $\ket{n} \equiv (S^-_{n+1} - S_n^-)\ket{F}$ depicted in Fig.~\ref{fig:1kstates}b.

Using Eq.~(\ref{allspin}), projectors of the Hamiltonian Eq.~(\ref{S1Hamiltonian}) on all the bonds except $\{n-1, n\}$, $\{n,n+1\}$ and $\{n+1,n+2\}$ contribute a total energy of $E = L - 3$ to the state $\ket{n}$. Since the projector on the bond $\{n,n+1\}$ vanishes (Eq.~(\ref{projectorvanishing2}), the only non-trivial projector actions are $P^{(2,1)}_{n-1,n}$ and $P^{(2,1)}_{n+1,n+2}$. Using Eq.~(\ref{singlethopping}) the scattering equation for $\ket{n}$ can be written as
\begin{equation}
     H\ket{n} = (L-1)\ket{n} + \frac{1}{2}(\ket{n-1} + \ket{n+1}).
\label{1singlethamil}
\end{equation}
Thus $\ket{1_k} = \sum_n{e^{i k n}\ket{n}}$ is an exact state with periodic boundary conditions for all $L$
\begin{eqnarray}
    H \ket{1_k} &=& (L-1)\ket{1_k} + \frac{1}{2}\sum_n{e^{i k n}(\ket{n-1} + \ket{n + 1})} \nn \\
    &=& (L - 1 + \cos(k))\ket{1_k}.
\end{eqnarray}
We thus have states for all momenta $\boldsymbol{k \neq 0}$ with $\boldsymbol{s = L - 1}$ and $\boldsymbol{E = L - 1 + \cos(k)}$. Note that these states, while rather trivial, have a very simple structure (observed in the entanglement spectrum) but non-rational energies.
\subsection{{\bf \boldsymbol{$2_n$} states}}\label{sec:2nstates}
\begin{figure}
    \centering
        \setlength{\unitlength}{1pt}
        \begin{picture}(180,10)(0,0)
        \linethickness{0.8pt}
        \multiput(0,10)(25,0){8}{\circle{4}} 
        \multiput(6,10)(25,0){8}{\circle{4}} 
        \multiput(3,10)(25,0){8}{\circle{14}}
        \thicklines
        \multiput(0,5)(25,0){2}{\vector(0,1){12}} 
        \put(6,5){\vector(0,1){12}} 
        \multiput(33,10)(100,0){2}{\line(1,0){15}}
        \multiput(45,10)(100,0){2}{\vector(1,0){0}}
        \multiput(56,5)(25,0){3}{\vector(0,1){12}}
        \multiput(75,5)(25,0){3}{\vector(0,1){12}}
        \multiput(156,5)(25,0){2}{\vector(0,1){12}} 
        \put(175,5){\vector(0,1){12}} 
        \put(25,-5){m}
        \put(45,-5){m+1}
        \put(120,-5){m+4}
        \put(145,-5){m+5}
        \put(-28, 10){(a)}
        \end{picture}\\[7mm]
        
        \begin{picture}(180,10)(0,0)
        \linethickness{0.8pt}
        \multiput(0,10)(25,0){8}{\circle{4}} 
        \multiput(6,10)(25,0){8}{\circle{4}} 
        \multiput(3,10)(25,0){8}{\circle{14}}
        \thicklines
        \multiput(0,5)(25,0){3}{\vector(0,1){12}} 
        \multiput(6,5)(25,0){2}{\vector(0,1){12}} 
        \multiput(58,10)(75,0){2}{\line(1,0){15}}
        \multiput(70,10)(75,0){2}{\vector(1,0){0}}
        \multiput(81,5)(25,0){2}{\vector(0,1){12}}
        \multiput(100,5)(25,0){2}{\vector(0,1){12}}
        \multiput(156,5)(25,0){2}{\vector(0,1){12}} 
        \put(175,5){\vector(0,1){12}} 
        \put(25,-5){m}
        \put(45,-5){m+1}
        \put(120,-5){m+4}
        \put(145,-5){m+5}
        \put(-28, 10){(b)}
        \end{picture}\\[7mm]
        
        \begin{picture}(180,10)(0,0)
        \linethickness{0.8pt}
        \multiput(0,10)(25,0){8}{\circle{4}} 
        \multiput(6,10)(25,0){8}{\circle{4}} 
        \multiput(3,10)(25,0){8}{\circle{14}}
        \thicklines
        \multiput(0,5)(25,0){2}{\vector(0,1){12}} 
        \put(6,5){\vector(0,1){12}} 
        \multiput(33,10)(25,0){2}{\line(1,0){15}}
        \multiput(45,10)(25,0){2}{\vector(1,0){0}}
        \multiput(81,5)(25,0){4}{\vector(0,1){12}}
        \multiput(100,5)(25,0){4}{\vector(0,1){12}}
        \multiput(156,5)(25,0){2}{\vector(0,1){12}} 
        \put(175,5){\vector(0,1){12}} 
        \put(25,-5){m}
        \put(45,-5){m+1}
        \put(70,-5){m+2}
        \put(-28, 10){(c)}
        \end{picture}\\[7mm]
        
        \begin{picture}(180,10)(0,0)
        \linethickness{0.8pt}
        \multiput(0,10)(25,0){8}{\circle{4}} 
        \multiput(6,10)(25,0){8}{\circle{4}} 
        \multiput(3,10)(25,0){8}{\circle{14}}
        \thicklines
        \multiput(0,5)(25,0){2}{\vector(0,1){12}} 
        \put(6,5){\vector(0,1){12}} 
        \multiput(33,10)(50,0){2}{\line(1,0){15}}
        \multiput(45,10)(50,0){2}{\vector(1,0){0}}
        \multiput(56,5)(25,0){1}{\vector(0,1){12}}
        \multiput(75,5)(25,0){1}{\vector(0,1){12}}
        \multiput(106,5)(25,0){4}{\vector(0,1){12}} 
        \multiput(125,5)(25,0){3}{\vector(0,1){12}} 
        \put(25,-5){m}
        \put(45,-5){m+1}
        \put(70,-5){m+2}
        \put(95,-5){m+3}
        \put(-28, 10){(d)}
        \end{picture}\\[7mm]
        
        \begin{picture}(180,10)(0,0)
        \linethickness{0.8pt}
        \multiput(0,10)(25,0){8}{\circle{4}} 
        \multiput(6,10)(25,0){8}{\circle{4}} 
        \multiput(3,10)(25,0){8}{\circle{14}}
        \thicklines
        \multiput(0,5)(25,0){1}{\vector(0,1){12}} 
        \put(6,5){\vector(0,1){12}} 
        \multiput(33,10)(100,0){1}{\line(1,0){15}}
        \multiput(45,10)(100,0){1}{\vector(1,0){0}}
        \multiput(81,5)(25,0){4}{\vector(0,1){12}}
        \multiput(75,5)(25,0){4}{\vector(0,1){12}}
        \multiput(156,5)(25,0){2}{\vector(0,1){12}}
        \qbezier(27,10)(40.5,22)(54,10)
        \put(44,16){\vector(1,0){0}}
        \put(175,5){\vector(0,1){12}} 
        \put(25,-5){m}
        \put(45,-5){m+1}
        \put(-28, 10){(e)}
        \end{picture} 
    \caption{Several configurations relevant for the $\ket{2_n}$ states (a) Two dimer configuration $\ket{m,m+4}$ (b) Scattering term $\ket{m+1,m+4}$ (c) Two dimer configuration $
    \ket{m, m+1}$ (d) Scattering term $\ket{m, m + 2}$. (e) Two dimer configuration $\ket{m, m}$}
    \label{fig:2singletconfig}
\end{figure}
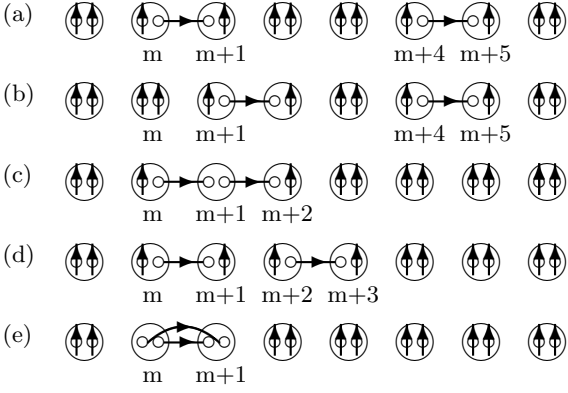
We now move on to the next simplest set of exact high energy states, those with two dimers on a ferromagnetic background. The appropriate basis states are $\ket{m,m+n}$, $n = 0, 1, \dots, L/2$, that denote the configuration with dimers on the bonds $\{m,m+1\}$ and $\{m+n,m+n+1\}$, examples of which are shown in Fig.~\ref{fig:2singletconfig}. Here, we deal with the cases $n > 2$ (\eg Fig.~\ref{fig:2singletconfig}a), $n = 1$ (Fig.~\ref{fig:2singletconfig}c) and $n = 2$ (Fig.~\ref{fig:2singletconfig}e) separately because the scattering rules in Appendix~\ref{scattering} (Eqs.~(\ref{singlethopping}) and (\ref{badrule})) are applied differently.
If $n > 2$, the terms in the Hamiltonian that contribute to the scattering are the four projectors around the two dimers. According to Eq.~(\ref{singlethopping}), the states scatter as 
\begin{eqnarray}
    H\ket{m,m+n} &=& (L-2)\ket{m,m+n} + \frac{1}{2}(\ket{m-1,m+n} \nonumber \\ &&+ \ket{m+1,m+n} + \ket{m,m+n+1} \nonumber \\ &&+ \ket{m,m+n-1}).
    \label{2singneigh}
\end{eqnarray}
An example of a configuration $\ket{m, m + 4}$ and a scattering term $\ket{m+1, m + 4}$ are shown in Figs.~\ref{fig:2singletconfig}a and \ref{fig:2singletconfig}b respectively. In Eq.~(\ref{2singneigh}), if $\ket{m, m + n}$ is site (bond) inversion symmetric, all the scattering terms are bond (site) inversion symmetric, possibly around different bonds (sites). For periodic boundary conditions and even $L$, they can thus be cancelled with a $k = \pi$ superposition. The exact state for $n > 2$ is then $\ket{2_n} = \sum_m{(-1)^m \ket{m,m+n}}$. However, note that if $L/2$ is odd, $\ket{2_{L/2}} = \sum_m{(-1)^m \ket{m, m + L/2}} = \sum_m{(-1)^{m + L/2} \ket{m + L/2, m + L}} = -\ket{2_{L/2}}$. $\ket{2_{L/2}}$ thus vanishes if $n = L/2$ and $L/2$ is odd. There are thus a set of $L/2 - 2$ states if $L/2$ is even and $L/2 -3$ states if $L/2$ is odd (we are treating $n = 0, 1, 2$ separately), all with $\boldsymbol{E = L-2}$, $\boldsymbol{s = L - 2}$ and $\boldsymbol{k = \pi}$.
When $n = 1$ (Fig.~\ref{fig:2singletconfig}c), the two dimers share a site and only two projectors of the Hamiltonian contribute in scattering. Using Eq.~(\ref{singlethopping}), we obtain
\begin{eqnarray}
    && H\ket{m,m+1} = (L-2)\ket{m,m+1} \nonumber \\
    && \qquad + \frac{1}{2}(\ket{m-1,m+1} + \ket{m,m+2}).
\label{fatsing}
\end{eqnarray}
Since $\ket{m, m+1}$ is site inversion symmetric (about site $m + 1$) and the scattering terms are bond inversion symmetric (\eg Fig.~\ref{fig:2singletconfig}d), $\ket{2_{n=1}} = \sum_m{(-1)^m \ket{m,m+1}}$ is an exact state with periodic boundary conditions and even $L$ with the quantum numbers $\boldsymbol{s = L - 2}$, $\boldsymbol{k = \pi}$ and $\boldsymbol{E = L - 2}$. 
If $n = 0$ (Fig.~\ref{fig:2singletconfig}e), the appropriate rule for scattering is given by Eq.~(\ref{double}). Hence, $\ket{m,m}$ scatters to 
\begin{eqnarray}
    H\ket{m,m} &=& (L-1)\ket{m,m} + \frac{1}{2}(\ket{m,m+1} + \ket{m-1,m})  \nonumber \\
    &&+ \frac{1}{6}(\ket{m-1,m-1} + \ket{m+1,m+1}). 
\end{eqnarray}
Since $\ket{m,m+1}$ (Fig.~\ref{fig:2singletconfig}c) is site inversion symmetric whereas $\ket{m,m}$ is bond inversion symmetric, $\ket{2_0} = \sum_m{(-1)^m \ket{m,m}}$ would give rise to the action
\begin{eqnarray}
    H\ket{2_0} &=& (L-1)\ket{2_{n=0}} - \frac{1}{3}\ket{2_{n=0}} \nn \\
    &=& (L - \frac{4}{3})\ket{2_{n=0}}.
\end{eqnarray}
Thus we have an exact state with with energy $\boldsymbol{E = L - 4/3}$, momentum $\boldsymbol{k = \pi}$ and spin $\boldsymbol{s = L - 2}$. 

If $n = 2$ (Fig.~\ref{fig:2singletconfig}d), from Eq.~(\ref{badrule}) we see that $P^{(2,1)}_{m+1,m+2}$ would act non-trivially on the configuration. This scatters $\ket{m, m + 2}$ into bond symmetric terms that cannot be cancelled with a $k = \pi$ superposition. Hence, $\ket{2_{n = 2}}$ is not an exact state.

To derive operator expressions for the $\ket{2_n}$ states, we note that $\ket{m, m + n}$ can be expressed in terms of spin variables by writing down its expression in terms of Schwinger bosons and using the dictionary Eq.~(\ref{spindict}). Noting that $S^-\ket{1} = \sqrt{2}\ket{0}$ and $S^-\ket{0} = \sqrt{2}\ket{\mm}$, this straightforward calculation leads to
\begin{equation}
    \begin{split}
    \ket{m, m + n} = 
    \threepartdef{\left((S^-_{m + 1} - S^-_m)\right.\\\left.(S^-_{m + n + 1} - S^-_{m + n})\right)\ket{F}}{n \neq 0,1,2.\\}{\left(S^-_{m+1}S^-_{m+2} + S^-_m S^-_{m+1} \right.\\
    \left.- S^-_m S^-_{m+2} - 2 (S^-_{m+1})^2\right)\ket{F}}{n = 1.\\}{2\left((S^-_{m+1})^2 + (S^-_m)^2 \right.\\
    \left.- S^-_m S^-_{m+1}\right)\ket{F}}{n = 0.}
    \end{split}
\label{2singop}
\end{equation}
Using Eq.~(\ref{2singop}), after simplification, the normalized states can be written as
\begin{equation}
    \begin{split}
    \widetilde{\ket{2_n}} = \sum_m{(-1)^m}{  \threepartdef{\frac{1}{2\sqrt{2L}} \left(S_m^-\right.\\\left.(S_{m+n+1}^- - S_{m+n-1}^-)\right)\widetilde{\ket{F}}}{n \neq 0,1\\}{\frac{1}{2\sqrt{5L}} \left(S_m^-(S_{m+2}^- - 2 S_{m}^-)\right)\widetilde{\ket{F}}}{n = 1\\}{\frac{1}{2\sqrt{L}}S_m^- S_{m+1}^-\widetilde{\ket{F}}}{n = 0.}}
    \end{split}
    \label{2nspin}
\end{equation}

\subsection{{\bf \boldsymbol{$2_k$} states}}
\begin{figure}
    \centering
        \setlength{\unitlength}{1pt}
        
        \begin{picture}(180,10)(0,0)
        \linethickness{0.8pt}
        \multiput(0,10)(25,0){8}{\circle{4}} 
        \multiput(6,10)(25,0){8}{\circle{4}} 
        \multiput(3,10)(25,0){8}{\circle{14}}
        \thicklines
        \multiput(0,5)(25,0){2}{\vector(0,1){12}} 
        \put(6,5){\vector(0,1){12}} 
        \multiput(33,10)(25,0){2}{\line(1,0){15}}
        \multiput(45,10)(25,0){2}{\vector(1,0){0}}
        \multiput(81,5)(25,0){4}{\vector(0,1){12}}
        \multiput(100,5)(25,0){4}{\vector(0,1){12}}
        \multiput(156,5)(25,0){2}{\vector(0,1){12}} 
        \put(175,5){\vector(0,1){12}} 
        \put(25,-5){n}
        \put(45,-5){n+1}
        \put(70,-5){n+2}
        \put(-28, 10){(a)}
        \end{picture}\\[7mm]

        \begin{picture}(180,10)(0,0)
        \linethickness{0.8pt}
        \multiput(0,10)(25,0){8}{\circle{4}} 
        \multiput(6,10)(25,0){8}{\circle{4}} 
        \multiput(3,10)(25,0){8}{\circle{14}}
        \thicklines
        \multiput(0,5)(25,0){2}{\vector(0,1){12}} 
        \multiput(6,5)(25,0){1}{\vector(0,1){12}} 
        \multiput(33,10)(75,0){2}{\line(1,0){15}}
        \multiput(45,10)(75,0){2}{\vector(1,0){0}}
        \multiput(56,5)(25,0){2}{\vector(0,1){12}}
        \multiput(75,5)(25,0){2}{\vector(0,1){12}}
        \multiput(131,5)(25,0){3}{\vector(0,1){12}} 
        \multiput(150,5)(25,0){2}{\vector(0,1){12}} 
        \put(25,-5){n}
        \put(45,-5){n+1}
        \put(95,-5){n+3}
        \put(120,-5){n+4}
        \put(-28, 10){(b)}
        \end{picture}\\[7mm]
        
        \begin{picture}(180,10)(0,0)
        \linethickness{0.8pt}
        \multiput(0,10)(25,0){8}{\circle{4}} 
        \multiput(6,10)(25,0){8}{\circle{4}} 
        \multiput(3,10)(25,0){8}{\circle{14}}
        \thicklines
        \multiput(0,5)(25,0){2}{\vector(0,1){12}} 
        \put(6,5){\vector(0,1){12}} 
        \multiput(33,10)(125,0){2}{\line(1,0){15}}
        \multiput(45,10)(125,0){2}{\vector(1,0){0}}
        \multiput(56,5)(25,0){4}{\vector(0,1){12}}
        \multiput(75,5)(25,0){4}{\vector(0,1){12}}
        \multiput(106,5)(25,0){2}{\vector(0,1){12}} 
        \multiput(125,5)(50,0){2}{\vector(0,1){12}} 
        \put(25,-5){n}
        \put(45,-5){n+1}
        \put(145,-5){n+5}
        \put(170,-5){n+6}
        \put(-28, 10){(c)}
        \end{picture}\\[7mm]
        
        \begin{picture}(180,10)(0,0)
        \linethickness{0.8pt}
        \multiput(0,10)(25,0){8}{\circle{4}} 
        \multiput(6,10)(25,0){8}{\circle{4}} 
        \multiput(3,10)(25,0){8}{\circle{14}}
        \thicklines
        \multiput(0,5)(25,0){2}{\vector(0,1){12}} 
        \put(6,5){\vector(0,1){12}} 
        \multiput(33,10)(50,0){2}{\line(1,0){15}}
        \multiput(45,10)(50,0){2}{\vector(1,0){0}}
        \multiput(56,5)(25,0){1}{\vector(0,1){12}}
        \multiput(75,5)(25,0){1}{\vector(0,1){12}}
        \multiput(106,5)(25,0){4}{\vector(0,1){12}} 
        \multiput(125,5)(25,0){3}{\vector(0,1){12}} 
        \put(25,-5){n}
        \put(45,-5){n+1}
        \put(70,-5){n+2}
        \put(95,-5){n+3}
        \put(-28, 10){(d)}
        \end{picture}\\[7mm]
        
        \begin{picture}(180,10)(0,0)
        \linethickness{0.8pt}
        \multiput(0,10)(25,0){8}{\circle{4}} 
        \multiput(6,10)(25,0){8}{\circle{4}} 
        \multiput(3,10)(25,0){8}{\circle{14}}
        \thicklines
        \multiput(0,5)(25,0){2}{\vector(0,1){12}} 
        \put(6,5){\vector(0,1){12}} 
        \multiput(33,10)(100,0){2}{\line(1,0){15}}
        \multiput(45,10)(100,0){2}{\vector(1,0){0}}
        \multiput(56,5)(25,0){3}{\vector(0,1){12}}
        \multiput(75,5)(25,0){3}{\vector(0,1){12}}
        \multiput(156,5)(25,0){2}{\vector(0,1){12}} 
        \put(175,5){\vector(0,1){12}} 
        \put(25,-5){n}
        \put(45,-5){n+1}
        \put(120,-5){n+4}
        \put(145,-5){n+5}
        \put(-28, 10){(e)}
        \end{picture}\\[2mm] 
    \caption{Typical configurations involved in the derivation of the $\ket{2_k}$ states. Two dimer configurations (a) $\ket{n,n+1}$ (b) $\ket{n,n+3}$ (c) $\ket{n, n+5}$. Scattering terms (d) $\ket{n, n + 2}$ (e) $\ket{n, n + 4}$}
    \label{fig:2ksinglet}
\end{figure}
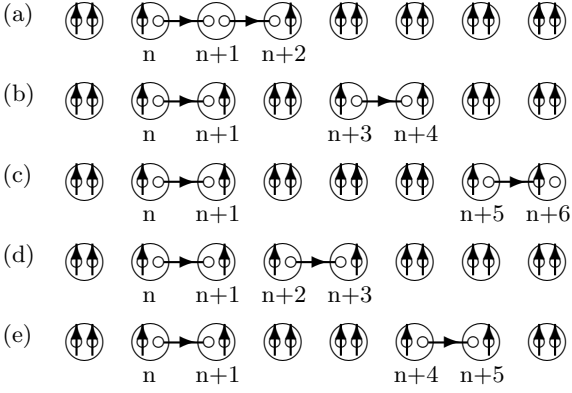
Using the scattering Eqs.~(\ref{2singneigh}) and (\ref{fatsing}), we obtain another set of exact states that can be constructed with two dimers on a ferromagnetic background are the $\ket{2_k}$ states. Consider
\begin{equation}
    \ket{2_k} = \sum_{m=0}^{L/2-1}{e^{i (k + \pi) m}\sum_n{e^{i k n}\ket{n,n + 2m + 1}}}.
\label{2kstate}
\end{equation}
In Eq.~(\ref{2kstate}), the choice of superpositions have been chosen such that scattering terms of a particular two dimer basis state are cancelled by those of other basis states. For example, $\ket{n, n+1}$ (Fig.~\ref{fig:2ksinglet}a) and $\ket{n, n + 3}$ (Fig.~\ref{fig:2ksinglet}b) produce the same scattering term $\ket{n, n + 2}$ (Fig.~\ref{fig:2ksinglet}d) according to Eqs.~(\ref{fatsing}) and (\ref{2singneigh}) respectively. Similarly the scattering term $\ket{n, n + 4}$(Fig.~\ref{fig:2ksinglet}e) obtained from both $\ket{n, n + 3}$ (Fig.~\ref{fig:2ksinglet}b) and $\ket{n, n + 5}$ (Fig.~\ref{fig:2ksinglet}c) but with opposite signs and hence vanishes. A detailed proof of the $\ket{2_k}$ state being exact can be found in Appendix~\ref{sec:2kexact}. We also show that these states have energy $E = L - 2$ and $L$ has to be even. 
While the state $\ket{2_k}$ can be proven to be exact for any momentum $k$, the state itself vanishes for certain momenta (see Eqs.~(\ref{2k1sub}) and (\ref{2k2sub}) in Appendix~\ref{sec:2kexact})
\begin{equation}
    \ket{2_k} = e^{i k \frac{L}{2}} e^{i \pi \left(\frac{L}{2} - 1\right)} \ket{2_k}.
\label{2kvanish}
\end{equation}
For the state not to vanish, the momenta should satisfy the relation $e^{i k \frac{L}{2}} e^{i \pi \left(\frac{L}{2} - 1\right)} = 1$. 

If $L = 4p$, $p$ integer, $e^{i k L/2} = -1$. Since $k = 2 \pi j/L$, where $j$ is an integer, the previous condition can be satisfied only for $j$ odd. So, for $L = 4p$, there is a state $\ket{2_k}$ at any odd momentum, a total of $2p$ states. If $L = 4p + 2$, we need $e^{i k L/2} = + 1$, restricting to even momenta, a total of $2p + 1$ states. Thus the $\ket{2_k}$ states lead to a total of $L/2$ exact states with $\boldsymbol{s = L - 2}$ and $\boldsymbol{E = L - 2}$. The operator expression for the normalized state $\widetilde{\ket{2_k}}$ directly obtained using Eqs.~(\ref{2kstate}) and (\ref{2singop}) reads
\begin{eqnarray}
    &\widetilde{\ket{2_k}} = \mathcal{N}\left(\sum_{m = 0}^{L/2-1}{\left(e^{i(k+\pi)m} \sum_n{ e^{i k n} S_n^- S_{n+2m+1}^-}\right)}\right. \nn \\
    &\left.- 2 e^{-i\frac{k}{2}} \text{secant}\left(\frac{k}{2}\right)\sum_n{e^{i k n} \left(S_n^-\right)^2}\right)\ket{F}
\end{eqnarray}
where $\mathcal{N}$ is the normalization constant.
\subsection{{\bf Special states}}\label{sec:specialstatesmain}
Apart from all the states above, we numerically observe a state that is at $s = L - 2$ that repeats only for particular values of $L$. The state $\ket{6_0}$ appears only for $L = 6p$, $p$ integer with energy $E = L - 1$. In this section, we exemplify a method to obtain such ``repeating states", i.e. states with an energy $E = L - \xi$ that appear periodically in $L$.

We work in the orthonormal, complete, spin basis. The highest weight state in the $s = L - 2$ can be obtained from $\ket{F}$, either by flipping one spin to $S_z = -1$ or flipping two spins to $S_z = 0$. Thus basis states of a fixed momentum $k$ can be labelled by $\ket{n_k}, n \in {0,1,...,L/2}$, where 
\begin{equation}
\ket{n_k} = \mathcal{N}\sum_{j=1}^{L}{e^{i k j} S_j^- S_{j+n}^- \ket{F}}
\label{spinket1}
\end{equation}
The Hamiltonian matrix in this basis constructed using Eqs.~(\ref{spin11}) - (\ref{spinm1m1}) and (\ref{spinket1}) is tridiagonal. Eigenvalues of the form $L - \xi$ that are periodic in $L$ can then be solved by a ``transfer" matrix method, as demonstrated for $k = 0$ in Appendix~\ref{sec:specialstates}. For $k = 0$, obtaining repeating states can be shown to reduce to solving the equation (see Eq.~(\ref{npqconstraintapp})) 
\begin{equation}
\frac{3}{2}\tan\left(\frac{y}{2}\right) = - \frac{\sin((r+2)y)}{\cos((r+3)y)}
\label{npqconstraint}
\end{equation}
with $r \in \mathbb{Z}_q$ and $y = 2\pi \frac{p}{q}$. As shown in Appendix~\ref{sec:specialstates}, a solution to Eq.~(\ref{npqconstraint}) with integer $p, q, r$, with $p \leq q/2$ and $q$ even corresponds to a state with energy $E = L - 4 \sin^2(\pi \frac{p}{q})$ that appears for every $L = 6 + 2 r + q m$, where $m \in \mathbb{Z}$.
Most of the states obtained by this method correspond to the $\ket{2_n}$ and $\ket{2_k}$ states that we have already discussed. For example, a solution of Eq.~(\ref{npqconstraint}) is $q = 4, p = 1, r = 0$ has $E = L - 2$ and is the $\ket{2_{k=0}}$ that appears for every $L = 4m + 2$. 

For $p, q, r \leq 1000$, we could find one new solution to Eq.~(\ref{npqconstraint}). It is $q = 6, p = 1, r = 0$. This state thus has energy $E = L - 1$, and since $q$ is even, it appears for $L = 6 p$, where $p$ is an integer. The exact state can be written as 
\begin{eqnarray}
&\widetilde{\ket{6_0}} = \mathcal{N}\sumal{i}{}{\sumal{n=0}{L/2}{\cos(\frac{n\pi}{3})S_i^- S_{i+n}^-\ket{F}}}
\label{60defn}
\end{eqnarray}
where $\mathcal{N}$ is a normalization constant.
We thus have an exact state with $\boldsymbol{E = L - 1}$, $\boldsymbol{k = 0}$ and $\boldsymbol{s = L -2}$.
In momentum space operators, the state can be written in a more elegant form.
\begin{equation}
    \ket{6_0} = (S_{\frac{\pi}{3}}^- S_{-\frac{\pi}{3}}^- + \sum_k{S_k^- S_{-k}^-})\ket{F}
\end{equation}
where the sum over $k$ runs over all the momenta.
The same exercise for any other momentum $k$ does not lead to any new states.
\subsection{{\bf Overlap with the tower of states}}
\begin{figure}
    \centering
        \setlength{\unitlength}{1pt}
        
        \begin{picture}(120,10)(0,0)
        \linethickness{0.8pt}
        \multiput(0,10)(25,0){6}{\circle{4}} 
        \multiput(6,10)(25,0){6}{\circle{4}} 
        \multiput(3,10)(25,0){6}{\circle{14}}
        \thicklines
        \multiput(0,5)(25,0){2}{\vector(0,1){12}} 
        \put(6,5){\vector(0,1){12}} 
        \multiput(33,10)(25,0){2}{\line(1,0){15}}
        \multiput(45,10)(25,0){2}{\vector(1,0){0}}
        \multiput(81,5)(25,0){3}{\vector(0,1){12}}
        \multiput(100,5)(25,0){2}{\vector(0,1){12}}
        \put(-28, 10){(a)}
        \end{picture}\\[6mm]

        \begin{picture}(120,10)(0,0)
        \linethickness{0.8pt}
        \multiput(0,10)(25,0){6}{\circle{4}} 
        \multiput(6,10)(25,0){6}{\circle{4}} 
        \multiput(3,10)(25,0){6}{\circle{14}}
        \thicklines
        \multiput(0,5)(25,0){2}{\vector(0,1){12}} 
        \multiput(6,5)(25,0){1}{\vector(0,1){12}} 
        \multiput(33,10)(75,0){2}{\line(1,0){15}}
        \multiput(45,10)(75,0){2}{\vector(1,0){0}}
        \multiput(56,5)(25,0){2}{\vector(0,1){12}}
        \multiput(75,5)(25,0){2}{\vector(0,1){12}}
        \multiput(131,5)(25,0){1}{\vector(0,1){12}} 
        \put(-28, 10){(b)}
        \end{picture}
        \caption{The only two possible dimer configurations up to translations in $\ket{S_{4}}$ for $L = 6$}
        \label{fig:overlap}
\end{figure}
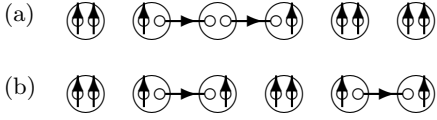
It is important to note that the tower of states $\{\ket{S_{2n}}\}$ obtained in Sec.~\ref{sec:towerofstates} Eq.~(\ref{evenspinsequence}) has some overlap with the high energy states we have presented. As we have seen earlier, $\ket{S_L} = \ket{F}$.  
Moreover, the $L/2 - 1$ magnon state $\ket{S_{L-2}}$ has a total spin $s = L - 2$.
As illustrated in Fig.~\ref{fig:S2Magnon}a, the operator $(S_n^+)^2$ always annihilates two dimers around a site $n$.
Since $N = L/2 - 1$ in Eq.~(\ref{evenspinsequence}) for $\ket{S_{L-2}}$, the state is a superposition of configurations with $L - 2$ dimers annihilated from the ground state configuration. 
Moreover, since $(S_n^+)^2$ annihilates the dimers on neighboring bonds $\{n - 1, n\}$ and $\{n, n+1\}$, for even $L$, the two remaining dimers always have an even (odd) number of bonds (sites) between them.
For example, the only possible configurations of $\ket{S_4}$ for $L = 6$ up to translation is shown in Fig.~\ref{fig:overlap}.  Thus, up to an overall constant, the appropriate two dimer basis state to describe the tower of states can be written as
\begin{equation}
\ket{n, n+2m +1} \sim \prod_{j=1}^m{(S_{n+2j}^+)^2} \prod_{j = m+1}^{L/2-1}{(S_{n+2j+1}^+)^2}\ket{G}
\label{towerbasis}
\end{equation}
where each $(S_i^+)^2$ annihilates dimers on bonds $\{i - 1, i\}$ and $\{i, i+1\}$. Thus, in Eq.~(\ref{towerbasis}), the dimers on bonds $\{n, n+1\}$ and $\{n + 2m + 1, n + 2m + 2\}$ are the only ones remaining. 
From Eq.~(\ref{evenspinsequence}), the exponent of (-1) in front of the configuration $\ket{n,n+2m+1}$ can be written as
\begin{eqnarray}
    &\sum_{j=1}^m{(n+2j)} + \sum_{j=m+1}^{L/2-1}{(n+2j+1)} \nonumber \\
    &= (\frac{L}{2} - 1)(\frac{L}{2}+1+n) - m.
\label{powerofm1}
\end{eqnarray}
Using Eqs.~(\ref{towerbasis}) and (\ref{powerofm1}), up to an overall phase, the state $\ket{S_{L-2}}$ can thus be written as 
\begin{equation}
\ket{S_{L-2}} = \sum_{m = 0}^{L/2-1}{\sum_n{(-1)^{(L/2-1)n + m}\ket{n,n+2m+1}}}
\label{overlapop}
\end{equation}
The two cases $L = 4p$ and $L = 4p+2$ are considered separately, $p$ being an integer. When $L = 4p+2$, $\ket{S_{L-2}}$ has $k = 0$, $(L/2 - 1)$ is even and when $L = 4p$, $\ket{S_{L-2}}$ has $k = \pi$, $L/2 -1$ is odd. From Eq.~(\ref{overlapop}), we obtain  
\begin{equation}
\ket{S_{L-2}} = \twopartdef{\ket{2_{k = 0}}}{L = 4p + 2.\\}{\sumal{m = 0}{L/2-1}{(-1)^m\ket{2_{n = 2m+1}}}}{L = 4p.}
\end{equation}  
\section{1D Spin-$\boldsymbol{S}$ AKLT Models with $\boldsymbol{S>1}$}\label{sec:spinsaklt}
AKLT models can be straightforwardly generalized to all dimensions and also to spins with different Lie algebras.\cite{kennedy1988two,arovas1988extended,greiter2007exact,greiter2007valence} In this section, we consider the generalization to spin-$S$ with $S$ being a positive integer. Such a model has been studied to explore the Haldane conjecture for $S = 2$.\cite{schollwock1996s, zang2010topological, jiang2010critical, pollmann2012symmetry} Particularly, it has been observed that odd integer spin chains are topological (due to the presence of half-integer dangling spins at the edge) whereas even integer spin chains are not.\cite{zang2010topological,pollmann2012symmetry} The generalization of the AKLT Hamiltonian that was used\cite{zang2010topological, jiang2010critical} is
\begin{equation}
    H^{(S)} = \sum_{i = 1}^L{\sum_{J = S + 1}^{2S}{\alpha_J P^{(J,S)}_{i,i+1}}}
\label{SHamiltonian}  
\end{equation}
with $\alpha_J \geq 0$ $\forall J$. As we will see later, the ground state is the same for all the Hamiltonians of the form Eq.~\eqref{SHamiltonian}. However, the entire energy spectrum is not the same. For example, the ferromagnetic state need not be the unique highest excited state unless
\begin{equation}
    \alpha_{2S} \geq \sum_{J = S + 1}^{2S-1}{\alpha_J}.
\end{equation}
As we did for the spin-1 AKLT model, we can write Hamiltonian Eq.~(\ref{SHamiltonian}) in terms of spin operators. The most general expression for $P^{(J,S)}_{i,j}$, is the projector onto total spin $J$ for two spin-$S$ on sites $i$ and $j$ and can be written in terms of the spin operators as
\begin{eqnarray}
    P^{(J,S)}_{i,j} &=& \prod_{s = 0, s \neq J}^{2S}{\left(\frac{(\vec{S}_i + \vec{S}_j)^2 - s(s+1)}{J(J+1) - s(s+1)}\right)}  \\
    &=& \prod_{s = 0, s \neq J}^{2S}{\left(\frac{ 2\vec{S}_i\cdot\vec{S}_j + 2S(S+1) - s(s+1)}{J(J+1) - s(s+1)}\right)}. \nonumber
\label{projector}
\end{eqnarray}
Since we are working with spin-$S$, each spin can be thought to be composed of $2S$ spin-1/2 Schwinger bosons, with the same algebra described in Appendix~\ref{dimers}. However, since the number of Schwinger bosons per site changes,
Eq.~(\ref{Si.Sj}) also changes to 
\begin{eqnarray}
    \vec{S_i}\cdot\vec{S_j} &=& -\frac{1}{2}c_{i j}^\dagger c_{i j} + \frac{1}{4}(a_i^\dagger a_i + b_i^\dagger b_i)(a_j^\dagger a_j + b_j^\dagger b_j) \nonumber \\
    &=& S^2 - \frac{1}{2}c_{i j}^\dagger c_{i j}
\label{SiSjspins}
\end{eqnarray}
Using Eq.~\eqref{SiSjspins} and Eq.~\eqref{dimernormexpand} for normal ordering, we obtain an expression for the spin-$S$ AKLT Hamiltonian,
\begin{equation}
    H^{(S)} = \sum_i{\left(1 + \sum_{j = 1}^{2S}{\gamma_j (c_{i,i+1}^\dagger)^j (c_{i,i+1})^j}\right)}
\label{SHamiltoniandimers}
\end{equation}
where the coefficients $\gamma_j$ depend on the coefficients  $\alpha_J$ in the projectors and the spin $S$, a closed form of which we could not obtain for a general $S$. We have the freedom to choose $(S - 1)$ $\alpha_J$ coefficients while retaining the standard ground state but changing the excitation spectrum. By choosing $\alpha_J = 1$ for all $J$, we observe that it is possible to set $\gamma_j = 0$, $1 \leq j \leq S - 1$. This is the only choice of $\{\alpha_J\}$ that satisfies the required condition. The Hamiltonian is then, 
\begin{equation}
    H^{(S)} = \sum_i{\left(1 + \sum_{j = S}^{2S}{\beta_j (c_{i,i+1}^\dagger)^j (c_{i,i+1})^j}\right)}
\label{SHamiltoniandimerspecial}
\end{equation}
As we will show later in this section, this choice of coefficients is crucial for us to have non-trivial exact states (including a tower of states) in the bulk of the spectrum. Since the algebra of dimers described in Appendix~\ref{dimers} is independent of the spin model we are working with, it holds here as well.

\subsection{Lower spectrum states}
We start our analysis of the spin-$S$ AKLT model with its ground state. Working in the spin basis, if $S$ dimers are formed between two spin-$S$ ($i$ and $i+1$) that have a total of $4S$ spin-1/2 Schwinger bosons, the maximum spin of both spin-$S$ combined cannot exceed $S$. Such a configuration must therefore be annihilated by all $P^{(J,S)}_{i,i+1}$ for $J > S$, and is thus the unique ground state $\ket{SG}$ of the Hamiltonian Eq.~(\ref{SHamiltonian}). This can be written as
\begin{equation}
    \ket{SG} = \prodal{i=1}{L}{({c^\dagger_{i,i+1}})^S}\vac.
\end{equation}
The cartoon picture for the ground state of the spin-2 AKLT model is shown in Fig.~\ref{fig:spin2gs}. The ground state with open boundary conditions is computed in Appendix~\ref{sec:obcspinSgs}.
\begin{figure}[t!]
    \setlength{\unitlength}{1pt}
    \begin{picture}(180,10)(0,0)
    \linethickness{0.8pt}
    \multiput(0,10)(25,0){8}{\circle{4}} 
    \multiput(6,10)(25,0){8}{\circle{4}} 
    \multiput(0,4)(25,0){8}{\circle{4}} 
    \multiput(6,4)(25,0){8}{\circle{4}} 
    \multiput(3,7)(25,0){8}{\circle{16}}
    \thicklines
    \multiput(-17,10)(25,0){9}{\line(1,0){15}} 
    \multiput(-5,10)(25,0){9}{\vector(1,0){0}}
    \multiput(-17,4)(25,0){9}{\line(1,0){15}} 
    \multiput(-5,4)(25,0){9}{\vector(1,0){0}}
    \end{picture}
    \caption{$S = 2$ AKLT ground state with two dimers between nearest neighbors. Spin-$S$ AKLT would have $S$ dimers. $\ket{2G}$ with periodic boundary conditions. }
    \label{fig:spin2gs}
\end{figure}
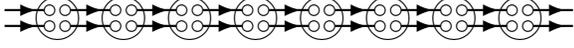
One might wonder if any low energy $s = 0$ magnons similar to the two Arovas magnon states discussed in Sec.~\ref{sec:spin1aklt} exist for the spin-$S$ AKLT models. We have used the method described in Sec.~\ref{sec:whatarethey} to detect exact eigenstates for $S = 2$ with $L \leq 10$ and $S = 3$ with $L \leq 8$. In both the systems, we find only one singlet exact state apart from the ground state, and it lies in the quantum number sector $s = 0$, $k = \pi$, $I = -1$ for even $L$, and has an energy $E = 2$. Since the Arovas B state in Sec.~\ref{sec:arovasb} is in the same sector, we need a similar configuration with short-ranged dimers. Numerically, we do not find an analogue of the Arovas A state for higher spin AKLT models with the chosen Hamiltonian given by Eq.~(\ref{SHamiltoniandimerspecial}) (i.e. with $\alpha_J = 1\; \forall J$ in Eq.~(\ref{SHamiltonian})). At the end of this section, we provide an intuitive explanation as to why this is the case. In Appendix~\ref{sec:S2ArovasA}, we prove that for $S = 2$, an Arovas A state cannot be obtained for the Hamiltonian Eq.~(\ref{SHamiltoniandimerspecial}) but an analogue can be constructed with another suitable choice of coefficients $\{\alpha_J\}$ in Eq.~(\ref{SHamiltonian}).
\begin{figure}[t!]
    \centering
        \setlength{\unitlength}{1pt}
        \begin{picture}(150,20)(0,0)
        \linethickness{0.8pt}
        \multiput(0,10)(25,0){7}{\circle{4}} 
        \multiput(6,10)(25,0){7}{\circle{4}} 
        \multiput(0,4)(25,0){7}{\circle{4}} 
        \multiput(6,4)(25,0){7}{\circle{4}} 
        \multiput(3,7)(25,0){7}{\circle{16}}
        \thicklines
        \multiput(-17,10)(25,0){2}{\line(1,0){15}} 
        \multiput(-5,10)(25,0){2}{\vector(1,0){0}}
        \multiput(-17,4)(25,0){8}{\line(1,0){15}} 
        \multiput(-5,4)(25,0){8}{\vector(1,0){0}}
        \multiput(58,10)(25,0){2}{\line(1,0){15}}
        \multiput(70,10)(25,0){2}{\vector(1,0){0}}
        \qbezier(33,10)(68.5,40)(104,10)
        \put(65.5, 25){\vector(1,0){0}}
        \qbezier(52,10)(87.5,40)(123,10)
        \put(97.5, 25){\vector(1,0){0}}
        \multiput(133,10)(25,0){2}{\line(1,0){15}}
        \multiput(145,10)(25,0){2}{\vector(1,0){0}}
        \put(75,-10){n}
        \put(47,-10){n-1}
        \put(95,-10){n+1}
        \put(120,-10){n+2}
        \put(22,-10){n-2}
        \put(-32,5){(a)}
        \end{picture}\\[10mm]
    
        \begin{picture}(150,20)(0,0)
        \linethickness{0.8pt}
        \multiput(0,15)(25,0){7}{\circle{4}} 
        \multiput(6,15)(25,0){7}{\circle{4}} 
        \multiput(0,9)(25,0){7}{\circle{4}} 
        \multiput(6,9)(25,0){7}{\circle{4}} 
        \multiput(3,12)(25,0){7}{\circle{16}}
        \thicklines
        \multiput(-17,15)(25,0){2}{\line(1,0){15}} 
        \multiput(-5,15)(25,0){2}{\vector(1,0){0}}
        \multiput(-17,9)(25,0){2}{\line(1,0){15}} 
        \multiput(-5,9)(25,0){2}{\vector(1,0){0}}
        \multiput(58,15)(25,0){2}{\line(1,0){15}}
        \multiput(70,15)(25,0){2}{\vector(1,0){0}}
        \multiput(58,9)(25,0){2}{\line(1,0){15}}
        \multiput(70,9)(25,0){2}{\vector(1,0){0}}
        \qbezier(33,15)(68.5,45)(104,15)
        \put(65.5, 30){\vector(1,0){0}}
        \qbezier(52,15)(87.5,45)(123,15)
        \put(97.5, 30){\vector(1,0){0}}
        \qbezier(33,9)(68.5,45)(104,9)
        \put(60, 24){\vector(1,0){0}}
        \qbezier(52,9)(87.5,45)(123,9)
        \put(107.5, 24){\vector(1,0){0}}
        \multiput(133,15)(25,0){2}{\line(1,0){15}}
        \multiput(145,15)(25,0){2}{\vector(1,0){0}}
        \multiput(133,9)(25,0){2}{\line(1,0){15}}
        \multiput(145,9)(25,0){2}{\vector(1,0){0}}
        \put(75,-5){n}
        \put(47,-5){n-1}
        \put(22,-5){n-2}
        \put(95,-5){n+1}
        \put(120,-5){n+2}
        \put(-32,10){(b)}
        \end{picture}\\[5mm]
        
        \begin{picture}(150,20)(0,0)
        \linethickness{0.8pt}
        \multiput(0,10)(25,0){7}{\circle{4}} 
        \multiput(6,10)(25,0){7}{\circle{4}} 
        \multiput(0,4)(25,0){7}{\circle{4}} 
        \multiput(6,4)(25,0){7}{\circle{4}} 
        \multiput(3,7)(25,0){7}{\circle{16}}
        \thicklines
        \multiput(-17,10)(50,0){1}{\line(1,0){15}} 
        \multiput(-5,10)(50,0){1}{\vector(1,0){0}}
        \multiput(-17,4)(25,0){1}{\line(1,0){15}} 
        \multiput(-5,4)(25,0){1}{\vector(1,0){0}}
        
        \multiput(8,10)(50,0){2}{\line(1,0){15}} 
        \multiput(20,10)(50,0){2}{\vector(1,0){0}}
        \multiput(8,4)(50,0){2}{\line(1,0){15}} 
        \multiput(20,4)(50,0){2}{\vector(1,0){0}}
        
        \multiput(83,4)(50,0){1}{\line(1,0){15}} 
        \multiput(95,4)(50,0){1}{\vector(1,0){0}}

        \multiput(108,4)(50,0){2}{\line(1,0){15}} 
        \multiput(120,4)(50,0){2}{\vector(1,0){0}}
        \multiput(108,10)(25,0){1}{\line(1,0){15}}
        \multiput(120,10)(25,0){1}{\vector(1,0){0}}
        \qbezier(33,10)(65.5,35)(98,10)
        \put(67, 23){\vector(1,0){0}}
        \qbezier(83,10)(115.5,35)(148,10)
        \put(122, 23){\vector(1,0){0}}
        \qbezier(52,10)(90.5,35)(129,10)
        \put(91, 23){\vector(1,0){0}}
        \qbezier(33,4)(90.5,-21)(148,4)
        \put(95,-9){\vector(1,0){0}}
        \qbezier(52,4)(90.5,-15)(129,4)
        \put(95, -5){\vector(1,0){0}}
        \multiput(158,10)(25,0){1}{\line(1,0){15}}
        \multiput(170,10)(25,0){1}{\vector(1,0){0}}
        \put(75,-18){n}
        \put(47,-18){n-1}
        \put(95,-18){n+1}
        \put(120,-18){n+2}
        \put(22,-18){n-2}
        \put(-32,5){(c)}
        \end{picture}\\[10mm]
    
        \begin{picture}(150,20)(0,0)
        \linethickness{0.8pt}
        \multiput(0,15)(25,0){7}{\circle{4}} 
        \multiput(6,15)(25,0){7}{\circle{4}} 
        \multiput(0,9)(25,0){7}{\circle{4}} 
        \multiput(6,9)(25,0){7}{\circle{4}} 
        \multiput(3,12)(25,0){7}{\circle{16}}
        \thicklines
        \multiput(-17,15)(25,0){3}{\line(1,0){15}} 
        \multiput(-5,15)(25,0){3}{\vector(1,0){0}}
        \multiput(-17,9)(25,0){3}{\line(1,0){15}} 
        \multiput(-5,9)(25,0){3}{\vector(1,0){0}}
        \multiput(83,15)(25,0){1}{\line(1,0){15}}
        \multiput(95,15)(25,0){1}{\vector(1,0){0}}
        \multiput(83,9)(25,0){1}{\line(1,0){15}}
        \multiput(95,9)(25,0){1}{\vector(1,0){0}}
        \qbezier(58,15)(87.5,45)(123,15)
        \put(97.5, 30){\vector(1,0){0}}
        \qbezier(58,9)(87.5,45)(123,9)
        \put(107.5, 24){\vector(1,0){0}}
        \multiput(133,15)(25,0){2}{\line(1,0){15}}
        \multiput(145,15)(25,0){2}{\vector(1,0){0}}
        \multiput(133,9)(25,0){2}{\line(1,0){15}}
        \multiput(145,9)(25,0){2}{\vector(1,0){0}}
        \qbezier(77,7)(90.5,-1)(104,7)
        \put(95.5, 3){\vector(1,0){0}}
        \qbezier(77,17)(90.5,25)(104,17)
        \put(95.5, 22){\vector(1,0){0}}
        \put(75,-5){n}
        \put(47,-5){n-1}
        \put(22,-5){n-2}
        \put(95,-5){n+1}
        \put(120,-5){n+2}
        \put(-32,10){(d)}
        \end{picture}\\[8mm]
    
        \begin{picture}(150,20)(0,0)
        \linethickness{0.8pt}
        \multiput(0,15)(25,0){7}{\circle{4}} 
        \multiput(6,15)(25,0){7}{\circle{4}} 
        \multiput(0,9)(25,0){7}{\circle{4}} 
        \multiput(6,9)(25,0){7}{\circle{4}} 
        \multiput(3,12)(25,0){7}{\circle{16}}
        \thicklines
        \multiput(-17,15)(25,0){2}{\line(1,0){15}} 
        \multiput(-5,15)(25,0){2}{\vector(1,0){0}}
        \multiput(-17,9)(25,0){4}{\line(1,0){15}} 
        \multiput(120,9)(50,0){2}{\vector(1,0){0}}
        \multiput(108,9)(50,0){2}{\line(1,0){15}} 
        \multiput(-5,9)(25,0){4}{\vector(1,0){0}}
        \multiput(58,15)(25,0){1}{\line(1,0){15}}
        \multiput(70,15)(25,0){1}{\vector(1,0){0}}
        \qbezier(33,15)(68.5,45)(104,15)
        \put(65.5, 30){\vector(1,0){0}}
        \qbezier(52,15)(87.5,45)(130,15)
        \put(97.5, 30){\vector(1,0){0}}
        \multiput(108,15)(25,0){1}{\line(1,0){15}}
        \multiput(120,15)(25,0){1}{\vector(1,0){0}}
        \multiput(158,15)(25,0){1}{\line(1,0){15}}
        \multiput(170,15)(25,0){1}{\vector(1,0){0}}
        \qbezier(82,15)(112.5,45)(148,15)
        \qbezier(82,10)(112.5,42)(148,10)
        \qbezier(101,7)(112.5,0)(131,7)
        \put(122.5, 30){\vector(1,0){0}}
        \put(122.5, 27){\vector(1,0){0}}
        \put(120, 4){\vector(1,0){0}}
        \put(75,-5){n}
        \put(47,-5){n-1}
        \put(95,-5){n+1}
        \put(120,-5){n+2}
        \put(22,-5){n-2}
        \put(-32,10){(e)}
        \end{picture}
    \caption{The spin-2 AKLT configurations (a) $\ket{BG_n}$ and (b) $\ket{B^2_n}$. Such configurations form the generalization of the Arovas B state in the spin-S AKLT model. (c) An example of a symmetric configuration that appears only in $\mathcal{C}_{B^2}^S$. (d) An example of a symmetric configuration that appears in both $\mathcal{C}_{BG}^S$ and $\mathcal{C}_{B^2}^S$. (e) An example of a non-symmetric scattering configuration $\ket{\zeta}$.}
    \label{fig:spin2arovas}
\end{figure}
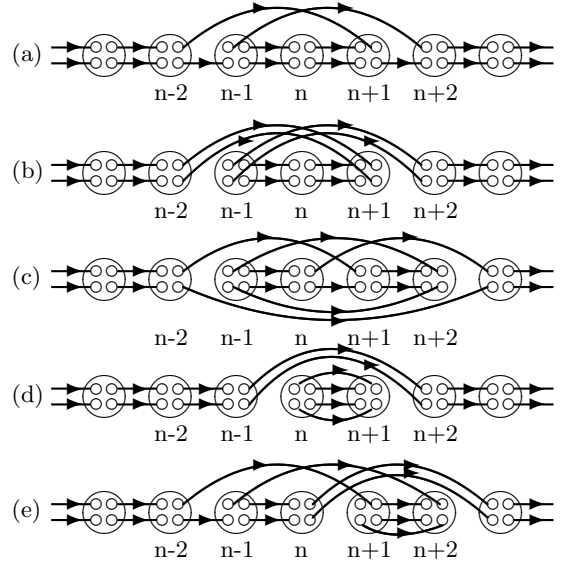

To exemplify the derivation of the generalized Arovas B state, we first focus on the spin-2 AKLT model. The corresponding Hamiltonian Eq.~(\ref{SHamiltoniandimerspecial}) reduces to
\begin{eqnarray}
    &H^{(2)} = \sumal{i}{}{ \left(1 - \frac{1}{84} {c^\dagger}^2_{i,i+1} {c}^2_{i,i+1} + \frac{1}{630} {c^\dagger}^3_{i,i+1} {c}^3_{i,i+1} \right.} \nn \\
    & \left. - \frac{1}{6720} {c^\dagger}^4_{i,i+1} {c}^4_{i,i+1} \right).
\label{spin2hamil}
\end{eqnarray}

One could form spin-2 basis states by gluing together two copies of spin-1 basis states, along with completely symmetrizing the spin-1s of the two copies. The spin-2 configuration $\ket{BG_n}$ (Fig.~\ref{fig:spin2arovas}a) is formed by gluing one one spin-1 Arovas B configuration $\ket{B_n}$ (Fig.~\ref{fig:ArovasBConfig}a) to the spin-1 ground state $\ket{G}$ (Fig.~\ref{fig:groundstate}) and $\ket{BG_n}$ (Fig.~\ref{fig:spin2arovas}b) by gluing two  $\ket{B_n}$s together. 
The scattering equations for the configurations $\ket{BG_n}$ and $\ket{B^2_n}$ with the choice of our Hamiltonian Eq.~(\ref{SHamiltoniandimerspecial}) are shown in Appendix~\ref{sec:S2Arovas} Eqs.~(\ref{spin2arovasbg}) and (\ref{spin2arovasb2}) respectively. We find that the scattering terms arising from $\ket{BG_n}$ and $\ket{B^2_n}$ fall into two types of configurations--symmetric with bond inversion symmetry and non-symmetric. While the bond inversion symmetric scattering terms are different for $\ket{BG_n}$ (set $\mathcal{C}_{BG}^S$) and $\ket{B^2_n}$ (set $\mathcal{C}_{B^2}^S$), the set of non-symmetric scattering terms are the same for both (set $\mathcal{C}^N$). The scattering equations assume the form (see Eqs.~(\ref{spin2arovasbg}) and (\ref{spin2arovasb2})) 
\begin{eqnarray}
    &H^{(2)}\ket{BG_n} = 2\ket{BG_n} + \sum_{\{\eta \in \mathcal{C}_{BG}^S\}}{\lambda^{BG}_\eta \ket{\eta}} \nonumber \\
    &+ x \sum_{\{\zeta \in \mathcal{C}^N\}}{\lambda_\zeta \ket{\zeta}}, \nn \\[10pt]
    &H^{(2)}\ket{B^2_n} = 2\ket{B^2_n} + \sum_{\{\eta \in \mathcal{C}_{B^2}^S\}}{\lambda^{B^2}_\eta \ket{\eta} } \nonumber \\
    &- 4\sum_{\{\zeta \in \mathcal{C}^N\}}{\lambda_\zeta \mathcal{T}^{\pm 1}\ket{\zeta}} + (1-x)\sum_{\{\zeta \in \mathcal{C}^N\}}{\lambda_\zeta \ket{\zeta}}
\label{BGscattering}
\end{eqnarray}
In Eq.~(\ref{BGscattering}), $\mathcal{T}$ is a translation operator that translates by one site to the right and $\lambda^{BG}_\eta, \lambda^{B^2}_\eta$ and $\lambda_\zeta$ are the scattering coefficients. We show an example of a configuration from each of the sets $\mathcal{C}_{B^2}^S$, $\mathcal{C}_{BG}^S$ and $\mathcal{C}^N$ in  Figs.~\ref{fig:spin2arovas}c, \ref{fig:spin2arovas}d and \ref{fig:spin2arovas}e respectively. In Eq.~(\ref{BGscattering}), since non-vanishing projectors of the Hamiltonian Eq.~(\ref{SHamiltoniandimerspecial}) act on on configurations $\ket{BG_n}$ and $\ket{B^2_n}$ symmetrically about site $n$ (on bonds $\{n-2,n-1\}$ and $\{n+1, n+2\}$ in Figs.~\ref{fig:spin2arovas}a and \ref{fig:spin2arovas}b), all the scattering terms that are bond inversion symmetric (\eg Fig.~\ref{fig:spin2arovas}c) appear in pairs that are related by a translation of an odd number of sites. Hence, they cancel with a momentum $\pi$ superposition of the configurations $\ket{BG_n}$ and $\ket{B^2_n}$ for an even system size $L$. Moreover, $\ket{B^2_n}$ and $\ket{BG_n}$ can be combined into
\begin{equation}
    \ket{2B_n} = 2\ket{BG_n} - \frac{1}{2}\ket{B^2_n},
\label{2Bconfig}
\end{equation}
such that the scattering equation for $\ket{2B_n}$ reads 
\begin{eqnarray}
H^{(2)}\ket{2B_n} &=& 2\ket{2B_n} + \sum_{\{\eta \in \mathcal{C}^S\}}{\lambda_\eta \ket{\eta} } \nonumber \\
    &&+ 2\sum_{\{\zeta \in \mathcal{C}^N\}}{\lambda_\zeta\left(\ket{\zeta} + \mathcal{T}^{\pm 1}\ket{\zeta}\right)}.
\label{spin2final}
\end{eqnarray}
where $\mathcal{C}^S = \mathcal{C}^S_{BG} \cup \mathcal{C}^S_{B^2}$. In Eq.~(\ref{spin2final}), the 
non-symmetric configurations also admit a momentum $\pi$ cancellation for $L$ even. 
Thus, $\ket{2B} = \sum_n{(-1)^n \ket{2B_n}}$ is an exact state for $L$ even with $\boldsymbol{s = 0}$, $\boldsymbol{k = \pi}$, and $\boldsymbol{E = 2}$. In terms of spin operators, using Eq.~(\ref{SiSjspins}), we find that the normalized state $\widetilde{\ket{2B}}$ can be written as
\begin{eqnarray}
    &\widetilde{\ket{2B}} = \mathcal{N}\sum_n{(-1)^n\left(-5 + \vec{S}_{n-1}\cdot\vec{S}_n + \vec{S}_n\cdot\vec{S}_{n+1} \right.}\nn \\ &\left.+\frac{1}{3}\{\vec{S}_{n-1}\cdot\vec{S}_n, \vec{S}_n\cdot\vec{S}_{n+1}\}\right)^2\widetilde{\ket{2G}}
\label{2bopexp}
\end{eqnarray}
where $\{,\}$ denotes the anti-commutator, $\widetilde{\ket{2G}}$ is the normalized ground state of the spin-2 AKLT Hamiltonian and $\mathcal{N}$ a normalization factor. 

The $j = 1$ term in general Hamiltonian Eq.~(\ref{SHamiltoniandimers}) scatters the state $\ket{B^2_n}$ into non-symmetric configurations (\eg Fig.~\ref{fig:spin2arovas}d) whereas it scatters $\ket{BG_n}$ into symmetric configurations; this precludes the possibility of cancellation of non-symmetric terms as earlier. We thus set $\beta_1 = 0$, justifying our choice of the Hamiltonian in Eq.~(\ref{SHamiltoniandimerspecial}).

Moving on to the spin-$S$ AKLT model, the set of configurations that can be derived from the Arovas B state $\ket{B_n}$ and the spin-1 ground state $\ket{G}$ is $\mathcal{S} = \{\ket{B^m G^{S-m}}\}$, $1 \leq m \leq S$, which is obtained by gluing $m$ spin-1 $\ket{B_n}$s with $S-m$ spin-1 $\ket{G}$s and completely symmetrizing the corresponding spins.
The derivation of the generalized exact states proceeds in a way similar to that of $S = 2$. 
Eq.~(\ref{SHamiltoniandimerspecial}) scatters the configurations in the set $\mathcal{S}$ into two types of configurations: symmetric under bond inversion symmetry and non-symmetric. Similar to the $S = 2$ case, we find that the non-symmetric scattering terms of $\ket{B^m G^{S-m}}$ are the same as the non-symmetric scattering terms of either of $\ket{B^{m + 1} G^{S - m - 1}}$ or $\ket{B^{m - 1} G^{S - m + 1}}$. For $S \leq 5$, we find that a $\ket{SB_n}$ can always be constructed from the configurations in the set $\mathcal{S}$ such that the scattering equation of $\ket{SB_n}$ reads
\begin{eqnarray}
    &H^{(S)}\ket{SB_n} = 2 \ket{SB_n} + \sum_{\eta \in\mathcal{C}^S}{\lambda_\eta\ket{\eta}} \nonumber \\
    & \qquad + \sum_{\zeta \in \mathcal{C}^N}{\lambda_\zeta (\ket{\zeta} + \mathcal{T}^{2 x_\zeta + 1}\ket{\zeta})}
\label{SBnscattering}
\end{eqnarray}
where $\mathcal{C}^S$ and $\mathcal{C}^N$ are the sets of bond inversion symmetric and non-symmetric configurations of all the configurations in $\mathcal{S}$ and $x_\zeta$ is an integer. In Eq.~(\ref{SBnscattering}), the bond inversion symmetric configurations appear in pairs and vanish for even $L$ under a momentum-$\pi$ superposition of $\ket{SB_n}$. The non-symmetric terms too appear in pairs, and hence vanish under the same conditions.
We have analytically derived the generalized Arovas B state up to $S = 5$. For generic $S$, we conjecture the following expression for $\ket{SB_n}$,
\begin{equation}
    \ket{SB_n} = \sum_{m=1}^S{\frac{(-1)^m}{m}\binom{S}{m}\ket{B^m G^{S-m}}}.
\label{SBnconfig}
\end{equation}
This reduces to $\ket{B_n}$ and $\ket{2B_n}$ of Eq.~(\ref{2Bconfig}) for $S = 1$ and $S = 2$ respectively. The normalized exact state is then
\begin{equation}
    \widetilde{\ket{SB}} = \mathcal{N}\sum_n{(-1)^n\ket{SB_n}},
\label{SBstate}
\end{equation}
where $\mathcal{N}$ is a normalization factor. This state has $\boldsymbol{s = 0}$, $\boldsymbol{k = \pi}$, $\boldsymbol{E = 2}$. In spite of the elegant form of Eq.~(\ref{SBnconfig}) in terms of dimers, we could not easily find a nice expression such as Eq.~(\ref{2bopexp}) for the state $\ket{SB}$ in terms of the spin operators.

As mentioned earlier, for our choice of the Hamiltonian Eq.~(\ref{SHamiltoniandimerspecial}), we do not find an analogue of the Arovas A state numerically for $S = 2$ or $S = 3$. Analytically, an obstacle encountered is that some of the scattering terms of  $\ket{A^m G^{S-m}}$ (the state obtained by gluing $m$ spin-1 Arovas A configurations with $S-m$ spin-1 ground states) are of the form of $\ket{A^n G^{S-n}}$. Such terms are bond inversion symmetric (for example, $\ket{A^2}$ shown in Fig.~\ref{fig:spin2arovas}d), thus precluding a cancellation with momentum $\pi$ (since the Arovas A configuration is also bond inversion symmetric). Moreover, for any superposition of the configurations $\{ \ket{A^m G^{S-m}} \}$, the scattering terms $\{\ket{A^n G^{S-n}}\}$ appear in superposition with a different set of coefficients, thus precluding the construction of an exact state. However, in Appendix~\ref{sec:S2ArovasA}, we show that for $S = 2$, a fine-tuning of the Hamiltonian yields an Arovas A configuration with a scattering equation similar to Eq.~(\ref{SBnscattering}), and hence an Arovas A exact excited state. 

%
%
\subsection{Mid-spectrum states}\label{sec:spinStower}
\begin{figure}[t!]
    \begin{subfigure}{}
    \setlength{\unitlength}{1pt}
    \begin{picture}(150,20)(0,0)
    \linethickness{0.8pt}
    \multiput(0,10)(25,0){7}{\circle{4}} 
    \multiput(6,10)(25,0){7}{\circle{4}} 
    \multiput(0,4)(25,0){7}{\circle{4}} 
    \multiput(6,4)(25,0){7}{\circle{4}} 
    \multiput(3,7)(25,0){7}{\circle{16}}
    \thicklines
    \multiput(-17,10)(25,0){3}{\line(1,0){15}} 
    \multiput(-5,10)(25,0){3}{\vector(1,0){0}}
    \multiput(-17,4)(25,0){3}{\line(1,0){15}} 
    \multiput(-5,4)(25,0){3}{\vector(1,0){0}}
    \multiput(108,10)(25,0){3}{\line(1,0){15}}
    \multiput(120,10)(25,0){3}{\vector(1,0){0}}
    \multiput(108,4)(25,0){3}{\line(1,0){15}}
    \multiput(120,4)(25,0){3}{\vector(1,0){0}}
    \multiput(56,5)(25,0){2}{\vector(0,1){12}} 
    \multiput(75,5)(25,0){2}{\vector(0,1){12}} 
    \multiput(56,-1)(25,0){2}{\vector(0,1){12}} 
    \multiput(75,-1)(25,0){2}{\vector(0,1){12}} 
    \put(75,-10){n}
    \put(47,-10){n-1}
    \put(95,-10){n+1}
    \put(-32,5){(a)}
    \end{picture}
    \end{subfigure}
    \begin{subfigure}{}
        \setlength{\unitlength}{1pt}
        \begin{picture}(147.5,25)(0,0)
        \linethickness{0.8pt}
        \multiput(0,10)(25,0){7}{\circle{4}} 
        \multiput(6,10)(25,0){7}{\circle{4}} 
        \multiput(0,4)(25,0){7}{\circle{4}} 
        \multiput(6,4)(25,0){7}{\circle{4}} 
        \multiput(3,7)(25,0){7}{\circle{16}}
        \thicklines
        \multiput(-17,10)(25,0){3}{\line(1,0){15}} 
        \multiput(-5,10)(25,0){3}{\vector(1,0){0}}
        \multiput(-17,4)(25,0){3}{\line(1,0){15}} 
        \multiput(-5,4)(25,0){3}{\vector(1,0){0}}
        \multiput(133,10)(25,0){2}{\line(1,0){15}}
        \multiput(145,10)(25,0){2}{\vector(1,0){0}}
        \multiput(133,4)(25,0){2}{\line(1,0){15}}
        \multiput(145,4)(25,0){2}{\vector(1,0){0}}
        \put(83,10){\line(1,0){15}}
        \put(95,10){\vector(1,0){0}}
        \put(83,4){\line(1,0){15}}
        \put(95,4){\vector(1,0){0}}
        \multiput(56,5)(50,0){2}{\vector(0,1){12}} 
        \multiput(75,5)(50,0){2}{\vector(0,1){12}} 
        \multiput(56,-1)(50,0){2}{\vector(0,1){12}} 
        \multiput(75,-1)(50,0){2}{\vector(0,1){12}} 
        \put(75,-10){n}
        \put(47,-10){n-1}
        \put(95,-10){n+1}
        \put(-32,5){(b)}
        \end{picture}
    \end{subfigure}
    \caption{(a) Spin-$4$ Magnon $\ket{2M_n}$ and (b) Scattering state $\ket{2N_n}$ for spin-2 AKLT model. A similar picture holds for spin-$S$ AKLT model.}
    \label{fig:Spin2SMagnon}
\end{figure}
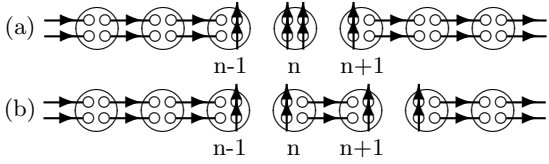
In this section, we derive the generalization of the tower of states described in Sec.~\ref{sec:towerofstates} for the spin-$S$ AKLT model. We start with a configuration gluing $S$ spin-1 $\ket{M_n}$ states to obtain a spin-$2S$ magnon $\ket{SM_n}$.
\begin{eqnarray}
    &\ket{SM_n} = \prod_{j = 1}^{n-2}(c_{j,j+1}^\dagger)^S (a_{n-1}^\dagger)^S (a_n^\dagger)^{2S} (a_{n+1}^{\dagger})^S \nonumber \\
    &\prod_{j = n+1}^{L} (c_{j,j+1}^\dagger)^S \vac.
\label{spin2Smagnondimer}
\end{eqnarray}
For example, $\ket{2M_n}$ for $S = 2$ is shown in Fig.~\ref{fig:Spin2SMagnon}a.
With $\beta_j = 0$ for $0 < j < S$, the Hamiltonian Eq.~(\ref{SHamiltoniandimerspecial}) does not have any terms $(c^\dagger)^s c^s$ for $s < S$. Due to Eq.~(\ref{nonscattering}), $\ket{2M_n}$ vanishes under the action of $(c^\dagger)^m c^m$ for $m > S$. Thus, using Eqs.~(\ref{highspinhopping}), the only term in the Hamiltonian that contributes to scattering is $(c^\dagger)^S c^S$. Thus, from Eq.~(\ref{highspinhopping}), the only scatterings term are those with $S$ dimers on the bonds $\{n-1,n\}$ or $\{n,n+1\}$, denoted by $\ket{SN_{n-1}}$ and $\ket{SN_n}$ respectively, and shown for $S = 2$ in Fig.~\ref{fig:Spin2SMagnon}b. With this, we find the scattering equation of $\ket{SM_n}$,
\begin{equation}
    H^{(S)} \ket{SM_n} = 2 \ket{SM_n} + \lambda_S (\ket{SN_{n-1}} + \ket{SN_n}).
\label{Mnscattering}
\end{equation}
In the above equation, the precise value of the coefficient $\lambda_S$ does not matter. For $S = 2$, $\lambda_{S = 2} = -2/7$ but it is hard to obtain a closed form for $\lambda$ in terms of $S$ since it involves the normal ordering recursion relations Eq.~(\ref{normrecrel}). From Eq.~(\ref{Mnscattering}), the exact state is
\begin{equation}
\ket{SS_2} = \sum_n{(-1)^n \ket{SM_n}}.
\end{equation}
In terms of spins, this can be expressed as
\begin{equation}
    \widetilde{\ket{SS_2}} = \mathcal{N} \sum_{n= 1}^L{(-1)^n (S^+_n)^{2S}}\widetilde{\ket{SG}}
\label{spin2smagnonopexp}
\end{equation}
where $\mathcal{N}$ is a normalization factor. Thus, we have an exact state for the spin-$S$ AKLT model that closely resembles the spin-2 magnon of the spin-1 AKLT model. This state has $\boldsymbol{E = 2}$, $\boldsymbol{k = \pi}$ and $\boldsymbol{s = 2S}$. Similar to the spin-2 magnon of the spin-1 AKLT model, the spin-$2S$ magnon generalizes for open boundary conditions as well, see Appendix~\ref{sec:obcspinStowerofstates}.
As for the spin-1 AKLT model tower of states, the state with $N$ spin-2$S$ magnons on the ground state is also exact for the spin-$S$ AKLT model. We denote the spin-2$S$ magnon creation operator as $\mcP = \sum_{n = 1}^L{(-1)^n (S_n^+)^{2S}}$. The tower of exact states is then
\begin{equation}
    \ket{SS_{2N}} = \mcP^N\ket{SG}.
\end{equation}
where $0 \leq N \leq L/2$. $\ket{S_{2N}}$ has $k = 0$ or $\pi$ depending on whether $N$ is odd or even, a total spin \boldsymbol{$s = 2SN$}, and an energy of \boldsymbol{$E = 2N$}. This tower of states connects the ground state to the ferromagnetic state. The proof proceeds exactly in the same way as for the spin-1 AKLT tower of states. As with the spin-1 AKLT tower of states in Sec.~\ref{sec:position}, we conjecture that the tower of states for any spin $S$ lies in the bulk of the energy spectrum, although it is hard to obtain any strong numerical evidence of this for $S > 1$.

\subsection{Upper spectrum states}
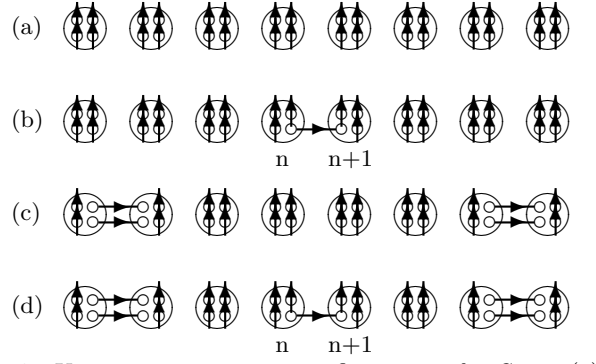
\begin{figure}[t!]
\centering
\begin{picture}(150,20)(0,0)
    \multiput(0,10)(25,0){8}{\circle{4}} 
    \multiput(6,10)(25,0){8}{\circle{4}} 
    \multiput(0,4)(25,0){8}{\circle{4}} 
    \multiput(6,4)(25,0){8}{\circle{4}} 
    \multiput(3,7)(25,0){8}{\circle{16}}
    \thicklines
    \multiput(6,5)(25,0){8}{\vector(0,1){12}} 
    \multiput(0,5)(25,0){8}{\vector(0,1){12}} 
    \multiput(0,-1)(25,0){8}{\vector(0,1){12}}
    \multiput(6,-1)(25,0){8}{\vector(0,1){12}}
    \put(-25,5){(a)}
\end{picture}\\[5mm]

\begin{picture}(150,20)(0,0)
    \multiput(0,10)(25,0){8}{\circle{4}} 
    \multiput(6,10)(25,0){8}{\circle{4}} 
    \multiput(0,4)(25,0){8}{\circle{4}} 
    \multiput(6,4)(25,0){8}{\circle{4}} 
    \multiput(3,7)(25,0){8}{\circle{16}}
    \thicklines
    \multiput(83,4)(25,0){1}{\line(1,0){15}}
    \multiput(95,4)(25,0){1}{\vector(1,0){0}}
    \multiput(6,5)(25,0){8}{\vector(0,1){12}} 
    \multiput(0,5)(25,0){8}{\vector(0,1){12}}
    \multiput(0,-1)(25,0){4}{\vector(0,1){12}}
    \multiput(6,-1)(25,0){3}{\vector(0,1){12}}
    \multiput(125,-1)(25,0){3}{\vector(0,1){12}}
    \multiput(106,-1)(25,0){4}{\vector(0,1){12}}
    \put(75,-10){n}
    \put(95,-10){n+1}
    \put(-25,5){(b)}
    \end{picture}\\[5mm]
    
    \begin{picture}(150,20)(0,0)
    \multiput(0,10)(25,0){8}{\circle{4}} 
    \multiput(6,10)(25,0){8}{\circle{4}} 
    \multiput(0,4)(25,0){8}{\circle{4}} 
    \multiput(6,4)(25,0){8}{\circle{4}} 
    \multiput(3,7)(25,0){8}{\circle{16}}
    \thicklines
    \multiput(31,5)(25,0){5}{\vector(0,1){12}}
    \multiput(181,5)(25,0){1}{\vector(0,1){12}}
    \multiput(31,-1)(25,0){5}{\vector(0,1){12}}
    \multiput(181,-1)(25,0){1}{\vector(0,1){12}}
    \multiput(0,5)(25,0){1}{\vector(0,1){12}}
    \multiput(50,5)(25,0){5}{\vector(0,1){12}}
    \multiput(0,-1)(25,0){1}{\vector(0,1){12}}
    \multiput(50,-1)(25,0){5}{\vector(0,1){12}}
    \put(-25,5){(c)}
    \multiput(8,10)(150,0){2}{\line(1,0){15}}
    \multiput(20,10)(150,0){2}{\vector(1,0){0}}
    \multiput(8,4)(150,0){2}{\line(1,0){15}}
    \multiput(20,4)(150,0){2}{\vector(1,0){0}}
    \end{picture}\\[5mm]
    
    \begin{picture}(150,20)(0,0)
    \linethickness{0.8pt}
    \multiput(0,10)(25,0){8}{\circle{4}} 
    \multiput(6,10)(25,0){8}{\circle{4}} 
    \multiput(0,4)(25,0){8}{\circle{4}} 
    \multiput(6,4)(25,0){8}{\circle{4}} 
    \multiput(3,7)(25,0){8}{\circle{16}}
    \thicklines
    \multiput(83,4)(25,0){1}{\line(1,0){15}}
    \multiput(95,4)(25,0){1}{\vector(1,0){0}}
    \multiput(0,5)(25,0){1}{\vector(0,1){12}}
    \multiput(50,5)(25,0){5}{\vector(0,1){12}}
    \multiput(31,5)(25,0){5}{\vector(0,1){12}}
    \multiput(181,5)(25,0){1}{\vector(0,1){12}}
    \multiput(0,-1)(25,0){1}{\vector(0,1){12}}
    \multiput(31,-1)(25,0){2}{\vector(0,1){12}}
    \multiput(125,-1)(25,0){2}{\vector(0,1){12}}
    \multiput(106,-1)(25,0){2}{\vector(0,1){12}}
    \multiput(50,-1)(25,0){2}{\vector(0,1){12}}
    \multiput(181,-1)(25,0){1}{\vector(0,1){12}}
    \put(75,-10){n}
    \put(95,-10){n+1}
    \put(-25,5){(d)}
    \multiput(8,10)(150,0){2}{\line(1,0){15}}
    \multiput(20,10)(150,0){2}{\vector(1,0){0}}
    \multiput(8,4)(150,0){2}{\line(1,0){15}}
    \multiput(20,4)(150,0){2}{\vector(1,0){0}}

    \end{picture}
    
\caption{Upper spectrum state configurations for $S = 2$ (a) The ferromagnetic state with $s = 2L$ and $E = L$. (b) An example of a non-scattering  configuration with $s = 2L - 1$, $E = L$. (c) A configuration that forms an exact state for $S = 2$ with $E = L - 2$, $s = 2L - 4$. (d) A dressed configuration that forms an exact state with $E = L - 2$, $s = 2L - 5$.}
\label{fig:nonscattering}
\end{figure}
In this section, we briefly comment on the structure of the simple upper spectrum excited states of the spin-$S$ AKLT model.
For the spin-$S$ AKLT model, there is no unique highest excited multiplet.
To see this, note that apart from a constant, the Hamiltonian Eq.~(\ref{SHamiltoniandimerspecial}) contains terms $(c_{ij}^\dagger)^m (c_{ij})^m$ for $m \geq S$. According to Eq.~(\ref{nonscattering}), all such terms vanish on any configuration with $s < S$ dimers on one bond. 
For example, the configuration shown in  Fig.~\ref{fig:nonscattering}b does not scatter and contributes one unit of energy, same as the fully ferromagnetic state (Fig.~\ref{fig:nonscattering}a).
We dub such configurations ``non-scattering".
Non-scattering configurations at least two bonds apart from each other lead to several excited states, all of which have $E = L$ using the specific set of $\alpha_J$ in the Hamiltonian Eq.~(\ref{SHamiltonian}) described previously. However, these non-scattering configurations have different total spins. 
In the previous section, we saw that the spin-$S$ AKLT tower of states can be obtained by replacing each dimer in spin-1 AKLT by $S$ dimers.
The construction worked because of the similarity of Eqs.~(\ref{singlethopping}) and (\ref{highspinhopping}) and the structure of the Hamiltonian Eq.~(\ref{SHamiltoniandimerspecial}).
Since Eq.~(\ref{singlethopping}) was crucial in obtaining the upper spectrum excited states for the spin-1 AKLT model, upper spectrum states of spin-$S$ AKLT models can be obtained similarly with Eq.~(\ref{highspinhopping}).
Thus, all the upper spectrum states of spin-1 AKLT model that used Eq.~(\ref{singlethopping}) in the scattering ($\ket{1_k}$, $\ket{2_n}$ and $\ket{2_k}$) have an analogue in the spin-$S$ AKLT model, replacing each dimer of the spin-1 model by $S$ dimers in the spin-$S$ model and spins $L - 1$ and $L - 2$ by $S L - S$ and $S L - 2S$ respectively.
For example, the configuration in Fig.~\ref{fig:nonscattering}c forms the $S = 2$ analogue of the $\ket{2_{n = 6}}$ state.
The exceptions to this rule are $\ket{2_{n = 0}}$, that uses a different scattering rule (Eq.~(\ref{double}) instead of Eq.~(\ref{singlethopping}) used for the other $\ket{2_n}$ states) and has no analogue in spin-$S$ and the special state $\ket{6_0}$.
Special states could be obtained for the $s = L S - 2$ sector the spin-$S$ AKLT too, where the Hamiltonian would be tri-diagonal.
In addition to the generalized versions of the spin-1 upper spectrum excited states, such states could always be dressed symmetrically with non-scattering configurations at least two bonds away from the scattering ones.
For example, the configuration shown in Fig.~\ref{fig:nonscattering}d is simply Fig.~\ref{fig:nonscattering}c dressed with a non-scattering configuration. Such a dressed configuration has the same energy as the parent configuration but a lower spin.
Thus, the upper spectrum of the spin-$S$ AKLT models are massively degenerate.
The states described above account for all of the rational energy states we observe numerically for $S = 2$. 
As mentioned earlier for spin-1 models, the quantum number sectors $(L-2) \leq s \leq L$ sectors are fully integrable.\cite{bibikov2016three} Similarly, for spin-$S$ models, it might be the case that all the spin sectors $S (L - 2) \leq s \leq SL$ are fully integrable. In such a case, the states we have described are some examples of states that have a simple analytical expression for a finite system. This is an interesting question for future work.
%
%
%
%
%
\section{Projection Principle}\label{sec:projection}
Spin-$S$ AKLT exact states have intriguing connections to the spin-1 AKLT exact states (\eg see Eq.~\eqref{SBnconfig}). Some of the exact spin-$S$ eigenstates can be constructed from other exact spin-$S'$ eigenstates with $S' < S$. This notion can be formalized via the {\it projection principle} (PP). Originally introduced for spin singlet groundstate wavefunctions\,\cite{greiter02jltp1029,greiter2007valence,rachel-09prb180420}, here we show that the PP can be applied to excited states as well. In particular, it can be applied to states with non-zero magnetization, i.e. non-singlet states. In the following, we demonstrate that the PP also holds beyond integer spin models. We briefly introduce the Majumdar--Ghosh model and its ground states, which we use to demonstrate the PP by constructing the AKLT ground state. Subsequently we discuss how it can be applied to exact excited eigenstates.
\subsection{Majumdar--Ghosh Model}
\label{sec:proj:mg}
Majumdar and Ghosh\,\cite{majumdar1969next} (MG) noticed in 1969 that on a  spin $S = \frac{1}{2}$ chain, the two valence bond solid or dimer states
\begin{eqnarray}
\big|{\text{MG}}^{\textrm{even}\rule{0pt}{5pt}\atop 
\textrm{(odd)}}\big\rangle \label{eq:mg1}
&=&
\prod_{i\ \textrm{even}\atop (i\ \textrm{odd})}  
c^\dag_{i,i+1}
\vac \\
&=&\left\{ \begin{array}{lc}
\ket{\hbox{\begin{picture}(102,8)(-4,-3)
\put(13,0){\makebox(0,0){\rule{10.pt}{ 0.8pt}}}
\put(41,0){\makebox(0,0){\rule{10.pt}{ 0.8pt}}}
\put(69,0){\makebox(0,0){\rule{10.pt}{ 0.8pt}}}
\put(6,0){\circle{4}}
\put(20,0){\circle{4}}
\put(34,0){\circle{4}}
\put(48,0){\circle{4}}
\put(62,0){\circle{4}}
\put(76,0){\circle{4}}
\put(90,0){\circle{4}}
\end{picture}}} 
&\quad\textrm{``even''}\quad\\
\ket{\hbox{\begin{picture}(102,8)(-4,-3)
\put(27,0){\makebox(0,0){\rule{10.pt}{ 0.8pt}}}
\put(55,0){\makebox(0,0){\rule{10.pt}{ 0.8pt}}}
\put(83,0){\makebox(0,0){\rule{10.pt}{ 0.8pt}}}
\put(6,0){\circle{4}}
\put(20,0){\circle{4}}
\put(34,0){\circle{4}}
\put(48,0){\circle{4}}
\put(62,0){\circle{4}}
\put(76,0){\circle{4}}
\put(90,0){\circle{4}}
\end{picture}}} 
&\quad\textrm{``odd''}\quad\rule{0pt}{15pt}
\end{array}\right.\rule{0pt}{25pt}
\label{eq:mg2}
\end{eqnarray}
where the product runs over all even sites $i$ for one state and over
all odd sites for the other, are exact zero-energy ground
states of the  Hamiltonian
\begin{equation}
H_{\text{MG}} = 
\sum_i \left(\vec{S}_i.\vec{S}_{i+1} +
\frac{1}{2}\vec{S}_i.\vec{S}_{i+2} +\frac{3}{8}\right).
 \label{eq:hmg}
\end{equation}
Note that Eq.~(\ref{eq:mg1}) implies that only one Schwinger boson operator ($a_i^\dagger \ket{\theta}_i = \ket{\uparrow}_i$  or $b_i^\dagger \ket{\theta}_i = \ket{\downarrow}_i$ where $\vac_i$ is the local vacuum of site $i$) is applied per lattice site $i$ leading to spin-1/2 states, unlike the spin-1 (spin-$S$) AKLT case where two ($2S$) Schwinger bosons act on each lattice site. The proof is simple.  We rewrite $H_{\textrm{MG}}=\frac{3}{4}\sum_i H_i$ with
\begin{equation}
H_i= P^{\left( \frac{3}{2},\frac{1}{2}\right)}_{i,i+1,i+2} = \frac{1}{3}\left[\bigl(\vec{S}_i +\vec{S}_{i+1} 
+\vec{S}_{i+2}\bigr)^2 -\frac{3}{4}\right],
\end{equation}
where $P^{\left( \frac{3}{2},\frac{1}{2}\right)}_{i,i+1,i+2}$ is the spin-1/2 projector onto total spin 3/2 of the three sites $i$, $i+1$, and $i+2$.
Hence, any state in which the total spin of three adjacent spins
$\frac{1}{2}$ is annihilated by $H_i$.  (The total spin can
only be $\frac{1}{2}$ or $\frac{3}{2}$, as
$\bf\frac{1}{2}\otimes\frac{1}{2}\otimes\frac{1}{2}=
\frac{1}{2}\oplus\frac{1}{2}\oplus\frac{3}{2}$.)  For the dimer states above, this is guaranteed as two of the three neighboring spins are in a singlet configuration, and 
$\bf 0\otimes\frac{1}{2}=\frac{1}{2}$.  Graphically, we may express
this as
\begin{equation}
H_i \ket{\hbox{\begin{picture}(40,8)(0,-3)
\put(13,0){\makebox(0,0){\rule{10.pt}{ 0.8pt}}}
\put(6,0){\circle{4}}
\put(20,0){\circle{4}}
\put(34,0){\circle{4}}
\end{picture}}}
= H_i \ket{\hbox{\begin{picture}(40,8)(0,-3)
\put(27,0){\makebox(0,0){\rule{10.pt}{ 0.8pt}}}
\put(6,0){\circle{4}}
\put(20,0){\circle{4}}
\put(34,0){\circle{4}}
\end{picture}}}=0.
\end{equation}
 As $H_i$ is positive definite, the two zero-energy eigenstates of
$H_{\textrm{MG}}$, shown in Eq.~(\ref{eq:mg2}) are also ground states. Note that the two ground states can be written as two translation invariant states with $k = 0$ and $k = \pi$ as 
\begin{eqnarray}
    \ket{\text{MG}^+} = \ket{\text{MG}^{{\textrm{even}}}} + \ket{\text{MG}^{\textrm{odd}}}, \nn \\
    \ket{\text{MG}^-} = \ket{\text{MG}^{\textrm{even}}} - \ket{\text{MG}^{\textrm{odd}}}.
\end{eqnarray}

\subsection{AKLT Ground state from the Projection Principle}
\label{sec:proj:aklt}

In the following, we slightly modify our cartoon pictures used previously for the AKLT states:
\begin{equation*}
\setlength{\unitlength}{1pt}
\begin{picture}(220,25)(0,-1)
\linethickness{0.8pt}
\multiput(0,10)(25,0){3}{\circle{4}} 
\multiput(6,10)(25,0){3}{\circle{4}} 
\multiput(3,10)(25,0){3}{\circle{14}}
\multiput(90,7)(25,0){3}{\circle{4}} 
\multiput(90,13)(25,0){3}{\circle{4}} 
\multiput(90,10)(25,0){3}{\circle{14}}
\multiput(178,7)(20,0){3}{\circle{4}}
\multiput(178,13)(20,0){3}{\circle{4}}
\thicklines
\put(-8,10){\line(1,0){6}}
\multiput(8,10)(25,0){2}{\line(1,0){15}} 
\multiput(20,10)(25,0){2}{\vector(1,0){0}} 
\put(58,10){\line(1,0){6}}
\put(81,7){\line(1,0){7}}
\multiput(92,13)(50,0){1}{\line(1,0){21}} 
\multiput(106,13)(50,0){1}{\vector(1,0){0}} 
\multiput(117,7)(50,0){1}{\line(1,0){21}} 
\multiput(131,7)(50,0){1}{\vector(1,0){0}} 
\put(142,13){\line(1,0){7}}
\put(170,7){\line(1,0){6}}
\multiput(180,13)(40,0){1}{\line(1,0){16}}
\multiput(200,7)(40,0){1}{\line(1,0){16}}
\put(220,13){\line(1,0){6}}
\put(73,11){\makebox(0,0)[c]{$\boldsymbol{\widehat{=}}$}}
\put(160,11){\makebox(0,0)[c]{$\boldsymbol{\widehat{=}}$}}
\end{picture}
\end{equation*}
Note that their physical interpretation remains unchanged. From the figure, note that the two spin 1/2 MG ground states of Eq.~\eqref{eq:mg1} can be glued together to form the spin-1 AKLT ground state Eq.~\eqref{groundstate},
\begin{eqnarray}
\ket{G} \, &=&
\prod_{i\ \textrm{even}}  c^\dag_{i,i+1} \cdot \prod_{i\ \textrm{odd}}  c^\dag_{i,i+1} \vac \nonumber \\
&=&
\setlength{\unitlength}{1pt}
\Big|
\begin{picture}(128,20)(-7,4)
\linethickness{0.8pt}
\multiput(0,10)(14,0){9}{\circle{4}}
\multiput(0,2)(14,0){9}{\circle{4}}
\thicklines
\multiput(2,10)(28,0){4}{\line(1,0){10}}
\multiput(16,2)(28,0){4}{\line(1,0){10}}
\thinlines
\put(37,-3){\line(1,0){10}}
\put(37,-3){\line(0,1){18}}
\put(37,15){\line(1,0){10}}
\put(47,-3){\line(0,1){18}}
\put(42,-3){\line(0,-1){7}}
\put(72,-15){\makebox(0,0)[c]{\small projection onto spin 1}}
\end{picture}\Big\rangle \label{PPMGGS}\\[10mm]
&\equiv&
\ket{\text{MG}^{\textrm{even}}} \odot |\text{MG}^{\textrm{odd}}\rangle. \label{MGe+MGo}
\end{eqnarray}
In terms of Schwinger bosons the projection onto the fully symmetric spin 1 subspace in Eq.~(\ref{PPMGGS}) is accomplished automatically due to their bosonic character. We can immediately generalize this construction to higher spin-$S$ AKLT ground states,
\begin{equation}
\ket{SG} = \ket{\text{MG}^{\textrm{even}}}^{\odot S} \odot |\text{MG}^{\textrm{odd}}\rangle^{\odot S}
\label{spinSGS}
\end{equation}
where we introduced
\begin{equation}
\ket{\psi}^{\odot n} \equiv \underbrace{\ket{\psi}\odot\ldots\odot \ket{\psi}}_{n-{\rm times}}.
\end{equation}
Applying Eq.~\eqref{MGe+MGo} to Eq.~(\ref{spinSGS}), we find that we can glue $S$ spin-1 AKLT ground states together to obtain one spin-$S$ AKLT ground state, $\ket{SG}=\ket{G}^{\odot S}$. The PP can thus be applied to obtain arbitrary spin-$S$ AKLT ground states.

\subsection{Projection Principle for Excited Singlet States}

We now establish that it is possible to combine MG states via PP to obtain the AKLT exact eigenstates that are obtained by a single-mode approximation, viz. the Arovas states and the spin-$2S$ magnon state. To be more precise, for the AKLT exact states that are of the form $\ket{\Xi_k} = \sum_n e^{ikn} \ket{\xi_n}$ (\eg for the Arovas A state of Eq.~(\ref{pisuparovas}),  $\ket{\xi_n} = \ket{A_n}$, $k = \pi$ and $\ket{\Xi_k} = \ket{A}$), we apply the projection principle to the configurations $\ket{\xi_n}$ but not directly to the eigenstates $\ket{\Xi_k}$. 

Certain exact excited states for the MG model have been obtained previously.\cite{caspers1982some,caspers1984majumdar} For instance, the state $\ket{\phi}$ of Eq.~(9) in Ref.~[\onlinecite{caspers1982some}] can be re-written as 
\begin{equation}
\ket{X_k} = \sum_n e^{i k n} \ket{x_n} \;\;  k = \frac{\pi}{2}, \frac{3\pi}{2},
\label{eq:mgsinglet}
\end{equation}
where $\ket{x_n}$ is defined as (see Appendix~\ref{mgdimer}, Eq.~(\ref{xkorig}))
        \begin{equation}
        \begin{picture}(160,20)(0,0)
        \put(-20,10){\makebox(0,0)[c]{$\ket{x_n} = \Big|$}}
        \linethickness{0.8pt}
        \multiput(0,10)(25,0){8}{\circle{4}} 
        \put(2,10){\line(1,0){21}}
        \put(77,10){\line(1,0){21}}
        \put(152,10){\line(1,0){21}}
        \qbezier(52,11)(87.5,30)(123,11)
        \put(75,3){\makebox(0,0)[c]{n}}
        \put(100,3){\makebox(0,0)[c]{n+1}}
        \put(182,10){\makebox(0,0)[c]{$\Big\rangle$}}
        \end{picture}
        \label{mgxn}
        \end{equation}
Careful analysis reveals that $\ket{A_n} = \ket{x_n} \odot \ket{\text{MG}^{\textrm{Parity(n)}}}$ 
resulting in 
\begin{equation}
\begin{picture}(190,20)(-20,0)
\put(-9,10){\makebox(0,0)[c]{$\Big|$}}
\linethickness{0.8pt}
\multiput(0,13)(25,0){8}{\circle{4}} 
\multiput(0,7)(25,0){8}{\circle{4}}
\put(2,13){\line(1,0){21}}
\put(77,13){\line(1,0){21}}
\put(152,13){\line(1,0){21}}
\multiput(27,7)(50,0){3}{\line(1,0){21}}
\qbezier(52,14)(87.5,32)(123,14)
\put(75,-3){\makebox(0,0)[c]{n}}
\put(100,-3){\makebox(0,0)[c]{n+1}}
\put(182,10){\makebox(0,0)[c]{$\Big\rangle$}}
\thinlines
\put(120,0){\line(1,0){10}}
\put(120,0){\line(0,1){18}}
\put(120,18){\line(1,0){10}}
\put(130,0){\line(0,1){18}}
\put(125,0){\line(0,-1){8}}
\put(132,-15){\makebox(0,0)[c]{\small projection onto spin 1}}
\end{picture}\\[5mm]
\end{equation}
is identical to the Arovas $A$ configuration $\ket{A_n}$ shown in Fig.\,\ref{fig:arovasA}a. Thus the PP can also be used for excited states. 

In the following, we show that all Arovas-type eigenstates can be constructed via projection principle. For instance, the Arovas $B$ state configuration reads $\ket{B_n} = \ket{x_{n-1}} \odot \ket{x_{n}}$, 
\begin{equation}
\begin{picture}(190,20)(-20,0)
        \put(-9,10){\makebox(0,0)[c]{$\Big|$}}
        \linethickness{0.8pt}
        \multiput(0,13)(25,0){8}{\circle{4}} 
        \multiput(0,7)(25,0){8}{\circle{4}}
        \put(2,13){\line(1,0){21}}
        \put(77,13){\line(1,0){21}}
        \put(152,13){\line(1,0){21}}
        \multiput(52,7)(75,0){2}{\line(1,0){21}}
        \qbezier(52,14)(87.5,26)(123,14)
        \qbezier(27,7)(62.5,-5)(98,7)
        \put(50,-5){\makebox(0,0)[c]{n-1}}
        \put(75,-5){\makebox(0,0)[c]{n}}
        \put(100,-5){\makebox(0,0)[c]{n+1}}
        \put(182,10){\makebox(0,0)[c]{$\Big\rangle$}}
        \thinlines
        \put(120,0){\line(1,0){10}}
        \put(120,0){\line(0,1){18}}
        \put(120,18){\line(1,0){10}}
        \put(130,0){\line(0,1){18}}
        \put(125,0){\line(0,-1){8}}
        \put(132,-18){\makebox(0,0)[c]{\small projection onto spin 1}}
        \end{picture}\\[5mm]
\end{equation}
As discussed before, the spin-2 Arovas configuration can be written as $\ket{2B_n} = 2 \ket{BG_n} - \frac{1}{2}\ket{B_n^2}$, see Eq.\,\eqref{2Bconfig}. In terms of the PP for spin-1 configuration, this can be formulated as
\begin{equation}
\ket{2B_n} = 2 \ket{G} \odot \ket{B_n} - \frac{1}{2} \ket{B_n}\odot \ket{B_n}
\end{equation}
where $\ket{G}$ and $\ket{B_n}$ are the spin-1 AKLT ground state and the Arovas B state respectively. 
This can be further generalized to express the spin-$S$ Arovas state configuration in terms of the PP as Eq.~(\ref{SBnconfig}),
\begin{equation}
\ket{SB_n} = \sum_{m=1}^S \frac{(-1)^m}{m} {S \choose m} \left[ \ket{G}^{\odot (S-m)} \, \odot \, \ket{B_n}^{\odot m} \right].
\tag{\ref{SBnconfig}}
\end{equation}
\subsection{Projection Principle for Non-Singlet States}
So far, we have applied the PP only to ground states and excited states which were spin singlets. Now we consider the simplest type of non-singlet states, and express them through the PP. Another MG eigenstate  which was found to be exact in Eq.~(8) of Ref.~[\onlinecite{caspers1982some}] can be rewritten as
\begin{equation}
    \ket{T_k} = \sum_n{e^{i k n} \ket{t_n}} \;\; k = \frac{\pi}{2}, \frac{3\pi}{2}
\label{eq:mgtriplet}
\end{equation} 
where (see Appendix~\ref{mgdimer}, Eq.~(\ref{tkorig}))
\begin{equation}
\begin{picture}(135,20)(-20,0)
\put(-20,10){\makebox(0,0)[c]{$\ket{t_n} = \big|$}}
\linethickness{0.8pt}
\multiput(0,10)(25,0){6}{\circle{4}} 
\multiput(2,10)(100,0){2}{\line(1,0){21}}
\put(133,10){\makebox(0,0)[c]{$\big\rangle$}}
\put(50,5){\vector(0,1){12}} 
\put(75,5){\vector(0,1){12}} 
\put(52,0){\makebox(0,0)[c]{n}}
\put(77,0){\makebox(0,0)[c]{n+1}}
\end{picture}\\[5mm]
\label{mgtn1}
\end{equation}
Using the PP we can readily express the magnon configuration $\ket{M_n}$ of Eq.~(\ref{npnmdimer}) for the spin-1 AKLT model as $\ket{M_n} = \ket{t_{n-1}} \odot \ket{t_{n}}$,
\begin{equation}
        \begin{picture}(190,20)(-20,0)
        \put(-9,10){\makebox(0,0)[c]{$\Big|$}}
        \linethickness{0.8pt}
        \multiput(0,14)(25,0){7}{\circle{4}} 
        \multiput(27,14)(100,0){2}{\line(1,0){21}}
        \put(75,9){\vector(0,1){12}} 
        \put(100,9){\vector(0,1){12}} 
        \multiput(0,3)(25,0){7}{\circle{4}} 
        \multiput(2,3)(100,0){2}{\line(1,0){21}}
        \put(50,-2){\vector(0,1){12}} 
        \put(75,-2){\vector(0,1){12}}
        \put(158,10){\makebox(0,0)[c]{$\Big\rangle$}}
        \put(52,-7){\makebox(0,0)[c]{n-1}}
        \put(77,-7){\makebox(0,0)[c]{n}}
        \put(102,-7){\makebox(0,0)[c]{n+1}}
        \thinlines
        \put(120,-2){\line(1,0){10}}
        \put(120,-2){\line(0,1){22}}
        \put(120,20){\line(1,0){10}}
        \put(130,-2){\line(0,1){22}}
        \put(125,-2){\line(0,-1){8}}
        \put(132,-20){\makebox(0,0)[c]{\small projection onto spin 1}}
        \end{picture}\\[6mm]
    \label{mgtn}
\end{equation}
leading to the spin-2 magnon state under momentum superpositions, $\ket{S_2} = \sum_n (-1)^n \ket{M_n}$. For higher spin-$S$ AKLT chains, magnon states with spin-$2S$ can be constructed from the PP. For instance, in the spin-2 AKLT chain we obtain the spin-4 magnon state via 
\begin{equation}
\ket{M_n}\odot\ket{M_n}=\Big|
    \setlength{\unitlength}{1pt}
    \begin{picture}(128,20)(15,4)
    \linethickness{0.8pt}
    \multiput(25,10)(25,0){5}{\circle{4}} 
    \multiput(31,10)(25,0){5}{\circle{4}} 
    \multiput(25,4)(25,0){5}{\circle{4}} 
    \multiput(31,4)(25,0){5}{\circle{4}} 
    \multiput(28,7)(25,0){5}{\circle{16}}
    \thicklines
    \put(23,10){\line(-1,0){6}}
    \put(23,4){\line(-1,0){6}}
    \multiput(33,10)(25,0){1}{\line(1,0){15}} 
    \multiput(45,10)(25,0){1}{\vector(1,0){0}}
    \multiput(33,4)(25,0){1}{\line(1,0){15}} 
    \multiput(45,4)(25,0){1}{\vector(1,0){0}}
    \multiput(108,10)(25,0){1}{\line(1,0){15}}
    \multiput(120,10)(25,0){1}{\vector(1,0){0}}
    \multiput(108,4)(25,0){1}{\line(1,0){15}}
    \multiput(120,4)(25,0){1}{\vector(1,0){0}}
    \multiput(56,5)(25,0){2}{\vector(0,1){12}} 
    \multiput(75,5)(25,0){2}{\vector(0,1){12}} 
    \multiput(56,-1)(25,0){2}{\vector(0,1){12}} 
    \multiput(75,-1)(25,0){2}{\vector(0,1){12}} 
    \put(133,10){\line(1,0){6}}
    \put(133,4){\line(1,0){6}}
    \put(75,-10){n}
    \put(47,-10){n-1}
    \put(95,-10){n+1}
      \end{picture}
      \Big\rangle.\\[7pt]
    \end{equation}
Since the MG model is $SU(2)$ symmetric, the ferromagnetic state (state with all spins having $S_z = +1/2$) is an exact eigenstate. The same symmetry dictates that the exact quasiparticle excitations around the ferromagnet are one-dimer configurations analogous to the $\ket{1_k}$ states of the spin-1 AKLT model. With these upper spectrum exact states of the MG model, the ferromagnetic and the one dimer configurations of the AKLT model (Fig.~\ref{fig:1kstates}), and the two dimer configurations (\eg Fig.~\ref{fig:2singletconfig}) can be trivially constructed via PP. 
\section{Conclusion}\label{sec:conclusion}
In this article, we have for the first time derived a tower exact eigenstates with a closed-form expression in non-integrable models, the spin-$S$ AKLT models. For that purpose, we first used finite size exact diagonalizations to look for states with a low rank of their reduced density matrix. These turned out to usually coincide with a rational or even an integer energy. For each of them, we have then derived an analytical formula. Apart from the tower of states from the ground state to the highest excited state, we have obtained several exact excited states in the low energy and the high energy spectrum. 

Our approach could potentially be applied to any non-integrable model, irrespective of its dimensionality. Numerically, for certain system sizes, we do see exact states of the Majumdar-Ghosh model (including the ones obtained by Caspers \ea\cite{caspers1982some}), spin-3/2 model\cite{rachel09epl37005} and of the spin-1 Heisenberg model. However, the exact states in those models do not include an infinite tower of states, as the AKLT models do. In the context of the Eigenstate Thermalization Hypothesis, some of the exact states we have obtained seem to be located in the bulk of the spectrum but still have non-generic entanglement properties,\cite{selftoappear} which questions the strong ETH. As we discuss in another article,\cite{selftoappear} those states have a low entanglement entropy, and it is unlikely that any of them would be thermal. Our results pave the way for the search of non-integrable models that provide some analytical insight on the Eigenstate Thermalization Hypothesis. They also suggest that a special class of ``semi-solvable" but non-integrable spin models could exist.\cite{turbiner2016one, bibikov2016three, bibikov2017bethe} These models would contain thermal and non-thermal eigenstates mixed together,\cite{shiraishi2017systematic, turner2018weak, turner2018quantum} something that is usually thought to be impossible in non-integrable models.
\section*{Acknowledgements}
We thank Paul Fendley for encouragement and initial discussions on this work, Elliot Lieb and David Huse for useful discussions. B. A. Bernevig wishes to thank Ecole Normale Superieure, UPMC Paris, and Donostia International Physics Center for their generous sabbatical hosting during some of the stages of this work. SR acknowledges funding from the DFG through SFB 1143. BAB acknowledges support for the analytic work Department of Energy de-sc0016239, Simons Investigator Award, the Packard Foundation, and the Schmidt Fund for Innovative Research. The computational part of the Princeton work was performed under NSF EAGER grant DMR-1643312, ONR-N00014-14-1-0330, ARO MURI W911NF-12-1-0461, NSF-MRSEC DMR-1420541.
\appendix
\section{Spin basis}\label{spins}
If all the spins are represented in the $S_z$ basis, and $\ket{S\ m_1, S\ m_2} \equiv \ket{m_1, m_2}$ represents the spins $i$ and $j$ having $S_z = m_1, m_2$, the rules for scattering by the action of the projector of two spin $S$ onto a spin-$J$ state (defined as $\ket{J\ J_z}_{JM}$) can be succinctly written as
\begin{eqnarray}
    &P^{(J,S)}_{ij} \ket{m_1, m_2} = \braket{J\ m_1 + m_2}{m_1, m_2} \ket{J\ m_1+m_2}_{JM} \nn \\
    &= \left(\braket{J\ m_1 + m_2}{m_1, m_2}\right. \nn \\
    &\sumal{m = 0}{m_1 + m_2}{\left(\braket{m, m_1 + m_2 - m}{J\ m_1 + m_2}\right.} \nn \\
    &\left.\left.\ket{m, m_1 + m_2 - m}\right)\right)
\label{projcg}
\end{eqnarray}
where  $\braket{J\ m_1 + m_2}{m_1, m_2}$ ($\braket{m, m_1 + m_2 - m}{J\, m_1 + m_2}$) is the Clebsh-Gordan coefficient for adding two spin-$S$ with $S_z = m_1$ and $S_z = m_2$ ($S_z = m$ and $S_z = m_1 + m_2 - m$) to obtain a spin-$J$ state with $S_z = m_1 + m_2$ ($\ket{J, m_1 + m_2}$).  When Eq.~(\ref{projcg}) is written down explicitly for $J = 2$ and $S = 1$ (spin-1 AKLT model), these yield
\begin{eqnarray}
&&P^{(2,1)}_{ij} \ket{1\ 1} = \ket{1\ 1} \label{spin11}\\
&&P^{(2,1)}_{ij} \ket{1\ 0} = P^{(2,1)}_{ij} \ket{0\ 1} = \frac{1}{2}(\ket{1\ 0} + \ket{0\ 1}) \label{spin10}\\
&&P^{(2,1)}_{ij} \ket{1\ \mm} = P^{(2,1)}_{ij} \ket{\mm\ 1} \nn \\
&& \hspace{19mm}= \frac{1}{6}(\ket{1\ \mm} + \ket{\mm\ 1}) + \frac{1}{3}\ket{0\ 0} \label{spin1m1}\\
&&P^{(2,1)}_{ij} \ket{0\ 0} = \frac{2}{3} \ket{0\ 0} + \frac{1}{3}(\ket{1\ \mm} + \ket{\mm\ 1}) \label{spin00}\\
&&P^{(2,1)}_{ij} \ket{0\ \mm} = P^{(2,1)}_{ij} \ket{\mm\ 0} = \frac{1}{2}(\ket{0\ \mm} + \ket{\mm\ 0}) \label{spin0m1}\\
&&P^{(2,1)}_{ij} \ket{\mm \mm} = \ket{\mm \mm}
\label{spinm1m1}
\end{eqnarray}

\section{Fundamental algebra of dimers}\label{dimers}
\subsection{Commutation relations}
We start by defining operators that create and annihilate up ($a^\dagger$, $a$) or down ($b^\dagger$, $b$) spin-1/2 Schiwinger bosons. By virtue of being bosons, they obey the boson commutation relations as
\begin{eqnarray}
    &&[a_i, a_j^\dagger] = \delta_{ij} \nonumber \\
    &&[a_i, a_j] = 0 \nonumber \\
    &&[b_i, b_j^\dagger] = \delta_{ij} \nonumber \\
    &&[b_i, b_j] = 0 \nonumber \\
    &&[a_i^\dagger, b_j] = 0 \nonumber \\
    &&[a_i, b_j] = 0
\label{bosoncomm}
\end{eqnarray}
where $i$ is the site index. We can then define dimer (singlet) annihilation and creation operators between sites $i$ and $j$ as
\begin{eqnarray}
    &&c_{ij} = a_i b_j - a_j b_i \nonumber \\
    &&c^\dagger_{ij} = a^\dagger_i b^\dagger_j - a^\dagger_j b^\dagger_i. 
\label{dimercreate}
\end{eqnarray}
Since bosons on each site are identical, note that a singlet (antisymmetric) state within each site vanishes ($c^\dagger_{ii} = 0$). Using Eq.~(\ref{bosoncomm}), one can derive the algebra for the dimer operators as
\begin{eqnarray}
    &[c_{mn}, c_{ij}^\dagger] = 2 \delta_{m i} \delta_{nj} - 2 \delta_{ni} \delta_{mj}  \nonumber \\ 
    & + (a_i^\dagger a_m + b_i^\dagger b_m) \delta_{nj} + (a_j^\dagger a_n + b_j^\dagger b_n) \delta_{mi} \nonumber \\ 
    & - (a_i^\dagger a_n + b_i^\dagger b_n) \delta_{mj} - (a_j^\dagger a_m + b_j^\dagger b_m) \delta_{ni} \nonumber \\
    &[c_{mn}, c_{ij}] = 0.
\label{dimeralgebra}
\end{eqnarray}
For $m = i$ and $n = j$, Eq.~(\ref{dimeralgebra}) can be written as
\begin{equation}
    [c_{ij}, c^\dagger_{ij}] = 2 + N_i + N_j
\label{dimernum}
\end{equation}
where $N_i = a_i^\dagger a_i + b_i^\dagger b_i$, the total number operator of the Schwinger bosons on site $i$. Commutation relations between the remaining operators can also be computed as
\begin{eqnarray}
    &&{[a_i^\dagger, c_{mn}]} = -\delta_{im}b_n + \delta_{in}b_m \nonumber \\
    &&{[b_i^\dagger, c_{mn}]} = -\delta_{in}a_m + \delta_{im}a_n \nonumber \\
    &&{[a_i, c_{mn}]} = 0 \nonumber \\
    &&{[b_i, c_{mn}]} = 0 \nonumber \\
    &&{[N_i, c_{mn}^\dagger]} = (\delta_{in} + \delta_{im})c_{mn}^\dagger \nonumber \\
    &&{[N_i, a_m^\dagger]} = \delta_{im}a_m^\dagger \nonumber \\
    &&{[N_i, b_m^\dagger]} = \delta_{im}b_m^\dagger.
\label{abcncomm}
\end{eqnarray}
Eqs.~(\ref{bosoncomm}), (\ref{dimeralgebra}), (\ref{dimernum}) and (\ref{abcncomm}) along with their Hermitian conjugates specify the entire algebra of all the objects we are working with.
\subsection{Normal ordering}\label{dimersnorm}
For our calculations, it is useful to know the expressions for the normal ordering of the operator $(c_{ij}^\dagger c_{ij})^n$. To compute this, we first need to know the commutation relation $[c_{ij}, (c_{ij}^\dagger)^n]$. Using the commutation relations in Eqs.~(\ref{dimernum}) and (\ref{abcncomm}), this can be easily computed to be
\begin{equation}
    [c_{ij}, (c_{ij}^\dagger)^n] = (n (N_i + N_j) + n (3 - n)) (c_{ij}^\dagger)^{n-1}
\label{dimermanycomm}
\end{equation}
We now work in a subspace of spin-S AKLT Hamiltonian basis states $\ket{\psi_S}$ that satisfy $N_i\ket{\psi_S} = 2 S \ket{\psi_S}$ for any site $i$. Using this fact and Eq.~(\ref{dimermanycomm}), we can expand $(c_{ij}^\dagger c_{ij})^n$ into a normal ordered form as
\begin{equation}
    (c_{ij}^\dagger c_{ij})^n = \sum_{m=0}^n{f(m,n) (c_{ij}^\dagger)^m (c_{ij})^m}.
\label{dimernormexpand}
\end{equation}
$f(m,n)$ is determined recursively with the relations
\begin{eqnarray}
    f(n, n) &=& f(n-1, n-1) \nonumber \\
    f(m, n) &=& f(m-1, n-1) \nonumber \\
    &+& ((4S + 1)m - m^2) f(m, n-1) \nonumber \\
    f(0, n) &=& \twopartdef{0}{n \geq 1}{1}{n = 0}.
\label{normrecrel}
\end{eqnarray}

\section{Overcompleteness of the dimer basis}\label{ortho}
In Sec.~\ref{sec:spin1aklt}, we mentioned that that dimer basis is overcomplete. We illustrate this property here with a few examples. For example, the following three different dimer configurations are not linearly independent, as shown in the following relation, where $i, j$ and $k$ are distinct:
\begin{eqnarray}
&(c_{ij}^\dagger a_k^\dagger + a_i^\dagger c_{jk}^\dagger)\vac = (a_i^\dagger b_j^\dagger a_k^\dagger - b_i^\dagger a_j^\dagger a_k^\dagger \nn \\
&+ a_i^\dagger a_j^\dagger b_k^\dagger - a_i^\dagger b_j^\dagger a_k^\dagger)\vac \nn \\
&= (a_i^\dagger a_j^\dagger b_k^\dagger - b_i^\dagger a_j^\dagger a_k^\dagger)\vac \nn \\
&= c_{ik}^\dagger a_j^\dagger \vac.
\label{overcomp1}
\end{eqnarray}
Written diagrammatically, Eq.~(\ref{overcomp1}) reads
\begin{equation}
\ket{
\begin{picture}(40,-30)(0,7.5)
\multiput(6,10)(14,0){3}{\circle{4}}
\thicklines
\multiput(8,10)(14,0){1}{\line(1,0){10}}
\multiput(17,10)(14,0){1}{\vector(1,0){0}}
\put(34,5){\vector(0,1){12}}
\put(6,-3){i}
\put(20,-3){j}
\put(34,-3){k}
\end{picture} } 
\, + \,
\ket{
\begin{picture}(40,-30)(0,7.5)
\multiput(6,10)(14,0){3}{\circle{4}}
\thicklines
\multiput(22,10)(14,0){1}{\line(1,0){10}}
\multiput(31,10)(14,0){1}{\vector(1,0){0}}
\put(6,5){\vector(0,1){12}}
\put(6,-3){i}
\put(20,-3){j}
\put(34,-3){k}
\end{picture} } 
\,=\,
\ket{
\begin{picture}(40,-30)(0,7.5)
\multiput(6,10)(14,0){3}{\circle{4}}
\thicklines
\qbezier(6,12)(20,30)(34,12)
\put(20,5){\vector(0,1){12}}
\multiput(24,21)(14,0){1}{\vector(1,0){0}}
\put(6,-3){i}
\put(20,-3){j}
\put(34,-3){k}
\end{picture} }. 
\end{equation}
Another example of linear dependence for dimer configurations on four sites is, where $i,j,k$ and $l$ are distinct,
\begin{eqnarray}
&(c_{ij}^\dagger c_{kl}^\dagger + c_{il}^\dagger c_{jk}^\dagger)\vac = ((a_i^\dagger b_j^\dagger - b_i^\dagger a_j^\dagger)(a_k^\dagger b_l^\dagger - b_k^\dagger a_l^\dagger) \nn \\
&+ (a_i^\dagger b_l^\dagger - b_i^\dagger a_l^\dagger)(a_j^\dagger b_k^\dagger - b_j^\dagger a_k^\dagger))\vac \nn \\
&= (a_i^\dagger b_k^\dagger - b_i^\dagger a_k^\dagger)(a_j^\dagger b_l^\dagger - b_j^\dagger a_l^\dagger)\vac \nn \\
&= c_{ik}^\dagger c_{jl}^\dagger\vac.
\label{overcomp2}
\end{eqnarray}
Written diagrammatically, Eq.~(\ref{overcomp2}) reads
\\
\begin{equation}
\ket{
\begin{picture}(55,-30)(0,7.5)
\multiput(6,10)(14,0){4}{\circle{4}}
\thicklines
\multiput(8,10)(28,0){2}{\line(1,0){10}}
\multiput(17,10)(28,0){2}{\vector(1,0){0}}
\put(6,-3){i}
\put(20,-3){j}
\put(34,-3){k}
\put(48,-3){l}
\end{picture} } 
\, + \,
\ket{
\begin{picture}(55,-30)(0,7.5)
\multiput(6,10)(14,0){4}{\circle{4}}
\thicklines
\qbezier(6,12)(27,30)(48,12)
\multiput(22,10)(14,0){1}{\line(1,0){10}}
\multiput(30,21)(14,0){1}{\vector(1,0){0}}
\multiput(31,10)(28,0){1}{\vector(1,0){0}}
\put(6,-3){i}
\put(20,-3){j}
\put(34,-3){k}
\put(48,-3){l}
\end{picture} } 
\,=\,
\ket{
\begin{picture}(55,-30)(0,7.5)
\multiput(6,10)(14,0){4}{\circle{4}}
\thicklines
\qbezier(6,12)(20,30)(34,12)
\qbezier(20,12)(34,30)(48,12)
\multiput(24,21)(14,0){2}{\vector(1,0){0}}
\put(6,-3){i}
\put(20,-3){j}
\put(34,-3){k}
\put(48,-3){l}
\end{picture} }. 
\end{equation}

\section{Dimer basis states and scattering rules for the spin-1 AKLT model}\label{scattering}
In this section, we derive rules for action of the projector $P^{(2,1)}_{ij}$ Eq.~(\ref{P2dimer}) on various configurations of dimers around sites $i$ and $j$. Since the projector is normal ordered, it is sufficient to determine the action of $(c_{ij})^m$ on the dimer configurations. A useful identity in simplifying dimer expressions is 
\begin{equation}
    c_{ij} c_{mi}^\dagger c_{jn}^\dagger\ket{\Theta} = -c_{mn}^\dagger\ket{\Theta}
\label{usefulidentity}
\end{equation}
where $i,j,m,n$ are assumed to be distinct and $\ket{\Theta}$, referred to as the local vacuum, is a state annihilated by any of annihilation operators involving sites $i,j,m,n$.

\subsection{Singlet basis states}\label{S1singlet}
\begin{figure}
    \centering
    \begin{picture}(100,-30)(15,7.5)
    \linethickness{0.8pt}
    \multiput(25,10)(25,0){4}{\circle{4}} 
    \multiput(31,10)(25,0){4}{\circle{4}} 
    \multiput(28,10)(25,0){4}{\circle{14}}
    \thicklines
    \multiput(33,10)(25,0){3}{\line(1,0){15}} 
    \multiput(45,10)(25,0){3}{\vector(1,0){0}} 
    \put(25,-5){m}
    \put(50,-5){i}
    \put(75,-5){j}
    \put(100,-5){n}
    \put(-25, 5){(a)}
    \end{picture}
    
    \begin{picture}(150,35)(-10,7.5)
    \linethickness{0.8pt}
    \multiput(0,10)(25,0){6}{\circle{4}} 
    \multiput(6,10)(25,0){6}{\circle{4}} 
    \multiput(3,10)(25,0){6}{\circle{14}}
    \thicklines
    \qbezier(8,10)(31,35)(54,10)
    \multiput(34.5,23)(69,0){2}{\vector(1,0){0}}
    \qbezier(77,10)(100,35)(123,10)
    \multiput(33,10)(50,0){2}{\line(1,0){15}}
    \multiput(45,10)(50,0){2}{\vector(1,0){0}}
    \put(0,-5){m}
    \put(25,-5){n}
    \put(75,-5){j}
    \put(50,-5){i}
    \put(100,-5){p}
    \put(125,-5){r}
    \put(-25,5){(b)}
    \end{picture}\\[3mm]
\caption{Two types of singlet configurations around a bond $\{i, j\}$.}
\label{singletconfigdiags}
\end{figure}
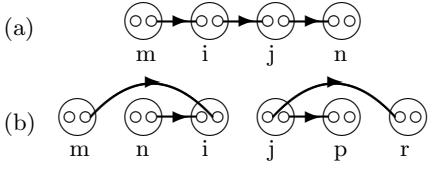
Since each dimer has $s = 0$ and $S_z = 0$, any basis state expressed in terms of only dimer creation operators ($c_{ij}^\dagger$s) is a singlet state. For the spin-1 AKLT model, there are two different possible configurations of dimers around two sites $i$ and $j$. They are $c_{ij}^\dagger c_{mi}^\dagger c_{jn}^\dagger \ket{\Theta}$ (Fig.~\ref{singletconfigdiags}a) and $c_{mi}^\dagger c_{ni}^\dagger c_{jp}^\dagger c_{jr}^\dagger \ket{\Theta}$ (Fig.~\ref{singletconfigdiags}b) where $\{m, n, p, r\}$ are distinct from $\{i, j\}$. Using Eq.~(\ref{dimeralgebra}), we can derive the action of $c_{ij}$ on the two different kinds of dimer configurations.
\begin{eqnarray}
    &c_{ij} c_{ij}^\dagger c_{mi}^\dagger c_{jn}^\dagger \ket{\Theta} = -c_{ij}^\dagger c_{mn}^\dagger \ket{\Theta} + 4 c_{mi}^\dagger c_{jn}^\dagger \ket{\Theta}  \label{singletidentities1}\\
    &c_{ij} c_{mi}^\dagger c_{ni}^\dagger c_{jp}^\dagger c_{jr}^\dagger \ket{\Theta} = -c_{mi}^\dagger c_{np}^\dagger c_{jr}^\dagger \ket{\Theta} - c_{ni}^\dagger c_{mp}^\dagger c_{jr}^\dagger \ket{\Theta}  \nonumber \\ & - c_{mi}^\dagger c_{jp}^\dagger c_{nr}^\dagger  \ket{\Theta} - c_{ni}^\dagger c_{jp}^\dagger c_{mr}^\dagger \ket{\Theta}
\label{singletidentities}
\end{eqnarray}
Similarly, the expressions for the action of $(c_{ij})^2$ on these configurations are
\begin{eqnarray}
    &(c_{ij})^2 c_{ij}^\dagger c_{mi}^\dagger c_{jn}^\dagger \ket{\Theta} = -6 c_{mn}^\dagger \ket{\Theta}\\
    &(c_{ij})^2 c_{mi}^\dagger c_{ni}^\dagger c_{jp}^\dagger c_{jr}^\dagger \ket{\Theta} = 2 (c_{mp}^\dagger c_{nr}^\dagger + c_{mr}^\dagger c_{np}^\dagger)\ket{\Theta}.
\label{singletidentities2}
\end{eqnarray}
Using Eqs.~(\ref{singletidentities1}) - (\ref{singletidentities2}), and the expression for the projector Eq.~(\ref{P2dimer}), the action of the projector on the configurations can be written as
\begin{eqnarray}
    &P^{(2,1)}_{ij} c_{ij}^\dagger c_{mi}^\dagger c_{jn}^\dagger \ket{\Theta} = 0 \label{projectorvanishing} \\
    &P^{(2,1)}_{ij}  c_{mi}^\dagger c_{ni}^\dagger c_{jp}^\dagger c_{jr}^\dagger \ket{\Theta}  =   c_{mi}^\dagger c_{ni}^\dagger c_{jp}^\dagger c_{jr}^\dagger \ket{\Theta}  \nonumber \\ & + \frac{1}{4} \left( c_{ij}^\dagger c_{mi}^\dagger c_{np}^\dagger c_{jr}^\dagger + c_{ij}^\dagger c_{ni}^\dagger c_{mp}^\dagger c_{jr}^\dagger  \right. \nonumber \\ & \left. + c_{ij}^\dagger c_{mi}^\dagger c_{jp}^\dagger c_{nr}^\dagger + c_{ij}^\dagger c_{ni}^\dagger c_{jp}^\dagger c_{mr}^\dagger \right) \ket{\Theta} \nonumber \\ &+ \frac{1}{12}\left((c_{ij}^\dagger)^2 c_{mr}^\dagger c_{np}^\dagger + (c_{ij}^\dagger)^2 c_{nr}^\dagger c_{mp}^\dagger \right)\ket{\Theta} 
\label{S1singletrules}
\end{eqnarray}
From Eq.~(\ref{projectorvanishing}), the projector vanishes on any singlet state that contains a $c_{ij}^\dagger$. This is heavily used in our calculations.
As shown in Fig.~\ref{diagrep}, these can be written as a set of diagrammatic rules for obtaining a set of configurations into which an initial configuration of dimers scatters.
\subsection{Non-singlet basis states}
\begin{figure}
    \begin{picture}(50,-30)(40,7.5)
    \linethickness{0.8pt}
    \multiput(50,10)(25,0){2}{\circle{4}} 
    \multiput(56,10)(25,0){2}{\circle{4}} 
    \multiput(53,10)(25,0){2}{\circle{14}}
    \thicklines
    \multiput(50,5)(25,0){2}{\vector(0,1){12}} 
    \multiput(56,5)(25,0){2}{\vector(0,1){12}} 
    \put(75,-5){j}
    \put(50,-5){i}
    \put(-25,5){(a)}
    \end{picture}
    
    \begin{picture}(100,25)(15,7.5)
    \linethickness{0.8pt}
    \multiput(25,10)(25,0){3}{\circle{4}} 
    \multiput(31,10)(25,0){3}{\circle{4}} 
    \multiput(28,10)(25,0){3}{\circle{14}}
    \thicklines
    \multiput(75,5)(25,0){1}{\vector(0,1){12}} 
    \multiput(56,5)(25,0){2}{\vector(0,1){12}} 
    \put(33,10){\line(1,0){15}}
    \put(45,10){\vector(1,0){0}}
    \put(75,-5){j}
    \put(50,-5){i}
    \put(25,-5){n}
    \put(-25,5){(b)}
    \end{picture}
    
    \begin{picture}(100,25)(15,7.5)
        \linethickness{0.8pt}
        \multiput(25,10)(25,0){4}{\circle{4}} 
        \multiput(31,10)(25,0){4}{\circle{4}} 
        \multiput(28,10)(25,0){4}{\circle{14}}
        \thicklines
        \multiput(33,10)(50,0){2}{\line(1,0){15}}
        \multiput(45,10)(50,0){2}{\vector(1,0){0}}
        \multiput(56,5)(25,0){1}{\vector(0,1){12}}
        \multiput(75,5)(25,0){1}{\vector(0,1){12}}
        \put(25,-5){m}
        \put(50,-5){i}
        \put(75,-5){j}
        \put(100,-5){n}
        \put(-25,5){(c)}
    \end{picture}
    
    \begin{picture}(50,35)(40,7.5)
    \linethickness{0.8pt}
    \multiput(50,10)(25,0){4}{\circle{4}} 
    \multiput(56,10)(25,0){4}{\circle{4}} 
    \multiput(53,10)(25,0){4}{\circle{14}}
    \thicklines
    \multiput(103.5,23)(69,0){1}{\vector(1,0){0}}
    \qbezier(77,10)(100,35)(123,10)
    \multiput(83,10)(50,0){1}{\line(1,0){15}}
    \multiput(95,10)(50,0){1}{\vector(1,0){0}}
    \multiput(50,5)(6,0){2}{\vector(0,1){12}}
    \put(75,-5){j}
    \put(50,-5){i}
    \put(100,-5){p}
    \put(125,-5){r}
    \put(-25, 5){(d)}
    \end{picture}
    
    \begin{picture}(100,35)(15,7.5)
    \linethickness{0.8pt}
    \multiput(25,10)(25,0){5}{\circle{4}} 
    \multiput(31,10)(25,0){5}{\circle{4}} 
    \multiput(28,10)(25,0){5}{\circle{14}}
    \thicklines
    \multiput(103.5,23)(69,0){1}{\vector(1,0){0}}
    \qbezier(77,10)(100,35)(123,10)
    \multiput(33,10)(50,0){2}{\line(1,0){15}}
    \multiput(45,10)(50,0){2}{\vector(1,0){0}}
    \multiput(56,5)(25,0){1}{\vector(0,1){12}}
    \put(25,-5){m}
    \put(75,-5){j}
    \put(50,-5){i}
    \put(100,-5){p}
    \put(125,-5){r}
    \put(-25, 5){(e)}
    \end{picture}
    
    \begin{picture}(50,30)(40,7.5)
    \linethickness{0.8pt}
    \multiput(50,10)(25,0){2}{\circle{4}} 
    \multiput(56,10)(25,0){2}{\circle{4}} 
    \multiput(53,10)(25,0){2}{\circle{14}}
    \thicklines
    \multiput(50,5)(25,0){1}{\vector(0,1){12}} 
    \multiput(81,5)(25,0){1}{\vector(0,1){12}} 
    \multiput(58,10)(25,0){1}{\line(1,0){15}} 
    \multiput(70,10)(25,0){1}{\vector(1,0){0}} 
    \put(75,-5){j}
    \put(50,-5){i}
    \put(-25, 5){(f)}
    \end{picture}
    
    \begin{picture}(100,25)(15,7.5)
    \linethickness{0.8pt}
    \multiput(25,10)(25,0){3}{\circle{4}} 
    \multiput(31,10)(25,0){3}{\circle{4}} 
    \multiput(28,10)(25,0){3}{\circle{14}}
    \thicklines
    \multiput(81,5)(25,0){1}{\vector(0,1){12}} 
    \multiput(33,10)(25,0){2}{\line(1,0){15}}
    \multiput(45,10)(25,0){2}{\vector(1,0){0}}
    \put(75,-5){j}
    \put(50,-5){i}
    \put(25,-5){m}
    \put(-25, 5){(g)}
    \end{picture}\\[2mm]
    
\caption{Types of non-singlet configurations around a bond $\{i,j\}$}
\label{nonsingletconfigdiag}
\end{figure}
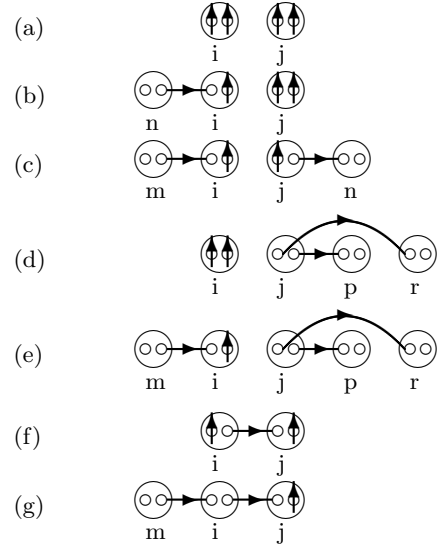

In the previous section we worked with singlet states, where all the configurations could be written only in terms of dimers. States with $s \neq 0$, because of $SU(2)$ symmetry, would appear in multiplets of $2 s + 1$ states. As mentioned in the main text, it is sufficient to focus on the highest weight states of each multiplet. If $s \neq 0$, the configurations in the highest weight multiplet would have free spin-$\uparrow$ ($a_i^\dagger$) on the chain. Once this possibility is allowed, several new configurations are possible. The distinct ones are ${a_i^\dagger}^2 {a_j^\dagger}^2\ket{\Theta}$ (Fig.~\ref{nonsingletconfigdiag}a), $c_{ni}^\dagger a_i^\dagger {a_j^\dagger}^2 \ket{\Theta}$ (Fig.~\ref{nonsingletconfigdiag}b), $c_{ni}^\dagger a_i^\dagger a_j^\dagger c_{jp}^\dagger\ket{\Theta}$ (Fig.~\ref{nonsingletconfigdiag}c), $(a_i^\dagger)^2 c_{jp}^\dagger c_{jr}^\dagger \ket{\Theta}$ (Fig.~\ref{nonsingletconfigdiag}d), $c_{mi}^\dagger a_i^\dagger c_{jp}^\dagger c_{jr}^\dagger\ket{\Theta}$ (Fig.~\ref{nonsingletconfigdiag}e), $c_{ij}^\dagger a_i^\dagger a_j^\dagger\ket{\Theta}$ (Fig.~\ref{nonsingletconfigdiag}f), and  $c_{ij}^\dagger c_{mi}^\dagger a_j^\dagger\ket{\Theta}$ (Fig.~\ref{nonsingletconfigdiag}g), where $\{m, n, p, r\}$ are distinct from $\{i, j\}$.
Analogous to Eqs.~(\ref{singletidentities1}) and (\ref{singletidentities}), we derive identities for the action of $c_{ij}$ on each of these configurations.
\begin{eqnarray}
    &c_{ij} {a_i^\dagger}^2 {a_j^\dagger}^2\ket{\Theta} = 0 \label{higherspinidentities1} \\
    &c_{ij} c_{ni}^\dagger a_i^\dagger {a_j^\dagger}^2 \ket{\Theta} = -2 a_n^\dagger a_i^\dagger a_j^\dagger \ket{\Theta} \label{higherspinidentities2} \\
    &c_{ij} c_{mi}^\dagger a_i^\dagger a_j^\dagger c_{jn}^\dagger\ket{\Theta} = - (c_{mi}^\dagger a_j^\dagger a_n^\dagger \nonumber \\ &+ a_m^\dagger a_i^\dagger c_{jn}^\dagger + a_i^\dagger a_j^\dagger c_{mn}^\dagger)\ket{\Theta} \\
    &c_{ij} (a_i^\dagger)^2 c_{jp}^\dagger c_{jr}^\dagger \ket{\Theta} = -2 (a_i^\dagger a_r^\dagger c_{jp}^\dagger + a_i^\dagger a_p^\dagger  c_{jr}^\dagger) \ket{\Theta} \\
    &c_{ij}c_{mi}^\dagger a_i^\dagger c_{jp}^\dagger c_{jr}^\dagger\ket{\Theta} = - (c_{mi}^\dagger c_{jp}^\dagger a_r^\dagger + c_{mi}^\dagger c_{jr}^\dagger a_p^\dagger \nonumber \\ 
    &+ c_{mr}^\dagger c_{jp}^\dagger a_i^\dagger + c_{mp}^\dagger c_{jr}^\dagger a_i^\dagger)\ket{\Theta} \label{higherspinidentities3}\\
    &c_{ij}c_{ij}^\dagger a_i^\dagger a_j^\dagger\ket{\Theta} = 4 a_i^\dagger a_j^\dagger \ket{\Theta} \\
    &c_{ij}c_{ij}^\dagger c_{mi}^\dagger a_j^\dagger\ket{\Theta} = -a_m^\dagger c_{ij}^\dagger \ket{\Theta} + 4 a_j^\dagger c_{ij}^\dagger \ket{\Theta}.\label{higherspinidentitieslast}
\end{eqnarray}
Similarly, the actions of $(c_{ij})^2$ are given by
\begin{eqnarray}
    &(c_{ij})^2 {a_i^\dagger}^2 {a_j^\dagger}^2\ket{\Theta} = 0 \label{higherspinidentities4} \\
    &(c_{ij})^2 c_{ni}^\dagger a_i^\dagger {a_j^\dagger}^2 \ket{\Theta} = 0 \label{higherspinidentities5} \\
    &(c_{ij})^2 c_{mi}^\dagger a_i^\dagger a_j^\dagger c_{jn}^\dagger\ket{\Theta} = 2 a_m^\dagger a_n^\dagger \ket{\Theta} \label{higherspinidentities10}\\
    &(c_{ij})^2 (a_i^\dagger)^2 c_{jp}^\dagger c_{jr}^\dagger \ket{\Theta} = 4 a_p^\dagger a_r^\dagger \ket{\Theta} \label{higherspinidentities9} \\
    &(c_{ij})^2 c_{mi}^\dagger a_i^\dagger c_{jp}^\dagger c_{jr}^\dagger\ket{\Theta} = 2 (c_{mp}^\dagger a_r^\dagger + c_{mr}^\dagger a_p^\dagger)\ket{\Theta}. \label{higherspinidentities6}\\
    &(c_{ij})^2 c_{ij}^\dagger a_i^\dagger a_j^\dagger\ket{\Theta} = 0
    \label{higherspinidentities7}\\
    &(c_{ij})^2 c_{ij}^\dagger c_{mi}^\dagger a_j^\dagger\ket{\Theta} = -6 a_m^\dagger \ket{\Theta}\label{higherspinidentities8}
\end{eqnarray}
Again, using the expression of the projector in Eq.~(\ref{P2dimer}) and Eqs.~(\ref{higherspinidentities1}) - (\ref{higherspinidentities8}), we derive the expression for the action of the projector $P^{(2,1)}_{ij}$ on each of the configurations.
\begin{equation}
    P^{(2,1)}_{ij} {a_i^\dagger}^2 {a_j^\dagger}^2\ket{\Theta} = {a_i^\dagger}^2 {a_j^\dagger}^2\ket{\Theta}
\label{allspin}
\end{equation}
\begin{eqnarray}
    &P^{(2,1)}_{ij} c_{ni}^\dagger a_i^\dagger {a_j^\dagger}^2 \ket{\Theta} =  c_{ni}^\dagger a_i^\dagger (a_j^\dagger)^2 \ket{\Theta} \nonumber \\ & + \frac{1}{2}a_n^\dagger a_i^\dagger c_{ij}^\dagger a_j^\dagger \ket{\Theta}
\label{singlethopping}
\end{eqnarray}
\begin{eqnarray}
    &P^{(2,1)}_{ij} c_{mi}^\dagger a_i^\dagger a_j^\dagger c_{jn}^\dagger\ket{\Theta} = c_{mi}^\dagger a_i^\dagger a_j^\dagger c_{jn}^\dagger\ket{\Theta} \nonumber \\ &+ \frac{1}{4}\left(c_{ij}^\dagger c_{mi}^\dagger a_j^\dagger a_n^\dagger + c_{ij}^\dagger a_m^\dagger a_i^\dagger c_{jn}^\dagger + c_{ij}^\dagger a_i^\dagger a_j^\dagger c_{mn}^\dagger\right)\ket{\Theta} \nonumber \\ &+ \frac{1}{12}\left((c_{ij}^\dagger)^2 a_n^\dagger a_m^\dagger\right)\ket{\Theta}
\label{badrule}
\end{eqnarray}
\begin{eqnarray}
    &P^{(2,1)}_{ij} (a_i^\dagger)^2 c_{jp}^\dagger c_{jr}^\dagger \ket{\Theta} = (a_i^\dagger)^2 c_{jp}^\dagger c_{jr}^\dagger \ket{\Theta} \nonumber \\ & +  \frac{1}{2}\left(a_i^\dagger a_r^\dagger c_{ij}^\dagger c_{jp}^\dagger + a_i^\dagger a_p^\dagger c_{ij}^\dagger c_{jr}^\dagger\right)\ket{\Theta} \nonumber \\ & + \frac{1}{6}\left(a_p^\dagger a_r^\dagger (c_{ij}^\dagger)^2 \right) \ket{\Theta}
\label{double}
\end{eqnarray}
\begin{eqnarray}
    &P^{(2,1)}_{ij}c_{mi}^\dagger a_i^\dagger c_{jp}^\dagger c_{jr}^\dagger\ket{\Theta} = c_{mi}^\dagger a_i^\dagger c_{jp}^\dagger c_{jr}^\dagger\ket{\Theta} + \frac{1}{4}\left(c_{ij}^\dagger c_{mi}^\dagger c_{jp}^\dagger a_r^\dagger \right. \nonumber \\ & \left. + c_{ij}^\dagger c_{mi}^\dagger c_{jr}^\dagger a_p^\dagger + c_{ij}^\dagger c_{mr}^\dagger c_{jp}^\dagger a_i^\dagger + c_{ij}^\dagger c_{mp}^\dagger c_{jr}^\dagger a_i^\dagger\right)\ket{\Theta} \nonumber \\ &+ \frac{1}{12}\left((c_{ij}^\dagger)^2 c_{mp}^\dagger a_r^\dagger + (c_{ij}^\dagger)^2 c_{mr}^\dagger a_p^\dagger\right)\ket{\Theta}
\label{useless}
\end{eqnarray}
\begin{equation}
    P^{(2,1)}_{ij} a_i^\dagger c_{ij}^\dagger a_{j}^\dagger\ket{\Theta} = 0
\label{projectorvanishing2}
\end{equation}
\begin{equation}
    P^{(2,1)}_{ij}c_{ij}^\dagger c_{mi}^\dagger a_j^\dagger\ket{\Theta} = 0
\label{projectorvanishing3}
\end{equation}
Eqs.~(\ref{projectorvanishing2}) and (\ref{projectorvanishing3}) along with Eq.~(\ref{projectorvanishing}) state that the projector $P^{(2,1)}_{ij}$ vanishes on all configurations that have a dimer on bond $\{i,j\}$. The actions of $P^{(2,1)}_{ij}$ on the various configurations are summarized diagrammatically in Fig.~\ref{diagrep}.

\section{Derivation of the \boldsymbol{$2_k$} states}\label{sec:2kexact}
We show that the $\ket{2_k}$ states given in Eq.~(\ref{2kstate}) are eigenstates of the spin-1 AKLT Hamiltonian. 
\begin{equation}
    \ket{2_k} = \sum_{m=0}^{L/2-1}{e^{i (k + \pi) m}\sum_n{e^{i k n}\ket{n,n + 2m + 1}}}.
\tag{\ref{2kstate}}
\end{equation}
Splitting Eq.~(\ref{2kstate}) into three parts, the $m = 0$, $m = L/2 -1$ and $m \in [1, L/2 - 2]$, we obtain
\begin{eqnarray}
    \ket{2_k} &=& \sum_n{e^{ikn}\ket{n, n+1}} + e^{i (k + \pi)(L/2 - 1)}\sum_n{e^{ikn}\ket{n, n - 1}} \nn \\
    &+& \sum_{m=1}^{L/2-2}{e^{i (k + \pi) m}\sum_n{e^{i k n}\ket{n,n + 2m + 1}}}.
\end{eqnarray}
The action of the Hamiltonian on $\ket{2_k}$ can be obtained using Eq.~(\ref{fatsing}) for the $m = 0, L/2 - 1$ terms in Eq.~(\ref{2kstate}) and Eq.~(\ref{2singneigh}) for the rest of the terms.
\begin{eqnarray}
    &H\ket{2_k} = (L-2)\ket{2_k} \nonumber \\ &+ \sum_n{e^{i k n} (\ket{n-1,n+1} + \ket{n,n+2})} \nonumber \\ &+ e^{i (k+\pi) (L/2 -1)}\sum_n{e^{i k n} (\ket{n-1,n+1} + \ket{n, n-2})}  \nonumber \\ &+ \sum_{m=1}^{L/2-2}{e^{i (k + \pi) m}} \sum_n{e^{i k n}}\nn \\ &(\ket{n,n + 2m} + \ket{n+1,n + 2m+1} \nn \\ &+ \ket{n,n + 2m + 2} + \ket{n-1,n + 2m+1} )
\label{2kexpand}
\end{eqnarray}
Simplifying the sum over $n$ corresponding to $m \in [1, L/2 - 2]$ in Eq.~(\ref{2kexpand}), one arrives at (by successively shifting $n$ and $m$),
\begin{eqnarray}
&\sum_{m=1}^{L/2-2}{e^{i(k+\pi)m} \sum_n{e^{i k n}(\ket{n, n+2m}(1+ e^{-i k})}} \nonumber \\ & +\ket{n,n+2m+2}(1+e^{i k})) \nonumber \\ 
&= \sum_n{e^{i k n} \sum_{m = 2}^{L/2-2}{e^{i (k + \pi)m} \ket{n,n+2m}(1 + e^{-i k}}} \nonumber \\ & +e^{-i(k+\pi)}(1+e^{i k})) +e^{i(k+\pi)} (1+e^{-i k})\ket{n,n+2} \nonumber \\ & +e^{i (k +\pi)(L/2-2)}(1+ e^{i k})\ket{n,n-2} \nonumber \\
&= \sum_n{e^{i k n} (-(1 + e^{i k})\ket{n,n+2}} \nonumber \\ &-e^{i(k+\pi)(L/2-1)} (1 + e^{-i k})\ket{n, n-2})
\label{2klastterm}
\end{eqnarray}
The sums in Eq.~(\ref{2kexpand}) corresponding to $m = 0$ and $m = L/2-1$ can be written as 
\begin{eqnarray}
&\sum_n{e^{i k n} (\ket{n-1,n+1} + \ket{n,n+2})} \nonumber \\ 
&+ e^{i (k+\pi) (L/2 -1)}\sum_n{e^{i k n} (\ket{n-1,n+1} + \ket{n, n-2})}  \nonumber \\
& = \sum_n{e^{i k n} ((1+e^{i k})\ket{n,n+2} }\nonumber \\
& +e^{i (k + \pi)(L/2 - 1)}(1 + e^{-i k})\ket{n,n-2})
\label{2kotherterms}
\end{eqnarray} 
Eq.~(\ref{2klastterm}) and (\ref{2kotherterms}) cancel with each other, thus showing that $\ket{2_k}$ is an exact state with energy $E = L-2$.  
However, as discussed earlier, the states vanish for certain momenta. We prove the relation
\begin{equation}
    \ket{2_k} = e^{i k \frac{L}{2}} e^{i \pi \left(\frac{L}{2} - 1\right)} \ket{2_k}.
\tag{\ref{2kvanish}}
\end{equation}
To derive Eq.~(\ref{2kvanish}), we make the variable substitution $n = n' - 2m - 1$. Eq.~(\ref{2kstate}) reduces to 
\begin{eqnarray}
    \ket{2_k} &=& \sum_{m=0}^{L/2-1}{e^{i (k + \pi) m} e^{-i k (2m + 1)}\sum_{n'}{e^{i k n'}\ket{n' - 2m - 1, n'}}}. \nn \\
    &=& \sum_{m=0}^{L/2-1}{e^{i \pi m - k m - k}\sum_{n'}{e^{i k n'}\ket{n' - 2m - 1, n'}}}.
\label{2k1sub}
\end{eqnarray}
In Eq.~(\ref{2k1sub}), the sum over $n$ was converted to a sum over $n'$ because it is over all the possible values of $n$ (for any given $m$). Furthermore, making the variable substitution $m = L/2 - 1 - m'$, we arrive at
\begin{eqnarray}
    \ket{2_k} &=& \sum_{m'=0}^{L/2-1}{e^{i \pi (\frac{L}{2} - 1- m') - i k (\frac{L}{2} - 1 - m') - i k}} \nn\\
    &&\sum_{n'}{e^{i k n'}\ket{n' - L + 2m' + 1, n'}} \nn \\
    &=& e^{i \pi \frac{L}{2}} e^{-i k (\frac{L}{2}-1)}\sum_{m'=0}^{L/2-1}{e^{i (k - \pi) m'}}\nn\\
    &&\sum_{n'}{e^{i k n'}\ket{n' + 2m' + 1, n'}} \nn \\
    &=& e^{i k \frac{L}{2}} e^{i \pi \left(\frac{L}{2} - 1\right)} \ket{2_k}
\label{2k2sub}
\end{eqnarray}
where in the last step, expressions have been simplified using the fact that $L/2$ and $m'$ are integers, and that the system is periodic.
\section{Special states in the high energy spectrum}\label{sec:specialstates}
We discuss the construction of the  special state $\ket{6_0}$, defined in Eq.~(\ref{60defn}); and in particular derive Eq.~(\ref{npqconstraint}). As discussed in Eq.~(\ref{spinket1}), we define the basis in the sector $s = L - 2, k = 0$ as $\ket{n} \equiv \ket{n_{k=0}}, n \in {0,1,...,L/2}$, where 
\begin{equation}
\ket{n} = \twopartdef{\frac{1}{\sqrt{L}}\sum_{i=1}^{L}{S_i^- S_{i+n}^- \ket{F}}}{n \leq L/2 - 1\\[5mm]}{\frac{1}{\sqrt{L/2}}\sum_{i=1}^{L}{S_i^- S_{i+n}^- \ket{F}}}{n = L/2}
\label{spinket}
\end{equation}
Note that this basis is orthonormal. In Eq.~(\ref{spinket}), each $\ket{n}$, $n \geq 1$ is a momentum superposition of spin configurations of the form 
\begin{equation}
    \ket{d_n} = \ket{1\ \dots\ 1\ \underbrace{0\ 1\ \dots\ 1\ }_{n} 0\ 1\ \dots 1} 
\label{exampconfigspecial}
\end{equation}
and $\ket{0}$ is a momentum superposition of the spin configuration
\begin{equation}
    \ket{d_0} = \ket{1\ \dots\ 1\ \mm\ 1\ \dots\ 1}.
\end{equation}
The action of $P^{(2,1)}_{i,i+1}$ on each pair of neighboring spins in the configuration Eq.~(\ref{exampconfigspecial}) is computed using Eqs.~(\ref{spin11}) - (\ref{spinm1m1}). For the configuration shown in Eq.~(\ref{exampconfigspecial}), the scattering equation for $n \geq 2$ reads (using Eqs.~(\ref{spin11}) and (\ref{spin10}))
\begin{equation}
    H \ket{d_n} = (L - 2) \ket{d_n} + \ket{d_{n - 1}} + \ket{d_{n + 1}}.
\end{equation}
Hence,
\begin{equation}
    H\ket{n} = (L - 2) \ket{n} + \ket{n - 1} + \ket{n + 1} \;\;\; 2 \leq n \leq L/2-2.
\label{nscattering}
\end{equation}
Using Eqs.~(\ref{spin11}), (\ref{spin10}) and (\ref{spin00}), the action of $H$ on $\ket{d_1}$ is given by
\begin{eqnarray}
    H\ket{d_1} &=& (L - 2)\ket{d_1} + \ket{d_2} + \frac{2}{3}\ket{d_1} + \frac{2}{3}\ket{d_0} \nn \\
    &=& (L - \frac{4}{3})\ket{d_1} + \ket{d_2} + \frac{2}{3}\ket{d_0}.
\end{eqnarray}
The scattering equation of $\ket{1}$ thus reads
\begin{equation}
    H\ket{1} = (L - \frac{4}{3})\ket{1} + \ket{2} + \frac{2}{3}\ket{0}. 
\end{equation}
Similarly, using Eqs.~(\ref{spin11}) and (\ref{spin1m1}), the scattering equation of $\ket{d_0}$ reads
\begin{eqnarray}
    H\ket{d_0} &=& (L - 2)\ket{d_0} + \frac{2}{3}\ket{d_0} + \frac{2}{3}\ket{d_1} \nn \\
    &=& (L - \frac{4}{3})\ket{d_0} + \frac{2}{3}\ket{d_1}. 
\end{eqnarray}
The scattering equation of $\ket{0}$ thus reads
\begin{equation}
    H\ket{0} = (L - \frac{4}{3})\ket{0} + \frac{2}{3}\ket{1}.
\end{equation}

The representation of the spin-1 AKLT Hamiltonian in the basis of $\{\ket{n}\}$ can thus be computed to be
\begin{equation}
H = 
\begin{blockarray}{cccccccc}
\ket{0} & \ket{1} & \dots & \dots & \dots & \dots & \ket{L/2} \\[2mm]
\begin{block}{(ccccccc)c}
    L - \frac{4}{3} & \frac{2}{3} & 0 & \dots & \dots & \dots & 0 & \ket{0} \\
    \frac{2}{3} & L - \frac{4}{3} & 1 & 0 & \dots & \dots & \vdots & \ket{1}\\
    0 & 1 & L -2 & 1 & 0 & \dots & \vdots & \vdots\\
    \vdots & \ddots & \ddots & \ddots & \ddots & \ddots & \vdots & \vdots \\
    \vdots & \ddots & \ddots & 1 & L - 2 & 1 & \vdots & \vdots \\
    \vdots & \ddots & \ddots & \ddots & 1 & L - 2 & \sqrt{2} & \vdots \\ 
    0 & \dots & \dots & \dots & \dots & \sqrt{2 }&  L - 2 & \ket{L/2} \\
\end{block}
\end{blockarray}
\label{spinL-2hamil}
\end{equation}

Any eigenstate with $E = L - \xi$ in this subspace decomposes as
\begin{equation}
\ket{\psi} = \sum_{n = 0}^{L/2}{x_n \ket{n}}
\label{guessstate}
\end{equation}
Acting the Hamiltonian on $\ket{\psi}$, we arrive at a set of equations
\begin{eqnarray}
&x_1 = (2 - \frac{3}{2}\xi)x_0 \nn \\
&x_2 = (\frac{3}{2}\xi^2 - 4\xi + 2)x_0 \nn \\
&x_3 = (2-\xi)x_2 - x_1 \nn \\
&\vdots \nn \\
&x_l = (2-\xi)x_{l-1} - x_{l-2} \nn \\
&\vdots \nn \\
&x_{L/2-1} = (\sqrt{2} x_{L/2} + x_{L/2 - 2})/(2-\xi) \nn \\
&x_{L/2} = \sqrt{2}x_{L/2-1}/(2-\xi)
\label{60eigenstateeqn}
\end{eqnarray}
Eqs.~(\ref{60eigenstateeqn}) can be written with a ``transfer matrix" as
\begin{equation}
\begin{pmatrix}
    x_{l+1} \\ x_l
\end{pmatrix}
= 
\begin{pmatrix}
    2 - \xi & -1 \\
    1 & 0 
\end{pmatrix}
\begin{pmatrix}
    x_l \\ x_{l-1}
\end{pmatrix}
\end{equation}
for $2 \leq l \leq L/2 - 2$.
Here the $2 \times 2$ matrix is the transfer matrix $M$.
The last two equations in Eq.~(\ref{60eigenstateeqn}) which signal the end of the transfer matrix, give
\begin{equation}
(2-\xi)x_{L/2-2} = (\xi^2 - 4\xi + 2)x_{L/2-1}.
\label{energytopconstraint}
\end{equation}
Since we are only looking for states that appear when the system size $L$ is a multiple of some integer $q$, we want the matrix $M$ to be idempotent. If $M^q = \mathds{1}$, we arrive at the constraint that both the eigenvalues of $M$ are $q$-th roots of unity. Since $M$ has real entries, the determinant of the matrix must be real, which imposes the constraint on the two eigenvalues to be conjugates of each other. Thus, $\lambda_{\pm} = e^{\pm i 2\pi \frac{p}{q}}$ for some $p, q$ such that ${\rm gcd}(p,q) = 1$. This  also means that ${\rm Tr}(M) (= 2 - \xi) =  2 \cos(2\pi\frac{p}{q})$. This imposes a constraint on the energies $L - \xi$ of the repeating states with the condition
\begin{equation}
\xi = 4 \sin^2\left(\pi \frac{p}{q}\right) \mbox{ for } p,q \in \mathbb{Z^+}
\label{energytopeqn}
\end{equation}
Using Eq.~(\ref{energytopeqn}), the identity with $y = 2 \pi \frac{p}{q}$
\begin{equation}
\begin{pmatrix}
    2 \cos(y) & -1 \\
    1 & 0
\end{pmatrix}^n
=
\frac{1}{\sin(y)}
\begin{pmatrix}
    \sin ((n+1)y) & -\sin (ny) \\
    \sin (ny) & -\sin ((n-1)y)
\end{pmatrix},
\end{equation}
the recurrence relation
\begin{equation}
\begin{pmatrix}
    x_{L/2-1} \\ x_{L/2-2}
\end{pmatrix}
= 
M^{L/2-3}
\begin{pmatrix}
    x_2 \\ x_1
\end{pmatrix},
\end{equation}
and the expressions for $x_1$ and $x_2$ in Eq.~(\ref{60eigenstateeqn}), the constraint Eq.~(\ref{energytopconstraint}) reduces to the equation
\begin{equation}
\frac{3}{2}\tan\left(\frac{y}{2}\right) = - \frac{\sin((r+2)y)}{\cos((r+3)y)}
\label{npqconstraintapp}
\end{equation}
where $r = L/2 - 3\;{\rm mod }\;q$.
If there is a solution to Eq.~(\ref{npqconstraintapp}) for integer $p, q, r$, we obtain a state with energy $E = L - 4 \sin^2(\pi \frac{p}{q})$, $s = L - 2$ that appears for every $L = 6 + 2r + 2 q m$, where $m \in \mathbb{Z}$.
In addition, if $q$ is even, and $r$ is a solution, $r + q/2$ is also a solution.
This means that the repeating state would appear for $L = 6 + 2 r + q m$, where $m \in \mathbb{Z}$.
Moreover, if $p$ is a solution, $q - p$ is also a solution that would give rise to the same transfer matrix (because the eigenvalues are the same).
So, we restrict ourselves to $p \leq q/2$.

From Eqs.~(\ref{guessstate}) and (\ref{spinket}), the operator expression for the unnormalized $\ket{\psi}$ is
\begin{equation}
    \ket{\psi} = \sum_{n = 0}^{L/2}{\sum_{i = 1}^{L}{x_n S^-_i S^-_{i+n}}}\ket{F}.
\label{specialopexp}
\end{equation}

\section{Dimer basis scattering rules for spin-$S$ AKLT basis states}\label{sec:highspindimer}
\begin{figure}[t!]
\centering
\begin{picture}(100,20)(0,0)
    \linethickness{0.8pt}
    \multiput(0,10)(25,0){5}{\circle{4}} 
    \multiput(0,10)(25,0){2}{\circle*{4}}
    \multiput(0,4)(25,0){2}{\circle*{4}}
    \multiput(6,10)(25,0){5}{\circle{4}} 
    \multiput(0,4)(25,0){5}{\circle{4}} 
    \multiput(6,4)(25,0){5}{\circle{4}} 
    \multiput(3,7)(25,0){5}{\circle{16}}
    \thicklines
    \multiput(33,4)(25,0){1}{\line(1,0){15}} 
    \multiput(45,4)(25,0){1}{\vector(1,0){0}}
    \multiput(33,10)(25,0){1}{\line(1,0){15}} 
    \multiput(45,10)(25,0){1}{\vector(1,0){0}}
    \multiput(83,4)(25,0){1}{\line(1,0){15}} 
    \multiput(95,4)(25,0){1}{\vector(1,0){0}}
    \multiput(83,10)(25,0){1}{\line(1,0){15}} 
    \multiput(95,10)(25,0){1}{\vector(1,0){0}}
    \qbezier(8,10)(31,30)(56,12)
    \multiput(35.5,22)(69,0){1}{\vector(1,0){0}}
    \qbezier(8,4)(31,-16)(56,2)
    \multiput(35.5,-7)(69,0){1}{\vector(1,0){0}}
    \qbezier(76,2)(90.5,-6)(105,2)
    \put(95.5, -2){\vector(1,0){0}}
    \qbezier(76,12)(90.5,24)(105,12)
    \put(95.5, 17){\vector(1,0){0}}
    \put(77,-10){j}
    \put(52,-10){i}
    \put(-32,5){(a)}
\end{picture}\\[8mm]
\begin{picture}(100,20)(0,0)
    \linethickness{0.8pt}
    \multiput(0,10)(25,0){2}{\circle*{4}}
    \multiput(0,4)(25,0){2}{\circle*{4}}
    \multiput(0,10)(25,0){5}{\circle{4}} 
    \multiput(6,10)(25,0){5}{\circle{4}} 
    \multiput(0,4)(25,0){5}{\circle{4}} 
    \multiput(6,4)(25,0){5}{\circle{4}} 
    \multiput(3,7)(25,0){5}{\circle{16}}
    \thicklines
    \multiput(33,4)(25,0){1}{\line(1,0){15}} 
    \multiput(45,4)(25,0){1}{\vector(1,0){0}}
    \multiput(33,10)(25,0){1}{\line(1,0){15}} 
    \multiput(45,10)(25,0){1}{\vector(1,0){0}}
    \multiput(83,4)(25,0){1}{\line(1,0){15}} 
    \multiput(95,4)(25,0){1}{\vector(1,0){0}}
    \multiput(83,10)(25,0){1}{\line(1,0){15}} 
    \multiput(95,10)(25,0){1}{\vector(1,0){0}}
    \qbezier(8,4)(31,-16)(56,2)
    \multiput(35.5,-7)(69,0){1}{\vector(1,0){0}}
    \qbezier(76,2)(90.5,-6)(105,2)
    \put(95.5, -2){\vector(1,0){0}}
    \qbezier(8,10)(55.5,30)(105,12)
    \put(60.5, 21){\vector(1,0){0}}
    \put(77,-10){j}
    \put(52,-10){i}
    \put(-32,5){(b)}
\end{picture}\\[8mm]
\begin{picture}(100,20)(0,0)
    \linethickness{0.8pt}
    \multiput(0,10)(25,0){2}{\circle*{4}}
    \multiput(31,10)(75,0){2}{\circle*{4}}
    \multiput(6,4)(100,0){2}{\circle*{4}}
    \multiput(0,4)(25,0){2}{\circle*{4}}
    \multiput(100,4)(25,0){1}{\circle*{4}}
    \multiput(0,10)(25,0){5}{\circle{4}} 
    \multiput(6,10)(25,0){5}{\circle{4}} 
    \multiput(0,4)(25,0){5}{\circle{4}} 
    \multiput(6,4)(25,0){5}{\circle{4}} 
    \multiput(3,7)(25,0){5}{\circle{16}}
    \thicklines
    \multiput(33,4)(25,0){1}{\line(1,0){15}} 
    \multiput(45,4)(25,0){1}{\vector(1,0){0}}
    \multiput(83,10)(25,0){1}{\line(1,0){15}} 
    \multiput(95,10)(25,0){1}{\vector(1,0){0}}
    \multiput(56,5)(25,0){1}{\vector(0,1){12}} 
    \multiput(75,5)(25,0){1}{\vector(0,1){12}} 
    \multiput(56,-1)(25,0){2}{\vector(0,1){12}} 
    \multiput(75,-1)(25,0){1}{\vector(0,1){12}} 
    \qbezier(8,10)(28,35)(48,10)
    \multiput(30.5,23)(69,0){1}{\vector(1,0){0}}
    \put(77,-10){j}
    \put(52,-10){i}
    \put(-32,5){(c)}
\end{picture}\\[5mm]
\begin{picture}(100,20)(0,0)
    \linethickness{0.8pt}
    \multiput(25,10)(25,0){4}{\circle{4}} 
    \multiput(31,10)(25,0){4}{\circle{4}} 
    \multiput(25,4)(25,0){4}{\circle{4}} 
    \multiput(31,4)(25,0){4}{\circle{4}} 
    \multiput(28,7)(25,0){4}{\circle{16}}
    \multiput(25,10)(25,0){1}{\circle*{4}}
    \multiput(31,10)(75,0){2}{\circle*{4}}
    \multiput(25,4)(25,0){1}{\circle*{4}}
    \multiput(100,10)(25,0){1}{\circle*{4}}
    \multiput(106,4)(25,0){1}{\circle*{4}}
    \thicklines
    \multiput(33,4)(50,0){2}{\line(1,0){15}} 
    \multiput(45,4)(50,0){2}{\vector(1,0){0}}
    \multiput(58,10)(25,0){1}{\line(1,0){15}} 
    \multiput(70,10)(25,0){1}{\vector(1,0){0}}
    \multiput(50,5)(25,0){1}{\vector(0,1){12}} 
    \multiput(81,5)(25,0){1}{\vector(0,1){12}} 
    \multiput(56,-1)(25,0){1}{\vector(0,1){12}} 
    \multiput(75,-1)(25,0){1}{\vector(0,1){12}} 
    \put(77,-10){j}
    \put(52,-10){i}
    \put(-32,5){(d)}
\end{picture}\\[5mm]
\begin{picture}(100,20)(0,0)
    \linethickness{0.8pt}
    
    \multiput(0,10)(25,0){4}{\circle{4}} 
    \multiput(6,10)(25,0){4}{\circle{4}} 
    \multiput(0,4)(25,0){4}{\circle{4}} 
    \multiput(6,4)(25,0){4}{\circle{4}} 
    \multiput(3,7)(25,0){4}{\circle{16}}
    \thicklines
    \multiput(33,4)(25,0){1}{\line(1,0){15}} 
    \multiput(45,4)(25,0){1}{\vector(1,0){0}}
    \multiput(0,10)(25,0){2}{\circle*{4}}
    \multiput(31,10)(75,0){1}{\circle*{4}}
    \multiput(6,4)(100,0){1}{\circle*{4}}
    \multiput(0,4)(25,0){2}{\circle*{4}}
    \multiput(56,5)(25,0){2}{\vector(0,1){12}} 
    \multiput(75,5)(25,0){1}{\vector(0,1){12}} 
    \multiput(56,-1)(25,0){2}{\vector(0,1){12}} 
    \multiput(75,-1)(25,0){1}{\vector(0,1){12}} 
    \qbezier(8,10)(28,35)(48,10)
    \multiput(30.5,23)(69,0){1}{\vector(1,0){0}}
    \put(77,-10){j}
    \put(52,-10){i}
    \put(-32,5){(e)}
\end{picture}\\[5mm]
\begin{picture}(100,20)(0,0)
    \linethickness{0.8pt}
    \multiput(0,10)(25,0){2}{\circle*{4}}
    \multiput(31,10)(75,0){1}{\circle*{4}}
    \multiput(6,4)(100,0){1}{\circle*{4}}
    \multiput(0,4)(25,0){2}{\circle*{4}}
    \multiput(0,10)(25,0){4}{\circle{4}} 
    \multiput(6,10)(25,0){4}{\circle{4}} 
    \multiput(0,4)(25,0){4}{\circle{4}} 
    \multiput(6,4)(25,0){4}{\circle{4}} 
    \multiput(3,7)(25,0){4}{\circle{16}}
    \thicklines
    \multiput(81,5)(25,0){1}{\vector(0,1){12}} 
    \put(6,5){\vector(0,1){12}}
    \multiput(50,5)(50,0){1}{\vector(0,1){12}} 
    \multiput(31,-1)(50,0){2}{\vector(0,1){12}} 
    \multiput(50,-1)(50,0){1}{\vector(0,1){12}} 
    \put(77,-10){j}
    \put(52,-10){i}
    \put(-32,5){(f)}
    \end{picture}
\caption{Scattering examples of $S = 2$ dimer configurations under the action of $c_{ij}$. Configurations of type (a) scatter to configurations of the type (b).  Configurations such as (c) and (d) are annihilated by $(c_{ij})^4$. Configuration (e) scatters to (f) under the action of $(c_{ij})^2$. The configurations of the filled small circles are irrelevant for the scattering due to $c_{ij}$.}
\label{fig:spinSscattering}
\end{figure}
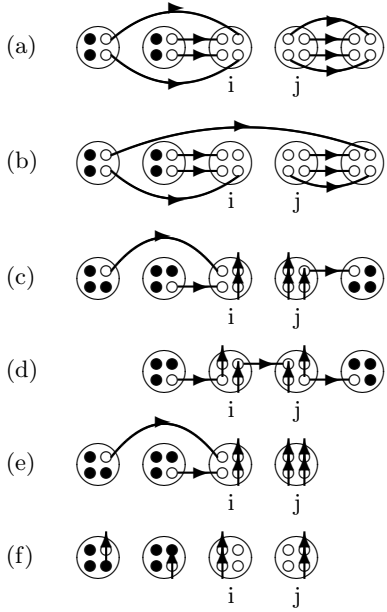
In this section, we give a brief overview on how terms in the Hamiltonian Eq.~\eqref{SHamiltoniandimerspecial} act on basis states of a spin-$S$ chain. To achieve this, since the Hamiltonian is normal ordered, it is sufficient to compute the actions of $(c_{ij})^m$ for $m = S, \dots, 2S$ on various configurations that can appear in a spin-$S$ chain. However, such expressions are lengthy in general and here we merely note the structure of the scattering terms. Note that the results in Appendices~\ref{spins} and \ref{dimers} are valid for any value of $S$.  
First we derive singlet scattering rules, analogous to the results in Appendix~\ref{S1singlet}. If the basis element does not contain any $c_{ij}^\dagger$, from Eqs.~(\ref{usefulidentity}) and (\ref{singletidentities}), the action of $c_{ij}$ on any basis state results in a sum of all possible configurations with one dimer annihilated on each site $i$ and $j$, and the resulting vacant sites on sites different from $i$ and $j$ connected. For example, the spin-2 singlet configuration shown in Fig.~\ref{fig:spinSscattering}a scatters to configurations such as the one shown in Fig.~\ref{fig:spinSscattering}b.
If the basis state has many dimers $(c_{ij}^\dagger)^n$, using Eqs.~(\ref{dimermanycomm}) and (\ref{singletidentities1}), the action of $c_{ij}$ gives rise to an additional term where the dimer $c_{ij}^\dagger$ is annihilated. 
With these observations, the action of $c_{ij}$ on a singlet basis state with no dimers $c_{ij}^\dagger$ and $N$ dimers connecting each of the sites $i$ and $j$ to other sites $\{p_l\}$ and $\{r_n\}$ respectively, can be written as
\begin{eqnarray}
    &c_{ij} \prodal{l = 1}{N}{c^\dagger_{p_l, i}}\prodal{n=1}{N}{c^\dagger_{j, r_n}}\ket{\Theta} = \nonumber \\
    &= - \sumal{l^\prime, n^\prime = 1}{N}{\prodal{l = 1, l \neq l^\prime }{N}{c^\dagger_{p_l, i}} \prodal{n = 1, n \neq n^\prime }{N}{c^\dagger_{j, r_n}} c^\dagger_{l^\prime, n^\prime}}\ket{\Theta}
\label{generalsingletrule}
\end{eqnarray}
In Eq.~(\ref{generalsingletrule}), on the first application of $c_{ij}$ on the spin-$S$ singlet state, $N = 2S$. Upon each application of $c_{ij}$, $N$ decreases by 1 (one dimer is annihilated on each of the sites $i$ and $j$).
The action of $(c_{ij})^m$ can be computed by consecutive applications of Eq.~(\ref{generalsingletrule}).  
The scattering configuration of $(c_{ij})^m$ acting on the original spin-$S$ singlet state would then be a sum of terms annihilating $m$ pairs of dimers, each pair with one connecting site $i$ and one connecting site $j$, and reconnecting the vacant sites in different possible ways.
The different ways to annihilate $m$ dimer pairs that lead to the same scattering term result in an overall $m!$ factor.
Morever, a factor of $(-1)^m$ appears because of the negative sign in Eq.~(\ref{generalsingletrule}).
Each $c_{ij}^\dagger$ in the original basis state gives rise to an additional scattering term where the $c_{ij}^\dagger$ is annihilated.
Though tedious to prove in general, we recover that $P^{(J,S)}_{ij}$, for $J > S$ vanishes on any configuration containing $(c_{ij}^\dagger)^S$, similar to Eq.~(\ref{projectorvanishing}).
To derive the non-singlet scattering rules, we need to consider basis elements that consist of some free spin-$\uparrow$ Schwinger bosons ($a^\dagger$).
Firstly, $c_{ij}$ annihilates any basis state that does not have dimers on sites $i$ or $j$, analogous to Eq.~\eqref{higherspinidentities1}.
From Eqs.~(\ref{higherspinidentities1}) to (\ref{higherspinidentitieslast}), observe that the action of $c_{ij}$ on other basis states results in a sum of terms with one free spin (or dimer, but not both spins) annihilated on each site $i$ and $j$ and the resulting vacant site(s) populated with a free spin (dimer). As earlier, each  $c_{ij}^\dagger$ in the original basis state gives rise to an additional scattering term with the $c_{ij}^\dagger$ annihilated. The action for $(c_{ij})^m$ can then be derived following the same procedure by repeated applications of $c_{ij}$s and accounting for the overcounting factors. 

The important conclusion from the above observations is that $(c_{ij})^m$ annihilates all configurations on bonds $\{i,j\}$ with a total of less than $m$ dimers on it or surrounding it, analogous to Eqs.~(\ref{higherspinidentities1}), (\ref{higherspinidentities4}), (\ref{higherspinidentities5}) and (\ref{higherspinidentities7}). For example, if a configuration has $N_i$ dimers connecting site $i$ to a site different from $j$, $N_j$ dimers connecting site $j$ to a site different from $i$ and $N_{ij}$ dimers connecting sites $i$ and $j$, the action of $(c_{ij})^m$ can be written as
\begin{eqnarray}
    &(c_{ij})^m \prodal{l = 1}{N_i}{c^\dagger_{p_l,i}}\prodal{n = 1}{N_j}{c^\dagger_{j,r_n}} (a^\dagger_i)^{2S - N_i} (c_{ij}^\dagger)^{N_{ij}} (a^\dagger_j)^{2S - N_j}\ket{\Theta} = 0  \nonumber \\ &\;\;\; \text{if} \;\;\; N_i + N_j + N_{ij} < m
\label{nonscattering}
\end{eqnarray}
where the sites $\{p_l\}$ and $\{r_n\}$ are assumed to be distinct from $j$ and $i$ respectively. For example, the spin-2 configurations shown in Figs.~\ref{fig:spinSscattering}c and \ref{fig:spinSscattering}d are annilated by $(c_{ij})^4$ since they have a total of three dimers on and around the bond $\{i,j\}$. 
Another useful configuration that we have used in our calculations is if $N_i + N_j = m$ and $N_{ij} = 0$. As discussed above, since each $c_{ij}$ annilates a dimer connected to either of the sites $i$ or $j$,  up to an overall constant, the scattering equation reads
\begin{eqnarray}
    &(c_{ij})^m \prodal{l = 1}{N_i}{c^\dagger_{p_l,i}}\prodal{n = 1}{N_j}{c^\dagger_{j,r_n}} (a^\dagger_i)^{2S - N_i} (a^\dagger_j)^{2S - N_j}\ket{\Theta} \nonumber \\ & \sim (a_i^\dagger a_j^\dagger)^{2S - m}\prodal{l = 1}{N_i}{a^\dagger_{p_l}} \prodal{n = 1}{N_j}{a^\dagger_{r_n}} \ket{\Theta}    \;\; \text{if} \;\;\; N_i + N_j = m.\nn\\[0.5mm]
\label{highspinhopping}
\end{eqnarray}
This is analogous to Eqs.~\eqref{higherspinidentities2}, \eqref{higherspinidentities10} and \eqref{higherspinidentities9}. In Eq.~(\ref{highspinhopping}), all the dimers connected to sites $i$ and $j$ are annihilated.
For example, the spin-2 configuration shown in Fig.~\ref{fig:spinSscattering}e scatters to the one shown in Fig.~\ref{fig:spinSscattering}f under the action of $(c_{ij})^2$. 
Thus the term $(c^\dagger_{ij})^m (c_{ij})^m$ in a Hamiltonian acting on such a configuration results in a configuration where the $m$ dimers around bond $\{i,j\}$ move on to bond $\{i,j\}$. 

\section{Scattering equations of the $\boldsymbol{S = 2}$ Arovas configurations}\label{sec:S2Arovas}
\begin{figure}[t!]
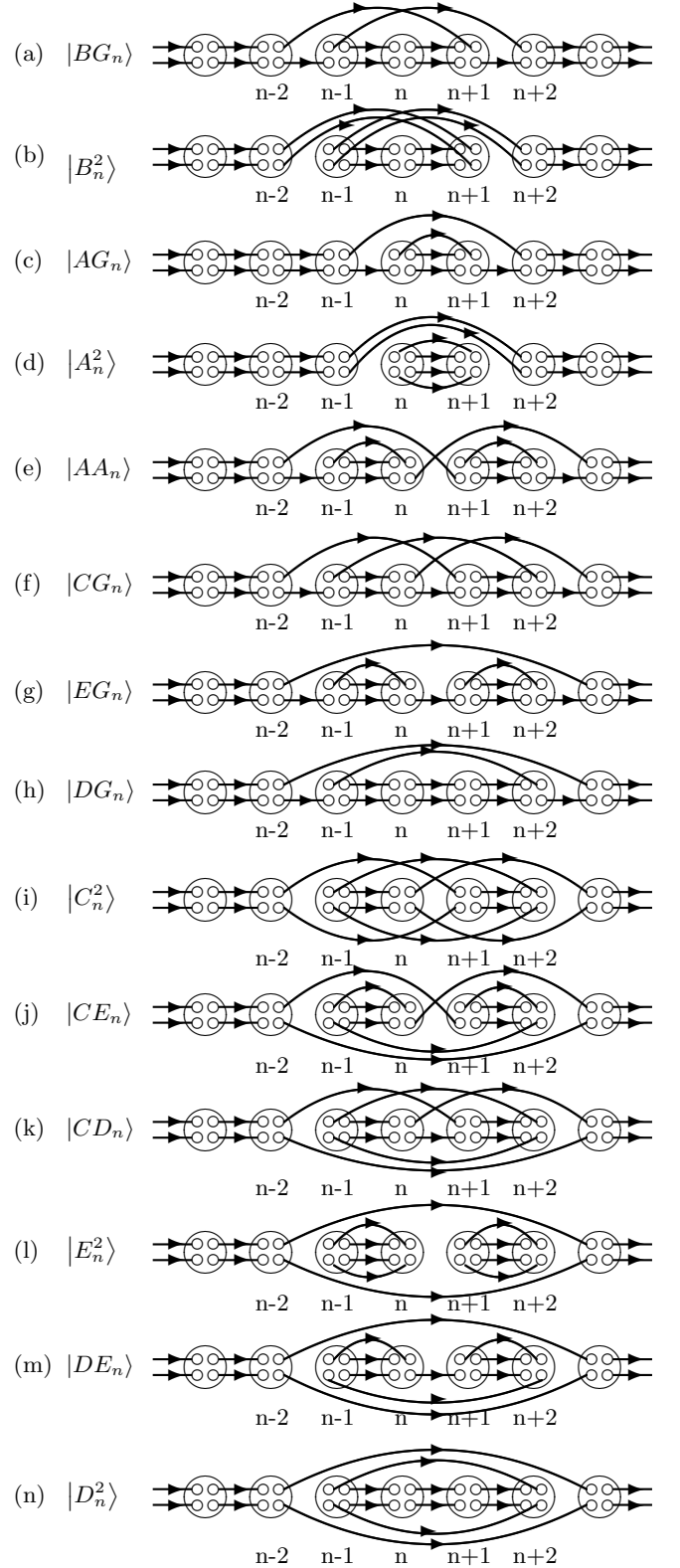

    \centering
\\[5mm]
    \caption{Symmetric scattering configurations that appear in the scattering equations of the $S =2$ Arovas configurations. All these configurations are obtained gluing two $S = 1$ Arovas scattering configurations, shown in Figs.~\ref{fig:arovasA} and \ref{fig:ArovasBConfig}.}
    \label{fig:spin2arovassym}
\end{figure}
\subsection{Arovas B configurations}\label{sec:S2ArovasB}
\begin{figure}[t!]
    \begin{picture}(105,20)(0,0)
        \linethickness{0.8pt}
        \multiput(0,15)(25,0){7}{\circle{4}} 
        \multiput(6,15)(25,0){7}{\circle{4}} 
        \multiput(0,9)(25,0){7}{\circle{4}} 
        \multiput(6,9)(25,0){7}{\circle{4}} 
        \multiput(3,12)(25,0){7}{\circle{16}}
        \thicklines
        \multiput(-17,15)(25,0){2}{\line(1,0){15}} 
        \multiput(-5,15)(25,0){2}{\vector(1,0){0}}
        \multiput(-17,9)(25,0){4}{\line(1,0){15}} 
        \multiput(120,9)(50,0){2}{\vector(1,0){0}}
        \multiput(108,9)(50,0){2}{\line(1,0){15}} 
        \multiput(-5,9)(25,0){4}{\vector(1,0){0}}
        \multiput(58,15)(25,0){1}{\line(1,0){15}}
        \multiput(70,15)(25,0){1}{\vector(1,0){0}}
        \qbezier(33,15)(68.5,45)(104,15)
        \put(65.5, 30){\vector(1,0){0}}
        \qbezier(52,15)(87.5,45)(130,15)
        \put(97.5, 30){\vector(1,0){0}}
        \multiput(108,15)(25,0){1}{\line(1,0){15}}
        \multiput(120,15)(25,0){1}{\vector(1,0){0}}
        \multiput(158,15)(25,0){1}{\line(1,0){15}}
        \multiput(170,15)(25,0){1}{\vector(1,0){0}}
        \qbezier(82,15)(112.5,45)(148,15)
        \qbezier(82,10)(112.5,42)(148,10)
        \qbezier(101,7)(112.5,0)(131,7)
        \put(122.5, 30){\vector(1,0){0}}
        \put(122.5, 27){\vector(1,0){0}}
        \put(120, 4){\vector(1,0){0}}
        \put(75,-5){n}
        \put(47,-5){n-1}
        \put(95,-5){n+1}
        \put(120,-5){n+2}
        \put(22,-5){n-2}
        \put(-70,10){(a)}
        \put(-50,10){$\ket{\zeta^1_n}$}
    \end{picture}\\[8mm]
    
    \begin{picture}(105,20)(0,0)
        \linethickness{0.8pt}
        \multiput(0,15)(25,0){7}{\circle{4}} 
        \multiput(6,15)(25,0){7}{\circle{4}} 
        \multiput(0,9)(25,0){7}{\circle{4}} 
        \multiput(6,9)(25,0){7}{\circle{4}} 
        \multiput(3,12)(25,0){7}{\circle{16}}
        \thicklines
        \multiput(-17,15)(50,0){2}{\line(1,0){15}} 
        \multiput(-5,15)(50,0){2}{\vector(1,0){0}}
        \multiput(-17,9)(50,0){2}{\line(1,0){15}} 
        \multiput(120,9)(50,0){2}{\vector(1,0){0}}
        \multiput(108,9)(50,0){2}{\line(1,0){15}} 
        \multiput(-5,9)(50,0){2}{\vector(1,0){0}}
        \multiput(83,15)(25,0){1}{\line(1,0){15}}
        \multiput(95,15)(25,0){1}{\vector(1,0){0}}
        \multiput(83,9)(25,0){4}{\line(1,0){15}}
        \multiput(95,9)(25,0){4}{\vector(1,0){0}}
        \qbezier(27,15)(65.5,45)(104,15)
        \put(62.5, 30){\vector(1,0){0}}
        \qbezier(58,15)(87.5,45)(124,15)
        \put(100.5, 30){\vector(1,0){0}}
        \multiput(133,15)(25,0){1}{\line(1,0){15}}
        \multiput(145,15)(25,0){1}{\vector(1,0){0}}
        \multiput(158,15)(25,0){1}{\line(1,0){15}}
        \multiput(170,15)(25,0){1}{\vector(1,0){0}}
        \qbezier(7,15)(37.5,45)(73,15)
        \qbezier(7,10)(37.5,42)(73,10)
        \qbezier(26,7)(37.5,0)(56,7)
        \put(47.5, 30){\vector(1,0){0}}
        \put(47.5, 27){\vector(1,0){0}}
        \put(45, 4){\vector(1,0){0}}
        \put(75,-5){n}
        \put(47,-5){n-1}
        \put(95,-5){n+1}
        \put(120,-5){n+2}
        \put(22,-5){n-2}
        \put(-70,10){(b)}
        \put(-50,10){$\ket{\widetilde{\zeta^1}_n}$}
        \end{picture}\\[5mm]
        
        \begin{picture}(105,20)(0,0)
        \linethickness{0.8pt}
        \multiput(0,10)(25,0){7}{\circle{4}} 
        \multiput(6,10)(25,0){7}{\circle{4}} 
        \multiput(0,4)(25,0){7}{\circle{4}} 
        \multiput(6,4)(25,0){7}{\circle{4}} 
        \multiput(3,7)(25,0){7}{\circle{16}}
        \thicklines
        \multiput(-17,10)(50,0){1}{\line(1,0){15}} 
        \multiput(-5,10)(50,0){1}{\vector(1,0){0}}
        \multiput(-17,4)(25,0){1}{\line(1,0){15}} 
        \multiput(-5,4)(25,0){1}{\vector(1,0){0}}
        \multiput(8,10)(25,0){3}{\line(1,0){15}} 
        \multiput(20,10)(25,0){3}{\vector(1,0){0}}
        \multiput(8,4)(50,0){2}{\line(1,0){15}} 
        \multiput(20,4)(50,0){2}{\vector(1,0){0}}
        \multiput(108,4)(50,0){2}{\line(1,0){15}} 
        \multiput(120,4)(50,0){2}{\vector(1,0){0}}
        \multiput(108,10)(25,0){1}{\line(1,0){15}}
        \multiput(120,10)(25,0){1}{\vector(1,0){0}}
        
        \qbezier(83,10)(112.5,40)(148,10)
        \put(122.5, 24){\vector(1,0){0}}
        \qbezier(52,2)(65.5,-7)(79,2)
        \put(70,-3){\vector(1,0){0}}
        \qbezier(102,10)(115.5,25)(129,10)
        \put(120,18){\vector(1,0){0}}
        \multiput(158,10)(25,0){1}{\line(1,0){15}}
        \multiput(170,10)(25,0){1}{\vector(1,0){0}}
        
        \qbezier(33,4)(90.5,-24)(148,4)
        \put(95,-10){\vector(1,0){0}}
        
        \qbezier(101,2)(112.5,-7)(131,2)
        \put(120, -3){\vector(1,0){0}}
        \put(75,-16){n}
        \put(47,-16){n-1}
        \put(95,-16){n+1}
        \put(120,-16){n+2}
        \put(22,-16){n-2}
        \put(-70,5){(c)}
        \put(-50,5){$\ket{\zeta^2_n}$}
        \end{picture}\\[12mm]
        
        \begin{picture}(105,20)(0,0)
        \linethickness{0.8pt}
        \multiput(0,10)(25,0){7}{\circle{4}} 
        \multiput(6,10)(25,0){7}{\circle{4}} 
        \multiput(0,4)(25,0){7}{\circle{4}} 
        \multiput(6,4)(25,0){7}{\circle{4}} 
        \multiput(3,7)(25,0){7}{\circle{16}}
        \thicklines
        \multiput(-17,10)(50,0){1}{\line(1,0){15}} 
        \multiput(-5,10)(50,0){1}{\vector(1,0){0}}
        \multiput(-17,4)(25,0){1}{\line(1,0){15}} 
        \multiput(-5,4)(25,0){1}{\vector(1,0){0}}
        \multiput(33,10)(50,0){3}{\line(1,0){15}} 
        \multiput(45,10)(50,0){3}{\vector(1,0){0}}
        \multiput(33,4)(50,0){3}{\line(1,0){15}} 
        \multiput(45,4)(50,0){3}{\vector(1,0){0}}
        \multiput(158,4)(50,0){1}{\line(1,0){15}} 
        \multiput(170,4)(50,0){1}{\vector(1,0){0}}
        \multiput(108,10)(25,0){1}{\line(1,0){15}}
        \multiput(120,10)(25,0){1}{\vector(1,0){0}}
        
        \qbezier(8,10)(37.5,40)(73,10)
        \put(47.5, 24){\vector(1,0){0}}
        \qbezier(77,2)(90.5,-7)(104,2)
        \put(95,-3){\vector(1,0){0}}
        \qbezier(27,10)(40.5,25)(54,10)
        \put(45,18){\vector(1,0){0}}
        \multiput(158,10)(25,0){1}{\line(1,0){15}}
        \multiput(170,10)(25,0){1}{\vector(1,0){0}}
        
        \qbezier(8,4)(65.5,-24)(123,4)
        \put(70,-10){\vector(1,0){0}}
        
        \qbezier(26,2)(37.5,-7)(56,2)
        \put(45, -3){\vector(1,0){0}}
        \put(75,-16){n}
        \put(47,-16){n-1}
        \put(95,-16){n+1}
        \put(120,-16){n+2}
        \put(22,-16){n-2}
        \put(-70,5){(d)}
        \put(-50,5){$\ket{\widetilde{\zeta^2}_n}$}
        \end{picture}\\[12mm]
        
        \begin{picture}(105,20)(0,0)
        \linethickness{0.8pt}
        \multiput(0,15)(25,0){7}{\circle{4}} 
        \multiput(6,15)(25,0){7}{\circle{4}} 
        \multiput(0,9)(25,0){7}{\circle{4}} 
        \multiput(6,9)(25,0){7}{\circle{4}} 
        \multiput(3,12)(25,0){7}{\circle{16}}
        \thicklines
        \multiput(-17,15)(25,0){2}{\line(1,0){15}} 
        \multiput(-5,15)(25,0){2}{\vector(1,0){0}}
        \multiput(-17,9)(25,0){4}{\line(1,0){15}} 
        \multiput(120,9)(50,0){2}{\vector(1,0){0}}
        \multiput(108,9)(50,0){2}{\line(1,0){15}} 
        \multiput(-5,9)(25,0){4}{\vector(1,0){0}}
        \multiput(58,15)(25,0){2}{\line(1,0){15}}
        \multiput(70,15)(25,0){2}{\vector(1,0){0}}
        \qbezier(52,15)(87.5,37)(130,15)
        \put(97.5, 30){\vector(1,0){0}}
        \multiput(108,15)(25,0){1}{\line(1,0){15}}
        \multiput(120,15)(25,0){1}{\vector(1,0){0}}
        \multiput(158,15)(25,0){1}{\line(1,0){15}}
        \multiput(170,15)(25,0){1}{\vector(1,0){0}}
        \qbezier(82,7)(112.5,-18)(148,7)
        \qbezier(101,7)(112.5,0)(131,7)
        \qbezier(33,15)(90.5,45)(148,15)
        \put(95,25){\vector(1,0){0}}
        \put(122, -6){\vector(1,0){0}}
        \put(120, 4){\vector(1,0){0}}
        \put(75,-15){n}
        \put(47,-15){n-1}
        \put(95,-15){n+1}
        \put(120,-15){n+2}
        \put(22,-15){n-2}
        \put(-70,10){(e)}
        \put(-50,10){$\ket{\zeta^3_n}$}
        \end{picture}\\[10mm]
        
        \begin{picture}(105,20)(0,0)
        \linethickness{0.8pt}
        \multiput(0,15)(25,0){7}{\circle{4}} 
        \multiput(6,15)(25,0){7}{\circle{4}} 
        \multiput(0,9)(25,0){7}{\circle{4}} 
        \multiput(6,9)(25,0){7}{\circle{4}} 
        \multiput(3,12)(25,0){7}{\circle{16}}
        \thicklines
        \multiput(-17,15)(50,0){4}{\line(1,0){15}} 
        \multiput(-5,15)(50,0){4}{\vector(1,0){0}}
        \multiput(-17,9)(50,0){4}{\line(1,0){15}} 
        \multiput(120,9)(25,0){3}{\vector(1,0){0}}
        \multiput(108,9)(25,0){3}{\line(1,0){15}} 
        \multiput(-5,9)(50,0){4}{\vector(1,0){0}}
        \multiput(58,15)(25,0){1}{\line(1,0){15}}
        \multiput(70,15)(25,0){1}{\vector(1,0){0}}
        \qbezier(27,15)(62.5,37)(105,15)
        \put(72.5, 30){\vector(1,0){0}}
        \multiput(158,15)(25,0){1}{\line(1,0){15}}
        \multiput(170,15)(25,0){1}{\vector(1,0){0}}
        \qbezier(7,7)(37.5,-18)(73,7)
        \qbezier(26,7)(37.5,0)(56,7)
        \qbezier(8,15)(65.5,45)(123,15)
        \put(70,26){\vector(1,0){0}}
        \put(47, -6){\vector(1,0){0}}
        \put(45, 4){\vector(1,0){0}}
        \put(75,-15){n}
        \put(47,-15){n-1}
        \put(95,-15){n+1}
        \put(120,-15){n+2}
        \put(22,-15){n-2}
        \put(-70,10){(f)}
        \put(-50,10){$\ket{\widetilde{\zeta^3}_n}$}
        \end{picture}\\[10mm]
        
        \begin{picture}(105,20)(0,0)
        \linethickness{0.8pt}
        \multiput(0,15)(25,0){7}{\circle{4}} 
        \multiput(6,15)(25,0){7}{\circle{4}} 
        \multiput(0,9)(25,0){7}{\circle{4}} 
        \multiput(6,9)(25,0){7}{\circle{4}} 
        \multiput(3,12)(25,0){7}{\circle{16}}
        \thicklines
        \multiput(-17,15)(25,0){3}{\line(1,0){15}} 
        \multiput(-5,15)(25,0){3}{\vector(1,0){0}}
        \multiput(-17,9)(25,0){4}{\line(1,0){15}} 
        \multiput(120,9)(50,0){2}{\vector(1,0){0}}
        \multiput(108,9)(50,0){2}{\line(1,0){15}} 
        \multiput(-5,9)(25,0){4}{\vector(1,0){0}}
        \multiput(83,15)(25,0){1}{\line(1,0){15}}
        \multiput(95,15)(25,0){1}{\vector(1,0){0}}
        
        \qbezier(58,15)(87.5,45)(130,15)
        \put(97.5, 30){\vector(1,0){0}}
        \multiput(108,15)(25,0){1}{\line(1,0){15}}
        \multiput(120,15)(25,0){1}{\vector(1,0){0}}
        \multiput(158,15)(25,0){1}{\line(1,0){15}}
        \multiput(170,15)(25,0){1}{\vector(1,0){0}}
        \qbezier(76,15)(112.5,45)(148,15)
        \qbezier(82,10)(112.5,42)(148,10)
        \qbezier(101,7)(112.5,0)(131,7)
        \put(122.5, 30){\vector(1,0){0}}
        \put(122.5, 27){\vector(1,0){0}}
        \put(120, 4){\vector(1,0){0}}
        \put(75,-5){n}
        \put(47,-5){n-1}
        \put(95,-5){n+1}
        \put(120,-5){n+2}
        \put(22,-5){n-2}
        \put(-70,10){(g)}
        \put(-50,10){$\ket{\zeta^4_n}$}
    \end{picture}\\[8mm]
    
    \begin{picture}(105,20)(0,0)
        \linethickness{0.8pt}
        \multiput(0,15)(25,0){7}{\circle{4}} 
        \multiput(6,15)(25,0){7}{\circle{4}} 
        \multiput(0,9)(25,0){7}{\circle{4}} 
        \multiput(6,9)(25,0){7}{\circle{4}} 
        \multiput(3,12)(25,0){7}{\circle{16}}
        \thicklines
        \multiput(-17,15)(50,0){2}{\line(1,0){15}} 
        \multiput(-5,15)(50,0){2}{\vector(1,0){0}}
        \multiput(-17,9)(50,0){2}{\line(1,0){15}} 
        \multiput(120,9)(50,0){2}{\vector(1,0){0}}
        \multiput(108,9)(50,0){2}{\line(1,0){15}} 
        \multiput(-5,9)(50,0){2}{\vector(1,0){0}}
        \multiput(108,15)(25,0){1}{\line(1,0){15}}
        \multiput(120,15)(25,0){1}{\vector(1,0){0}}
        \multiput(58,15)(25,0){1}{\line(1,0){15}}
        \multiput(70,15)(25,0){1}{\vector(1,0){0}}
        \multiput(83,9)(25,0){4}{\line(1,0){15}}
        \multiput(95,9)(25,0){4}{\vector(1,0){0}}
        \qbezier(27,15)(65.5,45)(98,15)
        \put(62.5, 30){\vector(1,0){0}}
        \multiput(133,15)(25,0){1}{\line(1,0){15}}
        \multiput(145,15)(25,0){1}{\vector(1,0){0}}
        \multiput(158,15)(25,0){1}{\line(1,0){15}}
        \multiput(170,15)(25,0){1}{\vector(1,0){0}}
        \qbezier(7,15)(37.5,45)(79,15)
        \qbezier(7,10)(37.5,42)(73,10)
        \qbezier(26,7)(37.5,0)(56,7)
        \put(47.5, 30){\vector(1,0){0}}
        \put(47.5, 27){\vector(1,0){0}}
        \put(45, 4){\vector(1,0){0}}
        \put(75,-5){n}
        \put(47,-5){n-1}
        \put(95,-5){n+1}
        \put(120,-5){n+2}
        \put(22,-5){n-2}
        \put(-70,10){(h)}
        \put(-50,10){$\ket{\widetilde{\zeta^4}_n}$}
        \end{picture}\\[5mm]
\caption{Non-symmetric scattering configurations that appear in the scattering equation of the $S = 2$ Arovas configurations.}
\label{fig:spin2arovasnonsym}
\end{figure}
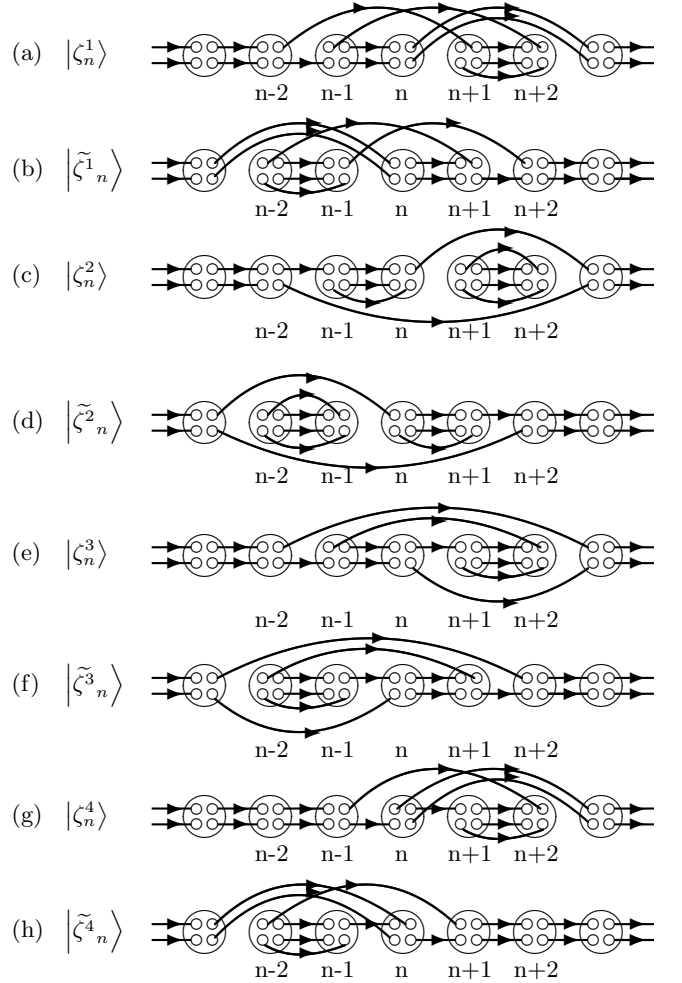

We start with the Arovas B configurations $\ket{BG_n}$ (Fig.~\ref{fig:spin2arovas}a) and $\ket{B^2_n}$ (Fig.~\ref{fig:spin2arovas}b). Applying the $S = 2$ AKLT Hamiltonian Eq.~(\ref{spin2hamil}) on $\ket{BG_n}$ and $\ket{B^2_n}$ generates non-trivial scattering terms due to the action of the projectors on the bonds $\{n-2,n-1\}$ and $\{n+1,n+2\}$. The scattering equations of $\ket{BG_n}$ and $\ket{B^2_n}$ read
\begin{eqnarray}
    &H^{(2)}\ket{BG_n} = 2\ket{BG_n} - \frac{1}{3}\ket{2G} + \frac{1}{3}\left(\ket{AG_{n-1}} + \ket{AG_n}\right)\nn \\
    &- \frac{2}{15}\left(\ket{AG_{n-2}} + \ket{AG_{n+1}}\right) + \frac{2}{15}\left(\ket{AA_{n-1}} + \ket{AA_n}\right) \nn \\
    &+ \frac{2}{3}\left(\ket{CG_{n-1}} + \ket{CG_n}\right) + \frac{2}{15}\left(\ket{EG_{n-1}} + \ket{EG_n}\right) \nn \\
    &+ \frac{1}{3}\left(\ket{DG_{n-1}} + \ket{DG_n}\right) - \frac{1}{60}\left(\ket{A^2_{n-2}} + \ket{A^2_{n+1}}\right) \nn \\
    &+ \frac{1}{15}\left(\ket{\zeta^1_n} + \ket{\widetilde{\zeta^1}_n}\right) + \frac{1}{30}\left(\ket{\zeta^2_n} + \ket{\widetilde{\zeta^2}_n}\right) \nn \\
    &+ \frac{2}{15}\left(\ket{\zeta^3_n} + \ket{\widetilde{\zeta^3}_n}\right)\label{spin2arovasbg} \\
    &H^{(2)}\ket{B^2_n} = 2\ket{B^2_n} - \frac{2}{21}\ket{2G} + \frac{4}{21}\left(\ket{AG_{n-1}} + \ket{AG_n}\right)\nn \\
    &- \frac{1}{21}\left(\ket{A^2_{n-1}} + \ket{A^2_n}\right) + \frac{16}{105}\left(\ket{AA_{n-1}} + \ket{AA_n}\right) \nn \\
    &- \frac{8}{105}\left(\ket{AG_{n-2}} + \ket{AG_{n+1}}\right) + \frac{8}{21}\left(\ket{CG_{n-1}} + \ket{CG_n}\right) \nn \\
    &+ \frac{16}{105}\left(\ket{EG_{n-1}} + \ket{EG_{n}}\right) - \frac{4}{21}\left(\ket{DG_{n-1}} + \ket{DG_n}\right) \nn \\
    & - \frac{1}{70}\left(\ket{A^2_{n-2}} + \ket{A^2_{n+1}}\right) - \frac{1}{21}\left(\ket{C^2_{n-1}} + \ket{C^2_n}\right) \nn \\
    &- \frac{16}{105}\left(\ket{CE_{n-1}} + \ket{CE_{n}}\right) - \frac{2}{105}\left(\ket{CD_{n-1}} + \ket{CD_n}\right) \nn \\
    &- \frac{1}{70}\left(\ket{E^2_{n-1}} + \ket{E^2_{n}}\right) - \frac{8}{105}\left(\ket{DE_{n-1}} + \ket{DE_n}\right) \nn \\
    &- \frac{1}{21}\left(\ket{D^2_{n-1}} + \ket{D^2_n}\right)
    - \frac{4}{21}\left(\ket{\zeta^1_{n-1}} + \ket{\widetilde{\zeta^1}_{n+1}}\right) \nn \\
    &- \frac{8}{105}\left(\ket{\zeta^2_{n-1}} + \ket{\widetilde{\zeta^2}_{n+1}}\right) 
    - \frac{8}{21}\left(\ket{\zeta^3_{n-1}} + \ket{\widetilde{\zeta^3}_{n+1}}\right)\nn \\
    &+ \frac{8}{105}\left(\ket{\zeta^1_{n}} + \ket{\widetilde{\zeta^1}_{n}}\right) 
    + \frac{2}{35}\left(\ket{\zeta^2_{n}} + \ket{\widetilde{\zeta^2}_{n}}\right)\nn \\
    &+ \frac{16}{105}\left(\ket{\zeta^3_{n}} + \ket{\widetilde{\zeta^3}_{n}}\right)
\label{spin2arovasb2}
\end{eqnarray}
where the symmetric and non-symmetric scattering configurations are defined in Figs.~\ref{fig:spin2arovassym} and \ref{fig:spin2arovasnonsym} respectively. The scattering equation for $\ket{2B_n} = 2 \ket{BG_n} - \frac{1}{2}\ket{B^2_n}$ (see Eq.~(\ref{2Bconfig})) is thus
\begin{eqnarray}
    &H^{(2)}\ket{2B_n} = 2\ket{2B_n} - \frac{13}{21}\ket{2G} + \frac{4}{7}\left(\ket{AG_{n-1}} + \ket{AG_n}\right)\nn \\
    &+ \frac{1}{42}\left(\ket{A^2_{n-1}} + \ket{A^2_n}\right) + \frac{4}{21}\left(\ket{AA_{n-1}} + \ket{AA_n}\right) \nn \\
    &- \frac{8}{35}\left(\ket{AG_{n-2}} + \ket{AG_{n+1}}\right) + \frac{8}{7}\left(\ket{CG_{n-1}} + \ket{CG_n}\right) \nn \\
    &+ \frac{4}{21}\left(\ket{EG_{n-1}} + \ket{EG_{n}}\right) + \frac{16}{21}\left(\ket{DG_{n-1}} + \ket{DG_n}\right) \nn \\
    & - \frac{11}{420}\left(\ket{A^2_{n-2}} + \ket{A^2_{n+1}}\right) + \frac{1}{42}\left(\ket{C^2_{n-1}} + \ket{C^2_n}\right) \nn \\
    &+ \frac{8}{105}\left(\ket{CE_{n-1}} + \ket{CE_{n}}\right) + \frac{1}{105}\left(\ket{CD_{n-1}} + \ket{CD_n}\right) \nn \\
    &+ \frac{1}{140}\left(\ket{E^2_{n-1}} + \ket{E^2_{n}}\right) + \frac{4}{105}\left(\ket{DE_{n-1}} + \ket{DE_n}\right) \nn \\
    &+ \frac{1}{42}\left(\ket{D^2_{n-1}} + \ket{D^2_n}\right)
    + \frac{2}{21}\left(\ket{\zeta^1_{n-1}} + \ket{\zeta^1_{n}}\right) \nn \\
    &+ \frac{4}{105}\left(\ket{\zeta^2_{n-1}} + \ket{\zeta^2_{n}}\right) 
    + \frac{4}{21}\left(\ket{\zeta^3_{n-1}} + \ket{\zeta^3_{n}}\right)\nn \\
    &+ \frac{2}{21}\left(\ket{\widetilde{\zeta^1}_{n}} + \ket{\widetilde{\zeta^1}_{n+1}}\right) 
    + \frac{4}{105}\left( \ket{\widetilde{\zeta^2}_{n}}+ \ket{\widetilde{\zeta^2}_{n+1}}\right)\nn \\
    &+ \frac{4}{21}\left( \ket{\widetilde{\zeta^3}_{n}}+ \ket{\widetilde{\zeta^3}_{n+1}}\right).
\label{spin2arovas2b}
\end{eqnarray}
Note that Eq.~(\ref{spin2arovas2b}) is of the form Eq.~(\ref{spin2final}) and all the scattering terms can be cancelled by a momentum $\pi$ superposition.
\subsection{Arovas A configurations}\label{sec:S2ArovasA}
In this section, we show that an exact state constructed from the spin-1 Arovas A configuration does not exist for our choice of Hamiltonian Eq.~(\ref{spin2hamil}). However, we construct two spin-2 Hamiltonians, that have such an exact excited state.
The Arovas A configurations that can be obtained for $S = 2$ are $\ket{A^2_n}$ (Fig.~\ref{fig:spin2arovassym}a) and $\ket{AG_n}$ (Fig.~\ref{fig:spin2arovassym}b). A general $S = 2$ AKLT Hamiltonian that has the ground state shown in Fig.~\ref{fig:spin2gs} reads
\begin{equation}
    H^{(2)}_x = \sum_i{\left(P^{(4,2)}_{i,i+1} + x P^{(3,2)}_{i,i+1}\right)}
\label{genspin2proj}
\end{equation}
where the projectors $P^{(J,S)}_{ij}$ are defined in Eq.~(\ref{projector}).
In terms of the dimers, $H^{(2)}_x$ can be written as
\begin{eqnarray}
    &H^{(2)}_x = \sum_i{\left(1 + \frac{x - 1}{8}c^\dagger_{i,i+1} c_{i,i+1} + \frac{3 - 7x}{336} {{{c^\dagger}^2_{i,i+1}}} {c^2_{i,i+1}}\right.} \nn \\
    &\left.+ \frac{21 x- 5}{10080}{c^\dagger}^3_{i,i+1} {c^3_{i,i+1}} + \frac{1 - 7x}{40320}{{c^\dagger}^4_{i,i+1}} {c^4_{i,i+1}}\right).
\label{genspin2hamil}
\end{eqnarray}
Note that our choice of Hamiltonian in Eq.~(\ref{spin2hamil}) to obtain the tower of states corresponds to $H^{(2)}_{x = 1}$. The scattering equations of $\ket{A^2_n}$ and $\ket{AG_n}$ under $H^{(2)}_x$ read
\begin{eqnarray}
    &H^{(2)}_x \ket{A^2_n} = 2\ket{A^2_n} + \frac{3-7x}{7}\ket{G} + 2(1-x)\ket{AG_{n}} \nn \\
    &+ \frac{5 - 21x}{35}\left(\ket{AG_{n-1}} + \ket{AG_{n+1}}\right) + \frac{6 - 14x}{7}\left(\ket{BG_n} + \ket{BG_{n+1}}\right)\nn \\
    &+\frac{1 - 7x}{70}\left(\ket{A^2_{n-1}} + \ket{A^2_{n+1}}\right) + \frac{3 - 7x}{14}\left(\ket{B^2_n}+\ket{B^2_{n+1}}\right) \nn \\
    &+ \frac{5 - 21x}{35}\left(\ket{\zeta^4_n} + \ket{\widetilde{\zeta^4}_n}\right) + (1-x)\left(\ket{\zeta^4_{n-1}} + \ket{\widetilde{\zeta^4}_{n+1}}\right) \nn \\
\end{eqnarray}
\begin{eqnarray}
    &H^{(2)}_x \ket{AG_n} = 2\ket{AG_n} + x\ket{G} - 2(1-x)\ket{AG_{n}} \nn \\
    &+ \frac{2x}{5}\left(\ket{AG_{n-1}} + \ket{AG_{n+1}}\right) + x\left(\ket{BG_n} + \ket{BG_{n+1}}\right)\nn \\
    &+ \frac{x}{20}\left(\ket{A^2_{n-1}} + \ket{A^2_{n+1}}\right) + \frac{x}{5}\left(\ket{\zeta^4_n} + \ket{\widetilde{\zeta^4}_n}\right)
\end{eqnarray}
where the symmetric and non-symmetric scattering configurations are defined in Fig.~\ref{fig:spin2arovassym} and Fig.~\ref{fig:spin2arovasnonsym} respectively. An exact state similar to the Arovas A state of the spin-1 AKLT model can be constructed from a superposition $\ket{2A_n(y)}$ of $\ket{A^2_n}$ and $\ket{AG_n}$, defined as
\begin{equation}
    \ket{2A_n(y)} = \ket{A^2_n} + y \ket{AG_n}
\end{equation}
only if non-symmetric configurations $\ket{\zeta^4_{n-1}}$ and $\ket{\zeta^4_n}$ appear in the scattering equation with the same coefficient, thus enabling their cancellation at momentum $\pi$. Hence, we want
\begin{equation}
    \frac{y x}{5} + \frac{5 - 21x}{35} = 1 - x,
\end{equation}
leading to 
\begin{equation}
    y = \frac{30 - 14x}{7x}.
\label{yconstraint}
\end{equation}
The scattering equation of $\ket{2A_n(y = \frac{30 - 14 x}{7x})}$ then reads
\begin{eqnarray}
    &H^{(2)}_x \ket{2A_n(y = \frac{30 - 14x}{7x})} = 2\ket{2A_n(y = \frac{30 - 14x}{7x})} + \frac{33-21 x}{7}\ket{G} \nn \\
    &- \frac{42 x^2 - 102 x + 60}{7x}\ket{AG_{n}}
    + \frac{65 - 49 x}{35}\left(\ket{AG_{n-1}} + \ket{AG_{n+1}}\right) \nn \\
    &+ \frac{36 - 28 x}{7}\left(\ket{BG_n} + \ket{BG_{n+1}}\right)\nn \\
    &+  \frac{8 - 7x}{35}\left(\ket{A^2_{n-1}} + \ket{A^2_{n+1}}\right) + \frac{3 - 7x}{14}\left(\ket{B^2_n}+\ket{B^2_{n+1}}\right).\nn \\
    &+ (1-x)\left(\ket{\zeta^4_{n-1}} + \ket{\zeta^4_n}\right) + (1-x) \left(\ket{\widetilde{\zeta^4}_n} + \ket{\widetilde{\zeta^4}_{n+1}}\right).\nn \\
\end{eqnarray}
With a momentum $\pi$ superposition of $\ket{2A_n(y = \frac{30 - 14 x}{7x})}$ and an even system size $L$, all the scattering terms except $\ket{A_n}$ and $\ket{AG_n}$ cancel. The scattering equation for the state $\ket{2A_x}$, defined as
\begin{equation}
    \ket{2A_x} = \sum_{n}{(-1)^n \ket{2A_n(y = \frac{30 - 14 x}{7x})}}
\end{equation}
thus reads
\begin{eqnarray}
    &H^{(2)}_x\ket{2A_x} = 2\ket{2A_x} - \frac{112 x^2 - 380x + 300}{35 x}\ket{AG} \nn \\
    &+ \frac{14x - 16}{35}\ket{A^2}
\label{2Ascattering}
\end{eqnarray}
where
\begin{equation}
    \ket{AG} = \sum_n{(-1)^n \ket{AG_n}}, \;\;\; \ket{A^2} = \sum_n{(-1)^n \ket{A^2_n}}.
\end{equation}
For $\ket{2A_x}$ to be an exact excited state, the coefficients of $\ket{AG}$ and $\ket{A^2}$ in Eq.~(\ref{2Ascattering}) should be in the ratio $y$, determined in Eq.~(\ref{yconstraint}). That is,
\begin{equation}
    \frac{112 x^2 - 380 x + 300}{35 x} = \frac{30 - 14x}{7x} \times \frac{14x-16}{35}.
\label{xconstraint}
\end{equation}
Solving Eq.~(\ref{xconstraint}), we obtain
\begin{equation}
    x = \frac{9}{7}\;\;\; {\rm or} \;\;\; x = \frac{15}{7}.
\label{xvalues}
\end{equation}
With these values of $x$, the Hamiltonians read
\begin{eqnarray}
    &H^{(2)}_{x = 9/7}\ket{2A_{x=9/7}} = \frac{72}{35}\ket{2A_{x=9/7}} \nn \\ &H^{(2)}_{x=15/7}\ket{2A_{x=15/7}} = \frac{12}{5}\ket{2A_{x=15/7}}.
\end{eqnarray}
The exact Arovas A eigenstates of the Hamiltonian Eq.~(\ref{genspin2proj}) with $x = 9/7$ and $x = 15/7$ are
\begin{equation}
    \ket{2A_{x = 9/7}} = \ket{A^2} + \frac{4}{3}\ket{AG}\;\; {\rm and} \;\; \ket{2A_{x = 15/7}} = \ket{A^2}
\end{equation}
with $\boldsymbol{s = 0}$, $\boldsymbol{k = \pi}$ and energies $\boldsymbol{E_{x = 9/7} = 72/35}$ and $\boldsymbol{E_{x = 15/7} = 12/5}$ respectively.

As expected, in both the Hamiltonians $H_{x = 9/7}$ and $H_{x = 15/7}$, we do not find an exact state analogous to the Arovas B state or a tower of states. It is likely that the Arovas A state can be generalized for a spin-$S$ Hamiltonian of the form of Eq.~(\ref{SHamiltonian}) by enforcing consistency conditions similar to Eqs.~(\ref{yconstraint}) and (\ref{xconstraint}) on the coefficients $\{\alpha_J\}$ in Eq.~(\ref{SHamiltonian}).

\section{Representing Majumdar-Ghosh exact states in terms of dimers}\label{mgdimer}
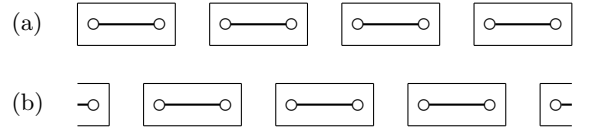
\begin{figure}
\centering
\begin{picture}(160,8)(-4,-3)
\multiput(6,0)(25,0){8}{\circle{4}}
\multiput(0,-8)(50,0){4}{\line(0,1){16}}
\multiput(37,-8)(50,0){4}{\line(0,1){16}}
\multiput(0,8)(50,0){4}{\line(1,0){37}}
\multiput(0,-8)(50,0){4}{\line(1,0){37}}
\put(-25,-2){(a)}
\thicklines
\multiput(8,0)(50,0){4}{\line(1,0){21}}
\end{picture}\\[7mm]

\begin{picture}(160,8)(-4,-3)
\multiput(6,0)(25,0){8}{\circle{4}}
\multiput(25,-8)(50,0){4}{\line(0,1){16}}
\multiput(12,-8)(50,0){4}{\line(0,1){16}}
\multiput(25,8)(50,0){3}{\line(1,0){37}}
\multiput(25,-8)(50,0){3}{\line(1,0){37}}
\put(12,8){\line(-1,0){12}}
\put(12,-8){\line(-1,0){12}}
\put(175,8){\line(1,0){12}}
\put(175,-8){\line(1,0){12}}
\put(-25,-2){(b)}
\thicklines
\multiput(33,0)(50,0){3}{\line(1,0){21}}
\put(4,0){\line(-1,0){4}}
\put(183,0){\line(1,0){4}}
\end{picture}\\[3mm]
\caption{Cells in the Majumdar-Ghosh model (a) ``Even" ground state (b) ``Odd" ground state}
\label{fig:mgcells}
\end{figure}
Ref.~[\onlinecite{caspers1982some}] introduces the Majumdar-Ghosh model in terms of the basic unit of a cell, a set of neighboring sites (Fig.~\ref{fig:mgcells}). Since each site has a spin-1/2, each cell can either be a singlet or a triplet. The ground state is the state where all the cells are singlets. The two equivalent choices of cells, shown in Fig.~\ref{fig:mgcells}a and Fig.~\ref{fig:mgcells}b, accounts for its doubly degenerate ground state Eq.~(\ref{eq:mg2}). We label the cells from $1$ to $N$ and the sites from $1$ to $L$. Thus, the cell $p$ comprises of sites $2p - 1$ and $2p$.

A singlet exact state $\ket{\phi}$ was first introduced in Eq.~(9) in Ref.~[\onlinecite{caspers1982some}]. If $N$ is the number of cells ($L = 2N$ number of sites), the exact state is defined as
\begin{equation}
    \ket{\phi} = \frac{1}{\sqrt{N}}\sum_{p=1}^N{(-1)^p \ket{0\,0}_{p,p+1}}
    \label{casperssingleteqn}
\end{equation}
where $\ket{0\,0}_{p,p+1}$ is a state with total spin 0 formed by the spin triplets of neighboring cells $p$ and $p + 1$. $\ket{0\,0}_{p, p+1}$ is defined as
\begin{equation}
    \ket{0\,0}_{p,p+1} = \frac{\ket{t_1\,t_{\mm}}_{p,p+1} + \ket{t_{\mm}\,t_1}_{p,p+1} - \ket{t_0\,t_0}_{p,p+1}}{\sqrt{3}}
    \label{neighsing}
\end{equation}
where $\ket{t_a\,t_b}_{p,p+1}$ is the state where the cells $p$ and $p + 1$ are comprised of spin triplet configurations with $S_z = a$ and $S_z = b$ and the rest of the cells are spin singlets. Thus,
\begin{eqnarray}
\ket{t_1}_{p} &=& a_{2p-1}^\dagger a_{2p}^\dagger \vac \nn \\
\ket{t_0}_{p} &=& \frac{(a_{2p - 1}^\dagger b_{2p}^\dagger + b_{2p - 1}^\dagger a_{2p}^\dagger)}{\sqrt{2}}\vac_p \nn \\ 
\ket{t_{\mm}}_{p} &=& b_{2p-1}^\dagger b_{2p}^\dagger \vac_p. 
\label{tripletexp}
\end{eqnarray}
where $\vac_p$ is the local vacuum of the cell $p$ (sites $2p - 1$ and $2p$), $a_i^\dagger \ket{\theta}_i = \ket{\uparrow}_i$ and $b_i^\dagger \ket{\theta}_i = \ket{\downarrow}_i$.
Expanding Eq.~(\ref{neighsing}) in terms of spin-1/2s using Eq.~(\ref{tripletexp}), we obtain
\begin{equation}
    \ket{0\,0}_{p,p+1} = \frac{\ket{x_{2p}} - \ket{{\rm MG}^{\text{odd}}}}{\sqrt{3}}
\end{equation}
where $\ket{x_n}$ and $\ket{{\rm MG}^{\text{odd}}}$ are defined in Eqs.~\eqref{mgxn} and \eqref{eq:mg2} respectively. Note that a momentum $\pi$ over the cells in Eq.~(\ref{casperssingleteqn}) corresponds to a momentum $\pi/2$ or $3\pi/2$ over the sites. Thus, up to an overall normalization factor, the singlet exact state $\ket{\phi}$ of Eq.~(\ref{casperssingleteqn}) can be written as $\ket{X_k}$ of Eq.~(\ref{eq:mgsinglet}) as 
\begin{eqnarray}
    \ket{X_k} &=&\sum_{n = 1}^{L}{e^{i k n} (\ket{x_n} - \ket{{\rm MG}^{\text{Parity(n)}}})} \nn \\
    &=& \sum_{n = 1}^{L}{e^{i k n}\ket{x_n}}, \;\;\; k = \frac{\pi}{2}, \frac{3 \pi}{2}
\label{xkorig}
\end{eqnarray}
where $L = 2N$. In the last equality we have used the fact that the sum over ground states vanishes with momentum $k = \pi/2, 3\pi/2$. 

Similarly, a triplet exact state $\ket{\psi_{1M}}$ was first introduced in Eq.~(8) of Ref.~[\onlinecite{caspers1982some}], as 
\begin{equation}
    \ket{\psi_{1M}} = \frac{1}{\sqrt{N}}\sum_{p = 1}^N{(-1)^p \ket{\Phi_p(M)}}.
\end{equation}
where $\ket{\Phi_p(M)}$ is the configuration where the cell $p$ comprises of a spin triplet with $S_z = M$ whereas the other cells are spin singlets. Using Eq.~(\ref{tripletexp}) to express $\ket{\Phi_p(M)}$ in terms of spin-1/2s, we obtain
\begin{equation}
    \ket{\Phi_p(M = 1)} = \ket{t_{2p - 1}}.
\end{equation}
The exact state up to an overall normalization factor can be written as
\begin{equation}
    \ket{T_k} = \sum_{n = 1}^L{e^{i k n} \ket{t_n}} \;\; k = \frac{\pi}{2}, \frac{3 \pi}{2}
\label{tkorig}
\end{equation}
where $\ket{t_n}$ is defined in Eq.~(\ref{mgtn1}).

\section{Exact excited states with Open Boundary Conditions}\label{sec:obc}
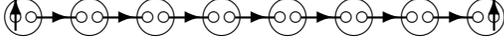
\begin{figure}[t!]
\begin{picture}(180,20)(0,0)
\linethickness{0.8pt}
\multiput(0,10)(25,0){8}{\circle{4}} 
\multiput(6,10)(25,0){8}{\circle{4}} 
\multiput(3,10)(25,0){8}{\circle{14}}
\thicklines
\multiput(8,10)(25,0){7}{\line(1,0){15}} 
\multiput(20,10)(25,0){7}{\vector(1,0){0}} 
\put(0,5){\vector(0,1){12}}
\put(181,5){\vector(0,1){12}}
\end{picture}
\caption{Triplet ground state $\ket{G^O}$ with open boundary conditions. The spins at the edge represent the two dangling spin-1/2s with $S_z = 1/2$ each.}
\label{fig:groundstateobc}
\end{figure}
We briefly discuss the extension of some of the exact states we have obtained to spin chains with Open Boundary Conditions (OBC). 
\subsection{Spin-1 AKLT Ground state}\label{sec:groundstateobc}
We start with the spin-1 AKLT chain. The spin-1 AKLT Hamiltonian for open boundary conditions $H^O$ excludes the term $P^{(2,1)}_{L,1}$ from the Hamiltonian $H$ defined for periodic boundary conditions in Eq.~(\ref{S1Hamiltonian}). $H^O$ then reads
\begin{equation}
    H^O = \sum_{i=1}^{L-1}{P^{(2,1)}_{i, i+1}}.
\label{aklthamiltonianobc}
\end{equation}
The symmetries of the Hamiltonian $H^O$ are the same as that of $H$ except for the momentum quantum number $k$ since translation symmetry is explicitly broken.
As discussed in Sec.~\ref{sec:groundstate}, with open boundary conditions, the ground state comprises of dangling spin-1/2 degrees of freedom at the edge (Fig.~\ref{fig:groundstateobc}). There are four degenerate ground states: three with $s = 1$, where the two dangling spin-1/2s form a triplet state with $S_z = 1, 0, -1$ and one with $s = 0$, where the dangling spin-1/2s form a singlet. The $s = 0$ ground state is thus identical to the ground state of the AKLT Hamiltonian with PBC, i.e., $\ket{G}$ of Eq.~\eqref{groundstate} (Fig.~\ref{fig:groundstate}). In the Schwinger boson notation, the $S_z = 1$ (highest weight) state of the triplet ground state is written as
\begin{equation}
    \ket{G^O} = a_1^\dagger a_L^\dagger\prod_{i = 1}^{L-1}{{c_{i,i+1}^\dagger}}\vac.
\label{groundstateobc}
\end{equation}

\subsection{Spin-2 Magnon}\label{sec:spin2magnonobc}
\begin{figure}[t!]
\begin{picture}(150,20)(0,0)
        \linethickness{0.8pt}
        \multiput(0,10)(25,0){7}{\circle{4}} 
        \multiput(6,10)(25,0){7}{\circle{4}} 
        \multiput(3,10)(25,0){7}{\circle{14}}
        \thicklines
        \multiput(58,10)(25,0){4}{\line(1,0){15}} 
        \multiput(70,10)(25,0){4}{\vector(1,0){0}}
        \multiput(0,5)(25,0){3}{\vector(0,1){12}} 
        \multiput(6,5)(25,0){2}{\vector(0,1){12}}
        \put(156,5){\vector(0,1){12}}
        \put(-32, 7.5){(a)}
        \end{picture}\\[5mm]
    
        \begin{picture}(150,20)(0,0)
        \linethickness{0.8pt}
        \multiput(0,10)(25,0){7}{\circle{4}} 
        \multiput(6,10)(25,0){7}{\circle{4}} 
        \multiput(3,10)(25,0){7}{\circle{14}}
        \thicklines
        \multiput(83,10)(25,0){3}{\line(1,0){15}} 
        \multiput(95,10)(25,0){3}{\vector(1,0){0}}
        \put(33,10){\line(1,0){15}}
        \put(45,10){\vector(1,0){0}}
        \multiput(6,5)(50,0){2}{\vector(0,1){12}}
        \multiput(0,5)(25,0){2}{\vector(0,1){12}}
        \multiput(75,5)(81,0){2}{\vector(0,1){12}}
        \put(-32, 7.5){(b)}
        \end{picture}
    \caption{(a) Spin-2 magnon state $\ket{M_2^O}$ for the chain with open boundary conditions. (b) $\ket{N_2^O}$, the only scattering state of $\ket{M_2^O}$.}
    \label{fig:S2Magnonobc}
\end{figure}
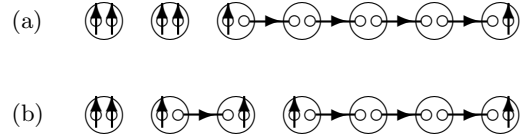
In an AKLT spin chain with OBC, spin-2 magnons $\ket{M^O_n} \equiv (-{S_n^+}^2/2)\ket{G^O}$ (similar to $\ket{M_n}$ in Eq.~(\ref{npnmdimer}) for PBC) can be defined on all the spins except the two edge spins. However, $\ket{M_1^O} = \ket{M_L^O} = 0$ since $(S_n^+)^2$ annihilates all states with $S_z \neq \mm$ on spin $n$ and $S_z \geq 0$ on the edge spins in the ground state $\ket{G^O}$ (see Fig.~\ref{fig:groundstateobc}). Moreover, since the projector $P^{(2,1)}_{L, 1}$ is absent in the Hamiltonian $H_O$ Eq.~(\ref{aklthamiltonianobc}), from Eq.~(\ref{singlethopping}) we obtain that the configurations $\ket{M_2^O}$ (Fig.~\ref{fig:S2Magnonobc}a) and $\ket{M_{L - 1}^O}$ scatter only to magnons $\ket{N_2^O}$ (Fig.~\ref{fig:S2Magnonobc}b) and $\ket{N_{L - 2}^O}$ respectively. The scattering equations of the spin-2 magnons $\ket{M_n^O}$, $2 \leq n \leq L - 1$ thus read
\begin{equation}
    H^O \ket{M_n^O} = 2\ket{M_n^O} + \frac{1}{2}\threepartdef{\ket{N_2^O}}{n = 2}{\ket{N_{L - 2}^O}}{n = L - 1}{(\ket{N_{n-1}^O}+ \ket{N_n^O})}{n \neq 2, L - 2}
    \label{Mnobcscattering}
\end{equation}
The scattering equation of the state $\ket{S_2^O} = \sum_{n = 1}^L{(-1)^n \ket{M_n^O}}$ is then given by
\begin{eqnarray}
    H^O \ket{S_2^O} &=& 2\ket{S_2^O} + \frac{1}{2}\left(\ket{N_2^O} + (-1)^{L - 1}\ket{N_{L-2}^O}\right) \nn \\ 
    &&+\frac{1}{2}\sum_{n = 3}^{L-2}{(-1)^n(\ket{N_{n - 1}^O} + \ket{N_n^O})} \nn \\[2mm]
    &=& 2\ket{S_2^O}.
\end{eqnarray}
Thus, $\ket{S_2^O}$ is an exact excited state of the AKLT chain with triplet open boundary conditions for {\it all} system sizes $L$. This state has an energy $\boldsymbol{E = 2}$ and spin $\boldsymbol{s = 3}$. The operator expression for the state is given by
\begin{equation}
    \widetilde{\ket{S_2^O}} = \mathcal{N}\sum_{n = 1}^L{(-1)^n {S_n^+}^2 \ket{G^O}}
\end{equation}
where $\mathcal{N}$ is a normalization constant.
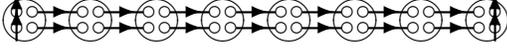
\begin{figure}[t!]
\vspace{10mm}
\begin{picture}(180,10)(0,0)
    \linethickness{0.8pt}
    \multiput(0,10)(25,0){8}{\circle{4}} 
    \multiput(6,10)(25,0){8}{\circle{4}} 
    \multiput(0,4)(25,0){8}{\circle{4}} 
    \multiput(6,4)(25,0){8}{\circle{4}} 
    \multiput(3,7)(25,0){8}{\circle{16}}
    \thicklines
    \multiput(8,10)(25,0){7}{\line(1,0){15}} 
    \multiput(20,10)(25,0){7}{\vector(1,0){0}}
    \multiput(8,4)(25,0){7}{\line(1,0){15}} 
    \multiput(0,5)(181,0){2}{\vector(0,1){12}}
    \multiput(0,-1)(181,0){2}{\vector(0,1){12}}
    \multiput(20,4)(25,0){7}{\vector(1,0){0}}
\end{picture}
\caption{The ground state $\ket{2G^O}$ of the spin-2 AKLT Hamiltonian with open boundary conditions. The two free spin-1/2s at each edge depict the dangling spin-1s. The ground state of the spin-$S$ AKLT model would have $S$ free Schwinger bosons on the edge spins, equivalent to one spin-$S/2$.}
\label{fig:spin2gsobc}
\end{figure}

\subsection{Spin-1 AKLT Tower of states}\label{sec:obctowerofstates}
A tower of states similar to the one discussed in Sec.~\ref{sec:towerofstates} can be constructed from the OBC spin-2 magnons. The state with $N$ spin-2 magnons with triplet open boundary conditions reads
\begin{equation}
    \ket{S_{2N}^O} = \sum_{\{l_j\}}{(-1)^{\sum_{j=1}^N{l_j}}\ket{M_{l_1}^O, M_{l_2}^O \dots M_{l_N}^O}}.
    \label{nmagnonstateobc} 
\end{equation}
Here $\{l_j\}$ satisfies the constaints $1 \leq j \leq N$, $l_{j+1} > l_j + 1$, $1 < l_j < L$ because the spin-2 magnons vanish on the edge spins with OBC. As shown for PBC in Sec.~\ref{sec:towerofstates}, the configurations $\ket{M_{l_1}^O, \dots, M_{l_k}^O \dots, M_{l_N}^O}$ and $\ket{M_{l_1}^O, \dots, M_{q_k}^O \dots, M_{l_N}^O}$ where $q_k = l_k \pm 1$ share a unique scattering configuration $\ket{M_{l_1}^O, \dots,N_{p_k}^O,\dots,M_{l_N}^O}$ where $p_k = l_k - 1$ or $p_k = l_k$. In the case of OBC, all configurations that share a scattering configuration have opposite signs in Eq.~(\ref{nmagnonstateobc}). Thus, all the scattering terms of $\ket{S_{2N}^O}$ vanish. Moreover, unlike PBC, there is no constraint on the system size $L$ in OBC since configurations with $q_k = l_1 = 1$ do not share a scattering term with $q_k = l_N = L$. Thus for OBC, $\ket{S_{2N}^O}$ is an exact state for {\it all} $L$ with $\boldsymbol{E = 2N}$, $\boldsymbol{s = 2N + 1}$. For odd $L$, $\ket{S_{L -1}^O}$ is a state with no dimers, and hence the highest excited ferromagnetic state with $E = L - 1$ and $s = L$.
\\

\subsection{Spin-$\boldsymbol{S}$ AKLT ground state}\label{sec:obcspinSgs}

Similar to the spin-1 AKLT Hamiltonian for OBC in Eq.~(\ref{aklthamiltonianobc}), the spin-$S$ AKLT Hamiltonian for OBC is defined as
\begin{equation}
    {H^{(S)}}^O = \sum_{i = 1}^{L-1}{\sum_{J = S + 1}^{2S}{P^{(J,S)}_{i,i+1}}}
\label{SHamiltonianOBC}  
\end{equation}
For OBC, the ground state of the spin-$S$ AKLT model has $S$ free Schwinger bosons on the edge spins, or, equivalently, one dangling spin-$S/2$. The highest weight ground state with $s = S$ (shown for $S = 2$ in Fig.~\ref{fig:spin2gsobc}) can be written as
\begin{equation}
    \ket{SG^O} = (a_1^\dagger)^S (a_L^\dagger)^S \prodal{i=1}{L-1}{({c^\dagger_{i,i+1}})^S}\vac.
\end{equation}
\subsection{Spin-$\boldsymbol{S}$ AKLT Tower of states}\label{sec:obcspinStowerofstates}
For OBC with the edge spins having $s = S$, the spin-$2S$ magnon $\ket{SM^O_n}$ is proportional to $(S_n^+)^{2S}\ket{SG^O}$, similar to the magnon $\ket{SM_n}$ defined for PBC in Eq.~(\ref{spin2Smagnondimer}). As discussed in Sec.~\ref{sec:spinStower}, the scattering equation of the spin-$2S$ magnon Eq.~(\ref{Mnscattering}) is similar to that of the spin-2 magnon described in Eq.~(\ref{npnmhamilaction}). This similarity holds for OBC too.
\begin{eqnarray}
    &{H^{(S)}}^O \ket{SM_n^O} = 2\ket{SM_n^O} \nn \\
    &+ \lambda_S\threepartdef{\ket{SN_2^O}}{n = 2}{\ket{SN_{L - 2}^O}}{n = L - 1}{(\ket{SN_{n-1}^O}+ \ket{SN_n^O})}{n \neq 2, L - 2}
    \label{SMnobcscattering}
\end{eqnarray}
where $\lambda_S$ is the same proportionality constant defined in Eq.~(\ref{spin2Smagnondimer}). The generalization of the OBC spin-$S$ tower of states then follows by the same  arguments as those in Sec.~\ref{sec:obctowerofstates}. The spin-$S$ tower of states thus exists for {\it all} $L$ has a spin $\boldsymbol{s = (2N + 1)S}$ and energy $\boldsymbol{E = 2N}$. 

\bibliography{aklt_exactstates}

\begin{thebibliography}{76}%
\makeatletter
\providecommand \@ifxundefined [1]{%
 \@ifx{#1\undefined}
}%
\providecommand \@ifnum [1]{%
 \ifnum #1\expandafter \@firstoftwo
 \else \expandafter \@secondoftwo
 \fi
}%
\providecommand \@ifx [1]{%
 \ifx #1\expandafter \@firstoftwo
 \else \expandafter \@secondoftwo
 \fi
}%
\providecommand \natexlab [1]{#1}%
\providecommand \enquote  [1]{``#1''}%
\providecommand \bibnamefont  [1]{#1}%
\providecommand \bibfnamefont [1]{#1}%
\providecommand \citenamefont [1]{#1}%
\providecommand \href@noop [0]{\@secondoftwo}%
\providecommand \href [0]{\begingroup \@sanitize@url \@href}%
\providecommand \@href[1]{\@@startlink{#1}\@@href}%
\providecommand \@@href[1]{\endgroup#1\@@endlink}%
\providecommand \@sanitize@url [0]{\catcode `\\12\catcode `\$12\catcode
  `\&12\catcode `\#12\catcode `\^12\catcode `\_12\catcode `\%12\relax}%
\providecommand \@@startlink[1]{}%
\providecommand \@@endlink[0]{}%
\providecommand \url  [0]{\begingroup\@sanitize@url \@url }%
\providecommand \@url [1]{\endgroup\@href {#1}{\urlprefix }}%
\providecommand \urlprefix  [0]{URL }%
\providecommand \Eprint [0]{\href }%
\providecommand \doibase [0]{http://dx.doi.org/}%
\providecommand \selectlanguage [0]{\@gobble}%
\providecommand \bibinfo  [0]{\@secondoftwo}%
\providecommand \bibfield  [0]{\@secondoftwo}%
\providecommand \translation [1]{[#1]}%
\providecommand \BibitemOpen [0]{}%
\providecommand \bibitemStop [0]{}%
\providecommand \bibitemNoStop [0]{.\EOS\space}%
\providecommand \EOS [0]{\spacefactor3000\relax}%
\providecommand \BibitemShut  [1]{\csname bibitem#1\endcsname}%
\let\auto@bib@innerbib\@empty
\bibitem [{\citenamefont {Basko}\ \emph {et~al.}(2006)\citenamefont {Basko},
  \citenamefont {Aleiner},\ and\ \citenamefont {Altshuler}}]{basko2006metal}%
  \BibitemOpen
  \bibfield  {author} {\bibinfo {author} {\bibfnamefont {D.}~\bibnamefont
  {Basko}}, \bibinfo {author} {\bibfnamefont {I.}~\bibnamefont {Aleiner}}, \
  and\ \bibinfo {author} {\bibfnamefont {B.}~\bibnamefont {Altshuler}},\
  }\href@noop {} {\bibfield  {journal} {\bibinfo  {journal} {Annals of
  Physics}\ }\textbf {\bibinfo {volume} {321}},\ \bibinfo {pages} {1126}
  (\bibinfo {year} {2006})}\BibitemShut {NoStop}%
\bibitem [{\citenamefont {Pal}\ and\ \citenamefont {Huse}(2010)}]{pal2010many}%
  \BibitemOpen
  \bibfield  {author} {\bibinfo {author} {\bibfnamefont {A.}~\bibnamefont
  {Pal}}\ and\ \bibinfo {author} {\bibfnamefont {D.~A.}\ \bibnamefont {Huse}},\
  }\href@noop {} {\bibfield  {journal} {\bibinfo  {journal} {Physical Review
  B}\ }\textbf {\bibinfo {volume} {82}},\ \bibinfo {pages} {174411} (\bibinfo
  {year} {2010})}\BibitemShut {NoStop}%
\bibitem [{\citenamefont {Imbrie}(2016)}]{imbrie2016many}%
  \BibitemOpen
  \bibfield  {author} {\bibinfo {author} {\bibfnamefont {J.~Z.}\ \bibnamefont
  {Imbrie}},\ }\href@noop {} {\bibfield  {journal} {\bibinfo  {journal}
  {Journal of Statistical Physics}\ }\textbf {\bibinfo {volume} {163}},\
  \bibinfo {pages} {998} (\bibinfo {year} {2016})}\BibitemShut {NoStop}%
\bibitem [{\citenamefont {Nandkishore}\ and\ \citenamefont
  {Huse}(2015)}]{nandkishore2014many}%
  \BibitemOpen
  \bibfield  {author} {\bibinfo {author} {\bibfnamefont {R.}~\bibnamefont
  {Nandkishore}}\ and\ \bibinfo {author} {\bibfnamefont {D.~A.}\ \bibnamefont
  {Huse}},\ }\href@noop {} {\bibfield  {journal} {\bibinfo  {journal} {Annu.
  Rev. Condens. Matter Phys.}\ }\textbf {\bibinfo {volume} {6}},\ \bibinfo
  {pages} {15} (\bibinfo {year} {2015})}\BibitemShut {NoStop}%
\bibitem [{\citenamefont {Luitz}\ \emph {et~al.}(2015)\citenamefont {Luitz},
  \citenamefont {Laflorencie},\ and\ \citenamefont {Alet}}]{luitz2015many}%
  \BibitemOpen
  \bibfield  {author} {\bibinfo {author} {\bibfnamefont {D.~J.}\ \bibnamefont
  {Luitz}}, \bibinfo {author} {\bibfnamefont {N.}~\bibnamefont {Laflorencie}},
  \ and\ \bibinfo {author} {\bibfnamefont {F.}~\bibnamefont {Alet}},\
  }\href@noop {} {\bibfield  {journal} {\bibinfo  {journal} {Physical Review
  B}\ }\textbf {\bibinfo {volume} {91}},\ \bibinfo {pages} {081103} (\bibinfo
  {year} {2015})}\BibitemShut {NoStop}%
\bibitem [{\citenamefont {Vosk}\ and\ \citenamefont
  {Altman}(2013)}]{vosk2013many}%
  \BibitemOpen
  \bibfield  {author} {\bibinfo {author} {\bibfnamefont {R.}~\bibnamefont
  {Vosk}}\ and\ \bibinfo {author} {\bibfnamefont {E.}~\bibnamefont {Altman}},\
  }\href@noop {} {\bibfield  {journal} {\bibinfo  {journal} {Physical Review
  Letters}\ }\textbf {\bibinfo {volume} {110}},\ \bibinfo {pages} {067204}
  (\bibinfo {year} {2013})}\BibitemShut {NoStop}%
\bibitem [{\citenamefont {Huse}\ \emph {et~al.}(2013)\citenamefont {Huse},
  \citenamefont {Nandkishore}, \citenamefont {Oganesyan}, \citenamefont {Pal},\
  and\ \citenamefont {Sondhi}}]{huse2013localization}%
  \BibitemOpen
  \bibfield  {author} {\bibinfo {author} {\bibfnamefont {D.~A.}\ \bibnamefont
  {Huse}}, \bibinfo {author} {\bibfnamefont {R.}~\bibnamefont {Nandkishore}},
  \bibinfo {author} {\bibfnamefont {V.}~\bibnamefont {Oganesyan}}, \bibinfo
  {author} {\bibfnamefont {A.}~\bibnamefont {Pal}}, \ and\ \bibinfo {author}
  {\bibfnamefont {S.}~\bibnamefont {Sondhi}},\ }\href@noop {} {\bibfield
  {journal} {\bibinfo  {journal} {Physical Review B}\ }\textbf {\bibinfo
  {volume} {88}},\ \bibinfo {pages} {014206} (\bibinfo {year}
  {2013})}\BibitemShut {NoStop}%
\bibitem [{\citenamefont {Canovi}\ \emph {et~al.}(2011)\citenamefont {Canovi},
  \citenamefont {Rossini}, \citenamefont {Fazio}, \citenamefont {Santoro},\
  and\ \citenamefont {Silva}}]{canovi2011quantum}%
  \BibitemOpen
  \bibfield  {author} {\bibinfo {author} {\bibfnamefont {E.}~\bibnamefont
  {Canovi}}, \bibinfo {author} {\bibfnamefont {D.}~\bibnamefont {Rossini}},
  \bibinfo {author} {\bibfnamefont {R.}~\bibnamefont {Fazio}}, \bibinfo
  {author} {\bibfnamefont {G.~E.}\ \bibnamefont {Santoro}}, \ and\ \bibinfo
  {author} {\bibfnamefont {A.}~\bibnamefont {Silva}},\ }\href@noop {}
  {\bibfield  {journal} {\bibinfo  {journal} {Physical Review B}\ }\textbf
  {\bibinfo {volume} {83}},\ \bibinfo {pages} {094431} (\bibinfo {year}
  {2011})}\BibitemShut {NoStop}%
\bibitem [{\citenamefont {Chandran}\ \emph {et~al.}(2014)\citenamefont
  {Chandran}, \citenamefont {Khemani}, \citenamefont {Laumann},\ and\
  \citenamefont {Sondhi}}]{chandran2014many}%
  \BibitemOpen
  \bibfield  {author} {\bibinfo {author} {\bibfnamefont {A.}~\bibnamefont
  {Chandran}}, \bibinfo {author} {\bibfnamefont {V.}~\bibnamefont {Khemani}},
  \bibinfo {author} {\bibfnamefont {C.}~\bibnamefont {Laumann}}, \ and\
  \bibinfo {author} {\bibfnamefont {S.}~\bibnamefont {Sondhi}},\ }\href@noop {}
  {\bibfield  {journal} {\bibinfo  {journal} {Physical Review B}\ }\textbf
  {\bibinfo {volume} {89}},\ \bibinfo {pages} {144201} (\bibinfo {year}
  {2014})}\BibitemShut {NoStop}%
\bibitem [{\citenamefont {Gogolin}\ \emph {et~al.}(2011)\citenamefont
  {Gogolin}, \citenamefont {M{\"u}ller},\ and\ \citenamefont
  {Eisert}}]{gogolin2011absence}%
  \BibitemOpen
  \bibfield  {author} {\bibinfo {author} {\bibfnamefont {C.}~\bibnamefont
  {Gogolin}}, \bibinfo {author} {\bibfnamefont {M.~P.}\ \bibnamefont
  {M{\"u}ller}}, \ and\ \bibinfo {author} {\bibfnamefont {J.}~\bibnamefont
  {Eisert}},\ }\href@noop {} {\bibfield  {journal} {\bibinfo  {journal}
  {Physical Review Letters}\ }\textbf {\bibinfo {volume} {106}},\ \bibinfo
  {pages} {040401} (\bibinfo {year} {2011})}\BibitemShut {NoStop}%
\bibitem [{\citenamefont {Rigol}(2009)}]{rigol2009breakdown}%
  \BibitemOpen
  \bibfield  {author} {\bibinfo {author} {\bibfnamefont {M.}~\bibnamefont
  {Rigol}},\ }\href@noop {} {\bibfield  {journal} {\bibinfo  {journal}
  {Physical Review Letters}\ }\textbf {\bibinfo {volume} {103}},\ \bibinfo
  {pages} {100403} (\bibinfo {year} {2009})}\BibitemShut {NoStop}%
\bibitem [{\citenamefont {Deutsch}(1991)}]{deutsch1991quantum}%
  \BibitemOpen
  \bibfield  {author} {\bibinfo {author} {\bibfnamefont {J.}~\bibnamefont
  {Deutsch}},\ }\href@noop {} {\bibfield  {journal} {\bibinfo  {journal}
  {Physical Review A}\ }\textbf {\bibinfo {volume} {43}},\ \bibinfo {pages}
  {2046} (\bibinfo {year} {1991})}\BibitemShut {NoStop}%
\bibitem [{\citenamefont {Srednicki}(1994)}]{srednicki1994chaos}%
  \BibitemOpen
  \bibfield  {author} {\bibinfo {author} {\bibfnamefont {M.}~\bibnamefont
  {Srednicki}},\ }\href@noop {} {\bibfield  {journal} {\bibinfo  {journal}
  {Physical Review E}\ }\textbf {\bibinfo {volume} {50}},\ \bibinfo {pages}
  {888} (\bibinfo {year} {1994})}\BibitemShut {NoStop}%
\bibitem [{\citenamefont {Rigol}\ \emph {et~al.}(2008)\citenamefont {Rigol},
  \citenamefont {Dunjko},\ and\ \citenamefont
  {Olshanii}}]{rigol2008thermalization}%
  \BibitemOpen
  \bibfield  {author} {\bibinfo {author} {\bibfnamefont {M.}~\bibnamefont
  {Rigol}}, \bibinfo {author} {\bibfnamefont {V.}~\bibnamefont {Dunjko}}, \
  and\ \bibinfo {author} {\bibfnamefont {M.}~\bibnamefont {Olshanii}},\
  }\href@noop {} {\bibfield  {journal} {\bibinfo  {journal} {Nature}\ }\textbf
  {\bibinfo {volume} {452}},\ \bibinfo {pages} {854} (\bibinfo {year}
  {2008})}\BibitemShut {NoStop}%
\bibitem [{\citenamefont {De~Luca}\ and\ \citenamefont
  {Scardicchio}(2013)}]{de2013ergodicity}%
  \BibitemOpen
  \bibfield  {author} {\bibinfo {author} {\bibfnamefont {A.}~\bibnamefont
  {De~Luca}}\ and\ \bibinfo {author} {\bibfnamefont {A.}~\bibnamefont
  {Scardicchio}},\ }\href@noop {} {\bibfield  {journal} {\bibinfo  {journal}
  {EPL (Europhysics Letters)}\ }\textbf {\bibinfo {volume} {101}},\ \bibinfo
  {pages} {37003} (\bibinfo {year} {2013})}\BibitemShut {NoStop}%
\bibitem [{\citenamefont {Gring}\ \emph {et~al.}(2012)\citenamefont {Gring},
  \citenamefont {Kuhnert}, \citenamefont {Langen}, \citenamefont {Kitagawa},
  \citenamefont {Rauer}, \citenamefont {Schreitl}, \citenamefont {Mazets},
  \citenamefont {Smith}, \citenamefont {Demler},\ and\ \citenamefont
  {Schmiedmayer}}]{gring2012relaxation}%
  \BibitemOpen
  \bibfield  {author} {\bibinfo {author} {\bibfnamefont {M.}~\bibnamefont
  {Gring}}, \bibinfo {author} {\bibfnamefont {M.}~\bibnamefont {Kuhnert}},
  \bibinfo {author} {\bibfnamefont {T.}~\bibnamefont {Langen}}, \bibinfo
  {author} {\bibfnamefont {T.}~\bibnamefont {Kitagawa}}, \bibinfo {author}
  {\bibfnamefont {B.}~\bibnamefont {Rauer}}, \bibinfo {author} {\bibfnamefont
  {M.}~\bibnamefont {Schreitl}}, \bibinfo {author} {\bibfnamefont
  {I.}~\bibnamefont {Mazets}}, \bibinfo {author} {\bibfnamefont {D.~A.}\
  \bibnamefont {Smith}}, \bibinfo {author} {\bibfnamefont {E.}~\bibnamefont
  {Demler}}, \ and\ \bibinfo {author} {\bibfnamefont {J.}~\bibnamefont
  {Schmiedmayer}},\ }\href@noop {} {\bibfield  {journal} {\bibinfo  {journal}
  {Science}\ }\textbf {\bibinfo {volume} {337}},\ \bibinfo {pages} {1318}
  (\bibinfo {year} {2012})}\BibitemShut {NoStop}%
\bibitem [{\citenamefont {Kemp}\ \emph {et~al.}(2017)\citenamefont {Kemp},
  \citenamefont {Yao}, \citenamefont {Laumann},\ and\ \citenamefont
  {Fendley}}]{kemp2017long}%
  \BibitemOpen
  \bibfield  {author} {\bibinfo {author} {\bibfnamefont {J.}~\bibnamefont
  {Kemp}}, \bibinfo {author} {\bibfnamefont {N.~Y.}\ \bibnamefont {Yao}},
  \bibinfo {author} {\bibfnamefont {C.~R.}\ \bibnamefont {Laumann}}, \ and\
  \bibinfo {author} {\bibfnamefont {P.}~\bibnamefont {Fendley}},\ }\href@noop
  {} {\bibfield  {journal} {\bibinfo  {journal} {Journal of Statistical
  Mechanics : Theory and Experiment}\ ,\ \bibinfo {pages} {063105}} (\bibinfo
  {year} {2017})}\BibitemShut {NoStop}%
\bibitem [{\citenamefont {Else}\ \emph {et~al.}(2017)\citenamefont {Else},
  \citenamefont {Fendley}, \citenamefont {Kemp},\ and\ \citenamefont
  {Nayak}}]{else2017prethermal}%
  \BibitemOpen
  \bibfield  {author} {\bibinfo {author} {\bibfnamefont {D.~V.}\ \bibnamefont
  {Else}}, \bibinfo {author} {\bibfnamefont {P.}~\bibnamefont {Fendley}},
  \bibinfo {author} {\bibfnamefont {J.}~\bibnamefont {Kemp}}, \ and\ \bibinfo
  {author} {\bibfnamefont {C.}~\bibnamefont {Nayak}},\ }\href@noop {}
  {\bibfield  {journal} {\bibinfo  {journal} {Physical Review X}\ }\textbf
  {\bibinfo {volume} {7}},\ \bibinfo {pages} {041062} (\bibinfo {year}
  {2017})}\BibitemShut {NoStop}%
\bibitem [{\citenamefont {Bahri}\ \emph {et~al.}(2015)\citenamefont {Bahri},
  \citenamefont {Vosk}, \citenamefont {Altman},\ and\ \citenamefont
  {Vishwanath}}]{bahri2015localization}%
  \BibitemOpen
  \bibfield  {author} {\bibinfo {author} {\bibfnamefont {Y.}~\bibnamefont
  {Bahri}}, \bibinfo {author} {\bibfnamefont {R.}~\bibnamefont {Vosk}},
  \bibinfo {author} {\bibfnamefont {E.}~\bibnamefont {Altman}}, \ and\ \bibinfo
  {author} {\bibfnamefont {A.}~\bibnamefont {Vishwanath}},\ }\href@noop {}
  {\bibfield  {journal} {\bibinfo  {journal} {Nature Communications}\ }\textbf
  {\bibinfo {volume} {6}} (\bibinfo {year} {2015})}\BibitemShut {NoStop}%
\bibitem [{\citenamefont {Bauer}\ and\ \citenamefont
  {Nayak}(2013)}]{bauer2013area}%
  \BibitemOpen
  \bibfield  {author} {\bibinfo {author} {\bibfnamefont {B.}~\bibnamefont
  {Bauer}}\ and\ \bibinfo {author} {\bibfnamefont {C.}~\bibnamefont {Nayak}},\
  }\href@noop {} {\bibfield  {journal} {\bibinfo  {journal} {Journal of
  Statistical Mechanics: Theory and Experiment}\ ,\ \bibinfo {pages} {P09005}}
  (\bibinfo {year} {2013})}\BibitemShut {NoStop}%
\bibitem [{\citenamefont {Beugeling}\ \emph {et~al.}(2014)\citenamefont
  {Beugeling}, \citenamefont {Moessner},\ and\ \citenamefont
  {Haque}}]{beugeling2014finite}%
  \BibitemOpen
  \bibfield  {author} {\bibinfo {author} {\bibfnamefont {W.}~\bibnamefont
  {Beugeling}}, \bibinfo {author} {\bibfnamefont {R.}~\bibnamefont {Moessner}},
  \ and\ \bibinfo {author} {\bibfnamefont {M.}~\bibnamefont {Haque}},\
  }\href@noop {} {\bibfield  {journal} {\bibinfo  {journal} {Physical Review
  E}\ }\textbf {\bibinfo {volume} {89}},\ \bibinfo {pages} {042112} (\bibinfo
  {year} {2014})}\BibitemShut {NoStop}%
\bibitem [{\citenamefont {Kim}\ \emph {et~al.}(2014)\citenamefont {Kim},
  \citenamefont {Ikeda},\ and\ \citenamefont {Huse}}]{kim2014testing}%
  \BibitemOpen
  \bibfield  {author} {\bibinfo {author} {\bibfnamefont {H.}~\bibnamefont
  {Kim}}, \bibinfo {author} {\bibfnamefont {T.~N.}\ \bibnamefont {Ikeda}}, \
  and\ \bibinfo {author} {\bibfnamefont {D.~A.}\ \bibnamefont {Huse}},\
  }\href@noop {} {\bibfield  {journal} {\bibinfo  {journal} {Physical Review
  E}\ }\textbf {\bibinfo {volume} {90}},\ \bibinfo {pages} {052105} (\bibinfo
  {year} {2014})}\BibitemShut {NoStop}%
\bibitem [{\citenamefont {Garrison}\ and\ \citenamefont
  {Grover}(2018)}]{garrison2015does}%
  \BibitemOpen
  \bibfield  {author} {\bibinfo {author} {\bibfnamefont {J.~R.}\ \bibnamefont
  {Garrison}}\ and\ \bibinfo {author} {\bibfnamefont {T.}~\bibnamefont
  {Grover}},\ }\href@noop {} {\bibfield  {journal} {\bibinfo  {journal}
  {Physical Review X}\ }\textbf {\bibinfo {volume} {8}},\ \bibinfo {pages}
  {021026} (\bibinfo {year} {2018})}\BibitemShut {NoStop}%
\bibitem [{\citenamefont {Steinigeweg}\ \emph {et~al.}(2013)\citenamefont
  {Steinigeweg}, \citenamefont {Herbrych},\ and\ \citenamefont
  {Prelov{\v{s}}ek}}]{steinigeweg2013eigenstate}%
  \BibitemOpen
  \bibfield  {author} {\bibinfo {author} {\bibfnamefont {R.}~\bibnamefont
  {Steinigeweg}}, \bibinfo {author} {\bibfnamefont {J.}~\bibnamefont
  {Herbrych}}, \ and\ \bibinfo {author} {\bibfnamefont {P.}~\bibnamefont
  {Prelov{\v{s}}ek}},\ }\href@noop {} {\bibfield  {journal} {\bibinfo
  {journal} {Physical Review E}\ }\textbf {\bibinfo {volume} {87}},\ \bibinfo
  {pages} {012118} (\bibinfo {year} {2013})}\BibitemShut {NoStop}%
\bibitem [{\citenamefont {Ikeda}\ \emph {et~al.}(2013)\citenamefont {Ikeda},
  \citenamefont {Watanabe},\ and\ \citenamefont {Ueda}}]{ikeda2013finite}%
  \BibitemOpen
  \bibfield  {author} {\bibinfo {author} {\bibfnamefont {T.~N.}\ \bibnamefont
  {Ikeda}}, \bibinfo {author} {\bibfnamefont {Y.}~\bibnamefont {Watanabe}}, \
  and\ \bibinfo {author} {\bibfnamefont {M.}~\bibnamefont {Ueda}},\ }\href@noop
  {} {\bibfield  {journal} {\bibinfo  {journal} {Physical Review E}\ }\textbf
  {\bibinfo {volume} {87}},\ \bibinfo {pages} {012125} (\bibinfo {year}
  {2013})}\BibitemShut {NoStop}%
\bibitem [{\citenamefont {Pozsgay}(2014)}]{pozsgay2014failure}%
  \BibitemOpen
  \bibfield  {author} {\bibinfo {author} {\bibfnamefont {B.}~\bibnamefont
  {Pozsgay}},\ }\href@noop {} {\bibfield  {journal} {\bibinfo  {journal}
  {Journal of Statistical Mechanics: Theory and Experiment}\ ,\ \bibinfo
  {pages} {P09026}} (\bibinfo {year} {2014})}\BibitemShut {NoStop}%
\bibitem [{\citenamefont {Affleck}\ \emph
  {et~al.}(1987{\natexlab{a}})\citenamefont {Affleck}, \citenamefont {Kennedy},
  \citenamefont {Lieb},\ and\ \citenamefont {Tasaki}}]{aklt1987rigorous}%
  \BibitemOpen
  \bibfield  {author} {\bibinfo {author} {\bibfnamefont {I.}~\bibnamefont
  {Affleck}}, \bibinfo {author} {\bibfnamefont {T.}~\bibnamefont {Kennedy}},
  \bibinfo {author} {\bibfnamefont {E.~H.}\ \bibnamefont {Lieb}}, \ and\
  \bibinfo {author} {\bibfnamefont {H.}~\bibnamefont {Tasaki}},\ }\href@noop {}
  {\bibfield  {journal} {\bibinfo  {journal} {Physical Review Letters}\
  }\textbf {\bibinfo {volume} {59}},\ \bibinfo {pages} {799} (\bibinfo {year}
  {1987}{\natexlab{a}})}\BibitemShut {NoStop}%
\bibitem [{\citenamefont {Affleck}\ \emph {et~al.}(1988)\citenamefont
  {Affleck}, \citenamefont {Kennedy}, \citenamefont {Lieb},\ and\ \citenamefont
  {Tasaki}}]{affleck1988valence}%
  \BibitemOpen
  \bibfield  {author} {\bibinfo {author} {\bibfnamefont {I.}~\bibnamefont
  {Affleck}}, \bibinfo {author} {\bibfnamefont {T.}~\bibnamefont {Kennedy}},
  \bibinfo {author} {\bibfnamefont {E.~H.}\ \bibnamefont {Lieb}}, \ and\
  \bibinfo {author} {\bibfnamefont {H.}~\bibnamefont {Tasaki}},\ }in\
  \href@noop {} {\emph {\bibinfo {booktitle} {Condensed Matter Physics and
  Exactly Soluble Models}}}\ (\bibinfo  {publisher} {Springer},\ \bibinfo
  {year} {1988})\ pp.\ \bibinfo {pages} {253--304}\BibitemShut {NoStop}%
\bibitem [{\citenamefont {Haldane}(1983{\natexlab{a}})}]{haldane1983nonlinear}%
  \BibitemOpen
  \bibfield  {author} {\bibinfo {author} {\bibfnamefont {F.}~\bibnamefont
  {Haldane}},\ }\href@noop {} {\bibfield  {journal} {\bibinfo  {journal}
  {Physical Review Letters}\ }\textbf {\bibinfo {volume} {50}},\ \bibinfo
  {pages} {1153} (\bibinfo {year} {1983}{\natexlab{a}})}\BibitemShut {NoStop}%
\bibitem [{\citenamefont {Haldane}(1983{\natexlab{b}})}]{haldane1983continuum}%
  \BibitemOpen
  \bibfield  {author} {\bibinfo {author} {\bibfnamefont {F.~D.~M.}\
  \bibnamefont {Haldane}},\ }\href@noop {} {\bibfield  {journal} {\bibinfo
  {journal} {Physics Letters A}\ }\textbf {\bibinfo {volume} {93}},\ \bibinfo
  {pages} {464} (\bibinfo {year} {1983}{\natexlab{b}})}\BibitemShut {NoStop}%
\bibitem [{\citenamefont {Fan}\ \emph {et~al.}(2004)\citenamefont {Fan},
  \citenamefont {Korepin},\ and\ \citenamefont
  {Roychowdhury}}]{fan2004entanglement}%
  \BibitemOpen
  \bibfield  {author} {\bibinfo {author} {\bibfnamefont {H.}~\bibnamefont
  {Fan}}, \bibinfo {author} {\bibfnamefont {V.}~\bibnamefont {Korepin}}, \ and\
  \bibinfo {author} {\bibfnamefont {V.}~\bibnamefont {Roychowdhury}},\
  }\href@noop {} {\bibfield  {journal} {\bibinfo  {journal} {Physical Review
  Letters}\ }\textbf {\bibinfo {volume} {93}},\ \bibinfo {pages} {227203}
  (\bibinfo {year} {2004})}\BibitemShut {NoStop}%
\bibitem [{\citenamefont {Katsura}\ \emph {et~al.}(2007)\citenamefont
  {Katsura}, \citenamefont {Hirano},\ and\ \citenamefont
  {Hatsugai}}]{katsura2007exact}%
  \BibitemOpen
  \bibfield  {author} {\bibinfo {author} {\bibfnamefont {H.}~\bibnamefont
  {Katsura}}, \bibinfo {author} {\bibfnamefont {T.}~\bibnamefont {Hirano}}, \
  and\ \bibinfo {author} {\bibfnamefont {Y.}~\bibnamefont {Hatsugai}},\
  }\href@noop {} {\bibfield  {journal} {\bibinfo  {journal} {Physical Review
  B}\ }\textbf {\bibinfo {volume} {76}},\ \bibinfo {pages} {012401} (\bibinfo
  {year} {2007})}\BibitemShut {NoStop}%
\bibitem [{\citenamefont {Korepin}\ and\ \citenamefont
  {Xu}(2010)}]{korepin2010entanglement}%
  \BibitemOpen
  \bibfield  {author} {\bibinfo {author} {\bibfnamefont {V.~E.}\ \bibnamefont
  {Korepin}}\ and\ \bibinfo {author} {\bibfnamefont {Y.}~\bibnamefont {Xu}},\
  }\href@noop {} {\bibfield  {journal} {\bibinfo  {journal} {International
  Journal of Modern Physics B}\ }\textbf {\bibinfo {volume} {24}},\ \bibinfo
  {pages} {1361} (\bibinfo {year} {2010})}\BibitemShut {NoStop}%
\bibitem [{\citenamefont {Xu}\ \emph {et~al.}(2008)\citenamefont {Xu},
  \citenamefont {Katsura}, \citenamefont {Hirano},\ and\ \citenamefont
  {Korepin}}]{xu2008entanglement}%
  \BibitemOpen
  \bibfield  {author} {\bibinfo {author} {\bibfnamefont {Y.}~\bibnamefont
  {Xu}}, \bibinfo {author} {\bibfnamefont {H.}~\bibnamefont {Katsura}},
  \bibinfo {author} {\bibfnamefont {T.}~\bibnamefont {Hirano}}, \ and\ \bibinfo
  {author} {\bibfnamefont {V.~E.}\ \bibnamefont {Korepin}},\ }\href@noop {}
  {\bibfield  {journal} {\bibinfo  {journal} {Journal of Statistical Physics}\
  }\textbf {\bibinfo {volume} {133}},\ \bibinfo {pages} {347} (\bibinfo {year}
  {2008})}\BibitemShut {NoStop}%
\bibitem [{\citenamefont {Santos}\ \emph {et~al.}(2011)\citenamefont {Santos},
  \citenamefont {Korepin},\ and\ \citenamefont {Bose}}]{santos2011negativity}%
  \BibitemOpen
  \bibfield  {author} {\bibinfo {author} {\bibfnamefont {R.~A.}\ \bibnamefont
  {Santos}}, \bibinfo {author} {\bibfnamefont {V.}~\bibnamefont {Korepin}}, \
  and\ \bibinfo {author} {\bibfnamefont {S.}~\bibnamefont {Bose}},\ }\href@noop
  {} {\bibfield  {journal} {\bibinfo  {journal} {Physical Review A}\ }\textbf
  {\bibinfo {volume} {84}},\ \bibinfo {pages} {062307} (\bibinfo {year}
  {2011})}\BibitemShut {NoStop}%
\bibitem [{\citenamefont {Santos}\ and\ \citenamefont
  {Korepin}(2012)}]{santos2012entanglement}%
  \BibitemOpen
  \bibfield  {author} {\bibinfo {author} {\bibfnamefont {R.~A.}\ \bibnamefont
  {Santos}}\ and\ \bibinfo {author} {\bibfnamefont {V.~E.}\ \bibnamefont
  {Korepin}},\ }\href@noop {} {\bibfield  {journal} {\bibinfo  {journal}
  {Journal of Physics A: Mathematical and Theoretical}\ }\textbf {\bibinfo
  {volume} {45}},\ \bibinfo {pages} {125307} (\bibinfo {year}
  {2012})}\BibitemShut {NoStop}%
\bibitem [{\citenamefont {Schollw{\"o}ck}(2011)}]{schollwock2011density}%
  \BibitemOpen
  \bibfield  {author} {\bibinfo {author} {\bibfnamefont {U.}~\bibnamefont
  {Schollw{\"o}ck}},\ }\href@noop {} {\bibfield  {journal} {\bibinfo  {journal}
  {Annals of Physics}\ }\textbf {\bibinfo {volume} {326}},\ \bibinfo {pages}
  {96} (\bibinfo {year} {2011})}\BibitemShut {NoStop}%
\bibitem [{\citenamefont {Perez-Garcia}\ \emph {et~al.}(2006)\citenamefont
  {Perez-Garcia}, \citenamefont {Verstraete}, \citenamefont {Wolf},\ and\
  \citenamefont {Cirac}}]{perez2006matrix}%
  \BibitemOpen
  \bibfield  {author} {\bibinfo {author} {\bibfnamefont {D.}~\bibnamefont
  {Perez-Garcia}}, \bibinfo {author} {\bibfnamefont {F.}~\bibnamefont
  {Verstraete}}, \bibinfo {author} {\bibfnamefont {M.~M.}\ \bibnamefont
  {Wolf}}, \ and\ \bibinfo {author} {\bibfnamefont {J.~I.}\ \bibnamefont
  {Cirac}},\ }\href@noop {} {\bibfield  {journal} {\bibinfo  {journal} {arXiv
  preprint quant-ph/0608197}\ } (\bibinfo {year} {2006})}\BibitemShut {NoStop}%
\bibitem [{\citenamefont {Or{\'u}s}(2014)}]{orus2014practical}%
  \BibitemOpen
  \bibfield  {author} {\bibinfo {author} {\bibfnamefont {R.}~\bibnamefont
  {Or{\'u}s}},\ }\href@noop {} {\bibfield  {journal} {\bibinfo  {journal}
  {Annals of Physics}\ }\textbf {\bibinfo {volume} {349}},\ \bibinfo {pages}
  {117} (\bibinfo {year} {2014})}\BibitemShut {NoStop}%
\bibitem [{\citenamefont {Pollmann}\ \emph {et~al.}(2010)\citenamefont
  {Pollmann}, \citenamefont {Turner}, \citenamefont {Berg},\ and\ \citenamefont
  {Oshikawa}}]{pollmann2010entanglement}%
  \BibitemOpen
  \bibfield  {author} {\bibinfo {author} {\bibfnamefont {F.}~\bibnamefont
  {Pollmann}}, \bibinfo {author} {\bibfnamefont {A.~M.}\ \bibnamefont
  {Turner}}, \bibinfo {author} {\bibfnamefont {E.}~\bibnamefont {Berg}}, \ and\
  \bibinfo {author} {\bibfnamefont {M.}~\bibnamefont {Oshikawa}},\ }\href@noop
  {} {\bibfield  {journal} {\bibinfo  {journal} {Physical Review B}\ }\textbf
  {\bibinfo {volume} {81}},\ \bibinfo {pages} {064439} (\bibinfo {year}
  {2010})}\BibitemShut {NoStop}%
\bibitem [{\citenamefont {Chen}\ \emph {et~al.}(2011)\citenamefont {Chen},
  \citenamefont {Gu},\ and\ \citenamefont {Wen}}]{chen2011classification}%
  \BibitemOpen
  \bibfield  {author} {\bibinfo {author} {\bibfnamefont {X.}~\bibnamefont
  {Chen}}, \bibinfo {author} {\bibfnamefont {Z.-C.}\ \bibnamefont {Gu}}, \ and\
  \bibinfo {author} {\bibfnamefont {X.-G.}\ \bibnamefont {Wen}},\ }\href@noop
  {} {\bibfield  {journal} {\bibinfo  {journal} {Physical Review B}\ }\textbf
  {\bibinfo {volume} {83}},\ \bibinfo {pages} {035107} (\bibinfo {year}
  {2011})}\BibitemShut {NoStop}%
\bibitem [{\citenamefont {Arovas}\ \emph {et~al.}(1988)\citenamefont {Arovas},
  \citenamefont {Auerbach},\ and\ \citenamefont
  {Haldane}}]{arovas1988extended}%
  \BibitemOpen
  \bibfield  {author} {\bibinfo {author} {\bibfnamefont {D.~P.}\ \bibnamefont
  {Arovas}}, \bibinfo {author} {\bibfnamefont {A.}~\bibnamefont {Auerbach}}, \
  and\ \bibinfo {author} {\bibfnamefont {F.}~\bibnamefont {Haldane}},\
  }\href@noop {} {\bibfield  {journal} {\bibinfo  {journal} {Physical Review
  Letters}\ }\textbf {\bibinfo {volume} {60}},\ \bibinfo {pages} {531}
  (\bibinfo {year} {1988})}\BibitemShut {NoStop}%
\bibitem [{\citenamefont {Majumdar}\ and\ \citenamefont
  {Ghosh}(1969)}]{majumdar1969next}%
  \BibitemOpen
  \bibfield  {author} {\bibinfo {author} {\bibfnamefont {C.~K.}\ \bibnamefont
  {Majumdar}}\ and\ \bibinfo {author} {\bibfnamefont {D.~K.}\ \bibnamefont
  {Ghosh}},\ }\href@noop {} {\bibfield  {journal} {\bibinfo  {journal} {Journal
  of Mathematical Physics}\ }\textbf {\bibinfo {volume} {10}},\ \bibinfo
  {pages} {1388} (\bibinfo {year} {1969})}\BibitemShut {NoStop}%
\bibitem [{\citenamefont {Pizorn}(2012)}]{pizorn2012universality}%
  \BibitemOpen
  \bibfield  {author} {\bibinfo {author} {\bibfnamefont {I.}~\bibnamefont
  {Pizorn}},\ }\href@noop {} {\bibfield  {journal} {\bibinfo  {journal} {arXiv
  preprint arXiv:1202.3336}\ } (\bibinfo {year} {2012})}\BibitemShut {NoStop}%
\bibitem [{\citenamefont {Thomale}\ \emph {et~al.}(2015)\citenamefont
  {Thomale}, \citenamefont {Rachel}, \citenamefont {Bernevig},\ and\
  \citenamefont {Arovas}}]{thomale2015entanglement}%
  \BibitemOpen
  \bibfield  {author} {\bibinfo {author} {\bibfnamefont {R.}~\bibnamefont
  {Thomale}}, \bibinfo {author} {\bibfnamefont {S.}~\bibnamefont {Rachel}},
  \bibinfo {author} {\bibfnamefont {B.~A.}\ \bibnamefont {Bernevig}}, \ and\
  \bibinfo {author} {\bibfnamefont {D.~P.}\ \bibnamefont {Arovas}},\
  }\href@noop {} {\bibfield  {journal} {\bibinfo  {journal} {Journal of
  Statistical Mechanics: Theory and Experiment}\ }\textbf {\bibinfo {volume}
  {2015}},\ \bibinfo {pages} {P07017} (\bibinfo {year} {2015})}\BibitemShut
  {NoStop}%
\bibitem [{\citenamefont {Zauner}\ \emph {et~al.}(2015)\citenamefont {Zauner},
  \citenamefont {Draxler}, \citenamefont {Vanderstraeten}, \citenamefont
  {Degroote}, \citenamefont {Haegeman}, \citenamefont {Rams}, \citenamefont
  {Stojevic}, \citenamefont {Schuch},\ and\ \citenamefont
  {Verstraete}}]{zauner2015transfer}%
  \BibitemOpen
  \bibfield  {author} {\bibinfo {author} {\bibfnamefont {V.}~\bibnamefont
  {Zauner}}, \bibinfo {author} {\bibfnamefont {D.}~\bibnamefont {Draxler}},
  \bibinfo {author} {\bibfnamefont {L.}~\bibnamefont {Vanderstraeten}},
  \bibinfo {author} {\bibfnamefont {M.}~\bibnamefont {Degroote}}, \bibinfo
  {author} {\bibfnamefont {J.}~\bibnamefont {Haegeman}}, \bibinfo {author}
  {\bibfnamefont {M.~M.}\ \bibnamefont {Rams}}, \bibinfo {author}
  {\bibfnamefont {V.}~\bibnamefont {Stojevic}}, \bibinfo {author}
  {\bibfnamefont {N.}~\bibnamefont {Schuch}}, \ and\ \bibinfo {author}
  {\bibfnamefont {F.}~\bibnamefont {Verstraete}},\ }\href@noop {} {\bibfield
  {journal} {\bibinfo  {journal} {New Journal of Physics}\ }\textbf {\bibinfo
  {volume} {17}},\ \bibinfo {pages} {053002} (\bibinfo {year}
  {2015})}\BibitemShut {NoStop}%
\bibitem [{\citenamefont {Zauner-Stauber}\ \emph {et~al.}(2018)\citenamefont
  {Zauner-Stauber}, \citenamefont {Vanderstraeten}, \citenamefont {Haegeman},
  \citenamefont {McCulloch},\ and\ \citenamefont
  {Verstraete}}]{zauner2018topological}%
  \BibitemOpen
  \bibfield  {author} {\bibinfo {author} {\bibfnamefont {V.}~\bibnamefont
  {Zauner-Stauber}}, \bibinfo {author} {\bibfnamefont {L.}~\bibnamefont
  {Vanderstraeten}}, \bibinfo {author} {\bibfnamefont {J.}~\bibnamefont
  {Haegeman}}, \bibinfo {author} {\bibfnamefont {I.}~\bibnamefont {McCulloch}},
  \ and\ \bibinfo {author} {\bibfnamefont {F.}~\bibnamefont {Verstraete}},\
  }\href@noop {} {\bibfield  {journal} {\bibinfo  {journal} {Physical Review
  B}\ }\textbf {\bibinfo {volume} {97}},\ \bibinfo {pages} {235155} (\bibinfo
  {year} {2018})}\BibitemShut {NoStop}%
\bibitem [{\citenamefont {Caspers}\ and\ \citenamefont
  {Magnus}(1982)}]{caspers1982some}%
  \BibitemOpen
  \bibfield  {author} {\bibinfo {author} {\bibfnamefont {W.}~\bibnamefont
  {Caspers}}\ and\ \bibinfo {author} {\bibfnamefont {W.}~\bibnamefont
  {Magnus}},\ }\href@noop {} {\bibfield  {journal} {\bibinfo  {journal}
  {Physics Letters A}\ }\textbf {\bibinfo {volume} {88}},\ \bibinfo {pages}
  {103} (\bibinfo {year} {1982})}\BibitemShut {NoStop}%
\bibitem [{\citenamefont {Arovas}(1989)}]{arovas1989two}%
  \BibitemOpen
  \bibfield  {author} {\bibinfo {author} {\bibfnamefont {D.~P.}\ \bibnamefont
  {Arovas}},\ }\href@noop {} {\bibfield  {journal} {\bibinfo  {journal}
  {Physics Letters A}\ }\textbf {\bibinfo {volume} {137}},\ \bibinfo {pages}
  {431} (\bibinfo {year} {1989})}\BibitemShut {NoStop}%
\bibitem [{\citenamefont {Poilblanc}\ \emph {et~al.}(1993)\citenamefont
  {Poilblanc}, \citenamefont {Ziman}, \citenamefont {Bellissard}, \citenamefont
  {Mila},\ and\ \citenamefont {Montambaux}}]{poilblanc1993poisson}%
  \BibitemOpen
  \bibfield  {author} {\bibinfo {author} {\bibfnamefont {D.}~\bibnamefont
  {Poilblanc}}, \bibinfo {author} {\bibfnamefont {T.}~\bibnamefont {Ziman}},
  \bibinfo {author} {\bibfnamefont {J.}~\bibnamefont {Bellissard}}, \bibinfo
  {author} {\bibfnamefont {F.}~\bibnamefont {Mila}}, \ and\ \bibinfo {author}
  {\bibfnamefont {G.}~\bibnamefont {Montambaux}},\ }\href@noop {} {\bibfield
  {journal} {\bibinfo  {journal} {Europhysics Letters}\ }\textbf {\bibinfo
  {volume} {22}},\ \bibinfo {pages} {537} (\bibinfo {year} {1993})}\BibitemShut
  {NoStop}%
\bibitem [{\citenamefont {Br{\'e}zin}\ and\ \citenamefont
  {Zee}(1993)}]{brezin1993universality}%
  \BibitemOpen
  \bibfield  {author} {\bibinfo {author} {\bibfnamefont {E.}~\bibnamefont
  {Br{\'e}zin}}\ and\ \bibinfo {author} {\bibfnamefont {A.}~\bibnamefont
  {Zee}},\ }\href@noop {} {\bibfield  {journal} {\bibinfo  {journal} {Nuclear
  Physics B}\ }\textbf {\bibinfo {volume} {402}},\ \bibinfo {pages} {613}
  (\bibinfo {year} {1993})}\BibitemShut {NoStop}%
\bibitem [{\citenamefont {Oganesyan}\ and\ \citenamefont
  {Huse}(2007)}]{oganesyan2007localization}%
  \BibitemOpen
  \bibfield  {author} {\bibinfo {author} {\bibfnamefont {V.}~\bibnamefont
  {Oganesyan}}\ and\ \bibinfo {author} {\bibfnamefont {D.~A.}\ \bibnamefont
  {Huse}},\ }\href@noop {} {\bibfield  {journal} {\bibinfo  {journal} {Physical
  Review B}\ }\textbf {\bibinfo {volume} {75}},\ \bibinfo {pages} {155111}
  (\bibinfo {year} {2007})}\BibitemShut {NoStop}%
\bibitem [{\citenamefont {Atas}\ \emph {et~al.}(2013)\citenamefont {Atas},
  \citenamefont {Bogomolny}, \citenamefont {Giraud},\ and\ \citenamefont
  {Roux}}]{atas2013distribution}%
  \BibitemOpen
  \bibfield  {author} {\bibinfo {author} {\bibfnamefont {Y.}~\bibnamefont
  {Atas}}, \bibinfo {author} {\bibfnamefont {E.}~\bibnamefont {Bogomolny}},
  \bibinfo {author} {\bibfnamefont {O.}~\bibnamefont {Giraud}}, \ and\ \bibinfo
  {author} {\bibfnamefont {G.}~\bibnamefont {Roux}},\ }\href@noop {} {\bibfield
   {journal} {\bibinfo  {journal} {Physical Review Letters}\ }\textbf {\bibinfo
  {volume} {110}},\ \bibinfo {pages} {084101} (\bibinfo {year}
  {2013})}\BibitemShut {NoStop}%
\bibitem [{\citenamefont {Affleck}\ \emph
  {et~al.}(1987{\natexlab{b}})\citenamefont {Affleck}, \citenamefont {Kennedy},
  \citenamefont {Lieb},\ and\ \citenamefont {Tasaki}}]{affleck1987rigorous}%
  \BibitemOpen
  \bibfield  {author} {\bibinfo {author} {\bibfnamefont {I.}~\bibnamefont
  {Affleck}}, \bibinfo {author} {\bibfnamefont {T.}~\bibnamefont {Kennedy}},
  \bibinfo {author} {\bibfnamefont {E.~H.}\ \bibnamefont {Lieb}}, \ and\
  \bibinfo {author} {\bibfnamefont {H.}~\bibnamefont {Tasaki}},\ }\href@noop {}
  {\bibfield  {journal} {\bibinfo  {journal} {Physical Review Letters}\
  }\textbf {\bibinfo {volume} {59}},\ \bibinfo {pages} {799} (\bibinfo {year}
  {1987}{\natexlab{b}})}\BibitemShut {NoStop}%
\bibitem [{\citenamefont {Li}\ and\ \citenamefont
  {Haldane}(2008)}]{lihaldane2008}%
  \BibitemOpen
  \bibfield  {author} {\bibinfo {author} {\bibfnamefont {H.}~\bibnamefont
  {Li}}\ and\ \bibinfo {author} {\bibfnamefont {F.~D.~M.}\ \bibnamefont
  {Haldane}},\ }\href {\doibase 10.1103/PhysRevLett.101.010504} {\bibfield
  {journal} {\bibinfo  {journal} {Physical Review Letters}\ }\textbf {\bibinfo
  {volume} {101}},\ \bibinfo {pages} {010504} (\bibinfo {year}
  {2008})}\BibitemShut {NoStop}%
\bibitem [{\citenamefont {Hastings}(2007)}]{hastings2007area}%
  \BibitemOpen
  \bibfield  {author} {\bibinfo {author} {\bibfnamefont {M.~B.}\ \bibnamefont
  {Hastings}},\ }\href@noop {} {\bibfield  {journal} {\bibinfo  {journal}
  {Journal of Statistical Mechanics: Theory and Experiment}\ ,\ \bibinfo
  {pages} {P08024}} (\bibinfo {year} {2007})}\BibitemShut {NoStop}%
\bibitem [{\citenamefont {Verstraete}\ and\ \citenamefont
  {Cirac}(2006)}]{verstraete2006matrix}%
  \BibitemOpen
  \bibfield  {author} {\bibinfo {author} {\bibfnamefont {F.}~\bibnamefont
  {Verstraete}}\ and\ \bibinfo {author} {\bibfnamefont {J.~I.}\ \bibnamefont
  {Cirac}},\ }\href@noop {} {\bibfield  {journal} {\bibinfo  {journal}
  {Physical Review B}\ }\textbf {\bibinfo {volume} {73}},\ \bibinfo {pages}
  {094423} (\bibinfo {year} {2006})}\BibitemShut {NoStop}%
\bibitem [{\citenamefont {Bibikov}(2016)}]{bibikov2016three}%
  \BibitemOpen
  \bibfield  {author} {\bibinfo {author} {\bibfnamefont {P.}~\bibnamefont
  {Bibikov}},\ }\href@noop {} {\bibfield  {journal} {\bibinfo  {journal}
  {Journal of Statistical Mechanics: Theory and Experiment}\ ,\ \bibinfo
  {pages} {033109}} (\bibinfo {year} {2016})}\BibitemShut {NoStop}%
\bibitem [{\citenamefont {Kiwata}\ and\ \citenamefont
  {Akutsu}(1994)}]{kiwata1994bethe}%
  \BibitemOpen
  \bibfield  {author} {\bibinfo {author} {\bibfnamefont {H.}~\bibnamefont
  {Kiwata}}\ and\ \bibinfo {author} {\bibfnamefont {Y.}~\bibnamefont
  {Akutsu}},\ }\href@noop {} {\bibfield  {journal} {\bibinfo  {journal}
  {Journal of the Physical Society of Japan}\ }\textbf {\bibinfo {volume}
  {63}},\ \bibinfo {pages} {3598} (\bibinfo {year} {1994})}\BibitemShut
  {NoStop}%
\bibitem [{\citenamefont {Kennedy}\ \emph {et~al.}(1988)\citenamefont
  {Kennedy}, \citenamefont {Lieb},\ and\ \citenamefont
  {Tasaki}}]{kennedy1988two}%
  \BibitemOpen
  \bibfield  {author} {\bibinfo {author} {\bibfnamefont {T.}~\bibnamefont
  {Kennedy}}, \bibinfo {author} {\bibfnamefont {E.~H.}\ \bibnamefont {Lieb}}, \
  and\ \bibinfo {author} {\bibfnamefont {H.}~\bibnamefont {Tasaki}},\
  }\href@noop {} {\bibfield  {journal} {\bibinfo  {journal} {Journal of
  Statistical Physics}\ }\textbf {\bibinfo {volume} {53}},\ \bibinfo {pages}
  {383} (\bibinfo {year} {1988})}\BibitemShut {NoStop}%
\bibitem [{\citenamefont {Greiter}\ \emph {et~al.}(2007)\citenamefont
  {Greiter}, \citenamefont {Rachel},\ and\ \citenamefont
  {Schuricht}}]{greiter2007exact}%
  \BibitemOpen
  \bibfield  {author} {\bibinfo {author} {\bibfnamefont {M.}~\bibnamefont
  {Greiter}}, \bibinfo {author} {\bibfnamefont {S.}~\bibnamefont {Rachel}}, \
  and\ \bibinfo {author} {\bibfnamefont {D.}~\bibnamefont {Schuricht}},\
  }\href@noop {} {\bibfield  {journal} {\bibinfo  {journal} {Physical Review
  B}\ }\textbf {\bibinfo {volume} {75}},\ \bibinfo {pages} {060401} (\bibinfo
  {year} {2007})}\BibitemShut {NoStop}%
\bibitem [{\citenamefont {Greiter}\ and\ \citenamefont
  {Rachel}(2007)}]{greiter2007valence}%
  \BibitemOpen
  \bibfield  {author} {\bibinfo {author} {\bibfnamefont {M.}~\bibnamefont
  {Greiter}}\ and\ \bibinfo {author} {\bibfnamefont {S.}~\bibnamefont
  {Rachel}},\ }\href@noop {} {\bibfield  {journal} {\bibinfo  {journal}
  {Physical Review B}\ }\textbf {\bibinfo {volume} {75}},\ \bibinfo {pages}
  {184441} (\bibinfo {year} {2007})}\BibitemShut {NoStop}%
\bibitem [{\citenamefont {Schollw{\"o}ck}\ \emph {et~al.}(1996)\citenamefont
  {Schollw{\"o}ck}, \citenamefont {Golinelli},\ and\ \citenamefont
  {Jolic{\oe}ur}}]{schollwock1996s}%
  \BibitemOpen
  \bibfield  {author} {\bibinfo {author} {\bibfnamefont {U.}~\bibnamefont
  {Schollw{\"o}ck}}, \bibinfo {author} {\bibfnamefont {O.}~\bibnamefont
  {Golinelli}}, \ and\ \bibinfo {author} {\bibfnamefont {T.}~\bibnamefont
  {Jolic{\oe}ur}},\ }\href@noop {} {\bibfield  {journal} {\bibinfo  {journal}
  {Physical Review B}\ }\textbf {\bibinfo {volume} {54}},\ \bibinfo {pages}
  {4038} (\bibinfo {year} {1996})}\BibitemShut {NoStop}%
\bibitem [{\citenamefont {Zang}\ \emph {et~al.}(2010)\citenamefont {Zang},
  \citenamefont {Jiang}, \citenamefont {Weng},\ and\ \citenamefont
  {Zhang}}]{zang2010topological}%
  \BibitemOpen
  \bibfield  {author} {\bibinfo {author} {\bibfnamefont {J.}~\bibnamefont
  {Zang}}, \bibinfo {author} {\bibfnamefont {H.-C.}\ \bibnamefont {Jiang}},
  \bibinfo {author} {\bibfnamefont {Z.-Y.}\ \bibnamefont {Weng}}, \ and\
  \bibinfo {author} {\bibfnamefont {S.-C.}\ \bibnamefont {Zhang}},\ }\href@noop
  {} {\bibfield  {journal} {\bibinfo  {journal} {Physical Review B}\ }\textbf
  {\bibinfo {volume} {81}},\ \bibinfo {pages} {224430} (\bibinfo {year}
  {2010})}\BibitemShut {NoStop}%
\bibitem [{\citenamefont {Jiang}\ \emph {et~al.}(2010)\citenamefont {Jiang},
  \citenamefont {Rachel}, \citenamefont {Weng}, \citenamefont {Zhang},\ and\
  \citenamefont {Wang}}]{jiang2010critical}%
  \BibitemOpen
  \bibfield  {author} {\bibinfo {author} {\bibfnamefont {H.-C.}\ \bibnamefont
  {Jiang}}, \bibinfo {author} {\bibfnamefont {S.}~\bibnamefont {Rachel}},
  \bibinfo {author} {\bibfnamefont {Z.-Y.}\ \bibnamefont {Weng}}, \bibinfo
  {author} {\bibfnamefont {S.-C.}\ \bibnamefont {Zhang}}, \ and\ \bibinfo
  {author} {\bibfnamefont {Z.}~\bibnamefont {Wang}},\ }\href@noop {} {\bibfield
   {journal} {\bibinfo  {journal} {Physical Review B}\ }\textbf {\bibinfo
  {volume} {82}},\ \bibinfo {pages} {220403} (\bibinfo {year}
  {2010})}\BibitemShut {NoStop}%
\bibitem [{\citenamefont {Pollmann}\ \emph {et~al.}(2012)\citenamefont
  {Pollmann}, \citenamefont {Berg}, \citenamefont {Turner},\ and\ \citenamefont
  {Oshikawa}}]{pollmann2012symmetry}%
  \BibitemOpen
  \bibfield  {author} {\bibinfo {author} {\bibfnamefont {F.}~\bibnamefont
  {Pollmann}}, \bibinfo {author} {\bibfnamefont {E.}~\bibnamefont {Berg}},
  \bibinfo {author} {\bibfnamefont {A.~M.}\ \bibnamefont {Turner}}, \ and\
  \bibinfo {author} {\bibfnamefont {M.}~\bibnamefont {Oshikawa}},\ }\href@noop
  {} {\bibfield  {journal} {\bibinfo  {journal} {Physical Review B}\ }\textbf
  {\bibinfo {volume} {85}},\ \bibinfo {pages} {075125} (\bibinfo {year}
  {2012})}\BibitemShut {NoStop}%
\bibitem [{\citenamefont {Greiter}(2002)}]{greiter02jltp1029}%
  \BibitemOpen
  \bibfield  {author} {\bibinfo {author} {\bibfnamefont {M.}~\bibnamefont
  {Greiter}},\ }\href@noop {} {\bibfield  {journal} {\bibinfo  {journal}
  {Journal of Low Temperature Physics}\ }\textbf {\bibinfo {volume} {126}},\
  \bibinfo {pages} {1029} (\bibinfo {year} {2002})}\BibitemShut {NoStop}%
\bibitem [{\citenamefont {Rachel}\ \emph {et~al.}(2009)\citenamefont {Rachel},
  \citenamefont {Thomale}, \citenamefont {F{\"u}hringer}, \citenamefont
  {Schmitteckert},\ and\ \citenamefont {Greiter}}]{rachel-09prb180420}%
  \BibitemOpen
  \bibfield  {author} {\bibinfo {author} {\bibfnamefont {S.}~\bibnamefont
  {Rachel}}, \bibinfo {author} {\bibfnamefont {R.}~\bibnamefont {Thomale}},
  \bibinfo {author} {\bibfnamefont {M.}~\bibnamefont {F{\"u}hringer}}, \bibinfo
  {author} {\bibfnamefont {P.}~\bibnamefont {Schmitteckert}}, \ and\ \bibinfo
  {author} {\bibfnamefont {M.}~\bibnamefont {Greiter}},\ }\href@noop {}
  {\bibfield  {journal} {\bibinfo  {journal} {Physical Review B}\ }\textbf
  {\bibinfo {volume} {80}},\ \bibinfo {pages} {180420(R)} (\bibinfo {year}
  {2009})}\BibitemShut {NoStop}%
\bibitem [{\citenamefont {Caspers}\ \emph {et~al.}(1984)\citenamefont
  {Caspers}, \citenamefont {Emmett},\ and\ \citenamefont
  {Magnus}}]{caspers1984majumdar}%
  \BibitemOpen
  \bibfield  {author} {\bibinfo {author} {\bibfnamefont {W.}~\bibnamefont
  {Caspers}}, \bibinfo {author} {\bibfnamefont {K.}~\bibnamefont {Emmett}}, \
  and\ \bibinfo {author} {\bibfnamefont {W.}~\bibnamefont {Magnus}},\
  }\href@noop {} {\bibfield  {journal} {\bibinfo  {journal} {Journal of Physics
  A: Mathematical and General}\ }\textbf {\bibinfo {volume} {17}},\ \bibinfo
  {pages} {2687} (\bibinfo {year} {1984})}\BibitemShut {NoStop}%
\bibitem [{\citenamefont {Rachel}(2009)}]{rachel09epl37005}%
  \BibitemOpen
  \bibfield  {author} {\bibinfo {author} {\bibfnamefont {S.}~\bibnamefont
  {Rachel}},\ }\href@noop {} {\bibfield  {journal} {\bibinfo  {journal}
  {Europhysics Letters}\ }\textbf {\bibinfo {volume} {86}},\ \bibinfo {pages}
  {37005} (\bibinfo {year} {2009})}\BibitemShut {NoStop}%
\bibitem [{\citenamefont {Moudgalya}\ \emph {et~al.}(2018)\citenamefont
  {Moudgalya}, \citenamefont {Regnault},\ and\ \citenamefont
  {Bernevig}}]{selftoappear}%
  \BibitemOpen
  \bibfield  {author} {\bibinfo {author} {\bibfnamefont {S.}~\bibnamefont
  {Moudgalya}}, \bibinfo {author} {\bibfnamefont {N.}~\bibnamefont {Regnault}},
  \ and\ \bibinfo {author} {\bibfnamefont {B.}~\bibnamefont {Bernevig}},\
  }\href@noop {} {\bibfield  {journal} {\bibinfo  {journal} {arXiv preprint
  arXiv:1806.09624}\ } (\bibinfo {year} {2018})}\BibitemShut {NoStop}%
\bibitem [{\citenamefont {Turbiner}(2016)}]{turbiner2016one}%
  \BibitemOpen
  \bibfield  {author} {\bibinfo {author} {\bibfnamefont {A.~V.}\ \bibnamefont
  {Turbiner}},\ }\href@noop {} {\bibfield  {journal} {\bibinfo  {journal}
  {Physics Reports}\ }\textbf {\bibinfo {volume} {642}},\ \bibinfo {pages} {1}
  (\bibinfo {year} {2016})}\BibitemShut {NoStop}%
\bibitem [{\citenamefont {Bibikov}(2018)}]{bibikov2017bethe}%
  \BibitemOpen
  \bibfield  {author} {\bibinfo {author} {\bibfnamefont {P.}~\bibnamefont
  {Bibikov}},\ }\href@noop {} {\bibfield  {journal} {\bibinfo  {journal}
  {Journal of Statistical Mechanics: Theory and Experiment}\ }\textbf {\bibinfo
  {volume} {2018}},\ \bibinfo {pages} {043108} (\bibinfo {year}
  {2018})}\BibitemShut {NoStop}%
\bibitem [{\citenamefont {Shiraishi}\ and\ \citenamefont
  {Mori}(2017)}]{shiraishi2017systematic}%
  \BibitemOpen
  \bibfield  {author} {\bibinfo {author} {\bibfnamefont {N.}~\bibnamefont
  {Shiraishi}}\ and\ \bibinfo {author} {\bibfnamefont {T.}~\bibnamefont
  {Mori}},\ }\href@noop {} {\bibfield  {journal} {\bibinfo  {journal} {Physical
  Review Letters}\ }\textbf {\bibinfo {volume} {119}},\ \bibinfo {pages}
  {030601} (\bibinfo {year} {2017})}\BibitemShut {NoStop}%
\bibitem [{\citenamefont {Turner}\ \emph
  {et~al.}(2018{\natexlab{a}})\citenamefont {Turner}, \citenamefont
  {Michailidis}, \citenamefont {Abanin}, \citenamefont {Serbyn},\ and\
  \citenamefont {Papi{\'c}}}]{turner2018weak}%
  \BibitemOpen
  \bibfield  {author} {\bibinfo {author} {\bibfnamefont {C.}~\bibnamefont
  {Turner}}, \bibinfo {author} {\bibfnamefont {A.}~\bibnamefont {Michailidis}},
  \bibinfo {author} {\bibfnamefont {D.}~\bibnamefont {Abanin}}, \bibinfo
  {author} {\bibfnamefont {M.}~\bibnamefont {Serbyn}}, \ and\ \bibinfo {author}
  {\bibfnamefont {Z.}~\bibnamefont {Papi{\'c}}},\ }\href@noop {} {\bibfield
  {journal} {\bibinfo  {journal} {Nature Physics}\ } (\bibinfo {year}
  {2018}{\natexlab{a}})}\BibitemShut {NoStop}%
\bibitem [{\citenamefont {Turner}\ \emph
  {et~al.}(2018{\natexlab{b}})\citenamefont {Turner}, \citenamefont
  {Michailidis}, \citenamefont {Abanin}, \citenamefont {Serbyn},\ and\
  \citenamefont {Papi{\'c}}}]{turner2018quantum}%
  \BibitemOpen
  \bibfield  {author} {\bibinfo {author} {\bibfnamefont {C.}~\bibnamefont
  {Turner}}, \bibinfo {author} {\bibfnamefont {A.}~\bibnamefont {Michailidis}},
  \bibinfo {author} {\bibfnamefont {D.}~\bibnamefont {Abanin}}, \bibinfo
  {author} {\bibfnamefont {M.}~\bibnamefont {Serbyn}}, \ and\ \bibinfo {author}
  {\bibfnamefont {Z.}~\bibnamefont {Papi{\'c}}},\ }\href@noop {} {\bibfield
  {journal} {\bibinfo  {journal} {arXiv preprint arXiv:1806.10933}\ } (\bibinfo
  {year} {2018}{\natexlab{b}})}\BibitemShut {NoStop}%
\end{thebibliography}%

\onecolumngrid
\begin{figure*}[p]
    \centering
    \begin{minipage}{\textwidth}
    \begin{equation}
    P^{(2,1)}_{ij}

    = 0
    \tag{\ref{projectorvanishing3}}
    \end{equation}
    \end{minipage}\\[5mm]
\caption{Diagrammatic representation of the action of the projector $P^{(2,1)}_{ij}$ on various configurations of dimers around bond $\{i,j\}$. The configurations of the filled small circles are not relevant to the scattering equation and are the same (in terms of the Schwinger bosons) on both sides of a given equation. The directions of the arrows are crucial; reversing an arrow contributes a factor of $(-1)$.}
\label{diagrep}
\end{figure*}
\twocolumngrid

\end{document}